\documentclass[twoside,12pt]{article}

\usepackage{graphics}
\usepackage{axodraw}
\usepackage{epsfig}
\usepackage{slashed}
\usepackage{amsmath,amsfonts}

\def\anti#1{\mathpalette{\@anti}{#1}#1}
\def\@anti#1#2{\sbox0{$#1#2$}
  \makebox[0pt][l]{$#1\kern.30\ht0\overline{\kern-.35\ht0\phantom{#2}}$}}

\newcommand{\lsim}{\mathrel{\rlap{\lower4pt\hbox{\hskip1pt$\sim$}}
    \raise1pt\hbox{$<$}}}         
\newcommand{\gsim}{\mathrel{\rlap{\lower4pt\hbox{\hskip1pt$\sim$}}
    \raise1pt\hbox{$>$}}}         
\newcommand{\be}{\begin{equation}}
\newcommand{\ee}{\end{equation}}
\newcommand{\bq}{\begin{eqnarray}}
\newcommand{\eq}{\end{eqnarray}}

\newcommand{\D}{\mathrm{d}}
\newcommand{\E}{\mathrm{e}}
\newcommand{\I}{\mathrm{i}}

\def\Vec#1{\mathpalette{\VVec}{#1}}                  
\def\VVec#1#2{\mbox{\boldmath$#1#2$\unboldmath}}

\newcommand{\newangle}{< \hspace{-.8ex} {\scriptscriptstyle )}}

\begin{document}

\title{\bf Transverse-Spin and Transverse-Momentum  Effects
\\ in High-Energy Processes}

\author{Vincenzo Barone$^{1,2}$, Franco Bradamante$^{3,4}$, 
Anna Martin$^{3,4}$ \\
 \\ 
$^1$ Di.S.T.A., Universit{\`a} del Piemonte 
Orientale, 15121 Alessandria, Italy \\
$^2$ INFN, Gruppo Collegato di Alessandria,  15121 Alessandria, Italy \\
$^3$ Dipartimento di Fisica, Universit{\`a} degli Studi di Trieste, 
34127 Trieste, 
Italy \\  
$^4$ INFN, Sezione di Trieste, 34127 Trieste, Italy }

\maketitle

\begin{abstract} 
The state of the art concerning transverse-spin and transverse-momentum 
phenomena in hard hadronic reactions is reviewed. 
An account is given of single-spin and azimuthal asymmetries 
in semiinclusive deep inelastic scattering, $e^+ e^-$ annihilation, 
Drell-Yan production, and hadroproduction. 
The ongoing experiments and the main theoretical frameworks 
are described in the first part of the paper. The second part 
is devoted to the experimental findings and their phenomenological 
interpretations. 
A brief discussion of the perspectives of future measurements is 
finally presented. \\
\end{abstract} 

\tableofcontents



\section{Introduction}
\label{intro}
For quite a long time the common lore in the hadron physics community  
has been that transverse polarisation effects are negligibly small 
in hard processes. In the last two decades 
a growing theoretical and experimental evidence 
has shown that this is not  the case and that 
transverse-spin phenomena are, on the contrary,  rather relevant 
in various  high-energy hadronic reactions. 

The prehistory of the subject started in the mid-70s, 
when substantial single-spin asymmetries (SSA's) 
were found in inclusive pion hadroproduction 
at the center-of-mass 
energies of the Argonne synchrotron (few GeV) 
\cite{Dick:1975ty,Klem:1976ui,Dragoset:1978gg}. 
At the same time, at Fermilab  
$\Lambda$ hyperons produced 
in unpolarised $pN$ collisions at $\sqrt{s} \simeq 24$ GeV 
and moderate transverse 
momenta $P_T$  (below 1.5 GeV) were found to possess a large transverse 
polarisation \cite{Bunce:1976yb}, a result 
subsequently confirmed at slightly higher 
$\sqrt{s}$ and $P_T$ \cite{Heller:1978ty}.    

These findings stimulated both experimental and theoretical work.
An experimental programme to investigate both longitudinal and 
transverse spin effects in high energy $pp$ and $\bar{p}p$ scattering 
was proposed in 1978
at FNAL~\cite{Auer:1978bj} and carried out more than 10 yeas later by the 
E704 Collaboration.
Measuring inclusive pion production in collisions of transversely polarised 
proton and antiproton beams with an hydrogen target at the center-of-mass 
energy $\sqrt{s} = 19.4$ GeV and for $P_{T}$ up to 2 GeV, the E704 
Collaboration found single-spin asymmetries as large as 40 \% in the forward 
region \cite{Adams:1991rw,Adams:1991cs,Adams:1991ru,Bravar:1996ki}.
More recently, the experimental collaborations STAR, PHENIX 
and BRAHMS, working at RHIC, have confirmed the early Fermilab 
findings on single-spin asymmetries in hadroproduction, 
pushing the frontier of the c.m. energy to $\sqrt{s} = 200$ GeV 
and covering wider kinematical ranges in $P_T$ and in the 
Feynman variable $x_F$ 
\cite{Adams:2003fx,Adams:2006uz,Adler:2003pb,Adler:2005in,Arsene:2007jd}.     

On the theoretical side, soon after the first experimental findings of 
transverse spin effects,
Kane, Pumplin and Repko proved in a famous paper \cite{Kane:1978nd}
that in collinear perturbative QCD (applicable to 
high $P_T$) SSA's are of the order of $\alpha_s (m_q/\sqrt{s})$ (where $m_q$ 
is the quark mass) and therefore vanish in the massless limit. 
Also, in other important theoretical works 
\cite{Efremov:1981sh,Efremov:1983eb,Ratcliffe:1985mp} 
it was shown that non-vanishing transverse single-spin asymmetries may 
arise in QCD only if one consider higher-twist contributions
(na{\" \i}vely expected to behave as a power of $(M/P_T)$, 
where $M$ is a hadronic scale. 

It took a while to realise that theory allows for transverse polarisation 
effects which are in some cases unsuppressed
(for a review see f.i. Ref.~\cite{Barone:2001sp}). 
In the early 90s various authors 
\cite{Artru:1989zv,Jaffe:1991kp,Jaffe:1992ra,Cortes:1991ja} 
rediscovered  the distribution of transversely polarised quarks in a 
transversely polarised nucleon first introduced by Ralston and Soper in 1979 
\cite{Ralston:1979ys}. 
This ``transversity'' distribution, usually denoted by $h_1(x)$ 
or by $\Delta_T q(x)$, is a leading-twist 
quantity that contributes dominantly to the double transverse asymmetry 
in Drell-Yan (DY) production. 
Due to its chiral-odd nature  
$h_1$ is not measurable in inclusive Deep Inelastic Scattering (DIS),
where transverse SSA's are prohibited by time-reversal invariance at lowest 
order in $\alpha_{\rm em}$ \cite{Christ:1966zz}. 
This argument, however, does not hold in semi-inclusive DIS (SIDIS), 
where at least one hadron in the final state is detected on top
of the scattered lepton.
In SIDIS processes no first principles forbid SSA's.
Various theoretical proposals were soon put forward to measure $h_1$
 \cite{Artru:1992jg,Artru:1993ad,Collins:1992kk,Collins:1993kq,Ji:1992ev,Jaffe:1996wp,Jaffe:1997hf}. 
In particular, Collins proposed a  mechanism, based on 
a spin asymmetry in the fragmentation of transversely polarised 
quarks into an unpolarised hadron (the ``Collins effect''), 
which involves a transverse-momentum dependent (TMD) fragmentation 
function, $H_1^{\perp}$.      
This mechanism was originally proposed as a 
``quark polarimeter'', and could be conveniently exploited to
measure the transversity function $h_1(x)$ in SIDIS.

In a different approach, one year before the publication of the
E704 results, Sivers had suggested that single-spin asymmetries could 
originate, at leading twist, from the intrinsic motion of quarks in the 
colliding hadrons \cite{Sivers:1989cc,Sivers:1990fh}.
The idea, in particular, was that there exists  
an azimuthal asymmetry of unpolarised quarks in a 
transversely polarised hadron (the so-called ``Sivers effect''),
and a new $T$-odd TMD distribution function, now commonly 
called Sivers function and usually denoted by $f_{1T}^{\perp}$,
was proposed to describe the partons in a transversely polarised
nucleon.

Originally this mechanism seemed to violate 
time-reversal ($T$) invariance \cite{Collins:1992kk} and it was
demonstrated that $f_{1T}^{\perp}$ had to be zero.
Brodsky, Hwang and Schmidt  \cite{Brodsky:2002cx,Brodsky:2002rv}
proved however by an explicit calculation   
 that final-state interactions in SIDIS, arising from 
gluon exchange between the struck quark and the nucleon remnant,  
or initial-state interactions in DY, 
produce a non-zero Sivers asymmetry. 
The situation was further clarified by Collins \cite{Collins:2002kn} who 
pointed out that, taking correctly into account the gauge links  
in the TMD distributions, time-reversal invariance does not imply a vanishing
$f_{1T}^{\perp}$, but rather a sign difference 
between the Sivers distribution measured in SIDIS and 
the same distribution measured in DY. 

The phenomenological analysis of the E704 results
 \cite{Anselmino:1998yz,Anselmino:1999pw}, however, showed that 
both the Collins and the Sivers effects are at work to
generate the observed asymmetries,
but a satisfactory theoretical description of the data is still 
missing today.

New experimental opportunities came out in the 90's, when it was 
realised that high energy SIDIS experiments were needed to investigate
the helicity structure of the nucleons.
In fact a major event had focused on the nucleon spin 
the attention of the high energy physics community. 
In 1988 the EMC Collaboration at CERN, scattering a high energy 
polarised muon beam on a transversely polarised proton target 
obtained a totally unexpected result, namely that the fraction 
of the nucleon spin carried by the quarks was small, even compatible 
with zero, within the accuracy of the measurement. 
From the inclusive cross-section difference for parallel and 
antiparallel spins one could extract a linear  combination of 
the quark helicity distributions $\Delta q$ (or $g_1$), defined as  
the difference of the quark  densities 
for quark spin parallel or antiparallel to the longitudinal nucleon spin. 
Using complementary information on the quark helicities derived from 
the weak decays of the hyperons it was possible to add up the quark 
helicities, thus obtaining $\Delta \Sigma$, the overall
quark contribution to the nucleon spin. 
The result, which came to be known as the ``spin crisis'' was at variance 
with the current paradigm, i.e. the
quark model and the beautifully simple explanation of the baryon 
magnetic moments. 
More than one thousand theoretical papers were 
written on the subject, and many experiment 
(SMC at CERN, E142, E143, E154 and E155 at SLAC) were proposed and executed 
to confirm the effect, to extend the result to the neutron, and to 
improve the accuracy of the measurement. The confirmation of this 
finding led to a growing attention to the other contributions to 
the proton spin, namely the gluon polarisation and the orbital 
angular momentum of both quarks and gluons,
as well as to a deeper look at the QCD description of the nucleon 
and to the relation between $h_1$ and $g_1$.
Thus, new generation experiments, well suited to investigate SIDIS
with both longitudinally and transversely
polarised targets, like COMPASS and HERMES, were proposed
and started their operations about 10 years ago.

In 2004 HERMES \cite{Airapetian:2004tw}
and COMPASS \cite{Alexakhin:2005iw,Ageev:2006da} presented the first 
data collected with transversely polarised proton and deuteron targets, 
which showed clear evidence of transverse SSA's on proton. 
One of the main advantages of SIDIS is that 
the Collins and Sivers effects,  as well as the other 
TMD effects, are not mixed, as in hadroproduction, but generate
different azimuthal asymmetries, which can be separately   
explored. 
Thus, the Collins and Sivers asymmetries could be extracted analysing 
the same data.

Another major step in the understanding of the Collins effect occurred
from the Belle Collaboration studies of the azimuthal correlation 
between the hadrons in the two jets created in $e^+e^-$ 
annihilations \cite{Abe:2005zx}.
In the process $e^+e^- \rightarrow q\bar{q}$ the transverse polarisations 
of the $q\bar{q}$ pair are correlated, thus the Collins effect
is expected to cause correlated azimuthal modulations
of the hadrons into which the $q$ and the $\bar{q}$ fragment.
The high precision of the Belle data provided very accurate measurements
of such modulations, and a combined analysis has allowed 
a first extraction of both the Collins function
and of the transversity distribution \cite{Anselmino:2007fs}.
The Sivers and Collins effects are by now theoretically 
well established and the overall picture is essentially in agreement
with the still limited set of results produced by
the SIDIS experiments. 

All this work on transverse polarisation effects
 eventually opened up the Pandora box of 
the transverse-momentum structure of hadrons. 
The importance of the intrinsic transverse momentum of quarks in
hadrons has been acknowledged since many years.
The transverse momentum of the quarks is responsible for the  large
azimuthal asymmetries of the hadrons produced in SIDIS processes
on unpolarised nucleons (the so-called Cahn effect).
In a similar way it is largely responsible for the
 azimuthal asymmetries observed in DY processes,
namely in the production of a lepton pair in hadron-hadron 
scattering at high energy. 
When the intrinsic transverse momentum of the quarks in the nucleon 
is taken into account, several new functions are needed to describe 
the transverse spin structure of the nucleon.
Transverse spin, in fact, couples naturally to the intrinsic transverse 
momentum of quarks, and the resulting correlations are expressed by various 
transverse-momentum dependent distribution and fragmentation functions, that
give rise to a large number of possible single-spin and azimuthal asymmetries 
\cite{Sivers:1989cc,Sivers:1990fh,Collins:1992kk,Kotzinian:1995dv,Tangerman:1994a1,Mulders:1995dh,Tangerman:1995hw}. 
Of particular interest are the correlations between the quark
transverse momentum and the nucleon spin, the quark spin and the
transverse momentum of the fragmenting hadron, and the quark
transverse spin and its transverse momentum in an unpolarised nucleon, 
which give rise to the Sivers function, the Collins function and the 
so-called Boer-Mulders function respectively.
All these three functions are (na{\"\i}vely) $T$-odd, and all three 
are responsible for transverse spin asymmetries in SIDIS.
In particular the Boer-Mulders function \cite{Boer:1997nt}, 
measures the transverse-spin asymmetries of quarks inside an 
unpolarised hadron, and contributes to the $\cos \phi$ and 
$\cos 2 \phi$ azimuthal modulations in the cross sections of 
unpolarised SIDIS and DY processes which have been observed since 
many years and are presently been accurately measured.

The TMD description of hard processes has been put on a firm basis 
by the proof of a non-collinear factorisation theorem for SIDIS and DY, 
in the low transverse momentum regime \cite{Ji:2004wu,Ji:2004xq}.   
On the other hand, it is known that twist-3 collinear effects, 
expressed by quark-gluon correlation functions, can also 
produce single-spin and azimuthal asymmetries 
\cite{Qiu:1991pp,Qiu:1991wg,Qiu:1998ia}. This mechanism 
works at high transverse momenta, $P_T \gg M$. Thus, there is an overlap 
region where both the collinear twist-3 factorisation 
and the non-collinear factorisation should be both valid.   
The relation between these two pictures, that is, between 
the $T$-odd TMD functions on one side and the multiparton correlators on the 
other side, has been clarified in a series of recent papers
\cite{Ji:2006ub,Ji:2006vf,Koike:2007dg}. These works have opened the 
way to the derivation of the evolution equations for the 
TMD functions, a longstanding problem in transverse-spin and 
transverse-momentum 
physics~\cite{Kang:2008ey,Zhou:2008mz,Braun:2009mi,Vogelsang:2009pj}. 

Before concluding these introductory remarks,  
a {\it caveat} is in order. This review is far from 
being exhaustive.
Transverse spin physics is in fact developing so fast 
that it is nearly impossible to cover all the results and 
the ongoing work. 
The following pages necessarily reflect the  
specific competence and the preference of the authors. 
Among the various processes involving the transverse 
spin and the transverse momentum structure of hadrons, 
we chose to focus on SIDIS. From a theoretical 
viewpoint these are the cleanest and best understood reactions. 
A related important 
process, that we will also treat, is hadron pair 
production in $e^+ e^-$ annihilation, which probes 
transverse-spin fragmentation functions. 
As we mentioned, a great wealth of data 
on transverse-spin phenomena come from inclusive hadroproduction. 
We will pay less attention to these processes,  
 because a recent review \cite{D'Alesio:2007jt}
is largely dedicated to their phenomenology.     
Generalised Parton Distributions will also be only mentioned
since they are nicely covered in a very recent and comprehensive
review~\cite{Burkardt:2008jw}.

A comment on the notation is in order.
The proliferation of distribution 
and fragmentation functions involved in transverse-spin 
and transverse-momentum phenomena makes the issue 
of notation and terminology a very problematic one. 
Throughout this paper we adopt for the 
distribution functions the Jaffe-Ji nomenclature \cite{Jaffe:1992ra}, 
extended  to transverse momentum dependent distributions by Mulders 
and collaborators \cite{Mulders:1995dh,Boer:1997nt} illustrated in detail 
in Section~\ref{tmds}.. 
Thus, $f_1(x)$, $g_1(x)$, and $h_1(x)$  are the unpolarised,
the helicity and the transversity distribution functions,
respectively, with the subscript 1 denoting leading-twist quantities. 
The main
disadvantage of this nomenclature is the use of $g_1$ to denote a quark
distribution function whereas the same notation is
universally adopted for one of the two structure functions
of polarised deep inelastic scattering. 
Other common names in the literature
are $q(x)$ for the unpolarised distribution, 
$\Delta q(x)$ for the helicity distribution, $\Delta_T q(x)$ 
for the transversity distribution (which is also called sometimes 
$\delta q$: here we reserve this name to the
tensor charge).   
The fragmentation functions are denoted by capital 
letters: $D$ (unpolarised), $G$  
(longitudinally polarised), $H$ (transversely 
polarised). Thus, $D_1$ is be the usual leading-twist unpolarised 
fragmentation function, $G_1$ the fragmentation 
function of longitudinally polarised quarks, $H_1$ 
the fragmentation function of transversely polarised quarks.  
Note that capital letters are also used for the gluon 
distribution functions \cite{Mulders:2000sh}
and for the generalised parton distributions 
\cite{Diehl:2003ny} and the reader should
be aware of this possible source of confusion. 

The outline of this paper is the following. 
In Section 2 a brief account of the technical features of the main
ongoing experiments is presented. 
Section 3 is devoted to the formal
aspects of the transverse-spin and transverse-momentum
structure of hadrons. 
In Section 4 we introduce the relevant 
processes and observables, and to the theoretical
frameworks that describe them. 
Section 5 illustrates the experimental
findings about single-spin and azimuthal asymmetries, and their
phenomenological interpretations. 
In Section 6 we discuss
the short- and mid-term perspectives of planned and proposed
measurements. 
Finally, Section 7 contains some concluding remarks.

\section{The experiments}

\subsection{DIS experiments}

Deep inelastic scattering as a tool to unveil the nucleon structure 
was invented in the late 60's at SLAC, when for the first time a high 
energy (1 GeV) electron accelerator became available, a wealth of e-N 
scattering data were collected, and eventually it became clear that 
scattering at large transverse momentum could be interpreted as elastic 
scattering off the nucleon constituents, the ``partons''. 
From the dependence of the cross-section on the energy and the momentum 
transfered to the nucleon it has been possible to identify the 
charged partons with the quarks, and assess the existence of the gluons, 
as carriers of half of the proton momentum. 
In these experiments, only the scattered electrons were detected with 
suitable magnetic spectrometers, and no coincidence experiments were 
possible, due to the small duty cycle of the intense electron beam. 
In the subsequent years the Linear Accelerator energy was gradually 
increased, to reach eventually 50 GeV in the most recent experiments, 
allowing to measure at larger and larger $Q^2$, and at smaller and 
smaller $x$. 
Higher energy experiments could be performed at CERN and at FNAL by 
constructing muon beams from $\pi$ and K decays: 
all these data and the neutrino-Nucleon data eventually led to the 
extraction of the parton distribution functions (PDF's) and of their $Q^2$ 
dependence.

In a second generation of experiments polarised lepton beams and polarised 
targets have been used. Sophisticated techniques have been developed to 
polarise the electron beam at SLAC: in the latest experiments (for instance 
E155), 85\% beam polarization was typically achieved by using as source 
the photoelectrons emitted by a gallium arsenide surface. 
The high energy muon beams on the other hand were naturally polarised in 
the weak decay process: by suitably choosing the ratio between muon and 
pion momentum (0.94 in the EMC experiment at the CERN SPS)
muon polarisation of $\sim$80\% are obtained. 
The goal of these experiments was the measurement of the structure 
functions $g_1$ and $g_2$, and to verify the Bjorken sum-rule. 
The beam was longitudinally polarised, while the targets were longitudinally 
polarised to measure $g_1$ and transversely polarised to measure $g_2$. 
A variety of solid targets have been used, butanol, ammonia or $^6$LiD, 
which are kept at very low temperatures ($< 0.1$ K) and in a strong magnetic 
field (typically 2.5 T, but for some small targets even 5 T). 
In these targets a dopant is added, which provides unpaired electrons 
which in the high magnetic field and at low temperature get polarised 
to almost 100\%. 
By irradiating the sample with microwave of proper frequency one can 
induce hyperfine transitions which flip the proton (or deuteron spin) 
to the preselected spin state (Dynamical Nuclear Polarisation, DNP, method).
Only the free protons (deuterons) are polarised, so that the
``figure--of--merit'' for a spin-asymmetry measurement, which
is $\sim 1/\sigma^2_{stat}$, is proportional
to $f^2 P_{tgt}^2$, where $f$ is the ratio between the polarisable nucleons
and the total number of nucleons in the target, and $P_{tgt}^2$ is the
free protons (deuterons) polarisation.
These experiments allowed to discover the ``spin crisis'', to extract the 
helicity quark distributions $g_1$, to provide first measurements of the 
quark contribution to the nucleon spin $\Delta \Sigma$, and to verify 
the validity of the Bjorken sum-rule.  

A third generation of experiments aiming at the study of the nucleon 
spin structure started in the past decade. 
They still use polarised beams and polarised targets, but complement the 
detection of the scattered lepton with the reconstruction and 
identification of the hadrons produced in the fragmentation of the struck 
quark, the so-called current jet. 
A suitable trigger system still allows these experiments to record DIS 
events, but the main objective of the measurement is SIDIS events. 
These coincidence experiments require the disposal of a continuous beam 
and of a large acceptance spectrometer with full particle identification. 

\subsection{SIDIS experiments}

\begin{figure}[tb]
\includegraphics[width=0.55\textwidth,bb=0 260 590 580]
{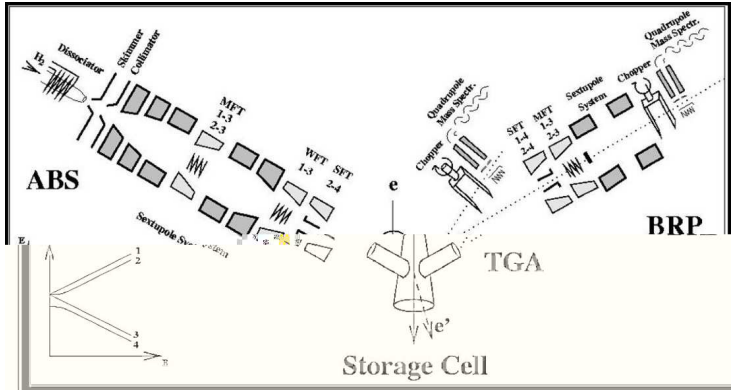}
\hspace{1cm}
\includegraphics[width=0.35\textwidth,bb=100 220 510 630]
{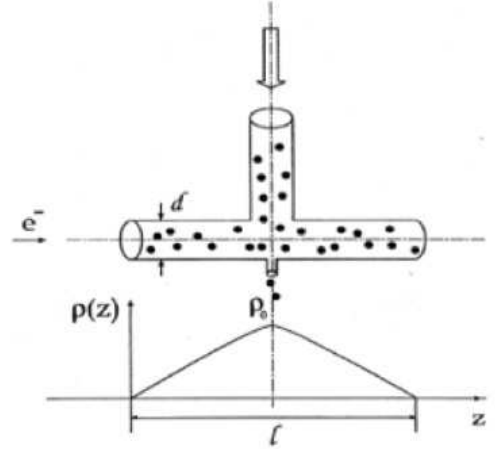}
\hfill
\caption{The HERMES polarized target.The atomic beam from the source
ABS is  focused in the sextupole system and diffuses in the
storage cell (right panel). Nuclear polarization is obtained by
inducing hyperfine transitions by RF-pumping. A Breit-Rabi polarimeter
(BRP) measures the beam polarization.
}
\label{fig:hermes_tgt} 
\end{figure}

At HERA the HERMES experiment was designed to utilise the circulating 
electron or positron beam. At the experiment, the stored beam (27.5 GeV 
and 40 mA) passes through a cell, a tube 60 cm long, coaxial with the beam, 
in which polarised atoms of hydrogen or deuterium are pumped in from an 
Atomic Beam Source. 
After diffusing in the storage cell, the atoms are pumped away by a huge 
pumping system before they diffuse into the electron beam pipe. 
Polarisation is achieved in the Atomic Beam Source by Stern-Gerlach 
filtering followed by radio-frequency transitions to the selected spin state. 
A schematic view of the target system is given in Fig. \ref{fig:hermes_tgt}.
%
This target system is particularly attractive when compared to the solid 
polarised targets because there is no dilution of the target polarisation 
due to the presence of the unpolarised nucleons bound in the other nuclei 
present in the material.
Of course, the target thickness cannot be increased at will, not to destroy 
the circulating electron beam, but densities of $10^{14}$ nucleons/cm$^2$ 
were regularly achieved. After the target, a large acceptance magnetic 
spectrometer based on a 1.3 Tm dipole magnet analysed all charged particles 
up  to 170 mrad in the horizontal plane and between 40 and 149 mrad in the 
vertical plane. 
The reduced acceptance in the vertical plane was due to the 
fact that since both the electron and the proton beam pipes of the HERA 
collider go through the middle plane of the magnet, its gap was divided 
into two identical sections by a horizontal septum plate that shields the 
electron and the proton beams from the dipole magnetic field. 
Charge particle tracking is 
provided by several micro-strip counters, multiwire proportional chambers, 
and drift chambers located before, inside and behind the magnet. 
Charged 
particle identification is provided by a RICH Cherenkov counter, while 
electron-hadron discrimination is achieved with a lead-glass calorimeter with 
a pre-shower hodoscope in front, and by a Transition Radiation detector. 
At the end of the spectrometer, a muon hodoscope located after an iron 
absorber helps the muon identification. 
The experiment took data with polarised targets until 2005. After an upgrade 
to implement the spectrometer with a recoil detector to investigate exclusive 
channels, it took data on unpolarised targets from 2008 to July 2009, when 
HERA ceased operation, and it has afterwards been dismantled.

\begin{figure}[tb]
\begin{center}
\includegraphics[angle=-90,width=0.7\textwidth]
{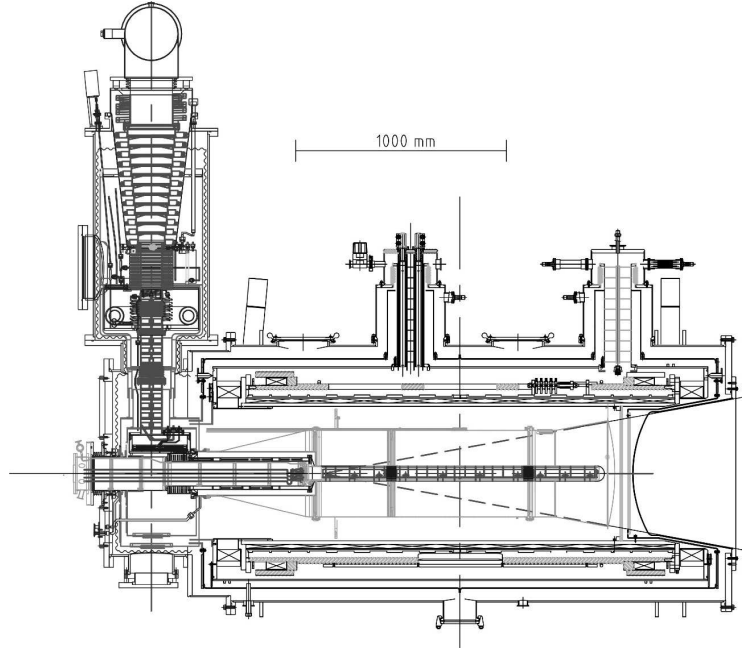}
\end{center}
\caption{Side view of the COMPASS polarised target. One can see the
three target cells inside the mixing chamber, the microwave cavity,
the solenoid coil, the correction coils, and the end compensation coil.
}
\label{fig:compass_target} 
\end{figure}

At CERN the COMPASS experiment has been assembled in the Hall 888, where 
the EMC and afterwards the SMC experiments were installed. 
Its physics 
program includes not only the investigation of the spin structure of the 
nucleon, but also the search of exotic light-quark hadronic states, like 
glueballs and  hybrids. 
For spectroscopy the experiment uses hadronic beams 
(mostly pions and protons), and data have been collected 
with a liquid hydrogen target and various nuclear targets in 2008 and 2009. 
In the following only the configuration which has 
been used for the study of the spin structure of the nucleon will be 
described, which uses the high energy muon beam at the CERN SPS.    
COMPASS has been designed to deal with a beam intensity five times larger 
than that of the previous EMC or SMC experiments. 
Typical intensities are 
$2\cdot10^8$ muons per spill (about 5 s every 18 s) at 160 GeV, the momentum 
at which most of the data have been collected. 
Since there is no problem of beam radiation in the target, the target length 
has been chosen as long as possible, within the boundaries of the complexity 
and cost of the cryogenic system. 
The target materials which have been used in so far 
are $^6$LiD ($f\simeq 0.4$) as a deuteron target, 
and NH$_3$ ($f\simeq 0.15$) as a proton target. The target 
system uses a solenoidal superconducting magnet, providing a highly 
homogeneous field of 2.5 T over a length of 130 cm along the axis, and is 
schematically shown in Fig.~\ref{fig:compass_target}. 
%
About one kg of material is contained in a 4 cm diameter cylinder, 
coaxial with the beam, over a length of 120 cm, distributed either over 
two cells (in 2002, 2003, and 2004), or over three cells (since 2006). Nearby 
cells are oppositely polarised, so that scattering data on the two 
orientation of the target are taken simultaneously to minimise systematic 
effects.  
A cryogenic system allows to keep the target at temperatures of about 
0.5 K, and to polarise it with the DNP method. 
Once high enough polarisation values are reached, the R-F is switched off, 
the temperature drops to less than 50 mK, the spins get frozen, and data taking 
can start. A set of two saddle coils allows to get a transverse field of up 
to 0.6 T which can be either used to adiabatically rotate the target 
polarisation 
from parallel to antiparallel to the beam, or to set it in the transverse 
mode, orthogonal to the beam direction.
At regular intervals, the polarisation orientation of the target cells are 
reversed by changing the frequency of the microwaves, so that possible 
effects due to the different acceptances of the different cells can be 
cancelled in the analysis. 
The experiment is still running, and in 2010 and 2011 will take more data 
with a NH$_3$ polarised proton target, 50\% of the time in the transverse 
polarisation mode and 50\% of the time in the longitudinal mode. 
Large angular 
and momentum acceptance is guaranteed by a two-stage magnetic spectrometer, 
60 m long, centred around two dipole magnets with 1 Tm and 4.4 Tm
bending power respectively.
A variety of tracking detectors ensures charged particles 
tracking from zero to $\sim$200 mrad scattering angle, and charged particle 
identification is provided by a RICH Cherenkov counter. Two hadronic 
calorimeters, two electromagnetic calorimeters and two muon 
filters complement the particle identification and allow the reconstruction 
of neutral pions. 

\begin{figure}[tb]
\begin{center}
\includegraphics[width=0.4\textwidth,bb=0 150 590 700]
{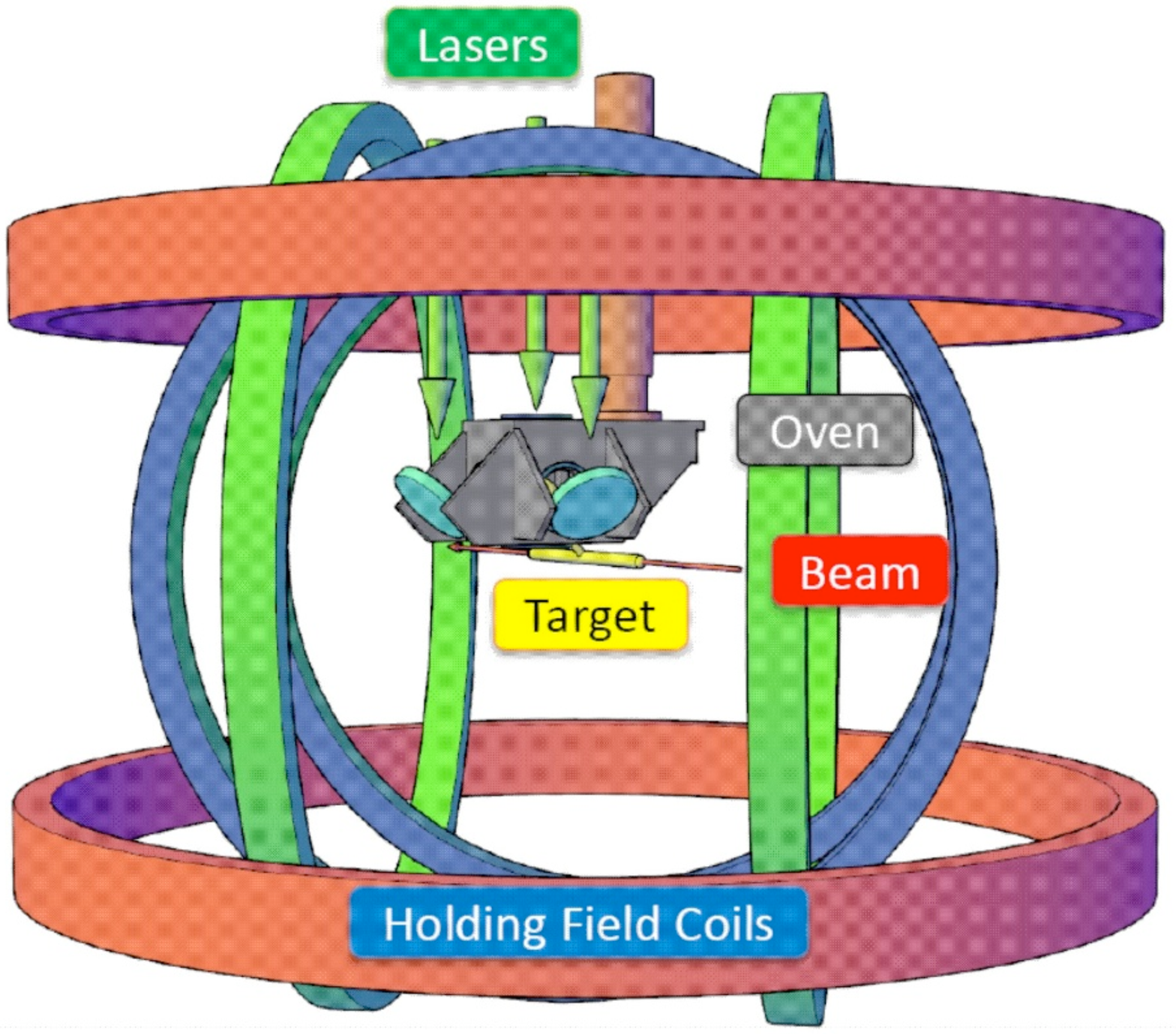}
\end{center}
\caption{An artistic view of the polarised $^3$He gas target used in Hall A of
Jlab for several experiments. For the E06-010 neutron transversity
experiment important upgrades have been made, including the isntallation
of a third pair of Helmoltz coil, enabling to orient the target spins
in any direction.
}
\label{fig:jlab_tgt} 
\end{figure}

At Jefferson Lab (JLab) many measurements of electron-nucleon scattering 
and  in particular of  DIS have been performed over the past 10 years 
using the electron beam of CEBAF (Continuous Electron Beam Accelerator
Facility), energies from 0.8 to 6 GeV, and polarised targets. 
In Hall A, the focus has been on measurements on a neutron target. 
Thanks to the very high beam current, a polarised $^3$He gas target 
could be used as a neutron target. 
The advantage of this target is that to 
first order one can think that its spin is carried entirely by the neutron, 
since the two protons have their spin anti-aligned. 
The $^3$He gas fills a pressurised glass vessel (10 atm, typically) and is 
mixed with Rubidium vapour whose electrons can be polarised via optical 
pumping with circularly polarised laser light. 
The $^3$He nuclei get polarised through spin exchange collisions with 
the Rubidium atoms. A schematic view of the target system is shown in Fig. 
\ref{fig:jlab_tgt}.
%
These targets can stand a much larger beam intensity than a solid target, 
without suffering of radiation damage. 
With a 15 $\mu$A electron beam on a 40 cm target at 10 atm a 
luminosity of $10^{36}$ cm$^{-2}$s$^{-1}$ is achieved. 
A series of high precision experiments have provided invaluable information 
on $g_1$ and $g_2$ for the neutron particularly at large $x$ values. 
The measurements have utilised a pair of 
high resolution magnetic spectrometers to measure the scattered electron, 
and by changing the angular settings and the momentum settings the 
structure functions could be precisely measured over a broad ($Q^2 - W$) 
plane. 
On the other hand, no SIDIS measurements were possible, due to the 
small angular acceptance of the spectrometers.  
 
Complementary measurements, using solid polarised targets (both Ammonia, 
and Deuterated ammonia) have been carried on in Hall B, using the CEBAF 
Large Angle Spectrometer (CLAS). 
The large acceptance of this spectrometer has allowed to detect also 
hadrons to study SIDIS and exclusive events, but the target geometry 
did not allow to put the polarisation orthogonal to the beam, so that 
no transversity measurements have been possible in so far. 

Recently, in 2009, the first transversity measurements have been performed at 
JLab, by the E06-010 collaboration, in Hall A. 
Another important feature of 
the $^3$He target is that the field necessary to hold the polarisation is 
low, so that it is easy with a set of three pairs of 
Helmoltz coils to rotate the 
polarisation in any direction, in particular to a direction 
orthogonal to the beam 
and measure SSA's on a transversely polarised target. 
The experiment has used 
the High Resolution Spectrometer to detect the hadrons in the SIDIS reaction 
($e \rightarrow e' \pi^{\pm}$), and a new large acceptance (64 msr) 
spectrometer, the BigBite 
spectrometer, to measure the electrons. Thus the first ever measurement of 
SIDIS on a transversely polarised neutron target should soon be available, to 
complement the HERMES and COMPASS proton and deuteron data. 

The HERMES, COMPASS  and JLab experiments have given an important 
contribution to precisely measuring $\Delta \Sigma$, the quark contribution 
to the nucleon helicity ($0.33 \pm 0.01 \pm 0.01$), 
and to verify with some precision the Bjorken sum-rule. 
Moreover HERMES and COMPASS  have provided first estimates of $\Delta G/G$, 
the gluon contribution to the nucleon spin, from the spin asymmetry of 
the cross section of pairs of hadrons. 
An important result obtained by the CLAS collaboration is the first 
evidence for a non-zero beam-spin azimuthal asymmetry in the 
semi-inclusive production of 
positive pions in the DIS kinematical regime. 
This effect, observed also by 
HERMES, is not leading twist, and should give information on quark-gluon 
correlation. 
Most relevant to this report, HERMES and COMPASS have been the first 
experiments to measure SIDIS events on transversely polarised targets, 
and the data they have collected are still the only ones in this field. 
HERMES took SIDIS data from 2002 to 2005 and new results from those data
are coming and will still come.
COMPASS took SIDIS data with the transversely polarised deuteron target
in 2002, 2003 and 2004, and with the transversely polarised proton target
in 2007, and more data will be collected in 2010.
JLab has carried on measurements with the transversely polarised He3 target 
in 2009, but no data exist yet at the time of this report.

\subsection{Hadroproduction experiments}

\begin{figure}[tb]
\begin{center}
\includegraphics[width=0.8\textwidth,bb=0 220 590 620]
{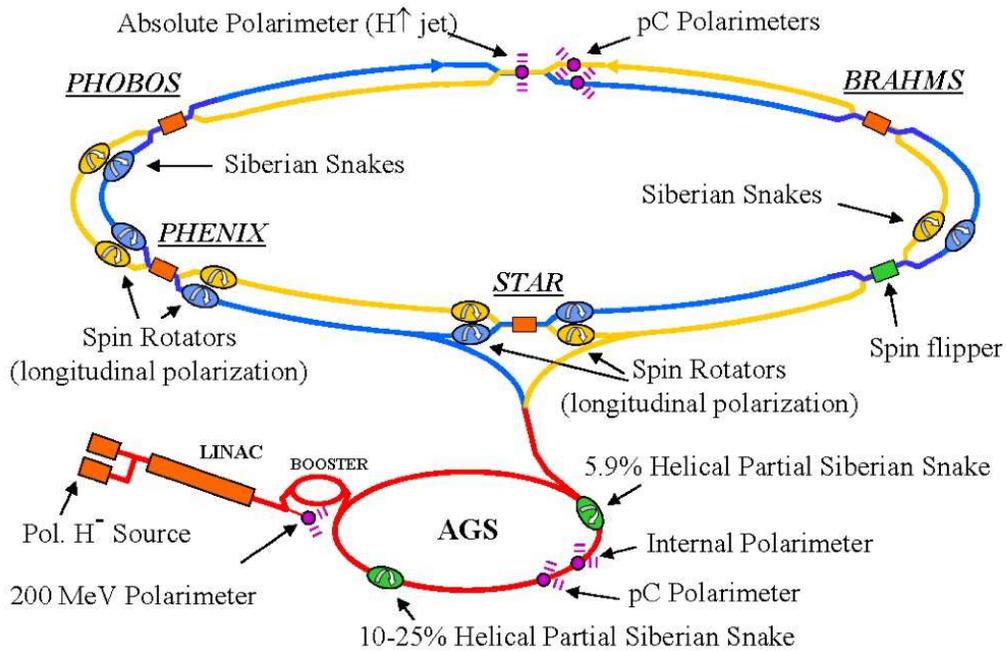}
\end{center}
\caption{
The Brookhaven hadron facility complex, which includes the AGS
Booster, the AGS, and RHIC. Two snakes per ring and four spin rotators
per each of the two large experiments (STAR and PHENIX) are also shown.
}
\label{fig:rhic} 
\end{figure}

A novel attack to transverse spin phenomena is provided by the Relativistic 
Heavy Ion Collider RHIC, at BNL, which has been designed to accelerate and 
store not only ion beams, but also polarised proton beams. 
As a high-energy polarised proton collider, RHIC by now is the 
world flagship facility for spin effects in hadron physics. 
The polarised proton beams are accelerated at the AGS to an energy of  
about 25 GeV, then transferred into the RHIC rings and accelerated up to 
the desired energy, typically in the range 100- 250 GeV. An overview of the 
accelerator complex is shown in Fig. \ref{fig:rhic}.
%
The polarisation in RHIC is maintained by two Siberian snakes, namely 
two sets of four helical dipole magnets which rotate the proton spin by 
180$^o$. 
The snakes are placed at two opposite points along the rings, so that 
the beam deflection between the two snakes is exactly 180$^o$ and the 
spin tune is thus equal to a half-integer and energy independent, thus 
cancelling the effect of  both imperfection and intrinsic resonances. 
In this way polarisation values of about 70\% have been obtained. 
Two more sets of the same helical dipole magnets before and after the 
two major collider experiments (PHENIX and STAR), for a total of 8 
spin ``rotators'',  allow to set the spin direction from vertical to 
parallel to the beam and then back to vertical, to allow the experiments 
to use either longitudinally polarised protons or transversely polarised 
protons. 
Polarisation-averaged 
cross-sections for pion production at $\sqrt{s}= 200$ GeV have already been 
measured at RHIC both at mid-rapidity and in the forward region and found to 
be well described by next-to-leading order (NLO) pQCD calculations. 

The major players at RHIC are the three experiments BRAHMS, PHENIX and STAR. 
Since no spin rotators are installed before and after the BRAHMS experiment, 
this experiment always took data with transversely polarised proton beams. 
The experimental set-up consisted of two movable spectrometer arms to measure 
charged hadrons over a large rapidity and transverse momentum range. The 
forward spectrometer consists of 4 dipole magnets, providing a total bending 
power of 9.2 Tm, it can be rotated between 2.3 and 15 degrees, and has 
particle identification capability thanks to a RICH Cerenkov counter. The 
second spectrometer uses a single dipole magnet (1.2 Tm bending power), can 
be rotated from 34 to 90 degrees relative to the beam, and covers a solid 
angle of approximately 5 msr. The experiment completed data taking in 2006.
The PHENIX experiment is one of the two large ongoing experiments 
investigating the proton spin structure at RHIC. The detector consists of two 
spectrometer arms at mid-rapidity ($\eta<0.35$) and two larger-rapidity 
spectrometer arms at $1.2<\eta<2.4$. The mid-rapidity spectrometers identify 
and track charged particles, and are equipped with electromagnetic 
calorimeters. Two more electromagnetic calorimeters cover the large rapidity 
region, $3.1<\eta<3.7$.
The experiment has been designed to detect rare probes, and it has a 
sophisticated triggering system and a fast data acquisition system.  
The second large ongoing experiment is the STAR experiment. The apparatus has 
cylindrical symmetry around the mean of the two beams directions, and 
consists of a Solenoid magnet, a large Time Projection Chamber (TPC), coaxial 
with the magnet, and a barrel electromagnetic calorimeter just outside of the 
TPC. More electromagnetic calorimetry is provided in the forward region by 
an endcap calorimeter and by two  pion 
detectors. 
Two more TPC's provide information on the charged 
particles in the angular range spanned by the forward calorimetry.

In the longitudinal mode both PHENIX and STAR have concentrated their 
experimental programme on the study of double-spin asymmetries in the 
production of mesons and jets at high $p_T$, to probe the gluon polarisation 
$\Delta G/G$, and first results have already been published, which confirm 
the smallness of $\Delta G/G$ observed in the SIDIS processes. 
In the transverse spin mode the large single spin asymmetries in meson 
(pion) production in proton-proton scattering already observed in the late 
1970's at lower energy at Argonne and at CERN, and in the 1990's by E704 at 
FNAL, have been confirmed to persist at the RHIC energies.
For inclusive pion SSA's the overall accumulated statistics by now is so 
large that data can be binned in $P_T$, in rapidity and in $x_F$. 
The behaviour of 
the SSA's as a function of the various kinematical variables is essential to 
constrain the parameters of the phenomenological models and assess the 
physical origin of the observed asymmetries. 

\subsection{Electron-positron collider experiments}
Experiments at electron-positron colliders play a special role in the 
extraction of the transversity PDF's.
It has been seen in the Introduction that the Collins conjecture, 
namely that in the 
hadronisation of a transversely polarised quark the hadrons of the 
jet might exhibit a left-right asymmetry relative to the plane defined 
by the quark momentum and the quark spin, 
might conveniently be exploited to measure the transversity distribution. 
To unfold the measured Collins asymmetry and extract the transversity 
distribution a knowledge of the Collins function is mandatory, 
but QCD tools are not capable to calculate either the 
quark distribution functions, or the fragmentation functions,
and the Collins function is no exception. 
However, such an effect might be detected 
in high energy $e^-e^+$ annihilations into two jets, and it was
unambiguously observed 
by the Belle Collaboration~\cite{Abe:2005zx}, analysing data collected 
at the asymmetric $e^-e^+$ KEKB storage rings, as will be explained
in Section.\ref{epluseminus}.
The Belle detector is a 
large-solid-angle magnetic spectrometer, based on a 
superconducting solenoidal magnet and many different tracking 
detectors, calorimeters and Cherenkov counters, which 
provide excellent particle identification. 
Particle identification is particularly important 
because the Collins effect which is needed for a global analysis 
with the SIDIS data is 
the effect occurring in the fragmentation of the light quarks. 
The fragmentation 
of charm quark or $b$-quark is expected to dilute the effect, 
because helicity is only 
conserved for nearly massless quarks, thus removal from the 
data sample of all the events 
associated with $c$- or $b$- quarks is very important.

\subsection{Drell-Yan experiments}

The measurement of the Drell-Yan cross-sections is 
not straightforward due to the smallness of the cross-section. 
Up to now only unpolarised hadrons have been used as either beams or 
targets, and the quark PDF's have been extracted from the 
coefficients of various angular modulation in the cross-section. 
 
The NA10 experiment~\cite{Falciano:1985eu,Guanziroli:1987rp} 
impinged a high intensity ($\sim 2 \cdot 10^9$ particles per burst, 
95\% $\pi^-$, corresponding to a mean intensity of $ 2 \cdot 10^8$/sec 
and a duty cycle of 30\%) negative beam from 
the CERN SPS unto a nuclear target, detecting muon pairs in a spectrometer 
whose analysing magnet had a hexagonal symmetry and 
produced a toroidal field. 
The experiment was run at 140 GeV, 194 GeV, and 286 
GeV beam momenta, and different targets were used: a tungsten target 
(either 5.6 or 12 cm long) at all three beam momenta, and a liquid 
deuterium target, 120 cm long, placed 2 m upstream of the tungsten target, 
for the 286 GeV runs. 
To reduce the flux in the spectrometer, the target was followed by a 
beam dump/hadron absorber, placed between the target and the core of 
the dump, at a distance (120 cm) such that there was no contamination 
in the data from muon-pairs created in the dump. 
After the absorber, the trajectories of the two muons were detected with 
two sets of multiwire proportional chambers, one upstream and one downstream 
of the spectrometer magnet. 
In the analysis, $J/\Psi$ and $\Upsilon$ events were eliminated from the 
muon-pair sample by suitable cuts in the pair invariant mass, so that 
the final sample of $\sim 0.3 \cdot 10^6$ events contained 
only DY events in the continuum. 
Very large modulations in $\cos 2\phi$, where $\phi$ is the
azimuthal angle between the hadron plane and the lepton plane,
have been measured, in strong disagreement with the prediction
of collinear QCD.

Similar results have been obtained at Fermilab by the experiment 
E615~\cite{Conway:1989fs}. 
The experiment was carried out in the Proton-West High intensity Area at 
Fermilab, and the measured reaction again was $\pi^- N \rightarrow
\mu^- \mu^+ X$.
The negative hadrons (93\% $\pi^-$) beam momentum was set at 252 GeV, 
and had an intensity of $ 2 \times 10^8$ particles/sec, with a duty cycle 
of 33\%. 
The experimental lay-out is similar to that of the 
CERN experiment, a nuclear target (a 20 cm long tungsten cylinder) was 
followed by a 
long (7.3 m) absorber, acting both as beam stopper and hadron absorber. 
The absorber was made of light material (beryllium-oxide bricks first, 
and graphite afterword), inserted between the pole faces of a  
dipole magnet, in order to sweep away from the spectrometer the low energy 
muons (corresponding to low-mass muon-pairs) and at the same time to focus 
the high mass pairs into the central part of the spectrometer. 
Low-Z material was chosen for the absorber in order to minimise the 
multiple scattering of the muons. 
In order to extend the measurements to very small muon angles with respect to 
the beam, the absorber was uniform in the plane transverse to the beam, 
and had no central plug of high-Z material, like the  uranium/tungsten  
core of the CERN experiment. 
After the absorber, a momentum analysing spectrometer consisting of a 
system of wire chambers upstream and downstream of a second dipole 
magnet was used to measure the muon-pair trajectories. 
The angular distributions have been measured in the invariant mass 
region $4.05< M < 8.55$ GeV, which is free from resonances. 

Recently the E866 collaboration at Fermilab has carried out 
measurements of the angular distributions of DY muon pairs produced 
by scattering 800 GeV protons on a deuterium 
and a proton target \cite{Zhu:2008sj}. 
The Collaboration uses the upgraded Meson-East magnetic pair spectrometer 
at Fermilab.
The primary proton beam, with an intensity of $\sim 2 \times 10^{12}$
protons/spill impinges over one of 
three identical 50.8 cm long target flasks containing either liquid 
hydrogen, liquid deuterium, or vacuum. 
A copper beam dump located inside the second dipole magnet absorbs 
the protons that passed through the target. 
Very much as in the other experiments, downstream of the 
beam dump an absorber wall removes all hadrons produced in the target 
and in the beam dump. 
The muon trajectories are detected by four tracking stations (drift 
chambers) and a momentum analysing dipole magnet. 
Extrapolating the tracks to a vertex in the target the 
parameters of each muon track are optimised, the invariant mass of 
the muon pair is evaluated, and the events from the $J/\Psi$ and 
$\Upsilon$ region are rejected from the DY final events sample. 
This experiment probes the DF's of the sea antiquarks, so it
provides information which is complementary to that of the
$\pi$ beam experiments.

To summarise, the study of the DY process is very promising and 
several new experiments are being proposed. 
At variance with all the 
past experiments, the new experiments will all use polarised beams and/or 
polarised targets.

\section{Transverse-spin and transverse-momentum structure \\ of hadrons}
\label{tmd}

Parton distribution functions (PDF's) and fragmentation functions (FF's)
incorporate a large part of the information 
on the internal structure of hadrons that can be probed in hard  
processes, that is, in strong-interaction 
processes characterised by at least one large momentum scale $Q$. 
We will start discussing the nature and the formalism of PDF's, and then 
extend our presentation to the FF's. 

Right after the first DIS experiments at SLAC, Feynman 
proposed the concept of parton distributions as the probability 
densities of finding a parton with a certain 
momentum fraction inside a nucleon \cite{Feynman:1972b1}. 
In its original formulation, Feynman's parton model was 
based on the observation that the time scale of the interaction 
between the virtual photon and the partons 
is  $\sim 1/Q$, hence much smaller than the time scale of the 
binding interactions of partons, which is $\sim 1/M$ ($M$ is the nucleon 
mass) in the target rest frame  and gets dilated in 
a reference system where the nucleon moves with a very 
large momentum (the ``infinite-momentum frame''). 
Thus, we can approximately assume that in DIS the electron 
interacts elastically with free partons, and define the PDF's  
as the single-particle momentum distributions 
of the nucleon's constituents. 

Focusing for definiteness on quarks, 
$f_1(x)$ denotes the number density of 
quarks carrying a fraction $x$ of the longitudinal 
momentum of the  nucleon. 
It is not difficult to introduce polarisation into 
this simple picture. 
Consider first a longitudinally polarised nucleon.  
The helicity distribution function $g_1(x)$ is defined as the helicity 
asymmetry of quarks in a longitudinally polarised nucleon, that is, 
the number density $f_+(x)$ of quarks with momentum fraction $x$ and 
polarisation parallel to that of the nucleon minus
the number density $f_-(x)$ of quarks with the same momentum fraction but 
antiparallel polarisation: $g_1 (x) = f_+(x) - f_-(x)$. 
In terms of $f_{\pm}$ the unpolarised distribution $f_1$ 
is simply the sum of the two probability densities: 
$f_1(x) = f_+(x) + f_-(x)$. 
The case of transverse polarisation  can be treated 
in a similar way: for a transversely polarised 
nucleon the transversity distribution $h_1(x)$ is defined as the 
number density of quarks with momentum fraction $x$ and polarisation parallel
to that of the hadron, minus the number density of quarks with the same
momentum fraction and antiparallel polarisation, that is,
denoting transverse polarisations by arrows, $h_1(x) 
= f_{\uparrow}(x) - f_{\downarrow}(x)$.  
In a basis of transverse polarisation states, $h_1$ too has a probabilistic
interpretation. In the helicity basis, in contrast, it has no simple meaning,
being related to an off-diagonal quark-hadron amplitude. 

Moving from this intuitive approach to quantum field theory, the 
PDF's admit a rigorous definition in terms of correlation functions 
of parton fields taken at two space-time points 
with a light-like separation \cite{Soper:1976jc,Soper:1979fq}
(a modern treatment is given in
Ref.~\cite{Collins:2003fm}). 
Since DIS probes the parton dynamics on the light-cone 
(see, e.g., Ref.~\cite{Jaffe:1996zw}), it is convenient 
to introduce here some notions concerning light-cone geometry. 
The light-cone components of a four vector $a^{\mu}$ 
are defined as $a^{\pm} = (a^0 \pm a^3)/\sqrt{2}$, 
and grouped in triplets of the form $a^{\mu} = (a^+, a^-, \Vec a_{T})$, 
where the transverse bi-vector is $\Vec a_{T} = (a^1, a^2)$. The 
norm of $a^{\mu}$ is given by $a^2 = 2 a^+ a^- - 
\Vec a_T^2$. 
It is customary to define two light-like vectors $n_+ = (1, 0, \Vec 0_T)$ and 
$n_- = (0,1, \Vec 0_T)$, sometimes called ``Sudakov vectors'', 
which identify the longitudinal direction and are such that 
$n_+ \cdot n_- = 1$. 
Any vector $a^{\mu}$ can be written as 
$a^{\mu} = a^+ n_+^{\mu} + a^- n_-^{\mu} 
+ a_T^{\mu}$, where $a_T^{\mu} = (0, 0, \Vec a_T)$. 
This is the four-dimensional generalisation 
of the familiar decomposition of a three-vector 
into longitudinal and transverse components with 
respect to a given direction. The reference frame  
of DIS (or SIDIS) is chosen so that 
the nucleon's momentum is purely longitudinal: 
$P^{\mu} \simeq P^+ n_+^{\mu}$, 
where the approximate equality means that we are 
neglecting the nucleon mass (a legitimate 
approximation in the deep inelastic limit). 
The infinite momentum frame corresponds to 
$P^+ \to \infty$. Dominant contributions to DIS 
are $\mathcal{O}(P^+)$, whereas subleading 
corrections are suppressed by inverse powers of $P^+$,  
or equivalently, in terms of the momentum transfer, 
by inverse powers  of $Q$. 

The field-theoretical expression of the quark number density $f_1(x)$ is 
(we postpone the QCD subtleties)
\be
f_1(x) \sim 
\int \! \D \xi^- \, \E^{\I x P^+ \xi^-} \langle N \vert 
\psi^{\dagger}_{(+)}(0) \psi^{\dagger}_{(+)}(\xi) 
\vert N \rangle \,, 
\label{f1_qft}
\ee
with $\xi^+ =0, \, \Vec \xi_T = \Vec 0_T$. 
 The peculiarity 
of eq.~(\ref{f1_qft}) is the appearance of the so-called 
``good'' components $\psi_{(+)}$ of the quark fields $\psi$. These admit  
the general decomposition  
$ \psi = \psi_{(+)} + \psi_{(-)}$, with  
$\psi_{({\pm})} = \frac{1}{2}\, \gamma^{\mp} \, \gamma^{\pm} \, \psi$.    
The  good components $\psi_{(+)}$ are the dominant 
ones in the infinite momentum frame, whereas the ``bad'' 
components $\psi_{(-)}$ are not
dynamically independent: using the equations of motion, they can be eliminated
in favour of ``good'' components and terms containing quark masses and gluon
fields. Due to the structure of eq.~(\ref{f1_qft}), one can insert 
between the quark fields a complete set of intermediate states $\vert X 
\rangle$, obtaining a modulus squared:   
\be
 f_1(x) \sim 
  \sum_X \, \delta \! \left(P^+ - x P^+ - P_X^+\right)
  | \langle N | \psi_{(+)}(0)| X \rangle|^2 \,. 
  \label{good12}
\ee
This expression confirms in a field-theoretical 
form the probabilistic interpretation 
of PDF's: $f_1(x)$ is the probability to extract from the nucleon 
$N$ a quark with longitudinal momentum $x P^+$, leaving an 
intermediate state $X$ of longitudinal momentum $P_X^+$. 
A similar  reasoning applies to polarised distributions and yields
\bq
g_1(x) 
&\sim&  
  \sum_X \, \delta \! \left(P^+ - x P^+ - P_X^+\right)
 \left \{  | \langle N | \mathcal{Q}_+ \psi_{(+)}(0)| X \rangle|^2
- | \langle N | \mathcal{Q}_- \psi_{(+)}(0)| X \rangle|^2 \right \}
\,, 
\label{good15} \\
h_1(x) 
&\sim&  
  \sum_X \, \delta \! \left(P^+ - x P^+ - P_X^+\right)
 \left \{  | \langle N | \mathcal{Q}_{\uparrow} \psi_{(+)}(0)| X \rangle|^2
- | \langle N | \mathcal{Q}_{\downarrow} \psi_{(+)}(0)| X \rangle|^2 \right \}
\,, 
\label{good16} 
\eq
where $\mathcal{Q}_{\pm}$ and $\mathcal{Q}_{\uparrow \downarrow}$ 
are the helicity and transversity projectors, respectively. 

If the quarks are perfectly collinear with the parent hadron, 
the three distribution functions we have mentioned so far, 
$f_1(x)$, $g_1(x)$, $h_1(x)$, exhaust the information on 
the internal dynamics of hadrons at leading twist, i.e., 
at zeroth order in $1/Q$  (for an operational definition  of 
twist, see Ref.~\cite{Jaffe:1996zw}). If instead we 
admit a non negligible quark transverse momentum, the number 
of distribution functions considerably increases. At leading twist, 
there are eight of them. In order to understand their origin and 
meaning, it is necessary to adopt a more systematic 
approach. 

\subsection{The quark correlation matrix}
\label{corrmatrix}

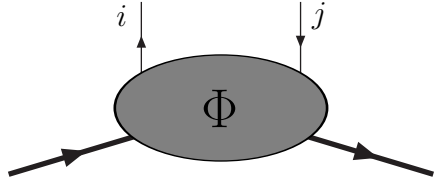
\begin{figure}
  \centering
  \begin{picture}(300,100)(0,0)
    \SetWidth{2}
    \ArrowLine(70,15)(120,30)
    \ArrowLine(180,30)(230,15)
    \SetWidth{0.5}
    \ArrowLine(120,50)(120,80)
    \ArrowLine(180,80)(180,50)
    \Text(115,75)[r]{$i$}
    \Text(185,75)[l]{$j$}
    \GOval(150,40)(20,40)(0){0.5}
    \Text(150,40)[]{\LARGE $\Phi$}
  \end{picture}
  \caption{The quark correlation matrix $\Phi$.}
  \label{phi}
\end{figure}

Quark distribution functions are contained in the 
correlation matrix $\Phi$ (Fig.~\ref{phi}), defined as 
\begin{equation}
  \Phi_{ij}(k, P, S) = \int \! \frac{\D^4\xi}{(2 \pi)^4} \, 
\E^{\I k{\cdot}\xi} \,
  \langle P,S | \bar \psi_j(0) \, \mathcal{W} [0, \xi]
\, \psi_i(\xi) | P,S \rangle \, .
  \label{df1}
\end{equation}
where $\vert P, S \rangle$ is the nucleon state 
of momentum $P^{\mu}$ and polarisation 
vector $S^{\mu}$, $i$ and $j$ are Dirac indices and a summation over colour is
implicit. 
The Wilson line $\mathcal{W}$, 
which guarantees the gauge invariance of the correlator, is a path-ordered 
exponential of the gluon field (see below) arising 
from multigluon final state interactions between the struck quark and 
the target spectators. 
The presence of this gauge link 
introduces in principle a path-dependence in $\Phi$, which in some cases 
turns out to be highly non trivial (Section~\ref{tmds}). 

Integrating $\Phi(k, P, S)$ over the quark momentum, 
with the condition $x = k^+/P^+$ that defines $x$ as the fraction of the 
longitudinal momentum of the nucleon carried by the quark, yields
\be
\Phi (x) = \int \! \D^4 k \; \Phi (k, P, S) \, \delta (k^+ - x P^+) 
= \int \frac{\D \xi^-}{2 \pi} \, 
\E^{\I x P^+ \xi^-} \, \langle P S \vert 
\anti\psi(0) \, \mathcal{W}^- [0, \xi]
\, \psi(\xi) | P,S \rangle \vert_{\xi^+ =0, \Vec \xi_T = \Vec 0}\,, 
\label{phix}
\ee
where the Wilson line $\mathcal{W}^- [0, \xi]$ 
connects $(0,0, \Vec 0_T)$ to $(0, \xi^-, \Vec 0_T)$ 
along the $n^-$ direction and reads ($\mathcal{P}$ denotes 
path ordering)
\be
  \mathcal{W}^- [0,\xi] =
  \mathcal{P} \, \exp \left( - \I g \int_0^{\xi^-} \! \D z^- \, 
A^+(0, z^-, \Vec 0_T) \right)\,. 
\label{w1}
\ee
In the light-cone gauge, $A^+ = 0$, the Wilson link reduces to 
unity and can be omitted. The situation is more complicated 
in the case of transverse-momentum distributions, 
which are defined in terms of field separations 
of the type $(0,\xi^-,\Vec \xi_{T})$: we shall return to this 
issue in Section~\ref{tmds}. 
  
$\Phi(x)$ contains the collinear (i.e., $x$-dependent, $k_T$-integrated) 
quark distribution functions. Introducing the longitudinal 
and transverse components of the 
the polarisation  vector of the nucleon, $S^{\mu} 
= (S_L/M) \, P^{\mu} + S_T^{\mu}$,  
the expression of $\Phi(x)$ at leading twist, that is 
at leading order in $P^+$, is 
\be
\Phi (x) = \frac{1}{2} 
\left \{ f_1 (x) \, \slashed n_+ + S_L \, g_1(x) \, \gamma_5 \slashed 
n_+ + h_1(x) \, \gamma_5 \, \frac{[ \slashed S_T, \slashed n_+]}{2}
\right \} \,. 
\label{phix2}
\ee
Here one sees the three distributions already introduced: 
the number density $f_1(x)$, the helicity distribution $g_1(x)$ 
and the transversity distribution $h_1(x)$,  
first identified by Ralston 
and Soper \cite{Ralston:1979ys}. 
The quark distributions can be extracted from (\ref{phix2}) 
by tracing $\Phi$ with some Dirac matrix $\Gamma$. We will use the 
notation
$ \Phi^{[\Gamma]}(x) \equiv \frac{1}{2}\, {\rm Tr} \left [ 
\Phi (x) \, \Gamma \right ]$.
The explicit 
expressions of the  leading-twist distributions are
(the transverse Dirac matrix $\gamma_T$ is either $\gamma^1$ 
or $\gamma^2$): 
\bq
& & f_1(x) = \Phi^{[\gamma^+]}(x) = 
\int \frac{\D \xi^-}{4 \pi} \, \E^{\I x P^+ \xi^-} \langle P, S \vert 
\bar \psi(0) \mathcal{W}^- [0, \xi] \gamma^+ \psi(\xi) \vert P, S \rangle 
\vert_{\xi^+ = 0, \, \Vec \xi_T = \Vec 0_T}\,, 
\label{f1} \\
& & g_1(x) = \Phi^{[\gamma^+ \gamma_5]}(x) = 
\int \frac{\D \xi^-}{4 \pi} \, \E^{\I x P^+ \xi^-} \langle P, S \vert 
\bar \psi(0) \mathcal{W}^- [0, \xi] 
\gamma^+ \gamma_5 \psi(\xi) \vert P, S \rangle 
\vert_{\xi^+ = 0, \, \Vec \xi_T = \Vec 0_T}\,, 
\label{g1} \\
& & h_1(x) = \Phi^{[\gamma^+ \gamma_T \gamma_5]}(x) = 
\int \frac{\D \xi^-}{4 \pi} \, \E^{\I x P^+ \xi^-} \langle P, S \vert 
\bar \psi(0) \mathcal{W}^- [0, \xi] \gamma^+ \gamma_T \gamma_5 
\psi(\xi) \vert P, S \rangle 
\vert_{\xi^+ = 0, \, \Vec \xi_T = \Vec 0_T}\,.
\label{h1}  
\eq
In QCD the operators appearing in (\ref{f1}-\ref{h1}) are 
ultraviolet divergent, so they have to be renormalised. This 
introduces a scale dependence into the distribution functions,  
$f_1 (x) \rightarrow f_1(x, \mu)$, etc.,  
which is governed 
by the renormalisation group equations, the well known 
DGLAP equations \cite{Dokshitzer:1977sgt,Gribov:1972rtt,Altarelli:1977zs}. 

\subsection{The transversity distribution}
\label{transversity}

The main  properties of the ``third'' parton density, 
the transversity distribution $h_1$, eq.~(\ref{h1}),
 are: \\
{\it i)} it 
is {\it chirally-odd} and therefore does not appear 
in the handbag diagram of inclusive DIS, which cannot 
flip the chirality; in order to measure 
$h_1$, the chirality must be flipped twice, so one always needs 
two hadrons, both in the initial 
state, or one in the initial state and one in the final state, 
and at least one of them must be transversely 
polarised ; \\
{\it ii)} there is no gluon 
transversity distribution: this would imply 
a helicity-flip gluon-nucleon amplitude, which does not  
exist since gluons have helicity $\pm 1$ and the nucleon 
cannot undergo an helicity change of two units. 

The DGLAP 
equations for $h_1$ have been worked out at leading order 
 \cite{Artru:1989zv}, and years later
at next-to-leading order 
\cite{Hayashigaki:1997dn,Kumano:1997qp,Vogelsang:1997ak}. 
There are two noteworthy features of the evolution of $h_1$: 
first of all, since there is no gluon transversity distribution, 
$h_1$ does not mix with gluons and evolves as a non-singlet 
density \cite{Artru:1989zv}; 
second, at low $x$, $h_1$ is suppressed by the evolution  
with respect to $g_1$ \cite{Barone:1997fh}. This has important consequences 
for those observables that involve $h_1$ at low $x$ and large $Q^2$, 
such as the Drell-Yan double transverse asymmetry at collider 
energies 
\cite{Barone:1997mj}.  

The transversity distribution satisfies 
a bound discovered by Soffer  \cite{Soffer:1994ww}: 
\be
\vert h_1(x) \vert \leq \frac{1}{2} 
\, [f_1(x) + g_1(x)]\,.  
\label{soff1}
\ee
This inequality, which is derived in the context 
of the parton model from the expressions of 
the distribution functions in terms of quark-nucleon forward 
amplitudes, is strictly preserved in leading-order QCD 
\cite{Barone:1997fh,Bourrely:1997bx}. At next-to-leading order, 
parton densities are not univoquely defined, but a regularisation scheme 
can be chosen such that the Soffer inequality is still valid 
\cite{Vogelsang:1997ak}. 

The integral of $\Phi(x)$ over $x$ gives the local matrix element 
$\langle P,S \vert \bar \psi(0) \psi(0) \vert P, S \rangle$, 
which can be parametrised in terms of the vector, axial and 
tensor charge of the nucleon. In particular, the tensor charge 
(that we call $\delta q$, for the flavour $q$) is given by
the matrix element of the operator $\bar \psi 
\I \sigma^{\mu \nu} \gamma_5 \psi$,   
\be
\langle P, S \vert \bar \psi_q(0) \I \sigma^{\mu \nu} 
\gamma_5 \psi_q(0) \vert P, S \rangle = 2 \delta q \, (S^{\mu} P^{\nu} 
- S^{\nu} P^{\mu})\,, 
\label{tcharge1}
\ee
and is related to the transversity distributions 
as follows 
\be
\int_0^1 \! \D x \, [h_1^q(x) - \bar h_1^q(x)] = \delta q\,. 
\label{tcharge2}
\ee
Note that, due to the charge-conjugation properties
of $\bar \psi \I \sigma^{\mu \nu} \gamma_5 \psi$, 
which is a $C$-odd operator, the tensor charge is the first 
moment of a flavour  non-singlet combination (quarks 
minus antiquarks). 

An important distinction 
between transverse spin and transverse polarisation
\cite{Jaffe:1991kp} is in order.   
The transverse spin operator, i.e. the generator 
of rotations, for a quark  
is $\Sigma_{T} = \gamma_5 \gamma_0 \gamma_{T}$,     
and  does not commute 
with the free quark Hamiltonian $H_0 = \alpha_z p_z$. 
Thus, there are no common eigenstates of $\Sigma_{T}$ 
and $H_0$: said otherwise, in a transversely polarised 
nucleon quarks cannot be in  a definite transverse spin state. 
The distribution related to $\Sigma_{T}$, called $g_T(x)$,  
is a twist-three quantity  that reflects a complicated quark-gluon 
dynamics with no partonic interpretation. 
On the other hand, the transversity distribution $h_1$ carries information 
about the transverse polarisation of quarks 
inside a transversely polarised nucleon. 
The transverse polarisation operator is 
$ \Pi_{T} = \frac{1}{2}\gamma_0 \Sigma_{T}$,  
and commutes with $H_0$, owing to the presence 
of an extra $\gamma_0$. Therefore, in a transversely 
polarised nucleon, quarks may exist in a definite 
transverse polarisation state, and a simple partonic picture applies to $h_1$. 

The argument above shows that 
the integral $\int \D x \, (h_1^q + \bar h_1^{q})$
does not represent the quark + antiquark 
contribution to the transverse spin of the nucleon. 
A transverse spin sum rule containing the 
first moment of $h_1 + \bar h_1$ has been 
derived in Ref.~\cite{Bakker:2004ib} within the parton model, 
but, in the light of what we have just said 
and of other general considerations, is subject
to some controversy (see the discussion in Ref.~\cite{Burkardt:2008jw}). 
A sum rule for the total angular momentum of transversely 
polarised quarks in an unpolarised hadron 
\cite{Burkardt:2005hp,Burkardt:2006ev}, 
involving the generalised parton distributions, 
will be introduced in Section~\ref{gpd}. 

\subsection{The transverse-momentum dependent (TMD) distribution 
functions}
\label{tmds}

\begin{figure}[t]
\begin{center}
\includegraphics[width=0.40\textwidth]
{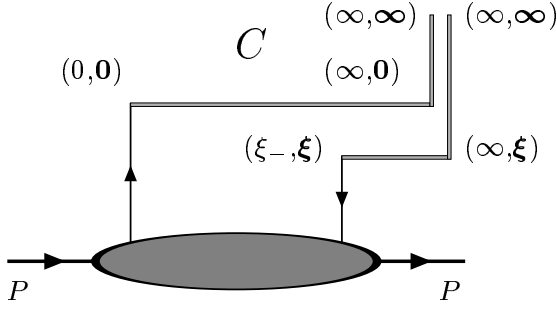}
\caption{\label{fig_links} 
The gauge-link structure of TMD distributions in SIDIS.
}
\end{center} 
\end{figure}

The intrinsic transverse motion of quarks, 
 is a  source of azimuthal and spin asymmetries in 
hadronic processes. 
Taking into account its transverse component, the quark momentum is given by
$k^{\mu} = x P^+ n_+^{\mu} + k_T^{\mu}$. 
As we will see later in this Section, at leading twist there 
are eight TMD distributions: three 
of them, once integrated over $\Vec k_T$, yield $f_1, g_1, h_1$; 
the remaining five are new and vanish upon $\Vec k_T$  
integration.   
Integrating $\Phi (k, P, S)$ over $k^+$ and $k^-$ 
only, one obtains 
the $\Vec k_T$-dependent correlation matrix
\be
  \Phi(x, \Vec k_{T}) = 
  \int \! \D k^+ \, \int \! \D k^-  \,
   \Phi (k, P, S) \, \delta(k^+ - xP^+)\,, 
\label{tmd1}
\ee
which contains the TMD distribution functions. The field-theoretical 
expression of $\Phi (x, \Vec k_T)$ \cite{Collins:1981uk} turns out to be
quite complicated due to the structure of the gauge link, which now 
connects two space-time points with a transverse separation. 
One 
has \cite{Ji:2002aa,Belitsky:2002sm}
\bq
\Phi(x, \Vec k_T) 
  &=&
  \int \! \frac{\D \xi^-}{2 \pi} \int \! 
\frac{\D^2\Vec\xi_T}{(2\pi)^2}
  \E^{\I x P^+ \xi^-} \, \E^{ - \I \Vec{k}_T{\cdot}\Vec\xi_T}
\nonumber \\
& & \times \, 
  \langle P,S
    \vert \bar \psi (0) \,\mathcal{W}^- [0, \infty]\,  
\mathcal{W}^T [0_T, \infty_T] \,  \mathcal{W}^T [\infty_T, \xi_T] \,  
\mathcal{W}^- [\infty, \xi]  \psi(\xi)
  \vert P,S \rangle \vert_{\xi^+ = 0}\,,    
  \label{tmd2}
\eq
with two longitudinal Wilson lines directed along $n^-$, 
from $(0,0,\Vec 0_T)$ to $(0, \infty, \Vec 0_T)$
and from $(0, \infty, \Vec \xi_T)$ to $(0, \xi^-, \Vec \xi_T)$, 
and two Wilson lines $\mathcal{W}^T$ at $\xi^- = \infty$ 
 containing the transverse 
gluon field $A_T^{\mu}$
(Fig.~\ref{fig_links}). 
This link structure, with the 
longitudinal Wilson lines $\mathcal{W}^-$ 
running to $\xi^- = + \infty$, applies  
to semi-inclusive deep inelastic scattering. 
In Drell-Yan processes, the Wilson line 
runs to $- \infty$ and this may change the sign of the distributions, 
as we will discuss later (Section~\ref{todd}). 

It is important to stress in eq. (\ref{tmd2}) the presence 
of the transverse links, 
which survive in the light-cone gauge $A^+ = 0$, enforcing 
gauge invariance under residual gauge transformations.  
These transverse links are responsible for the final-state 
or initial-state interactions 
that generate some TMD distributions otherwise forbidden by time-reversal 
invariance (the so-called $T$-odd distributions).  
In non singular gauges, on the contrary, the gauge potential 
vanishes at infinity and one is left with 
the longitudinal links. 
It is known that in this case there are 
light-cone logarithmic divergences arising in the limit 
$z^+ \to 0$ \cite{Collins:1981uk} due to 
contributions of virtual gluons with zero 
plus momentum, i.e., with infinitely negative rapidity. 
One way to avoid these singularities is 
to use Wilson lines slightly displaced from the light-like direction. 
This introduces a dependence of the TMD distributions on a new scalar 
quantity, $\zeta^2 = (2 P \cdot v)^2/v^2$ ($v$ is a vector 
slightly off the light-cone), acting as a rapidity cutoff. 
The light-cone divergences now appear as large logarithms of $\zeta$, which   
are resummed by the so-called Collins-Soper equation 
\cite{Collins:1981uk,Idilbi:2004vb}. 
A lucid presentation of this subject is contained 
in Ref.~\cite{Collins:2003fm}

Coming back to the quark correlator, at leading twist  
$\Phi(x, \Vec{k}_T)$ has the following structure 
\cite{Mulders:1995dh,Boer:1997nt}\footnote{In the ``Amsterdam classification'' 
of TMD distributions \cite{Mulders:1995dh} which we 
follow in this review the letters $f$, $g$, $h$
refer to unpolarised, longitudinally polarised, and transversely 
polarised distributions, respectively (as first proposed  
by Jaffe and Ji \cite{Jaffe:1991kp,Jaffe:1992ra}). 
The subscript 1 labels the leading twist. 
Subscripts $L$ and $T$ indicate that the parent
hadron is longitudinally or transversely polarised. A superscript
$\perp$ signals the presence of $k_T^i$ factors
 in the quark correlation function.}
\bq
\Phi (x, \Vec{k}_{T}) &=& 
\frac{1}{2} \left \{  f_1 
\, \slashed{n}_+ - f_{1T}^{\perp} \, 
\frac{\epsilon_T^{ij} k_{T i} S_{T j}}{M} 
\, \slashed{n}_+ +
\left ( S_L g_{1L} 
+ \frac{\Vec k_T \cdot \Vec S_T}{M} \, g_{1T} \right ) 
\gamma_5 \slashed{n}_+ \right. 
\nonumber \\
& & \;\;\;\; + \left. h_{1T} 
\frac{[\slashed{S}_T, \slashed{n}_+] \gamma_5}{2} 
+ \left (
S_L h_{1L}^{\perp} 
+ \frac{\Vec k_T \cdot \Vec S_T}{M} \, h_{1T}^{\perp}
\right ) 
 \frac{[\slashed{k}_T, \slashed{n}_+] \gamma_5}{2 M} 
+ \I h_1^{\perp}  \frac{[\slashed{k}_T, \slashed{n}_+]}{2 M} 
\right \}\,, 
\label{tmd100}
\eq
where $\epsilon_T^{ij}$ is the two-dimensional 
antisymmetric Levi-Civita tensor, with $\epsilon_T^{12} = 1$. 
By tracing $\Phi(x, \Vec k_T)$ with Dirac matrices, 
$ \Phi^{[\Gamma]} \equiv \frac{1}{2}{\rm Tr} \, (\Gamma \Phi)$, 
one gets
\bq
  \Phi^{[\gamma^+]} 
&=&  
f_1 (x, k_T^2)
 - \frac{\epsilon_T^{ij} k_{T i} S_{T j}}{M} 
f_{1T}^{\perp} (x, k_T^2) \,, 
\label{tmd110} \\
  \Phi^{[\gamma^+ \gamma_5]} 
  &=&  
S_L g_{1L}(x,  k_T^2) + \frac{\Vec k_T \cdot \Vec S_T}{M} 
g_{1T} (x,  k_T^2) 
\,, 
\label{tmd111} \\
  \Phi^{[\I \sigma^{i+} \gamma_5]} 
 & =&    
S_T^i h_1(x,  k_T^2) + S_L \frac{k_T^i}{M} h_{1L}^{\perp} 
\nonumber \\
& & - \frac{k_T^i k_T^j + \frac{1}{2} k_T^2 g_T^{ij}}{M^2} 
S_{T j} \, h_{1T}^{\perp}(x,  k_T^2) - 
\frac{\epsilon_T^{ij} k_{Tj}}{M} h_1^{\perp} (x,  k_T^2)\,, 
\;\;\;\; i = 1,2\,. 
\label{tmd112}
\eq 
where $h_1 \equiv h_{1T} + (k_T^2/2M) h_{1T}^{\perp}$.  
The three quantities $\Phi^{[\gamma^+]}$,  $\Phi^{[\gamma^+ \gamma_5]}$ and 
  $\Phi^{[\I \sigma^{i +} \gamma_5]}$ represent the probabilities 
of finding an unpolarised, a longitudinally polarised and 
a transversely polarised quark, respectively, with 
momentum fraction $x$ and transverse momentum $\Vec k_T$.  
In eqs.~(\ref{tmd110}-\ref{tmd112}) eight independent 
TMD distributions are present:   
$f_1, f_{1T}^{\perp}, g_1, g_{1T},  
h_1, h_{1L}^{\perp}, h_{1T}^{\perp}, h_1^{\perp}$. 
Upon integration over $\Vec k_T$, only three of these, 
$f_1(x, k_T^2), g_1(x,  k_T^2), h_1(x,  k_T^2)$, 
survive, yielding the $x$-dependent leading-twist distributions 
$f_1(x)$, $g_1(x)$, $h_1(x)$.    

From eq.~(\ref{tmd112}) one sees that 
the spin asymmetry of transversely polarised quarks inside a 
transversely polarised nucleon is given 
 not only by the unintegrated transversity  $h_1(x,  k_T^2)$, but also by 
the TMD distribution $h_{1T}^{\perp} (x,  k_T^2)$, 
which has been given the name of ``pretzelosity'', 
as it is somehow related to the non-sphericity 
of the nucleon shape \cite{Miller:2007ae} 
(for a review of the properties of $h_{1T}^{\perp}$, 
see Ref.~\cite{Avakian:2008dz}).   
Note that, due to the intrinsic transverse motion, 
quarks can also be transversely polarised in a
longitudinally polarised nucleon ($h_{1L}^{\perp}$), and longitudinally 
polarised in a transversely polarised nucleon ($g_{1T}$). 

\subsubsection{The $T$-odd couple: Sivers and Boer-Mulders distributions}
\label{todd}

From eq.~(\ref{tmd110}) the probability of finding an unpolarised  
quark with  longitudinal momentum fraction $x$ and transverse momentum 
$\Vec k_T$ inside a transversely polarised nucleon  is 
\be
f_{q/N^{\uparrow}} (x, \Vec k_T) 
= f_1(x, k_T^2) - \frac{(\hat {\Vec P} \times \Vec k_T) \cdot \Vec S_T}{M}
\,  f_{1T}^{\perp}(x, k_T^2) 
\, , 
\label{tmd114}
\ee
where $\hat {\Vec P} \equiv \Vec P/\vert \Vec P \vert$.  
Thus the azimuthal asymmetry is
\begin{equation}
f_{q/N^{\uparrow}}(x, \Vec k_T) - 
f_{q/N^{\uparrow}}(x, - \Vec k_T)
 = - 2 \, \frac{( \hat{\Vec P} \times \Vec k_T) \cdot \Vec S_T}{M} 
\, f_{1T}^{\perp} (x,  k_T^2)\,, 
\label{sivdef}
\end{equation}
which is proportional to the so-called 
Sivers function $f_{1T}^{\perp}$ \cite{Sivers:1989cc,Sivers:1990fh}.  
A non vanishing $f_{1T}^{\perp}$ signals that 
unpolarised quarks in a transversely polarised nucleon 
have a preferential motion direction: in particular, 
$f_{1T}^{\perp} >0$ means that in a nucleon moving 
along $+ \hat{z}$ with transverse polarisation in the 
$+ \hat{y}$ direction, unpolarised quarks tend to 
move to the right, i.e. towards $- \hat{x}$. 

Specularly, the distribution of transversely polarised 
quarks inside an unpolarised nucleon is \cite{Boglione:1999pz}
\be
f_{q^{\uparrow}/N}(x, \Vec k_T) = 
\frac{1}{2} \, \left [ 
f_1(x, k_T^2) - \frac{(\hat{\Vec P} \times \Vec k_T) 
\cdot \Vec S_{qT}}{M} \, h_1^{\perp}(x, k_T^2) \right ]\,, 
\label{bmdef}
\ee
and from this we get a spin asymmetry of the form
\begin{equation}
f_{q^{\uparrow}/N}(x, \Vec k_T) - 
f_{q^{\downarrow}/N}(x, \Vec k_T)
 = - \frac{( \hat{\Vec P} \times \Vec k_T) \cdot \Vec S_{qT}}{M} 
\, h_1^{\perp} (x, k_T^2)\,, 
\label{boerdef}
\end{equation} 
which is proportional to $h_1^{\perp}$, the 
Boer--Mulders distribution \cite{Boer:1997nt}. 
Positivity bounds for $f_{1T}^{\perp}$ and $h_1^{\perp}$  were derived 
in Ref.~\cite{Bacchetta:1999kz}. 
Note that in the literature (see Ref.~\cite{D'Alesio:2007jt}
and bibliography therein) one also 
encounters the notation 
\be 
\Delta^N f_{q/p^{\uparrow}} \equiv - \frac{2 \vert \Vec k_T \vert}{M} 
\, f_{1T}^{\perp q} \,, 
\;\;\;\;
\Delta^N f_{q^{\uparrow}/p} \equiv 
- \frac{\vert \Vec k_T \vert}{M} \, h_1^{\perp q}\,. 
\ee
The Sivers and Boer-Mulders functions
are associated with the time-reversal ($T$) odd correlations  
$(\hat{\Vec P} \times \Vec k_T  ) \cdot \Vec S_{T}$ and 
$(\hat{\Vec P} \times \Vec k_T ) \cdot \Vec S_{qT}$, 
hence the name of ``$T$-odd distributions''. To see 
the implications of time-reversal invariance one has to recall 
the operator definition of these distributions which,
in the case of the Sivers function, is: 
\be
f_{1T}^{\perp} (x, k_T^2)  \sim 
\int \! \D \xi^- 
\int \! \D^2 \Vec \xi_T \, 
\E^{\I x P^+ \xi^- - \I \Vec k_T \cdot  \Vec \xi_T}
  \langle P, S_T \vert \overline{\psi}(0) \gamma^+ 
\mathcal{W}[0, \xi] 
\psi(\xi) 
\vert P, S_T \rangle \vert_{\xi^+ = 0}
\label{sivfun1}
\ee
If the overall Wilson link $\mathcal{W}$ is na{\"\i}vely set to unity,   
the matrix element in (\ref{sivfun1})  changes sign 
under time reversal, and the Sivers function 
must therefore be zero \cite{Collins:1992kk}.   
On the other hand, a direct
calculation \cite{Brodsky:2002cx} in a spectator model 
shows that $f_{1T}^{\perp}$ is non vanishing: gluon exchange 
between the struck quark and the target remnant generates 
a non-zero Sivers asymmetry
(the presence of a quark transverse momentum smaller than
$Q$ ensures that this asymmetry is proportional to $M/k_T$, rather than to
$M/Q$, and therefore is a leading-twist observable).
The puzzle is solved by carefully considering the Wilson line in 
eq.~(\ref{sivfun1}) \cite{Collins:2002kn}. 
In fact $W[0, \xi]$ includes transverse links at infinity
that do not reduce to unity in the light-cone gauge \cite{Belitsky:2002sm}. 
Since time reversal changes  a future-pointing Wilson line 
 into a past-pointing Wilson line, $T$-invariance,  
rather than constraining $f_{1T}^{\perp}$ to zero,
gives a relation between processes that probe Wilson lines 
pointing in opposite time directions. 
In particular, since in SIDIS the Sivers asymmetry arises from the interaction 
between the spectator and the outgoing quark, whereas in Drell-Yan 
production it arises from the interaction between the spectator 
and an incoming quark, one gets
\be
f_{1T}^{\perp} (x, k_T^2)_{\rm SIDIS} = - 
f_{1T}^{\perp} (x, k_T^2)_{\rm DY} \,.
\label{signrev}
\ee  
A similar relation holds for the Boer-Mulders function $h_1^{\perp}$. 
Eq,~(\ref{signrev}) is an example of the ``time-reversal modified
universality'' of distribution functions in SIDIS, DY
production and $e^+ e^-$ annihilation studied in Ref.
 \cite{Collins:2004nx}.
The relation (\ref{signrev}) is a direct consequence of the gauge structure 
of parton distribution functions, 
and its experimental check would be extremely important.

Gauge link patterns of hadroproduction processes are more complicated and 
do not result in a simple sign flip of the distributions
\cite{Bomhof:2004aw,Bacchetta:2005rm,Bomhof:2006dp,Bomhof:2006ra,Bomhof:2007su,Bomhof:2007xt}. For these processes the authors of 
Refs.~\cite{Bacchetta:2005rm,Bomhof:2006dp} suggested that a  
factorisation scheme should hold with $k_T$ distributions containing 
process-dependent  Wilson lines.  This ``generalised TMD factorisation''  
evidently differs from the standard TMD factorisation, wherein the 
$k_T$ distributions are fully universal quantities. A recent study 
\cite{Rogers:2010dm}, however, has shown that even the generalised 
factorisation scheme is violated in hadroproduction of nearly 
back-to-back jets or hadrons, a process investigated experimentally 
by the STAR collaboration at RHIC \cite{Abelev:2007ii}.

The quark Sivers function 
has an exact gluonic counterpart, $f_{1T}^{\perp g}$, 
which represents the distribution of unpolarised gluons in 
a transversely polarised hadron. 
This function is called $G_T$ in Ref.~\cite{Mulders:2000sh}, where 
a complete classification of leading-twist 
$\Vec k_T$-dependent gluon distributions is presented.  
There is no gluonic equivalent of the Boer-Mulders function, 
but a somehow similar quantity is the distribution of 
linearly polarised gluons in an unpolarised hadron.  

A sum rule for the Sivers function
was derived in QCD by Burkardt \cite{Burkardt:2003yg,Burkardt:2004ur},  
who showed that the sum of all contributions to the 
average transverse momentum of unpolarised partons 
in a transversely polarised target
(that is, the average transverse momentum induced by the Sivers effect), 
must vanish: 
\be
\sum_{a = q, \bar q, g} 
\langle \Vec k_T^a \rangle \vert_{\rm Sivers} = 0\,. 
\label{burk1}
\ee
In terms of the Sivers function, the condition (\ref{burk1}) 
becomes \cite{Efremov:2004tp}
\be
\sum_{a = q, \bar q, g} 
\int_0^1 \D x \, f_{1T}^{\perp (1)\, a} (x) = 0 \,, 
\label{burk2}
\ee
where the first $k_T^2$-moment 
of $f_{1T}^{\perp}$ is given by 
\be
f_{1T}^{\perp (1)} (x) \equiv \int \D^2 \Vec k_T 
\, \frac{k_T^2}{2 M^2} \, f_{1T}^{\perp} (x,  k_T^2)\,.
\label{ktmom1}
\ee
Although some QCD aspects, such as   
ultraviolet divergences  and light-cone singularities, were not considered 
in the original derivation, eq.
(\ref{ktmom1}) is likely to be valid in general. 
In Ref.~\cite{Goeke:2006ef} it has been shown that 
the Burkardt sum rule is fulfilled for a quark 
target in perturbative QCD at one-loop order.  

From a phenomenological point of view, the importance 
of eq.~(\ref{burk2}) is that one can infer the size of 
the gluon Sivers function from fits to 
SIDIS observables involving the quark and antiquark 
Sivers functions \cite{Anselmino:2008sga,Arnold:2008ap}. 

\subsection{Higher-twist distributions and quark-gluon correlators}
\label{ht_tmd}

At twist three, suppressed by $1/Q$, i.e., by $1/P^+$ 
in the infinite momentum frame,  
with respect to leading twist, the quark correlator $\Phi(x)$ admits the 
general decomposition \cite{Jaffe:1991kp,Jaffe:1992ra}
\be
\Phi (x) \vert_{{\rm twist} \, 3}  = \frac{M}{2 P^+} 
\left \{ e (x) + 
g_T(x) \gamma_5 \slashed{S}_T  + 
S_L \, h_L (x) \, 
\frac{[\slashed{n}_+, \slashed{n}_-] \gamma_5}{2}  \right \}\,,  
\label{ht_tmd1}
\ee
displaying the three  distribution functions $e(x), g_T(x), h_L(x)$. 
In particular, $g_T(x)$ contributes to the polarised DIS 
structure function $g_2(x, Q^2)$ (see, e.g., Ref.~\cite{Barone:2003fy}). 
Higher-twist distributions do not have a probabilistic interpretation. 
They involve in fact both good and bad components of the quark fields, 
so the procedure leading to expressions such as 
eqs.~(\ref{good12}-\ref{good16}) cannot be applied. 

\begin{figure}[t]
  \centering
  \begin{picture}(300,110)(0,0)
    \SetWidth{2}
    \ArrowLine(70,15)(120,30)
    \ArrowLine(180,30)(230,15)
    \SetWidth{0.5}
    \ArrowLine(120,50)(120,80)
    \ArrowLine(180,80)(180,50)
    \Gluon(135,58)(135,80){3}{3.5}
     \Text(135,88)[l]{$x_g = x' - x$}
    \Text(65,15)[r]{$P$}
    \Text(235,15)[l]{$P$}
    \Text(115,65)[r]{$x$}
    \Text(185,65)[l]{$x'$}
    \GOval(150,40)(20,40)(0){0.5}
  \end{picture}
  \caption{The quark--gluon correlation matrix.}
  \label{phi_qg}
\end{figure}
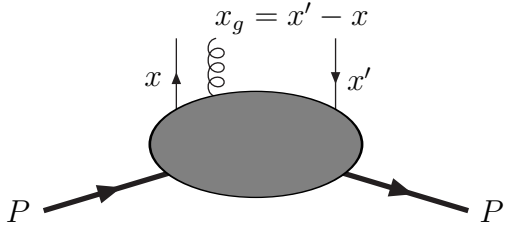

Higher-twist effects are a manifestation of quark-gluon correlations 
inside hadrons \cite{Ellis:1982wd,Ellis:1983cd}. 
At twist three there 
are four quark-gluon correlators, which depend on two momentum fractions, $x$ 
and $x'$ (see Fig.~\ref{phi_qg}): $G_F (x, x')$, $\tilde{G}_F(x, x')$, 
$H_F(x, x')$, $E_F(x, x')$.  
In the literature 
\cite{Qiu:1991pp,Qiu:1998ia,Ji:2006ub,Zhou:2008mz,Zhou:2009jm}, 
these correlators are also 
called $T_F$, $\tilde{T}_F$, $\tilde{T}_F^{(\sigma)}$, 
$T_F^{(\sigma)}$, respectively, with a normalisation varying 
from one paper to another. 
In QCD the quark-gluon correlation functions 
acquire a dependence on a scale $\mu$. 
The equations governing the evolution in $\mu$ 
have been recently written down and solved \cite{Kang:2008ey,Zhou:2008mz}. 

Using the QCD equations 
of motion and integrating over $x'$ one can show that 
\cite{Mulders:1995dh,Boer:2003cm}
\be
g_T(x) = \frac{g_{1T}^{(1)} (x)}{x} + \tilde{g}_T (x)\,, 
\;\;\; 
h_L (x) = - 2 \, 
\frac{h_{1L}^{\perp (1)}(x)}{x} + \tilde{h}_L(x)\,, 
\;\;\;
e(x) = \tilde{e}(x)\,,  
\ee
where 
$g_{1T}^{(1)}, 
h_{1L}^{\perp (1)}$ are the first transverse moments of 
$g_{1T}$ and $h_{1L}^{\perp}$, defined as in eq.~(\ref{ktmom1}), and  
the  tilde functions are genuinely twist-three 
distributions related to the quark-gluon correlators. 

Ignoring the contributions of tilde functions and 
of quark mass terms one gets the generalised Wandzura-Wilczek (WW) 
approximation, so called in analogy with the original WW relation
\cite{Wandzura:1977qf} between the polarised DIS 
structure functions $g_1(x,Q^2)$ and $g_2(x, Q^2)$
\cite{Anselmino:1994gn,Barone:2003fy,Accardi:2009au}.
The generalised WW approximation relates twist-three 
distributions to twist-two distributions. 
It has been investigated by various authors 
\cite{Tangerman:1994bb,Mulders:1995dh,Avakian:2007mv,Metz:2008ib}
and also applied in phenomenological analyses \cite{Kotzinian:2006dw}. 

Coming to the transverse motion of quarks,  
the structure of the $\Vec k_T$-dependent quark correlator 
$\Phi (x, \Vec k_T)$ at 
twist three has also been studied by various authors 
\cite{Mulders:1995dh,Boer:1997nt,Bacchetta:2004zf,Goeke:2005hb}.  
It is now known that there are 16 twist-three TMD distributions: 
$e$, $e_T^{\perp}$, $e_L$, $e_T$, 
$f_T$, $f_L^{\perp}$, $f_T^{\perp}$, $f^{\perp}$, 
$g_T$, $g_L^{\perp}$, $g_T^{\perp}$, $g^{\perp}$, 
$h_L$, $h_T$, $h$, $h_T^{\perp}$
in the classification of Ref. \cite{Bacchetta:2006tn}).   
Some of these functions, namely $g^{\perp}, e_{T}^{\perp}, f_T, f_T^{\perp}$, 
not identified in earlier studies, exist because the Wilson line 
in the quark correlator
provides an extra independent vector ($n_-$) for the 
Lorentz decomposition of $\Phi (x, \Vec k_T)$. 
If we integrate $\Phi(x, \Vec k_T)$  over $\Vec k_T$, the 
only non vanishing distributions are the three $T$-even functions   
in eq.~(\ref{ht_tmd1}). 

Concerning twist four, the integrated parton distributions were 
first identified in Refs.~\cite{Jaffe:1991kp,Jaffe:1992ra,Hoodbhoy:1993wx}. 
More recently, the complete expression of 
the $\Vec k_T$-dependent correlator $\Phi(x, \Vec k_T)$  
has been given  in Ref. \cite{Goeke:2005hb}, where it is shown 
that up to twist four there are in total 32 TMD distributions.   
The unintegrated correlation matrix $\Phi$ is also composed of 32 
Lorentz-scalar structures: 12 amplitudes associated  
to the four-vectors $k, P, S$ and 20 amplitudes associated to $n_-$.  
The number of distribution functions
being equal to the number of amplitudes of $\Phi$, 
all the TMD  distributions are independent and there are no 
general relations among them. In earlier studies    
\cite{Tangerman:1994bb,Mulders:1995dh}, some ``Lorentz-invariance 
relations'' (LIR's) were derived from an expansion of $\Phi$ 
that did not take into account  the amplitudes 
associated to the gauge link vector $n_-$. 
Two of these relations are: 
\be
g_T(x) = g_1 (x) + \frac{\D}{\D x} g_{1T}^{\perp (1)}(x)  \,, \;\;\;\;\;
h_L(x) = h_1 (x) - \frac{\D}{\D x} h_{1L}^{\perp (1)}(x)  \,.   
\label{lir}
\ee
The presence of the $n_-$-dependent amplitudes invalidate 
the LIR's, which are not valid in QCD \cite{Kundu:2001pk,Goeke:2003az}. 
However, they approximately hold in the generalised WW approximation
\cite{Metz:2008ib}. 

A general remark about higher-twist distributions is in order. 
While the distributions $e(x)$, $g_T(x)$, $h_L(x)$ -- or, to 
be precise, the corresponding quark-gluon correlators -- 
enter into the collinear twist-three factorisation theorem 
of QCD \cite{Qiu:1991pp,Qiu:1991wg}, the $k_T$-dependent 
higher-twist distributions are employed in factorisation 
formulae that lack a solid QCD foundation. 
Thus, they should rather be interpreted as a way to model subleading 
effects. 

\subsection{Generalised parton distributions}
\label{gpd}

The generalised parton distributions (GPD's), which are related 
to non-forward quark-quark (or gluon-gluon) correlators,  
emerge in the description of 
hard exclusive processes, such as deeply-virtual Compton scattering 
and exclusive meson production, characterised by a non-zero momentum 
transfer to the target nucleon 
\cite{Radyushkin:1996nd,Ji:1996ek,Goeke:2001tz,Diehl:2003ny,Belitsky:2005qn,Boffi:2007yc}. 
Here we will be mostly concerned 
with the relations existing between the GPD's and the transverse 
spin distributions (for more details, see Ref.~\cite{Meissner:2007rx}).  

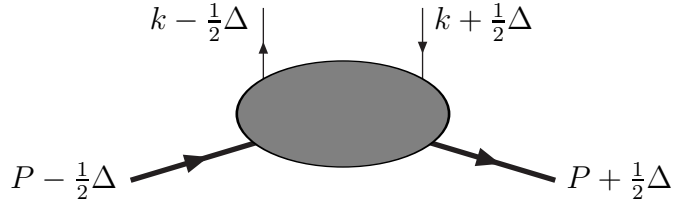
\begin{figure}
  \centering
  \begin{picture}(300,100)(0,0)
    \SetWidth{2}
    \ArrowLine(70,15)(120,30)
    \ArrowLine(180,30)(230,15)
    \SetWidth{0.5}
    \ArrowLine(120,50)(120,80)
    \ArrowLine(180,80)(180,50)
    \Text(115,75)[r]{$k - \frac{1}{2}\Delta$}
    \Text(185,75)[l]{$k + \frac{1}{2}\Delta$}
    \Text(65,15)[r]{$P - \frac{1}{2}\Delta$}
    \Text(235,15)[l]{$P + \frac{1}{2}\Delta$}
    \GOval(150,40)(20,40)(0){0.5}
  \end{picture}
  \caption{Kinematics of GPD's}
  \label{gpd_kin}
\end{figure}

The kinematics of GPD's is represented in fig.~\ref{gpd_kin} 
(we follow the conventions of \cite{Diehl:2003ny}). 
The  momenta of the incoming and the outgoing nucleon 
are $p = P - \frac{1}{2}\Delta$ and $p' = P + \frac{1}{2}\Delta$, 
respectively. 
The momentum transfer squared is $t = \Delta^2$. The GPD's 
depend on $t$ and on two light-cone momentum ratios: 
$x = k^+/P^+$ and  $\xi = - \Delta^+/2 P^+$.    
The variable $\xi$ is sometimes called ``skewness'', and the GPD's are also 
known as ``skewed parton distributions''. 
At leading twist, there are 8 GPD's \cite{Diehl:2001pm}:   
\be
H(x, \xi, t), \, E(x, \xi, t), \, 
\tilde{H}(x, \xi, t), \, \tilde{E}(x, \xi, t), \, 
H_T(x, \xi, t), \, E_T(x, \xi, t), \, \tilde{H}(x, \xi, t), \, 
\tilde{E}(x, \xi, t)\,. 
\label{gpd20}
\ee 
The first four  are chirally even 
and are related to the familiar form factors.  
Integrating $H$, $E$, $\tilde{H}$, $\tilde{E}$ over $x$, one gets 
in fact the Dirac, Pauli, axial 
and pseudoscalar form factors, respectively. 
The quantity 
\be
\int \! \D x \, E^q(x, 0, 0) = \kappa^q\,, 
\label{anmom}
\ee
is the contribution of the flavour $q$ to the 
anomalous magnetic moment of the nucleon, 
that is, to the Pauli form factor $F_2$ at $t=0$. 
The GPD's $H, \tilde{H}, H_T$, taken at $\xi = t =0$,  coincide   
with the integrated quark distributions $f_1, g_1, h_1$: 
\be
H(x, 0,0) = f_1(x)\,,
\; 
\tilde{H}(x,0,0) = g_1(x) \,, 
\;
H_T(x,0,0)   = h_1(x)\,.   
\label{gpd14}
\ee
The original interest in GPD's was prompted by 
Ji's sum rule relating the total angular momentum 
of quarks (in a nucleon with  polarisation vector $\Vec S$) 
to the second moment of $H$ and $E$ \cite{Ji:1996ek}:
\be
\langle J_q^i \rangle = S^i \int \! \D x \, 
x \, [H(x, 0, 0) + E (x, 0, 0)] \,. 
\label{jisr}
\ee
A similar decomposition for the 
angular momentum of quarks with transverse polarisation vector
$\Vec S_q$ in an unpolarised nucleon has been derived in Ref.
\cite{Burkardt:2005hp,Burkardt:2006ev} and is given by: 
\be
\langle J_q^i (\Vec S_q) \rangle = \frac{S_q^i}{4}  \int \! \D x \, 
x \, [H_T(x, 0, 0) + 2 \tilde{H}_T (x, 0, 0) + E_T (x, 0, 0)]\,. 
\label{busr}
\ee
Here $H_T (x,0,0)$ coincides with 
transversity, whereas the combination $2 \tilde{H}_T + E_T$ 
appears in the impact-parameter description of 
the Boer-Mulders effect (see Section~\ref{impact}). 
 
Note in conclusion that there are no direct and model-independent 
connections between the GPD's and the TMD distributions, as 
stressed in Refs.~\cite{Meissner:2007rx,Meissner:2008xs}.   
GPD's are instead directly 
related to the distribution functions in the impact-parameter 
space. 

\subsection{Distribution functions in the impact-parameter space} 
\label{impact} 

In the impact-parameter space one can get a more intuitive picture of   
some transverse spin and transverse momentum effects. 
To define the impact-parameter distributions (IPD's), one first introduces  
 nucleon states localised at a transverse position  
$\Vec R_{T}$, by means of an inverse  Peierls-Yoccoz projection: 
\be
\vert P^+, \Vec R_T ; S \rangle 
= \mathcal{N} \, \int \frac{\D^2 \Vec P_{T}}{(2 \pi)^2} 
\, \E^{- \I \Vec P_{T} \cdot \Vec R_{T}} 
\, \vert P, S \rangle\,,  
\label{ip1}
\ee
where $\mathcal{N}$ is a normalisation factor. 
The IPD's are light-cone 
correlations in these transverse-position nucleon eigenstates.  
For instance, the unpolarised IPD is given by 
\be
q(x, b_{T}^2) = 
\int \frac{\D z^-}{4 \pi} \, 
\E^{\I x P^+ z^-} \, 
\langle P^+, \Vec 0_{T}; S \vert 
\bar \psi (z_1) \,  
\mathcal{W}^- [z_1, z_2]
\gamma^+ \psi (z_2) \vert 
P^+, \Vec 0_{T}; S 
\rangle\,, 
\label{ip2}
\ee
with $z_{1,2} = (0^+, \mp \frac{1}{2}z^-, \Vec b_T)$. This 
is the number density of quarks with momentum fraction $x$ 
and transverse position $\Vec b_T$ inside an unpolarised
hadron.
The polarised IPD's  are obtained 
by inserting in the matrix element of eq.~(\ref{ip2}), instead of $\gamma^+$,  
the matrices $\gamma^+ \gamma_5$ and $ \I \sigma^{i +} \gamma_5$. 
 
IPD's are Fourier transforms not of the TMD distributions, but of the GPD's.  
The impact-parameter transform of a generic GPD $X$  
for $\xi = 0$ (which implies $\Delta^2 = -  \Delta_T^2$)  is defined as
\be
\mathcal{X}(x,b_T^2) = \int \frac{\D^2 \Vec \Delta_T}{(2 \pi)^2} 
\, \E^{- \I \Vec \Delta_T \cdot \Vec b_T} \, 
X (x, 0, -   \Delta_T^2)\,. 
\label{ip3}
\ee
It is straightforward to show \cite{Burkardt:2002hr} that the unpolarised 
IPD $q(x,  b_T^2)$ coincides with the Fourier transform 
of $H (x, 0, - \Vec \Delta_T)$, that is 
\be
q(x,  b_T^2) = \mathcal{H}_q(x,  b_T^2)\,. 
\label{ip4}
\ee
The impact-parameter density of unpolarised quarks 
in a transversely polarised nucleon ($N^{\uparrow}$) is 
\cite{Burkardt:2002hr,Diehl:2005jf,Meissner:2007rx}
\be
q_{N^{\uparrow}}(x, \Vec b_T) 
= \mathcal{H}_q (x, b_T^2) +  \frac{(\hat{\Vec P} 
\times \Vec b_T) \cdot \Vec S_T}{M} \mathcal{E}_q'(x, 
b_T^2) \,, 
\;\;\;\;\; {\rm with} \;\;
\mathcal{E}_q'(x, 
b_T^2) \equiv 
\frac{\partial}{\partial   b_T^2} 
\mathcal{E}_q (x,  b_T^2) \,,  
\label{ip5}
\ee
where $\mathcal{E}(x,  b_T^2)$ is the Fourier 
transform of $E (x, 0, -  \Delta_T^2)$. 
Notice the formal similarity with eq.~(\ref{tmd114}) 
and the correspondence $f_{1T}^{\perp} \leftrightarrow - \mathcal{E}'$.  . 
Due to the $\mathcal{E}_q'$ term, which can be regarded as the 
$\Vec b_T$-space analogue of the Sivers distribution,  
$q_{N^{\uparrow}}(x, \Vec b_T)$ is not axially 
symmetric and describes a spatial distortion of the quark distribution in 
the transverse plane. 
Final-state interactions can translate this position-space 
asymmetry into a momentum-space asymmetry. 
For instance, if the nucleon moves in the $+ \hat z$ directions and  is 
polarised in the $+ \hat x$ direction, a positive 
$\mathcal{E}_q'$ implies that quarks tend to be displaced 
in the $- \hat y$ direction, and final-state interactions,  
which is expected to be attractive on average, convert this transverse 
distortion into a momentum asymmetry in the $+ \hat y$ 
direction. 
This is the intuitive explanation of the Sivers effect in 
the impact-parameter picture \cite{Burkardt:2003yg,Burkardt:2003je}. 
A measure of the space distortion is given by the 
flavour dipole moment
\be
 d_q^i = \int \! \D x \int \! \D^2 \Vec b_T 
\,  b_T^i \, q_{N^{\uparrow}}(x, \Vec b_T)  
= - \frac{\epsilon_T^{ij} S_T^j}{2 M} \, \int \! 
\D x \, E_q(x, 0, 0) = - \frac{\epsilon_T^{ij} S_T^j}{2 M}
\, \kappa_q\,, 
\label{ip7}
\ee
where $\kappa^q$ is the contribution of the quark flavour $q$ 
to the anomalous magnetic moment of the nucleon, see eq.~(\ref{anmom}). 
The argument developed so far 
is summarised by the following qualitative relation between 
the Sivers function $f_{1T}^{\perp}$ and $\kappa^q$
\cite{Burkardt:2002ks,Burkardt:2003uw,Burkardt:2003je}
(any quantitative relation between these two 
quantities is necessarily model-dependent  
\cite{Meissner:2007rx,Meissner:2008xs})  
\be
f_{1T}^{\perp q} \sim  - \kappa^q\,,  
\label{ip8}
\ee
where the minus sign is a consequence of attractive final-state interactions 
that transform a preferential direction in the 
$\Vec b_T$-space into  the opposite direction in $\Vec k_T$. 
Eq.~(\ref{ip8}) leads to an immediate prediction: since the quark 
contributions to the anomalous magnetic moment of the proton $\kappa^p$, 
extracted from the experimental value of $\kappa^p$ using 
SU(2) flavour symmetry, are $\kappa^u \simeq 1.7, \kappa^d = - 2.0$, 
one expects $f_{1T}^{\perp u}< 0$ 
and $f_{1T}^{\perp d} > 0$. This prediction has been 
corroborated by the SIDIS experiments (see Section~\ref{siv_sidis}). 
 
\begin{figure}[t]
\begin{center}
\includegraphics[width=0.50\textwidth]
{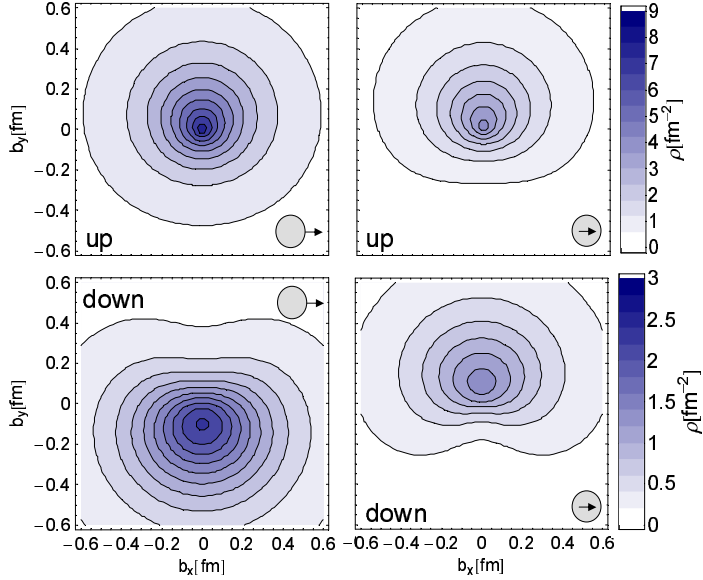}
\caption{\label{fig_ipd} 
First $x$-moments of the densities of unpolarised quarks 
in a transversely polarised nucleon (left) 
and transversely polarised quarks in an unpolarised 
nucleon (right) for $u$ (upper plots) and $d$ (lower plots) 
quarks. Quark spins (inner arrows) and nucleon 
spins (outer arrows) are oriented in the transverse plane 
as indicated. From Ref.~\cite{Gockeler:2006zu}. 
}
\end{center} 
\end{figure}

Consider now the case of transversely polarised 
quarks inside an unpolarised nucleon. 
Their impact-parameter distribution is \cite{Diehl:2005jf}
\be
q^{\uparrow}(x, \Vec b_T) 
= \frac{1}{2} 
\, \left \{ \mathcal{H}_q (x, b_T^2) +  \frac{(\hat{\Vec P} 
\times \Vec b_T) \cdot \Vec S_{qT}}{M} [ \mathcal{E}_{Tq}'(x,  
b_T^2) + 2 \tilde{H}_{Tq}'(x,  b_T^2) ] \right \}  \,, 
\label{ip9}
\ee
The term $\mathcal{E}' + 2 \tilde{H}_T'$  
is the analogue of the Boer-Mulders function 
in the $\Vec b_T$-space -- see eq.~(\ref{bmdef}). 
Again, we see that transverse spin (of quarks, in this case) 
causes a spatial distortion of the distribution,  
which is at the origin of the Boer-Mulders effect. 
One can repeat the same reasoning developed for the 
Sivers effect and introduce a transverse anomalous moment 
$\kappa_T^q$, defined by 
\be
\kappa_T^q \equiv \int \! \D x \, [E_{T}^q (x, 0, 0) + 2 \tilde{H}_{T}^q
(x, 0, 0)] \,. 
\label{ip10}
\ee
The Boer-Mulders function is expected to scale with this quantity, 
\be
h_{1}^{\perp q} \sim - \kappa_T^q\,. 
\label{ip11}
\ee
where the minus sign has the same meaning as before. 
Unfortunately, no data exist for $\kappa_T^q$. 
This quantity, however, and the impact-parameter distributions 
have been calculated in lattice QCD  \cite{Gockeler:2006zu,Brommel:2008zz} 
and are shown in Fig.~\ref{fig_ipd}). 
The result for $\kappa_T$ is: $\kappa_T^u = 3.0$, 
$\kappa_T^d = 1.9$. Thus, at variance with $f_{1T}^{\perp}$,  
we expect the  $u$ and $d$ components of $h_1^{\perp}$   
to have the same sign, and in particular to be both negative. 
Moreover, assuming simple proportionality 
between the ratio $h_1^{\perp}/f_{1T}^{\perp}$
and $\kappa/\kappa_T$, the $u$ component of $h_1^{\perp}$ 
should be approximately twice as large as the corresponding component 
of $f_{1T}^{\perp}$, while $h_1^{\perp d}$  and $f_{1T}^{\perp d}$ 
should have a comparable magnitude and opposite sign. 
These predictions are well supported by 
a phenomenological analysis of SIDIS data \cite{Barone:2009hw}
as will be shown in Section~\ref{siv_sidis}. 

\subsection{Model calculations of TMD distributions}
\label{models}

Models and other non-perturbative approaches like lattice 
calculations play a very important r{\^o}le when the experimental 
information about distribution functions is scarce 
or lacking at all.  
So, it is not surprising that a considerable effort has been made to 
compute the TMD distributions in various models of the nucleon 
and by lattice QCD. 
Here we will be not 
be able to give an exhaustive account of all this work still 
largely in progress and we will limit ourselves to a
general discussion. 
For a recent review of model results see Ref.~\cite{Avakian:2009jt}. 

The first calculation of TMD distributions was performed 
in a quark-diquark spectator model \cite{Jakob:1997wg}. 
This class of models, with various 
quark-diquark vertex functions, has been subsequently 
used by many authors. 
In particular,  using a 
simple scalar spectator model with gluon exchange 
is was shown explicitly  \cite{Brodsky:2002cx}
that the Sivers function is non vanishing.  
Since Wilson links, representing gluon 
insertions, are crucial in order to guarantee the existence 
of the $T$-odd distribution functions, these   
can only be computed in models containing gluonic degrees of freedom. 
Following Ref.~\cite{Brodsky:2002cx}, more refined calculations 
of the Sivers and Boer-Mulders functions were performed in spectator
models  with both scalar and axial-vector diquarks and various 
quark-diquark vertices 
\cite{Goldstein:2002vv,Gamberg:2003ey,Bacchetta:2003rz,Goeke:2006ef,Gamberg:2007wm,Bacchetta:2008af,Ellis:2008in}. 
Other models used to evaluate the $T$-odd functions include  the MIT bag model
\cite{Yuan:2003wk,Cherednikov:2006zn,Courtoy:2008dn,Courtoy:2009pc},  
the constituent quark model \cite{Courtoy:2008vi,Courtoy:2009pc} 
and a light-cone model \cite{Lu:2004au}. 
In Ref.~\cite{Cherednikov:2006zn} final state interactions 
were assumed to be induced by instanton effects. 

What emerges from models is that the Sivers function, although quite 
variable in magnitude, is  negative for $u$ quarks and positive for $d$ quark.
A different sign of $f_{1T}^{\perp d}$ is however 
found in the model of Ref.~\cite{Cherednikov:2006zn}.  
As for the Boer-Mulders function, the general prediction (with 
the exception of Ref.~\cite{Bacchetta:2003rz})  is that 
both the $u$ and the $d$ distributions are negative. 
These signs for $f_{1T}^{\perp}$ and $h_1^{\perp}$ 
are also expected in the impact-parameter picture
\cite{Burkardt:2002hr,Burkardt:2002ks,Burkardt:2003je,Burkardt:2003uw,Burkardt:2005hp}, 
in the large-$N_c$ approach \cite{Pobylitsa:2003ty}, which predicts 
the  isoscalar component of $f_{1T}^{\perp}$ and the isovector component 
of $h_1^{\perp}$  to be suppressed, and in chiral models \cite{Drago:2005gz}. 

Spectator models have been also used \cite{Afanasev:2006gw,Gamberg:2006ru}
to calculate $T$-odd twist-3 distributions, in particular 
$g^{\perp}$, which contributes to the longitudinal beam  
spin asymmetry in SIDIS. 

Models without gluonic degrees of freedom can be used 
to compute $T$-even TMD distributions only. These distributions  
have been calculated in a spectator model \cite{Bacchetta:2008af}, 
in light-cone quark models \cite{Pasquini:2008ax,Boffi:2009sh,She:2009jq}, 
in a covariant parton model with orbital motion \cite{Efremov:2009ze} 
and in the bag model \cite{Avakian:2008dz,Avakian:2010br}.  
In particular, Ref.~\cite{Avakian:2010br} presents a 
systematic study of leading and subleading 
twist TMD distributions and of the relations among them.

In any quark model without gluons, the Lorentz--invariance 
relations, obtained by neglecting the amplitudes 
of the quark-quark correlator related to the 
gauge link (Section~\ref{ht_tmd}), must obviously be valid.  
There are also a number of other relations  
that hold in some specific models like \cite{Avakian:2008dz}
\be
g_1(x,  k_T^2) - h_1 (x,  k_T^2) 
= \frac{ k_T^2}{2 M^2} \, h_{1T}^{\perp} (x,  k_T^2) \,.   
\label{models1}
\ee   
According to this relation, $h_{1T}^{\perp}$ can be interpreted as a measure 
of the relativistic effects in the nucleon, which   
are known to be responsible for the difference between the helicity and 
the transversity distributions \cite{Barone:1996un}. 
Other model-dependent relations involving the  
TMD distributions are listed in Refs.~\cite{Avakian:2009jt,Avakian:2010br}. 

Finally, one should keep in mind that  
models provide a dynamical picture of the nucleon at some fixed, very
low, scale $\mu^2 < 1$ GeV$^2$ 
\cite{Parisi:1976fz,Jaffe:1980ti,Schreiber:1991tc,Barone:1993iq}.  
The quark distributions that one gets are therefore 
valid at this unrealistic scale and must be evolved to the 
experimental scales. The evolution of the TMD distributions has been 
unknown until very recently and is therefore usually neglected or approximated 
in current phenomenological analyses.  
 
\subsection{Fragmentation functions} 
\label{frag}

In partially inclusive processes a parton hadronises
into a particle $h$ carrying away a fraction of the parton's momentum. 
In the following  it is supposed that the 
fragmentation process is initiated by a quark, as is the case in SIDIS 
and $e^+ e^-$ annihilation at leading order.  
The momentum of the fragmenting quark is indicated as $\kappa^{\mu}$ and 
$z$ is the fraction of its longitudinal component carried by the final hadron, 
$z = P_h^-/\kappa^-$. 
Since the hadron moves in the opposite direction with respect to the target 
nucleon, the minus components of momenta are 
the dominant ones. The fragmenting quark  has a transverse 
momentum $\Vec \kappa_T$ with respect to the final hadron. 
Conversely, 
the hadron has a transverse momentum $\Vec p_T = - z \Vec \kappa_T$ 
with respect to the quark. 

The fragmentation analogue of $\Phi_{ij} (x, \Vec k_T)$ is: 
\bq
\Xi_{ij} (z, z \Vec \kappa_T) 
& =& \frac{1}{2z} \, \sum_X \int \frac{\D \xi^+}{2 \pi} 
\int \frac{\D^2 \Vec \xi_T}{(2 \pi)^3}
\, \E^{\I P_h^- \xi^+/z} 
\, \E^{- \I \Vec \kappa_T \cdot \Vec \xi_T} 
\nonumber \\
& & \times
  \langle 0 | \mathcal{W}[+ \infty, \xi] \, \psi_i(\xi) | P_h, S_h; X \rangle
  \langle P_h, S_h; X | \anti\psi_j(0)\,  
\mathcal{W}[0, + \infty]| 0 \rangle 
\vert_{\xi^- = 0}\,,  
\label{sidis114}
\eq
where each Wilson line   includes a longitudinal 
link along $n_+$ and a transverse link at infinity~\cite{Bacchetta:2006tn}. 
In the case of fragmentation one has the same gauge structure in SIDIS 
and in $e^+ e^-$ annihilation, which means that 
there is no difference between the fragmentation functions 
of these processes, i.e. they are universal quantities 
in a full sense \cite{Collins:2004nx}. 
The integrated fragmentation correlator is given by   
\be
\Xi (z)= \int \! \D^2 \Vec p_T \, 
\Xi (z, \Vec p_T) = 
\frac{1}{2} \, \left \{
D_1 (z) \slashed{n}_- + S_L \, G_1 (z) \gamma_5 
\slashed{n}_- + H_1(z) \, \frac{[\slashed{S}_T, \slashed{n}_- ] 
\gamma_5}{2} \right \} \, + \, {\rm h.t.}\,,  
\ee
where ``h.t.'' denotes higher-twist terms. 
$D_1$, $G_1$ and $H_1$ 
are the integrated leading-twist fragmentation functions (FF's):  
$D_1$ is the ordinary unpolarised fragmentation function, $G_1$ 
is the analogue of the helicity distribution, $H_1$ is the analogue 
of the transversity 
distribution (it describes the fragmentation of a transversely 
polarised quark into a transversely polarised hadron).  

To compute azimuthal asymmetries the transverse-momentum 
dependent FF's are needed. 
For simplicity, we limit ourselves to listing the 
FF's of main phenomenological interest. 
The traces of the fragmentation matrix corresponding to 
unpolarised and transversely polarised quarks are \cite{Mulders:1995dh}
\bq
& & \Xi^{[\gamma^-]}(z, \Vec p_T)  = 
D_1 (z,  p_T^2)
+ \frac{\epsilon_{T ij} p_T^i S_{hT}^j}{z M_h} 
\, D_{1T}^{\perp} (z,  p_T^2) \,, 
\label{xig} \\
& & \Xi^{[\I \sigma^{i-} \gamma_5]}(z, \Vec p_T)
 = S_{hT}^i H_1(z,  p_T^2) + 
\frac{\epsilon_T^{ij} p_{Tj}}{z M_h} \, 
H_1^{\perp} (z,  p_T^2) + \ldots \, .
\label{xis}
\eq
$D_{1T}^{\perp}$ is analogous to the Sivers distribution function  
and describes the production of transversely polarised hadrons 
from unpolarised quarks. 
For this reason it is 
called ``polarising fragmentation function'' \cite{Anselmino:2000vs}.  

\subsubsection{The Collins function}
\label{collinsfunction}

The most noteworthy FF appearing in (\ref{xis}) is $H_{1}^{\perp}(z, k_T^2)$,  
the so-called Collins function, describing 
the fragmentation of a transversely polarised quark into an 
unpolarised hadron \cite{Collins:1992kk}. 
The resulting transverse-momentum asymmetry of hadrons is 
\be
 D_{h/q^{\uparrow}}(z, \Vec p_T) - 
D_{h/q^{\uparrow}}(z, - \Vec p_T) = 
2 \, \frac{(\hat{\Vec \kappa} \times \Vec p_T) \cdot \Vec S_{qT}'}{z M_h}
H_1^{\perp}(z,  p_T^2)\,, 
\ee
where $\Vec S_q'$ is the spin vector of the fragmenting quark. 
From the structure of the 
correlation $(\hat{\Vec \kappa} \times \Vec p_T) \cdot \Vec S_{qT}'$
one sees that a positive $H_1^{\perp}$ corresponds to a preference of the 
hadron to be emitted on the left side of the jet 
if the quark spin points upwards. Through this mechanism 
the transverse momentum 
of the produced hadron with respect to the jet direction 
 acts as a quark polarimeter.  

The Collins function satisfies a sum rule 
arising from the conservation of the intrinsic transverse 
momentum during quark fragmentation. This sum rule, 
discovered by Sch{\"a}fer and Teryaev \cite{Schafer:1999kn}, reads
\be
\sum_h \int \! \D z \, z \, H_1^{\perp (1) q}(z) = 0\,, 
\;\;\; {\rm with} \;\;\;
H_1^{\perp (1)}(z) \equiv z^2 \, \int \! \D^2 \Vec \kappa_T 
\frac{ \kappa_T^2}{2 M_h^2} \, H_1^{\perp} (z, 
z^2  \kappa_T^2)\,. 
\label{schaefer1}
\ee 

\begin{figure}[t]
\begin{center}
\vspace{-1.5cm}
\includegraphics[width=0.60\textwidth]
{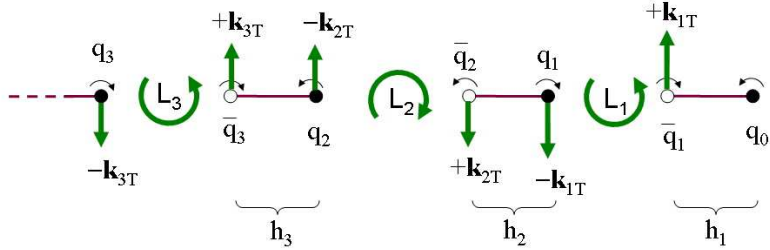}
\vspace{-1.5cm} 
\caption{\label{fig:string} 
The fragmentation process in the recursive string model~\cite{Artru:2010st}.
The $\hat{z}$ direction is along the string and the fragmenting quark $q_0$
is supposed to be polarised in the $+ \hat{y}$ direction, out of the page.
}
\end{center} 
\end{figure}

A simple qualitative explanation of the Collins effect is provided by the
``recursive string model'' of Ref. 
\cite{Artru:1993ad,Artru:1995bh,Artru:2010st},
which is illustrated in Fig.~\ref{fig:string}. 
Suppose that a quark $q_0$, 
polarised in the $+ \hat{y}$ direction, i.e. out of the page in the
figure, fragments into a pion with an antiquark 
$\bar q_1$ created by string breaking. 
If the $q_1 \bar q_1$ pair is in a ${}^3 P_0$ state, the orbital angular 
momentum of the pair is $L =1$, and the pion, inheriting the transverse 
momentum of $\bar q_1$, moves in the $+\hat{x}$ direction.    
The quark $q_1$, with the subleading pion that contains it, 
moves in the opposite direction. 
This model predicts 
opposite Collins asymmetries for $\pi^+$ and $\pi^-$ assuming $u$ dominance, 
and a positive (negative) sign for the favoured (unfavoured) 
Collins function. 
``Favoured'' refers to the fragmentation 
of a quark or an antiquark belonging to the valence component 
of the final hadron, 
 e.g. $u \to \pi^+$, $d \to \pi^-$, $\bar d \to \pi^+$, etc..  

The Collins function for pions has been computed in various 
fragmentation models
\cite{Bacchetta:2001di,Bacchetta:2002tk,Bacchetta:2003xn,Gamberg:2003eg,Gamberg:2003pz}. What is common to these approaches is that 
$H_1^{\perp}$ arises from the interference between a tree 
level amplitude and loop corrections that provide the necessary 
imaginary parts.  
The differences reside in the pion-quark couplings and in the nature 
of the virtual particles in the loops (pions or gluons). 
An assessment of model calculations 
of the Collins function is contained in Ref.~\cite{Amrath:2005gv}.

\section{Processes and observables related to transverse spin} 
\label{processes}

In this section we will present a general description of 
the processes probing the transverse-spin and transverse-momentum 
structure of hadrons, and of the relevant observables: 
single-spin asymmetries, double-spin asymmetries 
and unpolarised azimuthal asymmetries. 
The focus will be on two classes of reactions that have   
clear and well-established theoretical descriptions, namely 
SIDIS with the related process $e^+ e^-$ annihilation 
into hadron pairs,  and DY production.  
Hadroproduction will also be described, but in lesser 
detail. 
The last subsection contains  a sketchy 
presentation of other processes somewhat related to transverse spin. 

\subsection{Semi-inclusive deep inelastic scattering (SIDIS)}
\label{sidis}

SIDIS is the process $\ell (l)  +  N (P)  \rightarrow  \ell' (l')
 +  h (P_h)  +  X (P_X)$,  
where $\ell$ ($\ell'$) is the incoming (outgoing) lepton, 
$N$ the nucleon target, $h$ the detected hadron. 
The corresponding four-momenta are 
given in parentheses. In the following we will denote by $S_{\parallel}$ 
and $\Vec S_{\perp}$ the longitudinal and transverse component of the 
target spin vector, respectively, and by $\lambda_{\ell}$ 
the longitudinal polarisation of the incident lepton. 

SIDIS is usually described in terms of the invariant variables
$x_B = Q^2/2 \, P \cdot q$, $y =  P \cdot q/P \cdot l$, 
$z_h = P \cdot P_h/P \cdot q$, with $ q = l - l'$ and $Q^2 \equiv - q^2$.
In the deep inelastic limit, $Q^2$ is much larger than the 
mass $M$ of the nucleon and the mass $M_h$ of the final hadron. 
Hereafter mass corrections are neglected unless otherwise 
stated.  

\begin{figure}[t]
\begin{center}
\includegraphics[width=0.50\textwidth]
{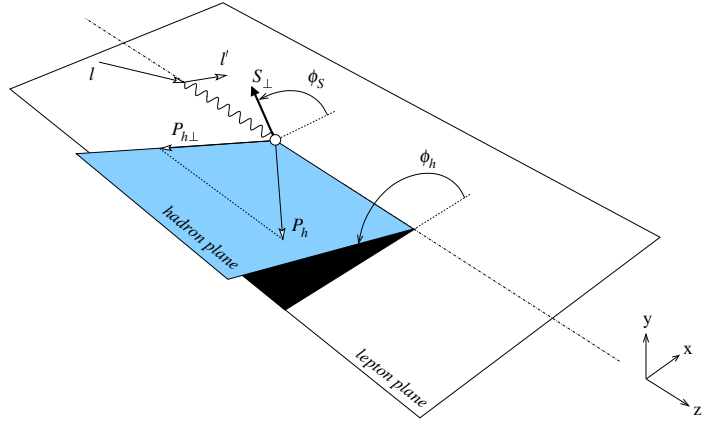}
\caption{\label{plane} Lepton and hadron planes in semi-inclusive 
deep inelastic scattering.
The reference frame is defined according to the convention
of Ref.~\cite{Bacchetta:2004jz}.
}
\end{center} 
\end{figure}

To parametrise the cross section in terms 
of structure functions the so-called ``$\gamma^* N$ collinear frame'' 
\cite{Bacchetta:2004jz}
is usually adopted. As shown in Fig.~\ref{plane} in this reference frame 
the virtual photon and the target nucleon are collinear
and directed along the z axis, with the 
photon moving in the positive $z$ direction, and 
the final hadron has  a transverse momentum $\Vec P_{h \perp}$.  
All azimuthal angles are referred to the lepton scattering plane: 
$\phi_h$ is the azimuthal angle of the hadron $h$, $\phi_S$ is 
the azimuthal angle of the nucleon spin vector $\Vec S_{\perp}$.  
The phase space of the process 
contains  another angle, $\psi$, which is the azimuthal 
angle of the outgoing lepton around the beam axis
with respect to an arbitrary fixed direction, 
which is chosen to be given by the target spin. 
Up to corrections of order $M^2/Q^2$ one has $\D \psi \simeq 
\D \phi_S$ \cite{Diehl:2005pc}. 

In this Section we consider the case of a spinless or unpolarised 
detected hadron, while leptoproduction of transversely polarised hadrons will 
be treated in Section~\ref{transvbar}. 
The SIDIS differential cross section 
in the six variables $x_B, y, z_h, \phi_S, 
P_{h \perp} \equiv   \vert \Vec P_{h \perp} \vert$ and $\phi_h$, is given by 
\be
\frac{\D^6 \sigma}{\D x_B  \D y  \D z_h  \D \phi_S  
\D \phi_h  \D   P_{h \perp}^2} = 
\frac{\alpha_{\rm em}^2}{ 8 Q^4} \frac{y}{z_h} 
\, L_{\mu \nu} W^{\mu \nu} \,, 
\label{sidis1} 
\ee
where $L_{\mu \nu}$ is the usual DIS leptonic tensor and 
$W^{\mu \nu}$ is the hadronic tensor 
\bq
W^{\mu \nu} = \frac{1}{(2 \pi)^4} 
\sum_X &\int & \! \frac{\D^3 \Vec P_X}{(2 \pi)^3 2 E_X}
\, (2 \pi)^4 \delta^4 (P + q - P_X - P_h) 
\nonumber \\
& & \times \langle P, S \vert J^{\mu}(0) \vert X; P_h, S_h 
\rangle \langle X; P_h, S_h \vert J^{\nu}(0) \vert P, S \rangle\,. 
\label{sidis2} 
\eq 
Neglecting for simplicity the $M^2 x^2/Q^2$ corrections
the complete SIDIS  cross section can be parametrised 
in terms of 18 structure functions as follows 
\cite{Diehl:2005pc,Bacchetta:2006tn};
 \bq
& & 
 \frac{\D^6 \sigma}{\D x_B  \D y  \D z_h   
\D \phi_h  \D  P_{h\perp}^2 \, \D \phi_S} = 
 \frac{\alpha_{\rm em}^2}{ x_B y Q^2} \left \{ 
(1 - y + \frac{1}{2} y^2) \, F_{UU, T} 
+ (1 - y) \, F_{UU, L} \right.
\nonumber \\
& & \hspace{2cm} + (2 - y) \sqrt{1 - y} \, \cos \phi_h F_{UU}^{\cos \phi_h} 
+ (1 - y) \, \cos 2 \phi_h \, F_{UU}^{\cos 2 \phi_h} 
 + \lambda_{\ell} \, y \sqrt{1 - y} \, \sin \phi_h \, 
F_{LU}^{\sin \phi_h} 
\nonumber \\
& & \hspace{2cm} + S_{\parallel} \left [ 
(2 - y) \sqrt{1 - y} \, \sin \phi_h \, F_{UL}^{\sin \phi_h} 
+ (1 - y) \, \sin 2 \phi_h \, F_{UL}^{\sin 2 \phi_h} \right ]
\nonumber \\
& & \hspace{2cm} + S_{\parallel} \, \lambda_{\ell} 
\, \left [ y (1 - \frac{1}{2} y) \, F_{LL} + y \sqrt{1 - y} \cos \phi_h 
\, F_{LL}^{\cos \phi_h} \right ]
\nonumber \\
& & \hspace{2cm} +  S_{\perp} \, \left [ 
\sin (\phi_h - \phi_S) \, \left ( (1 - y + \frac{1}{2} y^2) \, 
F_{UT, T}^{\sin (\phi_h - \phi_S)} 
+ (1 - y) \, F_{UT, L}^{\sin (\phi_h - \phi_S)} \right )
\right. 
\nonumber \\
& & \hspace{2cm} + (1 -y) \, \sin (\phi_h + \phi_S) \, F_{UT}^{\sin (\phi_h + 
\phi_S)} + (1 - y ) \, \sin (3 \phi_h - \phi_s) \, 
F_{UT}^{\sin (3 \phi_h - \phi_S)} 
\nonumber \\
& & \hspace{2cm} 
+ \left. (2 - y) \sqrt{1 - y} \, \sin \phi_S \, F_{UT}^{\sin \phi_S} 
+ (2 - y) \sqrt{1 - y} \, \sin (2 \phi_h - \phi_S) \, 
F_{UT}^{\sin (2 \phi_h - \phi_S)} \right ]
\nonumber \\
& & \hspace{2cm} +  S_{\perp} \,  \lambda_{\ell} 
\, \left [ y (1 - \frac{1}{2} y) \, \cos (\phi_h - \phi_S) \, 
F_{LT}^{\cos (\phi_h - \phi_S)} 
+ y \sqrt{1 - y} \, \cos \phi_S \, F_{LT}^{\cos \phi_S} 
\right. 
\nonumber \\
& & \hspace{2cm} + \left. \left. y \sqrt{1-y} \, \cos (2 \phi_h - \phi_S) \, 
F_{LT}^{\cos (2 \phi_h - \phi_S)} \right ] \right \}\,.  
\label{sidiscs}
\eq
The structure functions $F$ depend on $x_B, y, z_h$ and  $P_{h \perp}^2$. 
Their first and second subscript 
denote the polarisation of the beam and of the target, 
respectively ($U$ = unpolarised, 
$L$ = longitudinally polarised, $T$ = transversely polarised), 
whereas the third subscript refer to the polarisation of the virtual photon. 

If we integrate (\ref{sidiscs}) over $\Vec P_{h \perp}$,  
only 5 structure functions survive: $F_{UU,T}, F_{UU,L},  
F_{LL}, F_{LT}^{\cos \phi_S}$, and $F_{UT}^{\sin \phi_S}$. 
The first two, upon a further integration in $z$ and a sum over 
all outgoing hadrons,  
yield the unpolarised DIS structure functions $F_T (x_B, Q^2) = 
2 x F_1(x_B, Q^2)$ 
and $F_L (x_B, Q^2) = F_2 (x_B, Q^2) - 2 x_B F_1(x_B, Q^2)$. The second two 
lead to combinations of the structure functions $g_1(x_B, Q^2)$ 
and $g_2(x_B, Q^2)$ of longitudinally polarised DIS. The fifth one vanishes. 
The fact that 
\be
\sum_h \int \! \D z_h \, z_h \int \! \D^2 \Vec P_{h \perp} \, 
F_{UT}^{\sin \phi_S} = 0 \, 
\ee
is a consequence of time-reversal invariance \cite{Diehl:2005pc} 
and is another way to express the Christ-Lee theorem \cite{Christ:1966zz}, 
according to which there cannot be transverse spin asymmetries 
in inclusive DIS. 
On the contrary, in SIDIS no first principle forbids 
the existence of transverse spin asymmetries.

In the literature, the spin and azimuthal asymmetries of SIDIS 
are defined in two different ways: 
\begin{itemize} 
\item[-]
as the structure function ratios:  
\be
A_{XY}^{w(\phi_h, \phi_S)}(x, z, P_{h \perp}^2) \equiv 
\frac{F_{XY}^{w (\phi_h, \phi_S)}}{F_{UU}}\,, 
\label{asym_def1}
\ee
where $X$ and $Y$ label the polarisation of the beam 
and of the target respectively,  
$w(\phi_h, \phi_S)$ is a trigonometric function of 
its arguments, and  $F_{UU} \equiv (1 - y + \frac{1}{2} y^2) 
F_{UU,T} + (1 - y) F_{UU,L}$;
\item[-]
as the azimuthal moments of cross sections \cite{Bacchetta:2004jz}: 
\be
\mathcal{A}^{w (\phi_h, \phi_S)}(x_B, y, z_h, P_{h \perp}^2)
\equiv 2 \langle w(\phi_h, \phi_S) \rangle 
\equiv 2 
\frac{\int \D \phi_h \int \D \phi_S \, w (\phi_h, \phi_S) \, 
\D \sigma (\phi_h, \phi_S)}{\int \D \phi_h \int \D \phi_S  \, \D \sigma 
(\phi_h, \phi_S)}\,,
\ee
where  $\D \sigma$ is a shorthand notation for the 
fully differential cross section. 
\end{itemize}

Notice that the two definitions of asymmetries
differ for $y$-dependent factors. 


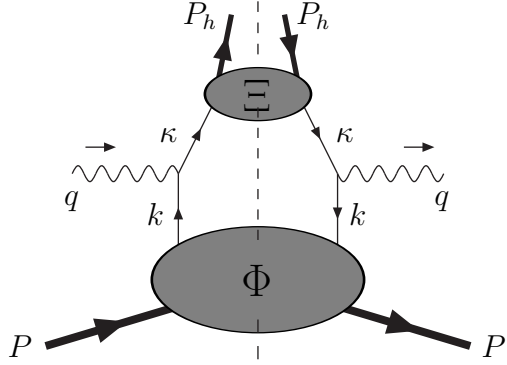
\begin{figure}
  \centering
  \begin{picture}(300,160)(0,20)
    \SetWidth{3}
    \ArrowLine(70,25)(120,40)
    \ArrowLine(180,40)(230,25)
    \SetWidth{0.5}
    \ArrowLine(120,60)(120,90)
    \ArrowLine(180,90)(180,60)
    \ArrowLine(120,90)(135,120)
    \ArrowLine(165,120)(180,90)
    \Photon(80,90)(120,90){3}{4.5}
    \Photon(220,90)(180,90){3}{4.5}
    \GOval(150,50)(20,40)(0){0.5}
    \SetWidth{2}
    \ArrowLine(135,125)(140,150)
    \ArrowLine(160,150)(165,125)
    \SetWidth{0.5}
    \GOval(150,120)(10,20)(0){0.5}
    \LongArrow(85,100)(95,100)
    \LongArrow(205,100)(215,100)
    \Text(65,25)[r]{$P$}
    \Text(235,25)[l]{$P$}
    \Text(80,80)[]{$q$}
    \Text(220,80)[]{$q$}
    \Text(135,150)[r]{$P_h$}
    \Text(165,150)[l]{$P_h$}
    \Text(115,75)[r]{$k$}
    \Text(185,75)[l]{$k$}
    \Text(120,105)[r]{$\kappa$}
    \Text(180,105)[l]{$\kappa$}
    \Text(150,120)[]{\Large$\Xi$}
    \Text(150,50)[]{\Large$\Phi$}
    \DashLine(150,20)(150,35){4}
    \DashLine(150,65)(150,115){4}
    \DashLine(150,135)(150,155){4}
  \end{picture}
  \caption{Diagram contributing to semi-inclusive DIS in the parton model.}
  \label{handbag2}
\end{figure}

\subsubsection{SIDIS in the parton model}
\label{sidis_parton}

In the parton model the virtual photon strikes a quark (or an antiquark), 
which successively fragments into a hadron $h$. 
The process is represented by the diagram in Fig.~\ref{handbag2}. 
We will take transverse momenta 
of quarks into account and refer to this description as the ``extended 
parton model''.

For the partonic description of SIDIS we 
work in a reference frame where the momenta of the target nucleon and 
of the outgoing hadron  are collinear and define
the longitudinal direction. 
In this ``$hN$ collinear frame'', one has $P^{\mu} = P^+ n_+^{\mu}$ 
and $P_h^{\mu}= P^- n_-^{\mu}$,  
whereas the virtual photon momentum acquires a transverse 
component $\Vec q_T$. 
The incoming quark momentum is 
$k^{\mu} = x P^{\mu} + k_T^{\mu}$, with $x = k^+/P^+$; 
the fragmenting quark momentum is $\kappa^{\mu} 
= P_h^{\mu}/z + \kappa_T^{\mu}$, with $z = P_h^-/\kappa^-$. 
Notice that the ``transverse'' quantities in the $hN$ frame 
(labelled by the subscript $T$)  differ 
from the ``perpendicular'' quantities in the $\gamma^*N$ frame (labelled 
by the subscript $\perp$)  by terms suppressed 
at least as $1/Q$. In particular, $\Vec q_T$ is related 
to $\Vec P_{h \perp}$ by $\Vec q_T = - \Vec P_{h \perp}/z_h$,  
up to $1/Q^2$ corrections.   

The hadronic tensor corresponding to Fig.~\ref{handbag2} is given by  
\bq
  W^{\mu\nu} &=&
 \sum_a e_a^2
  \int \!  \D^4 k 
  \int \!  \D^4 \kappa \;
  \delta^4(k + q - \kappa)
  \, {\rm Tr} \, 
  [ \Phi^a(k) \, \gamma^\mu \, 
\Xi^a (\kappa) \gamma^\nu ]\,, 
\nonumber \\
  &=&  2 z_h \, \sum_a e_a^2
  \int \!  \D^2\Vec{k}_T
  \int \!  \D^2\Vec\kappa_T \;
  \delta^2(\Vec{k}_T + \Vec{q}_T - \Vec\kappa_T)
  \, {\rm Tr} \, 
  [ \Phi^a (x_B, \Vec k_T) \, \gamma^\mu \, 
\Xi^a (z_h, \Vec \kappa_T) \gamma^\nu ] \,, 
  \label{sidis22a}
\eq
where the second expression has been obtained by working out the 
momentum-conservation delta function and neglecting $1/Q^2$ terms. 
In this case one has $z = z_h$ and $x = x_B$. 

Inserting the expressions of $\Phi$ and $\Xi$ into (\ref{sidis22a})
and contracting $W^{\mu \nu}$ with $L_{\mu \nu}$ leads to 
the SIDIS structure functions. With the following notation 
for the transverse momenta convolutions
\be
\mathcal{C} \, [w f D] = 
 \sum_a e_a^2 \, x \, \int \D^2 \Vec k_T \int \D^2 \Vec \kappa_T
\, \delta^2 ( \Vec k_T - \Vec \kappa_T - \Vec P_{h \perp}/z) 
\, w(\Vec k_T, \Vec \kappa_T) 
\, f^a (x,  k_T^2) D^a (z,  \kappa_T^2)\,,  
\label{convol}
\ee
the non vanishing structure functions at leading twist are
\cite{Mulders:1995dh,Boer:1997nt,Bacchetta:2006tn}
\bq
F_{UU,T} &= & \mathcal{C} \, \left [f_1 D_1 \right ] \,, 
\label{sf1} \\
F_{UU}^{\cos 2 \phi_h} &=& 
\mathcal{C} 
\left [ -\frac{2 (\hat{\Vec h} \cdot \Vec k_T) (\hat{\Vec h} 
\cdot \Vec \kappa_T) - \Vec k_T \cdot \Vec \kappa_T}{M M_h} 
\, h_1^{\perp} \, H_1^{\perp} \right ] \,, 
\label{sf3} \\
F_{UL}^{\sin 2 \phi_h} &=& 
\mathcal{C} 
\left [ -\frac{2 (\hat{\Vec h} \cdot \Vec k_T) (\hat{\Vec h} 
\cdot \Vec \kappa_T) - \Vec k_T \cdot \Vec \kappa_T}{M M_h} 
\, h_{1L}^{\perp} \, H_1^{\perp} \right ] \,, 
\label{sf4} \\
F_{LL} &=& \mathcal{C} \, \left [ g_{1L} D_1  \right ] 
\label{sf5} \\
F_{UT, T}^{\sin (\phi_h - \phi_S)} &=& 
\mathcal{C} \left [ - \frac{\hat{\Vec h} \cdot \Vec k_T}{M} \, 
f_{1T}^{\perp} D_1 \right ] \,, 
\label{sf6} \\
F_{UT}^{\sin (\phi_h + \phi_S)} &=& 
\mathcal{C} \left [ - \frac{\hat{\Vec h} \cdot \Vec \kappa_T}{M_h} \, 
h_1 H_1^{\perp} \right ] \,, 
\label{sf8} \\
F_{UT}^{\sin (3 \phi_h - \phi_S)} &=& 
\mathcal{C} 
\left [ \frac{ 2 (\hat{\Vec h} \cdot \Vec \kappa_T) (\Vec k_T 
\cdot \Vec \kappa_T) +  k_T^2 (\hat{\Vec h} \cdot \Vec \kappa_T) - 
4 (\hat{\Vec h} \cdot \Vec k_T)^2 (\hat{\Vec h} \cdot \Vec \kappa_T)}{2 
M^2 M_h} \, h_{1T}^{\perp} H_1^{\perp} \right ] \,, 
\label{sf9} \\
F_{LT}^{\cos (\phi_h - \phi_S)} &=& 
\mathcal{C} \left [ \frac{\hat{\Vec h} \cdot \Vec k_T}{M} \, 
g_{1T} D_1 \right ]\,. 
\label{sf9bis}
\eq 
 where $\hat{\Vec h} \equiv \Vec P_{h \perp}/\vert \Vec P_{h \perp} \vert$.
The structure function $F_{UU,T}$ gives the 
dominant contribution to the unpolarised cross section 
integrated over $\phi_h$.
 
The Collins term  $F_{UT}^{\sin (\phi_h + \phi_S)}$ 
couples the Collins function $H_1^{\perp}$ 
to the transversity distribution $h_1$, thus representing  
one of the privileged ways to access this quantity. 
Note that in the original paper \cite{Collins:1992kk} 
the Collins angle $\Phi_C$ was defined as 
the angle between the momentum of the produced 
hadron and the spin of the fragmenting quark, i.e. 
$\Phi_C \equiv \phi_h - \phi_{S_q'}$. 
In terms of the azimuthal angle of the target spin $\phi_S$, 
one has $\Phi_C = \phi_h + \phi_S - \pi$. 
On the other hand, 
according to the conventions of Ref.~\cite{Bacchetta:2004jz}, the Collins 
angle is defined as  $\Phi_C' \equiv \phi_h + \phi_S$. 
Thus, one gets different signs for the Collins asymmetry, 
depending on which definition of the Collins angle, either 
$\Phi_C$ or $\Phi_C'$, is adopted. 

Another leading-twist asymmetry source is 
the Sivers term  $F_{UT,T}^{\sin (\phi_h - \phi_S)}$,  
which couples the Sivers function $f_{1T}^{\perp}$  to the unintegrated 
unpolarised fragmentation function $D_1$.      
In the transversely polarised case, a further
angular modulation, of the type $\sin (3 \phi_h - \phi_S)$, 
involves the distribution function $h_{1T}^{\perp}$.  

In unpolarised SIDIS,  a leading-twist  
azimuthal asymmetry is generated by the 
structure function 
$F_{UU}^{\cos 2 \phi_h}$, which couples the 
Boer-Mulders distribution
$h_1^{\perp}$ to the Collins fragmentation function $H_1^{\perp}$. 

Going to twist three, i.e. to order $1/Q$, it turns out that the leading-twist 
structure functions (\ref{sf1}-\ref{sf9}) do not acquire 
any extra contribution, but there appear other 
non vanishing structure functions~\cite{Bacchetta:2006tn}.  
Among them, of particular 
phenomenological importance are those related to the $\cos \phi_h$ 
and $\sin \phi_h$ modulations. 
Ignoring, in the spirit of the parton model,  interaction-dependent 
terms, i.e. quark-gluon correlations, and quark mass contributions  
(the  generalised Wandzura--Wilczek approximation) 
one finds \cite{Bacchetta:2006tn}
\bq
F_{UU}^{\cos \phi_h} &=& \frac{2 M}{Q}   
\, \mathcal{C} \left [ - \frac{(\hat{\Vec h} \cdot \Vec \kappa_T) 
\,  k_T^2}{M_h M^2} \, h_1^{\perp} \, H_1^{\perp}
-   \frac{\hat{\Vec h} \cdot \Vec k_T}{M} \, f_1 D_1 \right ]  
\,, 
\label{sf10} \\
F_{UL}^{\sin \phi_h} &=& \frac{2 M}{Q}   
\, \mathcal{C} \left [ - \frac{(\hat{\Vec h} \cdot \Vec \kappa_T) 
\,  k_T^2}{M_h M^2} \, h_{1L}^{\perp} \, H_1^{\perp} \right ] \,. 
\label{sf11} 
\eq
In the same approximation one gets $F_{LU}^{\sin \phi_h} = 0$. 
Thus a deviation of the beam-spin $\sin \phi_h$ asymmetry  
from zero might signal the relevance of interaction effects 
in the nucleon. 
One should recall however that 
at high transverse momenta $F_{LU}^{\sin \phi_h}$ is non 
zero in next-to-leading order QCD.   
The term in $F_{UU}^{\cos \phi_h}$ containing the product 
of the unpolarised functions $f_1 D_1$  
is a purely kinematical contribution arising from the intrinsic 
transverse motion of quarks, with no relation to spin. 
This contribution was discovered longtime ago by 
Cahn \cite{Cahn:1978se,Cahn:1989yf}, and the corresponding azimuthal asymmetry 
is referred to as the ``Cahn effect''.  
A similar contribution emerges at twist four, that is  
at order $1/Q^2$, in the $\cos 2 \phi_h$ term: 
\be
F_{UU, {\rm Cahn}}^{\cos 2 \phi_h} = 
\frac{M^2}{Q^2} \, 
\mathcal{C} 
\left [ \frac{(2 (\hat{\Vec h} \cdot \Vec k_T)^2 -  k_T^2)}{M^2} 
 f_1 \, D_1 \right ] 
\,.  
\label{cahn_asym}
\ee

All the above parton-model results have been obtained using the 
most general decompositions 
of the correlation matrices $\Phi$ and $\Xi$, and 
inserting them into the SIDIS hadronic tensor $W^{\mu \nu}$. 
There is an alternative approach, which relies on the helicity formalism 
and expresses the cross section as a convolution of 
helicity amplitudes
of elementary subprocesses with partonic distribution and fragmentation 
functions, taking fully into account non collinear 
kinematics \cite{Anselmino:2005sh}.    
Considering for simplicity 
an unpolarised lepton beam and a spinless or unpolarised 
final hadron, and adopting the 
$\gamma^* N$ collinear frame, the basic factorisation formula 
for the SIDIS cross section in this picture is, up to order $1/Q$, 
\be
\D \sigma 
= \sum_{q_i} \sum_{\lambda_{q_i} \lambda'_{q_i}}
  \int \! \D^2\Vec{k}_{\perp} \, \int \! \D^2 \Vec p_{\perp} \,   
  \rho^{q_i}_{\lambda_{q_i}\lambda'_{q_i}} \, f_{q_i} (x, \Vec k_{\perp}) 
\D \hat{\sigma}_{\lambda_{q_i}\lambda'_{q_i}} \, 
  D^{h/q_f}(z, \Vec p_{\perp}) \, \delta^2 (z \Vec k_{\perp} 
- \Vec p_{\perp} - \Vec P_{h \perp} )\,, 
  \label{noncoll2}
\ee
where the $\lambda$'s are helicity indexes,  
$f_{q_i}(x,\Vec{k}_{\perp})$ is the probability of finding a quark $q_i$
with momentum fraction $x$ and transverse momentum $\Vec{k}_{\perp}$ inside the
target nucleon, $\rho^q_{\lambda_{q_i} \lambda'_{q_i}}$ 
is the helicity density matrix of the quark $q_i$, 
$D^{h/q_f}(z,\Vec p_{\perp})$ is the fragmentation function 
of the struck quark $q_f$ into the hadron $h$ with transverse
momentum $\Vec p_{\perp}$ with respect to the fragmenting quark,
and $\D \hat{\sigma}_{\lambda_{q_i}\lambda'_{q_i}} \sim 
 \hat{M}_{\lambda_{\ell '} \lambda_{q_f}; \lambda_{\ell} 
\lambda_{q_i}} {\hat M}^*_{\lambda_{\ell '} \lambda_{q_f}; \lambda_{\ell} 
\lambda'_{q_i}}$ is the cross section  of   
lepton-quark scattering $\ell q_i \to \ell' q_f$ at tree level. 
Note that, whereas in the collinear case 
the produced hadron is constrained to have $\Vec P_{h \perp} = 0$ 
 and the entire process takes place 
in the scattering plane, the intrinsic transverse momentum 
of quarks introduces  a non planar geometry.   
The elementary scattering amplitudes $\hat{M}$'s 
take into account this non collinear and out-of-plane kinematics.  
Neglecting  
 $\mathcal{O}(k_{\perp}^2/Q^2)$ contributions,  
 no Jacobian factors appear in eq.~(\ref{noncoll2})  and one has 
$x = x_B, z = z_h$.    

Despite their apparent dissimilarity, the two 
parton model approaches described so far, namely the approach based on 
quark correlation matrices and eq.~(\ref{sidis22a}) 
and the approach based on the  
helicity formalism and eq.~(\ref{noncoll2}), are equivalent 
as far as parton interactions are ignored.  
In other terms, all the leading-twist asymmetries listed  
in eqs.~(\ref{sf1}-\ref{sf9bis}) can be exactly reobtained from 
eq.~(\ref{noncoll2}) \cite{Anselmino:2010aa}, whereas 
at twist three the results of the two approaches 
are identical if one neglects the ``tilde'' distribution 
and fragmentation functions arising from quark-gluon 
correlations. 

\subsubsection{TMD factorisation in QCD}
\label{tmdfact}

So far, we have only considered the extended parton model, i.e. the 
parton model incorporating intrinsic transverse momenta. 
One may wonder whether the results we have presented have any 
solid QCD foundation. 
The answer to this question is positive, 
at least in a particular kinematic regime.  
Semi-inclusive processes are characterised by two scales, 
besides the confinement scale $\Lambda_{QCD}$: 
the momentum transfer $Q$ and the transverse momentum 
of the final hadron $P_{h \perp}$ or, equivalently, the 
transverse momentum $Q_T \equiv \vert \Vec q_T \vert$
of the virtual photon in the $hN$ collinear frame.  

Extending the pioneering work
on back-to-back jet production of Ref. \cite{Collins:1981uk}, 
a TMD factorisation theorem for SIDIS and DY
has been proven \cite{Ji:2004xq,Ji:2004wu,Collins:2007ph}.
The proof is valid
in the low transverse-momentum region $P_{h \perp} (Q_T) \ll Q$. 
In this framework the unpolarised SIDIS structure function 
is written as  
\bq
F_{UU,T} (x_B, z_h, Q^2, Q_T^2) 
&=& \sum_a e_a^2 x \, \int \! \D^2 \Vec k_T \, 
\int \! \D^2 \Vec \kappa_T \, \int \! \D^2 \Vec l_T
\, \delta^2 (\Vec k_T - \Vec \kappa_T + \Vec l_T 
+ \Vec q_T) \nonumber \\
& & \times 
  H(Q^2) \, 
f_1^a (x_B, k_T^2) 
\, D_1^a (z_h, k_T^2) 
\, U(l_T^2)\,. 
\label{tmdfact1}
\eq
For simplicity the dependence 
of the distribution functions on $\zeta^2 = (2 v \cdot P)^2/v^2$ 
and of the fragmentation function on 
$\zeta_h^2 = (2 \tilde v \cdot P)^2/\tilde{v}^2$
($v$ and $\tilde v$ are vectors off the light-cone) is omitted . 
The variables $\zeta$ and $\zeta_h$ serve to regulate 
the light-cone singularities, as explained in Section~\ref{tmds}. 
$H$ is a perturbative hard factor written as a series in 
powers of $\alpha_s$. The soft factor $U$ arises from the 
radiation of soft gluons (of transverse momentum $\Vec l_T$) 
and is a matrix element of Wilson lines in the QCD vacuum. 
Also not displayed in eq.~(\ref{tmdfact1}) 
is  the dependence of all quantities on 
the renormalisation scale $\mu$ and on the soft-gluon rapidity
cut-off $\rho = \sqrt{(2 v \cdot \tilde v)^2/ v^2 \tilde{v}^2}$. 
Of course, the physical observable $F$ does not depend on 
any of these regulators. 

The generalisation of eq.~(\ref{tmdfact1}) to 
the polarised structure functions, in particular 
to those generating transverse SSA's, has been 
proposed in Refs.~\cite{Ji:2006br,Bacchetta:2008xw}. 
The parton model expressions of Section~\ref{sidis_parton}  
are recovered at tree level, i.e. $\mathcal{O}(\alpha_s^0)$, since 
$H^{(0)} = 1$ and $U^{(0)}(\Vec l_T) = \delta^2 (\Vec l_T)$. 

\subsubsection{QCD description at high transverse momenta}
\label{highpt}

\begin{figure}
\begin{center}
\includegraphics[width=0.65\textwidth]
{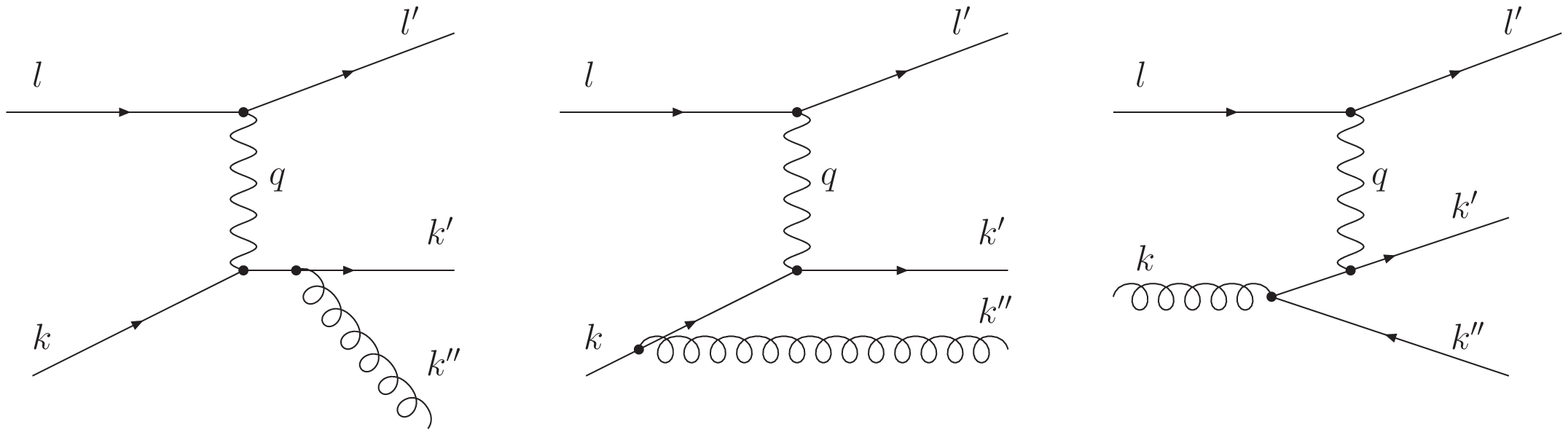}
\end{center}
\caption{\label{feynman} Feynman diagrams 
of the elementary processes contributing 
to SIDIS at first order 
 in $\alpha_s$.}
\end{figure}

At high transverse momenta, $Q_T \gg \Lambda_{QCD}$, 
SIDIS structure functions can be described in collinear QCD.   
The azimuthal angular dependence of hadrons 
in leptoproduction was proposed longtime ago 
as a test of perturbative QCD \cite{Georgi:1977tv}. 
In collinear factorisation, transverse momenta 
are generated by gluon radiation.  At first order in $\alpha_s$ 
the hard elementary processes shown in Fig.~\ref{feynman}
contribute to the four unpolarised SIDIS structure 
functions $F_{UU}$ 
and to the two double-longitudinal structure functions $F_{LL}$.  
Introducing the partonic variables $\hat x$ and $\hat z$, defined as 
$\hat x = Q^2/2k \cdot q  = x_B/x$, 
$\hat z = k \cdot k'/k \cdot q  = z_h/z$, 
where $k$ and $k'$ are the four-momenta of the incident and fragmenting
partons, respectively, and $x$ and $z$ are the usual light-cone momentum
fractions, i.e. $k = x P$ and $k' = P_h/z$, one has for 
the $F_{UU}$'s at leading order in $\alpha_s$ 
and leading twist \cite{Mendez:1978zx,Chay:1991nh,Bacchetta:2008xw}
\bq
& & F_{UU}(x, Q^2) = \frac{\alpha_s}{4 \pi^2 \, z^2 q^2} 
\sum_a e_a^2 x_B \, \int_{x_B}^1 \frac{\D \hat x}{\hat x} 
\int_{z_h}^1 \frac{\D \hat z}{\hat z} \, 
\delta \left ( \frac{Q_T^2}{Q^2} - \frac{(1 - \hat x) 
(1 - \hat z)}{\hat x \hat z} \right ) 
\nonumber \\
& & \hspace{0.2cm} \times  
\left [ f_1^a \left (\frac{x_B}{\hat x} \right ) 
D_1^a \left ( \frac{z_h}{\hat z} \right ) \, C_{UU}^{\gamma^* q \to q g}
+ f_1^a \left (\frac{x_B}{\hat x} \right ) 
D_1^g \left ( \frac{z_h}{\hat z} \right ) \, C_{UU}^{\gamma^* q \to g q}
+ f_1^g \left (\frac{x_B}{\hat x} \right ) 
D_1^a \left ( \frac{z_h}{\hat z} \right ) \, 
C_{UU}^{\gamma^* g \to q \bar q} \right ]\,,   
\label{highpt1}
\eq
and analogous formulae for the $F_{LL}$'s. 
The Wilson coefficients $C$ represent elementary cross sections 
and are listed in Ref.~\cite{Bacchetta:2008xw}. 

The structure function $F_{LU}^{\sin \phi_h}$ encountered 
in Section~\ref{sidis_parton}, which produces 
a beam-spin asymmetry and vanishes in the 
parton model, gets a non zero perturbative QCD contribution 
at leading twist and order $\alpha_s^2$ \cite{Hagiwara:1982cq,Ahmed:1999ix}.   

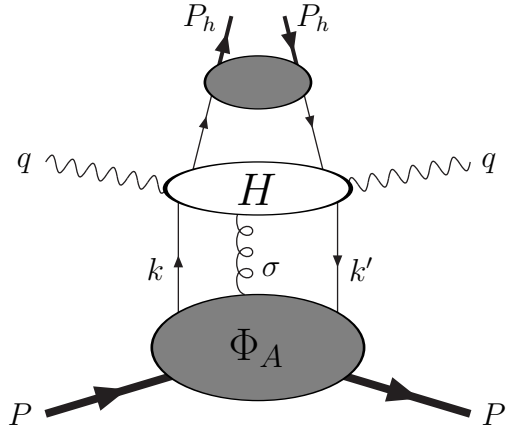
\begin{figure}
  \centering
  \begin{picture}(300,180)(0,20)
    \SetWidth{3}
    \ArrowLine(70,25)(120,40)
    \ArrowLine(180,40)(230,25)
    \SetWidth{0.5}
    \ArrowLine(120,60)(120,105)
    \ArrowLine(180,105)(180,60)

    \Photon(70,120)(120,110){3}{6.5}
    \Photon(230,120)(180,110){3}{6.5}
    \Gluon(145,70)(145,110){3}{4.5}

    \ArrowLine(125,115)(135,155)
    \ArrowLine(165,155)(175,115)

    \GOval(150,50)(20,40)(0){0.5}
    \GOval(150,110)(10,35)(0){1}

    \SetWidth{1.5}
    \ArrowLine(135,155)(140,175)
    \ArrowLine(160,175)(165,155)
    \SetWidth{0.5}
    \GOval(150,150)(10,20)(0){0.5}

    \Text(65,25)[r]{$P$}
    \Text(235,25)[l]{$P$}
    \Text(65,120)[r]{$q$}
    \Text(235,120)[l]{$q$}
    \Text(135,175)[r]{$P_h$}
    \Text(165,175)[l]{$P_h$}
    \Text(115,80)[r]{$k$}
    \Text(185,80)[l]{$k'$}
    \Text(152,80)[l]{$\sigma$}
    \Text(150,110)[]{\Large$H$}
    \Text(150,50)[]{\Large$\Phi_A$}
  \end{picture}

  \caption{General diagram contributing 
to SIDIS SSA's in the twist-three factorisation.}
  \label{fig_twist3}
\end{figure}

On the contrary, the transversely polarised structure functions $F_{UT}$, 
which vanish at leading twist in collinear factorisation,  
since there is no chirally-odd fragmentation  
function, emerge at twist three, as the result of 
quark-gluon correlations.  
Following the early work of 
Ref.\cite{Efremov:1981sh,Efremov:1983eb,Efremov:1984ip},
a twist-three collinear factorisation theorem valid at large transverse 
momenta was proven \cite{Qiu:1991pp,Qiu:1991wg,Qiu:1998ia}. 
In this approach the cross section 
for SIDIS with a transversely polarised target 
has the general form \cite{Ji:2006br,Eguchi:2006mc,Eguchi:2006qz}
 \be
\D \sigma \sim G_F (x, x') \otimes \D \hat \sigma \otimes 
D_1 (z) + h_1 (x) \otimes \D \hat \sigma' \otimes \hat E_F (z, z')\,, 
\label{twistthree10}
\ee
where the first term contains a quark-gluon correlation 
function for the transversely polarised nucleon 
and the ordinary unpolarised fragmentation function for the final 
hadron, whereas the second term combines the transversity distribution 
with a twist-three fragmentation function. 
Let us focus on the first contribution (twist-three 
effects in the initial state). 
The hadronic tensor can then be written as 
\be
W_{\mu \nu} (P, q, P_h) = \sum_a 
\int \! \frac{\D z}{z} \, w_{\mu \nu}^a (P, q, P_h/z) 
\, z D_1^a (z) \,, 
\label{twistthree11}
\ee
where the partonic tensor $w_{\mu \nu}$ 
contributing to the transversely polarised 
structure functions is (see Fig.~\ref{fig_twist3})
\be
w_{\mu \nu} (P, q, P_h) 
= \int \! \D^4 k \, \int \! \D^4 k' 
\, {\rm Tr} \, \left [\Phi_A (k, k') \, H_{\mu \nu}(k, k', q, P_h) 
\right ] \,. 
\label{twistthree12}
\ee
In this expression $\Phi_A$ is the quark-gluon correlation matrix
\be
  \Phi_{A}(k, k') = \int \! \frac{\D^4\xi}{(2 \pi)^4} \, 
\int \! \frac{\D^4 \eta}{(2 \pi)^4} \, 
\E^{\I k{\cdot}\xi} \, \E^{\I (k'-k) \cdot \eta}
  \langle P,S | \bar \psi (0) \, \mathcal{W}^- [0, \eta]
\, g A^+ (\eta) \, \mathcal{W}^- [\eta, \xi] \, 
 \psi(\xi) | P,S \rangle \, , 
\label{twistthree13}
\ee
and $H_{\mu \nu}$ represents the perturbatively 
calculable partonic hard scattering. 
By means of the collinear expansion \cite{Qiu:1991wg} one can get
\be
H(k, k') = H(x, x') + \left. \frac{\partial H}{\partial k_{\alpha}} 
\right \vert_{x, x'} (k_{\alpha} - x P_{\alpha}) 
+ \left. \frac{\partial H}{\partial k'_{\alpha}} 
\right \vert_{x, x'} (k'_{\alpha} - x' P_{\alpha}) \,, 
\ee
and finally end up with 
\be
w_{\mu \nu} = \I \, \int \! \D x \, \int \! \D x' 
\; {\rm Tr} \, \left [\Phi_F^{\alpha} (x, x') \, 
\frac{\partial H (x, x')}{\partial {k'}^{\alpha}} \right ]\,, 
\ee
where the quark-gluon correlation matrix $\Phi_F^{\alpha}$, defined as 
\be
\Phi_F^{\alpha}(x, x') = 
\int \frac{\D \xi^-}{2 \pi} \int \frac{\D \eta^-}{2 \pi} 
\E^{\I x P^+ \xi^-} \E^{\I (x' - x) P^+ \eta^-} 
\langle P, S \vert \bar \psi(0) \, \mathcal{W}^- [0, \eta] 
\, g F^{+\alpha} (\eta) \mathcal{W}^- [\eta, \xi] 
\, \psi(\xi) \vert P, S \rangle \,, 
\label{qgcorr1} 
\ee
contains the multiparton distributions $G_F, \tilde{G}_F, H_F, E_F$ 
introduced in Section~\ref{ht_tmd}. 
It is easy to verify that, due to the structure of $\Phi_F^{\alpha}$,  
the hadronic tensor receives contributions only 
from the imaginary part of the hard blob, arising 
from internal propagator poles. 

Considering for definiteness the Sivers contribution to 
the cross section, its explicit expression is \cite{Ji:2006br,Eguchi:2006mc} 
\bq
\D\sigma \vert_{\rm Siv} 
&\sim& \int \! \frac{\D x}{x} 
\, \int \! \frac{\D z}{z} \, \delta \left (\frac{Q_T^2}{Q^2} - 
\left (1 - \frac{x}{x_B} \right ) \left ( 1 - \frac{z}{z_h} \right )
\right ) \nonumber \\
& & \times 
\sum_a e_a^2 \left [ x \, \frac{\D G_F^a (x,x)}{\D x} 
\,  \hat \sigma_D + G_F^a (x,x) \,  \hat \sigma_G 
+ G_F^a (x, 0) \, \hat \sigma_F + G_F^a (x, x_B) \,  \hat \sigma_H 
 \right ] D_1^a (z) + \ldots
\label{siv_twist3}
\eq
The first two terms represent the so-called 
``soft-gluon pole'' contribution ($x_g = x' - x = 0$), 
the third term is the ``soft-fermion pole'' contribution 
($x' = 0$), the fourth term is the ``hard pole'' contribution ($x' = x_B$). 
The dots represent the contributions of $\tilde{G}_F$ and of the gluonic 
correlation functions.  

In the intermediate transverse-momentum region, i.e. 
$\Lambda_{QCD} \ll Q_T^2 (P_{h \perp}^2) \ll Q^2$, 
one expects that both the TMD and the twist-three pictures should hold. 
This has been explicitly verified in Refs.~\cite{Ji:2006ub,Koike:2007dg}. 
The output of these important works is a set of relations 
that connect the $T$-odd TMD distributions (Sivers and Boer-Mulders 
functions) on one side, with the quark-gluon correlations 
on the other side. 

\subsubsection{Leptoproduction of transversely polarised spin $1/2$ baryons} 
\label{transvbar}

The leptoproduction of spin $1/2$ baryons in another interesting channel to
access transversity.
If the nucleon target and the detected hadron are 
both transversely polarised, $\ell + 
N^{\uparrow} \rightarrow  \ell' + B^{\uparrow} + X$, 
the spin transfer between the initial and the final particle occurs
in collinear kinematics. 
The cross section  integrated over the transverse momentum gets in fact  
a double-spin term proportional to the product of the  
transversity distribution $h_1$ and the ``transversity'' 
fragmentation function $H_1$. 

This doubly polarised process has the advantage of being free 
from the theoretical complications related to the transverse 
motion of quarks: the ordinary collinear QCD
factorisation theorem applies and a perturbative 
study is possible, since we know the $Q^2$ evolution of both $h_1$ 
(Section~\ref{transversity}) and $H_1$ \cite{Stratmann:2001pt}. 
At leading order in QCD, the transverse polarisation 
$\mathcal{P}_T^{B}$ of the produced baryon is given by 
\be
\mathcal{P}_T^B 
=   \mathcal{P}_T^N \, \hat{D}_{NN} (y) \, 
\frac{\sum_a e_a^2 h_1^a(x_B, Q^2) H_1^{B/a}(z_h, Q^2)}{\sum_a e_a^2 
f_1^a(x_B, Q^2) 
D_1^{B/a}(z_h, Q^2)}\,,
\label{polar1}
\ee
where $\mathcal{P}_T^N$ is the nucleon polarisation and $\hat{D}_{NN}(y) 
= (1 - y)/(1 - y + y^2/2)$ is the spin transfer coefficient of quarks.   

In the case of $\Lambda$ hyperons, 
information on the spin transfer in the fragmentation 
process can be obtained from the $\Lambda$ polarisation extracted from the 
angular distribution in the weak $\Lambda \to p \, \pi^-$ decay.
The transverse polarisation of $\Lambda$'s produced in hard processes
with initial transversely polarised hadrons was studied long time ago in 
Refs.~\cite{Craigie:1980tc,Baldracchini:1981uq},  
and  reinvestigated more recently \cite{Artru:1990wq,Jaffe:1996wp}. 
Phenomenological analyses are presented in 
Refs.~\cite{Anselmino:2000ga,Ma:2001rm,Yang:2001sy}. 
The transversity fragmentation function $H_1^{\Lambda}$ 
is measurable in $e^+ e^-$ production 
of transversely polarised $\Lambda$ pairs. 

Transverse $\Lambda$ polarisation can also be observed
in SIDIS with an unpolarised target, by 
measuring asymmetries in the transverse momentum distribution of the hyperons. 
One contribution ($T$-odd in the final 
state) involves the quark density $f_1$ and  the polarising 
fragmentation function $D_{1T}^{\perp \Lambda}$. 
This mechanism for $\Lambda$ polarisation 
in SIDIS was studied in Ref.~\cite{Anselmino:2001js}. 
Another contribution ($T$-odd in the initial state)   
involves the Boer-Mulders function
$h_1^{\perp}$ coupled to the (unintegrated) fragmentation function 
$H_1^{\Lambda} (z, \Vec p_T^2)$. At high $P_{h \perp}$, 
this effect has a counterpart  
in the twist-three approach. The initial-state $T$-odd 
mechanisms for producing $\Lambda$ polarisation      
have been investigated in the context of 
twist-three factorisation by Zhou, Yuan and Liang
\cite{Zhou:2008fb}. 

\subsubsection{Two-hadron leptoproduction}
\label{twoparticle}

Another partially inclusive DIS reaction that can provide 
information on the transverse-spin structure of hadrons, 
and in particular on transversity, 
is two-particle leptoproduction from a transversely polarised target, 
$  \ell (l)  +  N^{\uparrow} (P)  \rightarrow
  \ell' (l')  +  h_1 (P_1)  +  h_2 (P_2)  +  X (P_X)$,  
with the two spinless final hadrons in the same jet. 
Two-hadron production in SIDIS has been proposed and studied by various authors
\cite{Collins:1994ax,Jaffe:1997hf,Bianconi:1999cd,Bianconi:1999uc} as a
process probing the transverse polarisation distribution
in combination with a dihadron fragmentation function (DiFF). 
 The idea is to look at an angular correlation 
between the spin of the fragmenting quark and the relative transverse momentum 
of the hadron pair, without involving  the transverse momenta of quarks. 
Integrating over the total transverse momentum of the final hadrons, 
one gets an asymmetry in the azimuthal 
angle between the two-hadron plane and the scattering plane.  
This asymmetry is determined by a fragmentation function usually called 
$H_1^{\newangle}$, which does not depend 
on the intrinsic transverse motion of quarks and 
arises from the interference between different channels of the 
fragmentation process into the  two-hadron system. 
Thus, all the difficulties 
related to non-collinearity are in this process avoided.  

The first authors who suggested resonance interference 
as a way to produce non-diagonal
fragmentation matrices of quarks were Cea et al. \cite{Cea:1988gw} in their
attempt to explain the observed transverse polarisation of $\Lambda$ hyperons
produced in $pN$ interactions \cite{Bunce:1976yb}.
The unpolarised dihadron fragmentation functions appeared 
for the first time in the context of jet calculus 
\cite{Konishi:1978yx,Konishi:1979cb}. 
The extension to the polarised case was discussed in 
Refs.~\cite{Collins:1994ax,Collins:1993kq,Artru:1995zu,Jaffe:1997hf} 
and a complete classification 
of the DiFF's was given at leading twist in Ref.~\cite{Bianconi:1999cd} 
and at twist-3 in Ref.~\cite{Bacchetta:2003vn}. 

\begin{figure}
\begin{center}
\includegraphics[width=0.50\textwidth]
{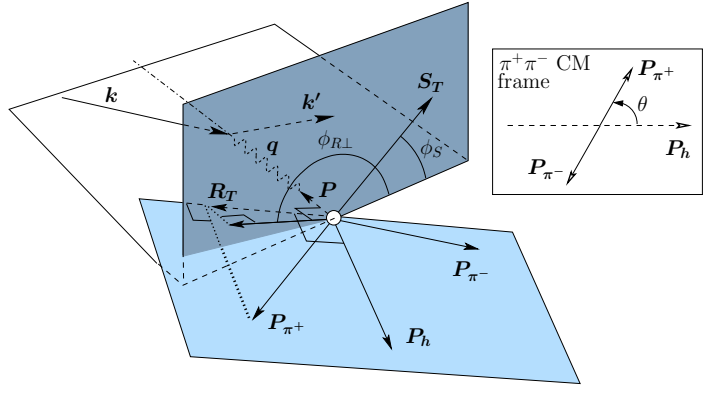}
\end{center}
\caption{\label{fig:2had} Kinematics of two-hadron leptoproduction.}
\end{figure}

The kinematics of the process 
in the $\gamma^* N$ frame is shown in Fig.~\ref{fig:2had}.  
We introduce the total momentum of the hadron pair $P_h = P_1 + P_2$ 
(with invariant mass $M_h^2 = P_h^2$), the relative momentum 
$R = (P_1 - P_2)/2$, and the variables 
 $z = z_1 + z_2 = P_1^-/\kappa^- + 
P_2^-/\kappa^- = P_h^-/\kappa^-$ (the light-cone fraction 
of the fragmenting-quark momentum carried 
by the hadron pair) and $\zeta = 2 R^-/P_h^-$ 
(which describes how the total momentum of the 
pair is split into the two hadrons). 
$\Vec R_T$ is the transverse component of $R$ 
with respect to $\Vec P_h$, and $\phi_R$ is the azimuthal 
angle of $\Vec R_T$ in the plane orthogonal to the $\gamma^*N$ 
axis, measured with respect to the scattering plane. The azimuthal 
angle of the target spin vector is $\phi_S$. 
Calling $\Vec \kappa_T$ the transverse momentum of the fragmenting quark 
with respect to $\Vec P_h$, the unpolarised and transverse-spin projections 
of the two-hadron fragmentation matrix $\Delta$ at leading twist are 
\bq
& &   \Delta^{[\gamma^-]} =
 D_1(z, \zeta, {\kappa}_T^2, 
R_T^2, \Vec \kappa_T{\cdot}\Vec{R}_T ) \, ,
  \label{twopart17}
\\
& &  \Delta^{[\I \sigma^{i-} \gamma_5]} =
  \frac1{M_1 + M_2}
  \left[
    \varepsilon_T^{ij} \kappa_{Tj} \,
    H_1^\perp(z, \zeta, {\kappa}_T^2, R_T^2, 
{\Vec\kappa}_T{\cdot}\Vec{R}_T)
  +   \varepsilon_T^{ij} R_{Tj} \,
    H_1^{\newangle}
      (z, \zeta, {\kappa}_T^2, R_T^2, 
{\Vec\kappa}_T{\cdot}\Vec{R}_T)
  \right]\,. 
  \label{twopart19}
\eq
These  are the probabilities for an unpolarised  quark and 
for a transversely polarised quark, respectively, 
to fragment into a hadron pair.
Upon integration over $\Vec \kappa_T$, the contribution 
of the Collins-type DiFF $H_1^{\perp}$ 
disappears, and the only remaining transverse-spin term  
is the one containing the (integrated) interference fragmentation 
function $H_1^{\newangle} (z, \zeta, M_h^2)$. 

It is convenient to consider   
the centre-of-mass frame of the two hadrons (Fig.~\ref{fig:2had}), 
where $R_T \equiv \vert \Vec R_T \vert 
= \vert \Vec R \vert \, \sin \theta$  and $\theta$ is the angle between  
the direction of the hadron emission and $\Vec P_h$ (in the $\gamma^* N$ 
frame).
For two alike hadrons of mass $m$, one has $\zeta = 2 \, 
(\vert \Vec R \vert / M_h) 
 \cos \theta$ and $\vert \Vec R \vert = \frac{1}{2} \sqrt{M_h^2 - 4 m^2}$.  
In terms of these variables, keeping only the unpolarised and the 
transversely polarised terms, the leading-twist partonic expression 
for the cross section of two-hadron leptoproduction reads 
\bq
\frac{\D^7 \sigma}{\D x_B  \D y  \D z_h 
 \D \phi_R  \D \phi_S \D M_h^2 \D \cos \theta} 
&=&  \frac{\alpha_{\rm em}^2}{2 \pi x y Q^2} 
 \sum_a e_a^2 \, x 
\, \left \{ \left (1 - y + \frac{y^2}{2} \right ) \, 
f_1^a (x_B) D_1^a (z_h, M_h^2, \cos \theta) 
\right. 
\nonumber \\
& & \hspace{-1cm} - \left. (1 - y) \, \frac{ S_{\perp} 
 \vert \Vec R \vert}{M_h} 
\sin \theta \sin (\phi_R + \phi_S) \, h_1^a (x_B) H_1^{\newangle \, a}  
(z_h,  
M_h^2, \cos \theta) \right \}\,.  
\label{twopart10} 
\eq
Here it is $z = z_h \equiv P \cdot P_h/P \cdot q$, a relation valid 
modulo $1/Q^2$ corrections. 
At twist three, there appear extra terms in the cross section: a   
$\cos \phi_R$ unpolarised contribution and a $\sin \phi_S$ transverse-spin 
contribution \cite{Bacchetta:2003vn}. 

The first model for the two-pion DiFF $H_1^{\newangle, sp}$
was presented in Ref.~\cite{Jaffe:1997hf}, where the phase difference 
between $s$ and $p$ waves was taken from $\pi \pi$ 
phase shifts in elastic scattering. 
The resulting fragmentation function (called 
$\delta \hat{q}_I$ in Ref.~\cite{Jaffe:1997hf}), 
changes sign around the $\rho$ mass. 
 
In a more recent model \cite{Bacchetta:2002ux} 
the fragmentation functions are expanded in Legendre 
polynomials of $\cos \theta$ keeping only the first few terms, corresponding 
to the lowest values of relative orbital momentum. 
This truncation is expected to be legitimate for not very large $M_h$. 
Thus one can write 
\bq
& & D_1 (z_h, M_h^2, \cos \theta) 
= D_{1}^o(z_h, M_h^2) + D_{1}^{sp}(z_h, M_h^2) \, \cos \theta 
+ D_1^{pp} (z_h, M_h^2) \frac{1}{4} (3 \cos^2 \theta - 1) \,, 
\label{diff1} \\
& & H_1^{\newangle} (z_h, M_h^2, \cos \theta) 
= H_{1}^{\newangle,   sp}(z_h, M_h^2) + H_1^{\newangle,   pp}
(z_h, M_h^2) \, \cos \theta 
\label{diff2} 
\eq
(remember that $H_1^{\newangle}$ multiplies a $\sin \theta$ factor in the 
cross section). 
The interpretation of these terms, signaled by their superscripts, is 
the following \cite{Bacchetta:2002ux}: $D_1^o$ is a diagonal 
component, receiving contributions from $s$ and $p$ waves
of the dihadron system separately (the ``background''); 
$D_1^{sp}$ and $H_1^{\newangle,   sp}$ originate 
from the interference of a $s$ wave and a $p$ wave; $D_1^{pp}$ 
and $H_1^{\newangle,  pp}$ arise from the interference of two $p$ waves. 
The main channels contributing to the fragmentation 
of a quark $q$ into a $\pi^+ \pi^-$ pair are: 1) incoherent 
fragmentation, $q \to \pi^+ \pi^- X$; 2) fragmentation 
via a $\rho$ resonance, $q \to \rho X  \to 
\pi^+ \pi^- X$; 3) fragmentation via a $\omega$ resonance decaying into 
three pions, $q \to \omega X \to \pi^+ \pi^- \pi^0 X$. 
Pions in channel 1 are expected to be 
mostly produced in $s$ wave; pions in channel 
2 come from the two-body decay of a vector meson and are in a 
relative $p$ wave; pions in channel 3 are prevalently in $p$ wave,  
but a fraction of them may also be in $s$ wave.  
The functions $D_1^{sp}$ and $H_1^{\newangle, sp}$ 
arise from the interference of channels 1-2 and 1-3 
\cite{Bacchetta:2006un}. 

\begin{figure}[t]
\begin{center}
\includegraphics[width=0.40\textwidth]
{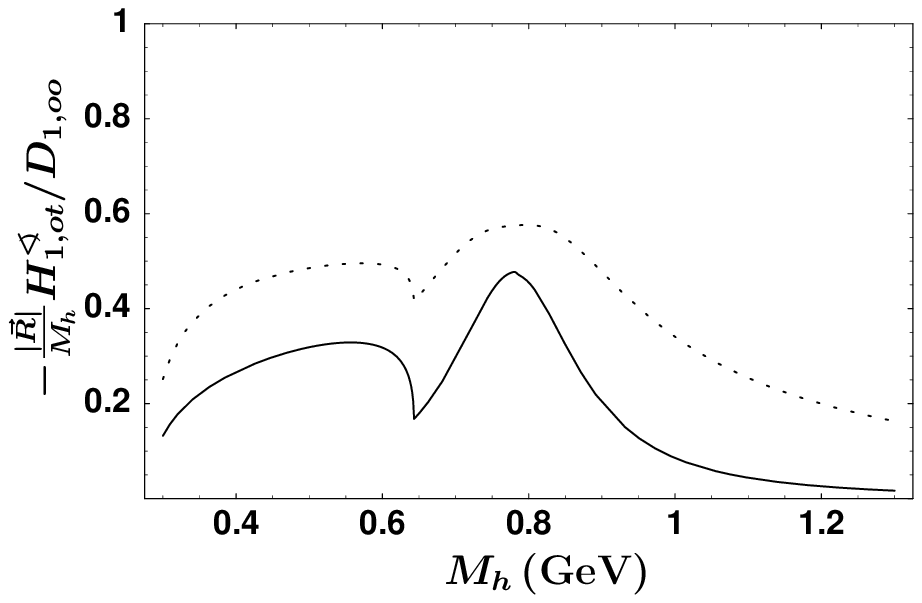}
\hspace{0.5cm}
\includegraphics[width=0.40\textwidth]
{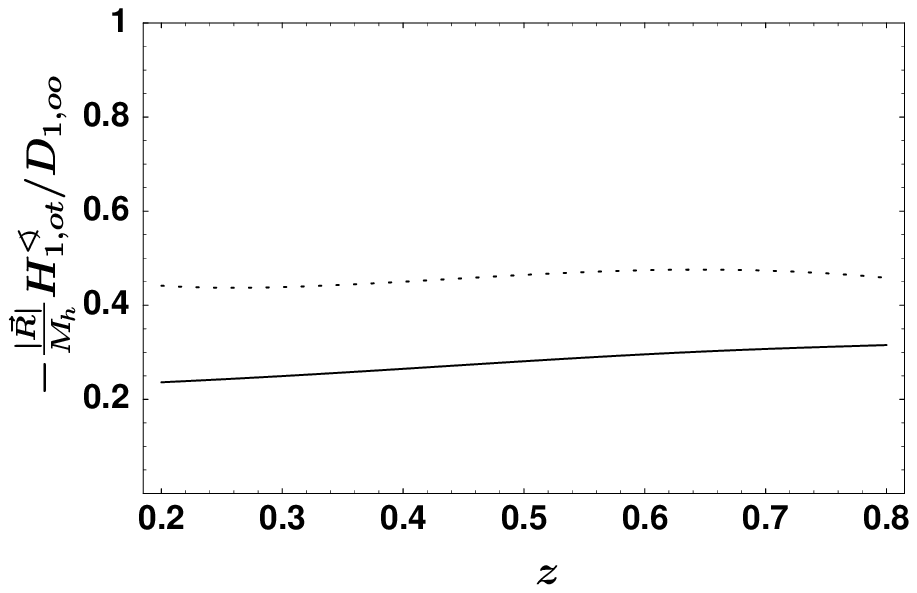}
\caption{\label{fig_mod_h1ang}  Model 
prediction for the ratio $(- \vert \Vec R \vert 
H_1^{\newangle, sp}/ (M_h D_1^o)$ as a function of $M_h$ (left) 
and $z$ (right). The dotted lines represent the positivity bounds. 
}
\end{center} 
\end{figure}

A model based on a more sophisticated 
analysis of the fragmentation channels \cite{Bacchetta:2006un} 
predicts a completely different behaviour for $H_1^{\newangle, sp}$, 
with  a peak at the $\rho$ mass and a broader maximum 
at the $\omega$ mass. Its size is about 30 \%  of the unpolarised 
fragmentation function as shown in Fig.~\ref{fig_mod_h1ang}. 

A different definition of the relative transverse momentum is  
proposed by Artru \cite{Artru:1993ad}, who 
uses the vector $\Vec r_{\perp}  = (z_2 \Vec P_{1 \perp} - 
z_1 \Vec P_{2 \perp})(z_1 + z_2)$, which is 
perpendicular to the $\gamma^* N$ axis, and its 
azimuthal angle $\phi_r$. 
The advantage of $\phi_r$ is that 
it is by construction invariant with respect to boosts 
along the $\gamma^* N$ direction, which is not the case 
of $\phi_R$ although the two angles are the same in
the $\gamma^* N$ frame.
Using the recursive fragmentation string model (see Section \ref{twoparticle})
the ``joint $p_T$ spectrum'' of the first and the second rank
hadrons has been calculated~\cite{Artru:2010st}.
By suitably integrating the spectrum, expressions
for both the Collins FF and the dihadron FF in principle
may be obtained.

\subsection{Inclusive production of hadron pairs in $e^+ e^-$ annihilation}
\label{epluseminus}

An independent source of information on the Collins fragmentation function
$H_1^\perp$ is inclusive two-hadron production in 
electron--positron collisions, 
$  e^+  +   e^- \, \rightarrow  h_1 +   h_2 +  X$,  
with the two hadrons (typically pions) in different hemispheres. 
We know that the Collins function $H_{1}^{\perp}$ produces 
a $\cos \phi$ modulation, where $\phi$ is the azimuthal angle 
between the plane containing the quark and the hadron momenta, and  the plane 
normal to $\Vec S_q$.  
Considering a single jet in $e^+ e^-$ hadron production,  the
Collins modulation would average to zero in a large event sample. 
Thus, in $e^+ e^-$ annihilation the Collins effect  
can only be observed in the combination of two fragmenting processes 
of a quark and an antiquark, resulting   
in the product of two Collins functions 
with an overall modulation of the type $\cos (\phi_1 + \phi_2)$, 
where $\phi_1$ and $\phi_2$ are the azimuthal angles of the 
final hadrons around the quark-antiquark axis, with respect 
to the $e^+ e^- \to q \bar q$ scattering plane.  

\begin{figure}[t]
\begin{center}
\includegraphics[width=0.35\textwidth,bb=0 220 590 600]
{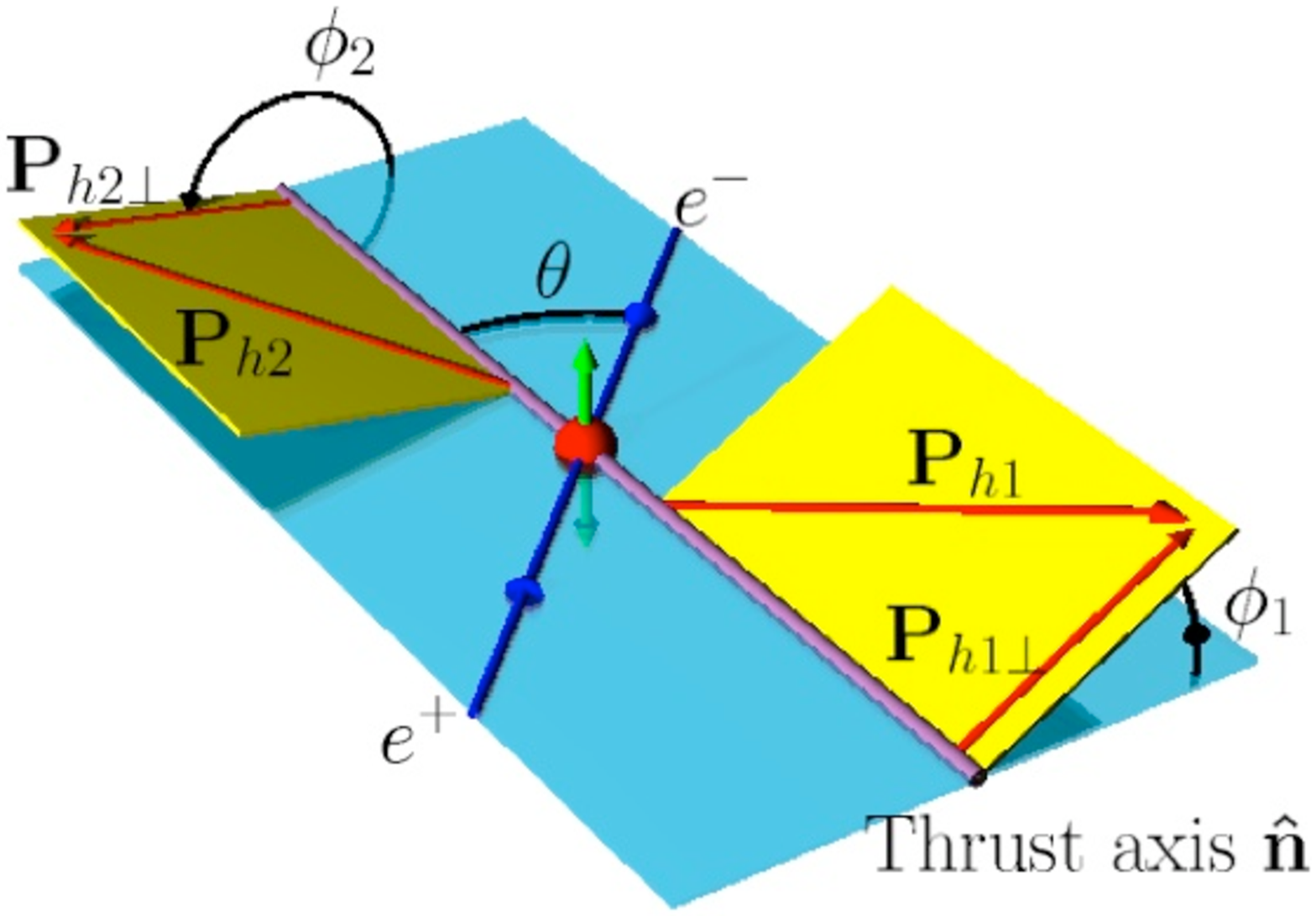}
\hspace{0.2cm}
\includegraphics[width=0.35\textwidth,bb=0 220 590 600]
{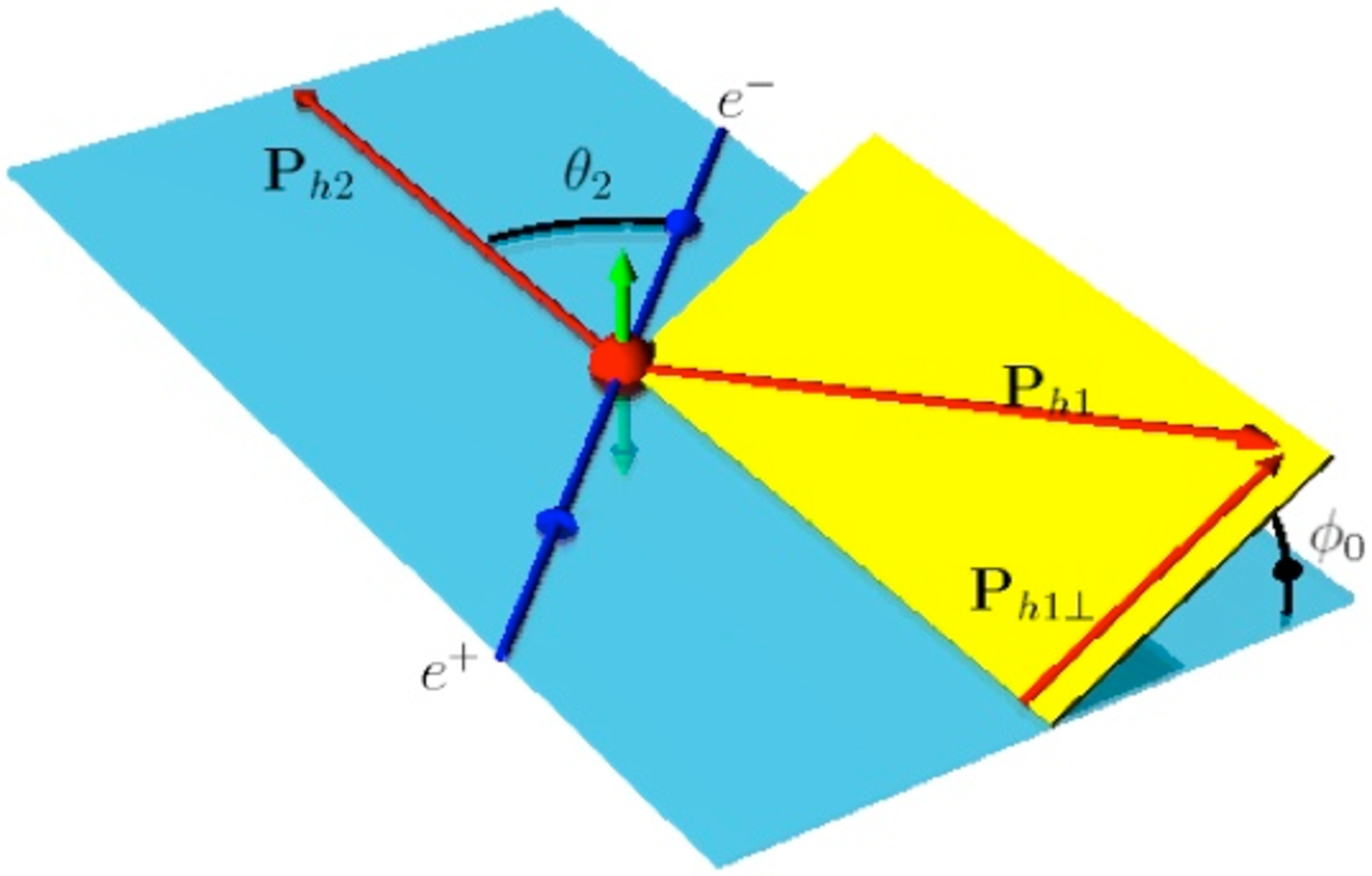}
\caption{\label{epem_geom} Left: Configuration of 
the process $e^+ e^- \to h_1 h_2 X$ in the jet frame, used
for the 
$\cos (\phi_1 + \phi_2)$ reconstruction of the 
Collins asymmetry. Right: The same process 
in the Gottfried-Jackson frame,  used for the the $\cos 2 \phi_0$ 
reconstruction of the Collins asymmetry. 
}
\end{center} 
\end{figure}

Two-hadron production in $e^+e^-$ collisions was studied in 
Refs.~\cite{Chen:1994ar,Bonivento:1995aa,Boer:1997qn,Boer:1997mf,Boer:1998aa,Boer:2008fr}. 
The tree-level differential cross section in the jet frame 
with respect to the quark-antiquark direction (Fig.~\ref{epem_geom}, 
left) reads
\bq
\frac{\D^6 \sigma}{\D \Omega \D z_1 
\D z_2  \D \phi_1 \D \phi_2}
&=& 
\frac{3 \alpha_{\rm em}^2}{4 s} 
\sum_{a=q, \bar q} e_a^2 z_1^2 \, z_2^2 
\left \{ (1 + \cos^2 \theta) \, 
D_1^{a[0]}(z_1) \overline{D}_1^{a[0]} (z_2) 
\right. 
\nonumber \\
& & + \left. \sin^2 \theta 
\, \cos (\phi_1 + \phi_2) \, 
H_{1}^{\perp a [1]} (z_1) \, 
\overline{H}_{1}^{\perp a [1]} (z_2) \right \}\,, 
\label{ee2} 
\eq
where $\D \Omega = \D \cos \theta \, \D \phi_{\ell}$  
($\theta$ is the angle between the lepton axis and 
the $q \bar q$ axis, in the $q \bar q$ centre-of-mass frame, 
whereas $\phi_{\ell}$ gives the orientation of the  
scattering plane around the $q \bar q$ axis), 
and we have introduced the one-dimensional moments 
\be
F^{[n]}(z) \equiv \int \D  p_T^2 
\, \left (\frac{ p_T}{M_h} \right )^n 
\, F(z, p_T^2) \,. 
\ee
Eq.~(\ref{ee2}) and the following results refer to the case 
of photon-mediated $e^+ e^-$ annihilation. $Z$ production 
and $\gamma^* Z$ interference effects have been investigated 
in Refs.~\cite{Boer:1997qn,Boer:2008fr}.  

From an experimental point of view, the quark-antiquark direction 
is not directly accessible, and is approximated by the 
dijet thrust axis $\hat{\Vec n}$, defined by 
\be
T = {\rm max} \; \frac{\sum_h \vert \Vec P_h \cdot \hat{\Vec n} \vert}{\sum_h 
\vert \Vec P_h \vert}\,,  
\label{ee3} 
\ee 
where the sum is over all detected particles. The resulting 
$\cos (\phi_1 + \phi_2)$ asymmetry is given by 
\be
a_{12} (\theta, z_1, z_2) = \frac{\sin^2 \theta}{1 + \cos^2 \theta}
\, 
\frac{\sum_a e_a^2 \left ( H_{1}^{\perp a [1]} (z_1) \, 
\overline{H}_{1}^{\perp a [1]} (z_2) \right )}{\sum_a e_a^2 
\left (D_1^{a[0]}(z_1) \overline{D}_1^{a[0]} \right )}\,.
\label{a12}
\ee 
We will refer to this method of extracting the Collins asymmetry 
by measuring azimuthal distributions around the 
thrust axis as to the ``$\cos (\phi_1 + \phi_2)$ method'' 
\cite{Bonivento:1995aa,Abe:2005zx,Seidl:2008xc,Boer:2008fr}.  

There is another way of reconstructing the asymmetry, 
the so-called ``$\cos 2 \phi_0$ method'', 
which is based on a different geometry and does not 
require the knowledge of the thrust axis. 
In this case one measures the hadron yields as a function of $\phi_0$, 
the angle between the plane containing the momentum 
of hadron 2 and the leptons, and the plane 
defined by the two hadron momenta \cite{Boer:1997qn,Boer:1997mf,Boer:2008fr}   
(this frame, shown in  Fig.~\ref{epem_geom} (right), 
is similar to the Gottfried-Jackson frame in Drell-Yan 
processes \cite{Gottfried:1964nx}). 
The corresponding cross section is
\bq
& & \frac{\D^6 \sigma}{\D \Omega \D z_1 \D z_2 
\D^2 \Vec q_T} = 
\frac{3 \alpha_{\rm em}^2}{4 Q^2} \sum_a 
e_a^2 z_1^2 z_2^2 
\nonumber \\
& & \hspace{1cm}
 \left \{ (1 + \cos^2 \theta) \, \mathcal{C} \,  
[D_1^a \overline{D}_1^a ]  
 +  \sin^2 \theta \, \cos 2 \phi_0 \, 
\mathcal{C} \left [ \frac{2 \hat{\Vec h} \cdot \Vec \kappa_{1T} 
\hat{\Vec h} \cdot \Vec \kappa_{2 T} - \Vec \kappa_{1T} 
\cdot \Vec \kappa_{2T}}{M_1 M_2} \, H_1^{\perp a} 
\overline{H}_1^{\perp a} \right ] \right \}  \,,  
\label{cos2phi0}
\eq
where $\hat{\Vec h} \equiv \Vec P_{1 \perp}/\vert \Vec P_{1\perp} \vert$ 
and $\Vec \kappa_{1T}, \Vec \kappa_{2T}$ are the transverse momenta 
of the two fragmenting quarks with respect to the hadron directions.
The $\cos 2 \phi_0$ asymmetry reads
\be
a_0(\theta, z_1, z_2) = 
\frac{\sin^2 \theta}{1 + \cos^2 \theta}
\, 
\frac{\sum_a e_a^2 \, \mathcal{C} 
\left [ (2 \hat{\Vec h} \cdot \Vec \kappa_{1T} \, 
\hat{\Vec h} \cdot \Vec \kappa_{2 T} - \Vec \kappa_{1T} 
\cdot \Vec \kappa_{2T}) \, H_1^{\perp a} 
\overline{H}_1^{\perp a} \right ]}{M_1 M_2 \, \sum_a e_a^2 \, 
\mathcal{C} \, 
[D_1^a \overline{D}_1^a ]} \,. 
\label{a0}
\ee

Electron-positron scattering can also allow accessing the
fragmentation function $H_1$ of transversely polarised baryons
\cite{Artru:1992jg,Chen:1994ar,Contogouris:1995xc}. 
In the case of $\Lambda$'s, the specific process 
is back-to-back $\Lambda \overline{\Lambda}$ inclusive production, 
$e^+ e^- \to \Lambda \, \overline{\Lambda} \, X$, 
with the hyperon and the anti-hyperon 
decaying into $p \pi^-$ and $\bar p \pi^+$, respectively. 
It was shown in Ref.~\cite{Chen:1994ar} that the  
cross section of this process contains an azimuthal 
modulation proportional of the type $\sin \theta 
\cos (\phi + \bar \phi) H_1^{\Lambda} (z_1) 
\overline{H}_1^{\overline{\Lambda}}(z_2) \propto H_1^{\Lambda}(z_1) 
H_1^{\Lambda}(z_2)$, where $\theta$ is the angle between 
the collision axis and the $q \bar q$ axis in 
the jet frame (Fig.~\ref{epem_geom}, left) and 
$\phi, \bar \phi$ are the azimuthal angles of the proton and 
antiproton in the same frame (with the $z$ direction given 
by the $q \bar q$ jet and the $x$ axis in the plane 
of the beams and jets. 
The $\cos (\phi + \bar \phi)$ asymmetry 
is determined by measuring the difference between 
the number of $p \bar p$ pairs 
on the same side of the scattering plane and the number 
of pairs on opposite sides. 

\begin{figure}[t]
\begin{center}
\includegraphics[width=0.50\textwidth]
{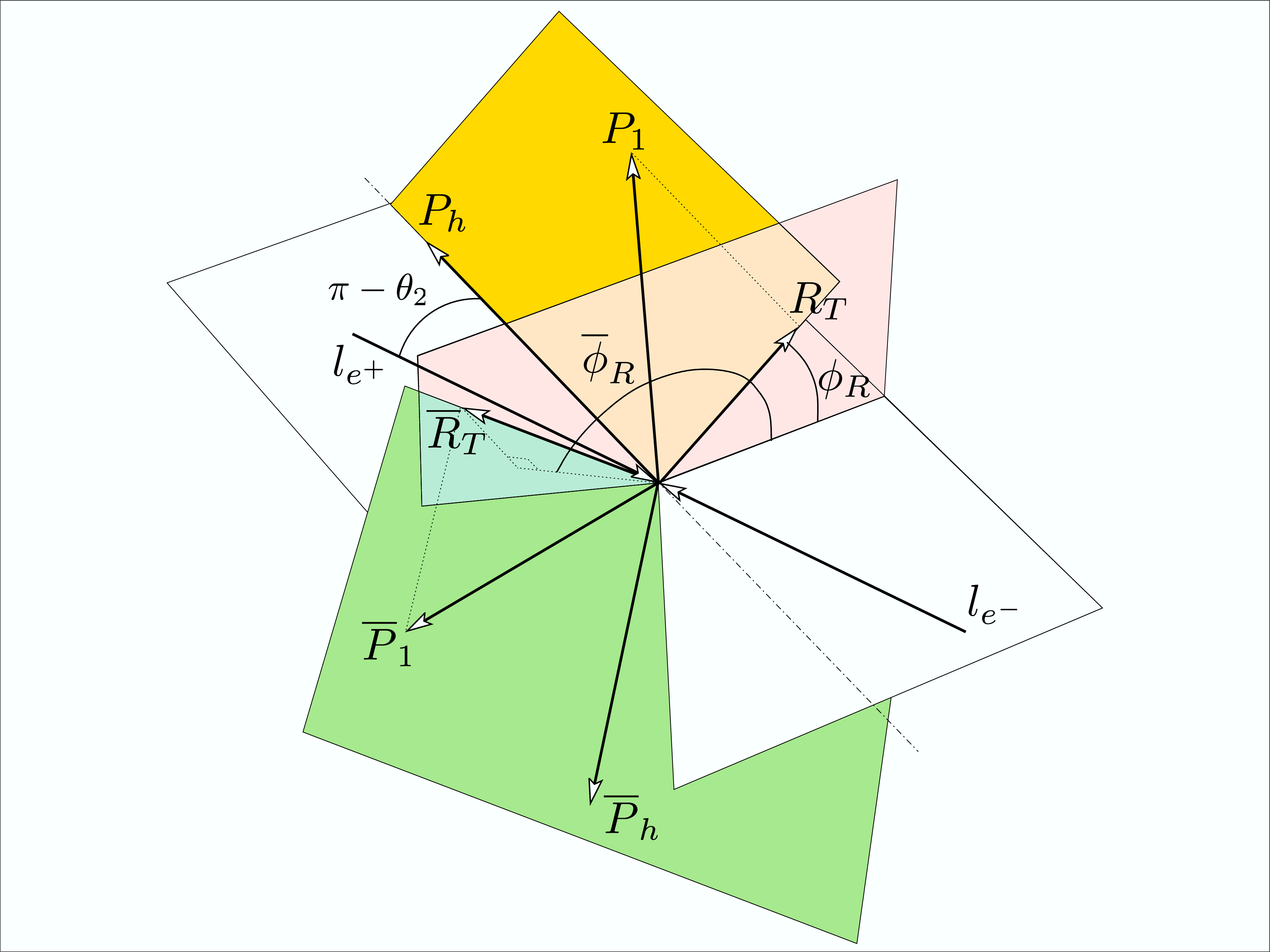}
\caption{\label{fig_4had}  Geometry of 
two hadron-pair production in $e^+e^-$ collisions.
}
\end{center} 
\end{figure}

Finally, the interference fragmentation function $H_1^{\newangle}$
can be extracted from the production of two hadron pairs 
in electron-positron annihilation: $e^+ e^- \to (h_1 h_2) (h_1' h_2') X$, 
where the particles in brackets belong to two back-to-back jets 
\cite{Artru:1995zu}.  
The (complicated) geometry of this process is shown in Fig.~\ref{fig_4had}. 
The observable quantity is the angular correlation of the production planes, 
expressed by the so-called Artru-Collins asymmetry. 
The kinematics of the process is described by doubling the 
variables introduced in Section~\ref{twoparticle}. 
If we call $\phi_R$ and $\bar \phi_R$ the azimuthal angles of the 
transverse relative momenta $\Vec R_T$ and $\bar{\Vec R_T}$ 
of the two hadron pairs, the Artru-Collins azimuthal asymmetry 
is the $\cos (\phi_R + \bar \phi_R)$ correlation. 

In $e^+ e^-$ annihilation, the interference fragmentation 
functions and  the Collins function  
are typically probed at much larger scales compared to SIDIS.  
However, the evolution of $H_1^{\newangle}$, 
differently from that of $H_1^{\perp}$,   
is known \cite{Ceccopieri:2007ip}. 
Therefore, a consistent combined analysis of dihadron production 
in $e^+ e^-$ annihilation and SIDIS is possible and 
may provide an alternative way to extract the transversity 
distributions \cite{Bacchetta:2008wb}. 

\subsection{Drell-Yan production} 
\label{sec:dy}

Drell-Yan (DY) dilepton production with various polarisations
of the two particles in the initial state
is a very rich source of knowledge  on the hadronic structure. 
The main advantage of this class 
of reactions is that they do not involve fragmentation 
functions, but only parton distributions. 
However, unless one considers antiproton--proton scattering,  
or pion-proton scattering, DY processes necessarily involve sea $\times$ 
valence products. 
This means that, while they provide direct information 
about antiquark distributions, which are less determined 
in SIDIS, their asymmetries are generally small. 

In principle, DY production 
with two transversely polarised hadrons is the cleanest 
reaction for studying the transversity distribution $h_1(x)$
and the pioneering works of Ref. \cite{Ralston:1979ys} 
and \cite{Pire:1983tv} were indeed devoted to this process. 
However, in order to observe sizable
double-spin asymmetries and extract $h_1$,  
we  probably have to wait for a new generation of 
experiments with polarised antiprotons \cite{Barone:2005pu}. 
On the other hand, unpolarised and singly-polarised 
DY processes can probe a large variety of  TMD's related to 
transverse spin, and are now attracting a wide
theoretical and experimental interest.   

\subsubsection{Kinematics and observables} 
\label{dy_kin}

Drell-Yan lepton-pair production is the process
$  A_1  (P_1)  +  A_2 (P_2)  \rightarrow
  \ell^+  (l)  +  \ell^-  (l')  +  X$, 
where $A_1$ and $A_2$ are hadrons  and $X$ is an undetected
system. The center-of-mass energy squared of this reaction is
$s=(P_1+P_2)^2\simeq2\,P_1{\cdot}P_2$, having neglected in the 
approximate equality  the hadron masses $M_1$ and $M_2$. 
The lepton pair originates from a virtual photon with four-momentum 
$q= l+ l'$. 
In contrast to DIS, $q$ is a time-like vector:
$Q^2=q^2>0$, and the invariant mass $M^2$ of the lepton pair 
coincides with $Q^2$.
The deep inelastic limit corresponds to $Q^2,s\to\infty$, with
$\tau \equiv Q^2/s$ fixed and finite.

\begin{figure}[t]
\begin{center}
\includegraphics[width=0.40\textwidth]
{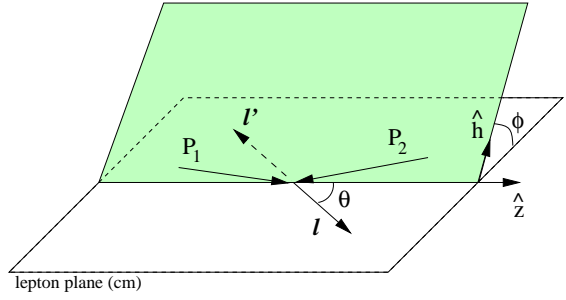}
\caption{\label{fig_dy_frame} 
The Collins-Soper frame.
The z-axis bisects the angle between  $\Vec P_2$ and $ - \Vec P_1$,
the momenta of the two initial state hadrons.
}
\end{center} 
\end{figure}

The DY cross section is usually expressed 
in a dilepton center-of-mass frame and can be written as 
\be
\frac{\D^6 \sigma}{\D^4 q \, \D \Omega} = 
\frac{\alpha_{\rm em}^2}{2 s Q^4} \, 
L_{\mu \nu} W^{\mu \nu} \,, 
\label{dycross}
\ee
where $L_{\mu \nu}$ is the familiar leptonic tensor
and $W^{\mu \nu}$ is the DY hadronic tensor.  
Among the infinite dilepton c.m. frames, related to each other 
by a rotation, the most often used is the Collins-Soper (CS) frame 
\cite{Collins:1977iv}, characterised by a z axis that bisects 
 the angle between $\Vec P_2$ and $ - \Vec P_1$ 
as shown in Fig.~\ref{fig_dy_frame}. 
Another common dilepton 
c.m. frame is the Gottfried--Jackson frame \cite{Gottfried:1964nx}, 
where the $z$ axis coincides with the direction of one of 
the colliding hadrons. 

In the unpolarised case the DY hadronic tensor contains 
four independent structure functions \cite{Lam:1978pu}. 
Using the classification of  Ref.~\cite{Arnold:2008kf}
the cross-section becomes
\be
\frac{\D^6 \sigma_{UU}}{\D^4 q \, \D \Omega} =  
\frac{\alpha_{\rm em}^2}{6 s Q^2} 
\, \left \{ (1 + \cos^2 \theta) \, W_{UU}^1 + \sin^2 \theta \, W_{UU}^2 
 +  \sin 2 \theta \, \cos \phi \, W_{UU}^{\cos \phi} 
+ \sin^2 \theta \cos 2 \phi \, W_{UU}^{\cos 2 \phi} \right \} \,. 
\label{dycross2} 
\ee
The double subscript refers to the polarisation states 
of the two colliding hadrons: $U$ = unpolarised, $L$ = longitudinally 
polarised, $T$ = transversely polarised. 
In literature \cite{Lam:1978pu}, the structure functions $W_{UU}^1, W_{UU}^2, 
W_{UU}^{\cos \phi}, W_{UU}^{\cos 2 \phi}$ are also 
called (apart  from a common factor), $W_T, W_L, W_{\Delta}, 
W_{\Delta \Delta}$, respectively. 
The angular distribution of leptons is often parametrised as 
\be
\frac{1}{N_{\rm tot}} \,  \frac{\D N}{\D \Omega} = 
\frac{3}{4 \pi} \frac{1}{\lambda + 3} 
\, \left ( 1 + \lambda \, \cos^2 \theta 
 +   \mu \, \sin 2 \theta \, \cos 2 \phi 
+ \frac{\nu}{2} \, \sin^2 \theta \, \cos 2 \phi \right )\,.   
\label{nomega} 
\ee
The three quantities $\lambda, \mu$, and $\nu$ are related to 
$W_{UU}^1, W_{UU}^2, W_{UU}^{\cos \phi}$, and $W_{UU}^{\cos 2 \phi}$ by  
\be
 \lambda = \frac{W_{UU}^1 - W_{UU}^2}{W_{UU}^1 + W_{UU}^2}\,, 
\;\;\;
 \mu = \frac{W_{UU}^{\cos \phi}}{W_{UU}^1 + W_{UU}^2}\,, 
\;\;\;
 \nu = \frac{2 \, W_{UU}^{\cos 2 \phi}}{W_{UU}^1 + W_{UU}^2}
\,. 
\label{lmn}
\ee
The so-called Lam-Tung relation \cite{Lam:1978pu,Lam:1978zr,Lam:1980uc}
\be
\lambda + 2 \nu = 1\,, 
\ee
which corresponds to $W_{UU}^2 = 2 W_{UU}^{\cos 2 \phi}$, 
is  valid at order $\alpha_s$ in collinear 
QCD \cite{Collins:1978yt} (see below) and is slightly violated 
at order $\alpha_s^2$ \cite{Mirkes:1994dp}.   

In the polarised case, a complete analysis 
of the DY hadronic tensor is contained in Ref.~\cite{Arnold:2008kf}, where 
it is shown that in single-polarised DY, first studied in 
\cite{Pire:1983tv}, the number of independent 
structure functions is 16, whereas the 
double-polarised hadronic tensor contains
28 structure functions, so that altogether the number 
of independent structure functions in DY is 48. 
The full cross section can be found in Ref.~\cite{Arnold:2008kf}. 
Limiting ourselves to unpolarised, single-transverse 
and double-transverse contributions, the cross section reads 
\bq
& & \frac{\D^6 \sigma}{\D^4 q \, \D \Omega} 
= \frac{\alpha_{\rm em}^2}{6 s Q^2} 
\left \{ \left [ (1 + \cos^2 \theta) \, W_{UU}^1 + 
\sin^2 \theta \, W_{UU}^2   
+ \sin 2 \theta \, \cos \phi 
\, W_{UU}^{\cos \phi} 
 +  \sin^2 \theta \, \cos 2 \phi 
\, W_{UU}^{\cos 2 \phi} \right ] \right.
\nonumber \\
& & \hspace{0.5cm} + S_{1T} \left [
\sin \phi_{S_1} \left ( (1 + \cos^2 \theta)\, W_{TU}^1
+ \sin^2 \theta \, W_{TU}^2 + \sin 2 \theta \, \cos \phi 
\, W_{TU}^{\cos \phi}  
+ \sin^2 \theta \, \cos 2 \phi \, W_{TU}^{\cos 2 \phi} 
\right ) \right. 
\nonumber \\  
& & \hspace{0.5cm} 
+ \left. \cos \phi_{S_1} \, (\sin 2 \theta \, \sin \phi \, W_{TU}^{\sin \phi}
+ \sin^2 \theta \, \sin 2 \phi \, W_{TU}^{\sin 2 \phi} ) \right ] 
+ ( 1 \leftrightarrow 2, T \leftrightarrow U) 
\nonumber \\
& & \hspace{0.5cm} + 
S_{1T} \, S_{2T} \, \left [ \cos (\phi_{S_1} + \phi_{S_2}) 
\, \left ( (1 + \cos^2 \theta) \, W_{TT}^1 
+ \sin^2 \theta \, W_{TT}^2 
\right. \right. 
\nonumber \\
& &  \hspace{0.5cm} +  
\left. \left. \sin 2 \theta \, \cos \phi 
\, W_{TT}^{\cos \phi} + \sin^2 \theta \, \cos 2 \phi 
\, W_{TT}^{\cos 2 \phi} \right ) \right.
\nonumber \\
& & \hspace{0.5cm} +  \cos (\phi_{S_1} - \phi_{S_2}) 
\, \left ( (1 + \cos^2 \theta) \, \overline{W}_{TT}^1 
+ \sin^2 \theta \, \overline{W}_{TT}^2 + \sin 2 \theta \, \cos \phi 
\, \overline{W}_{TT}^{\cos \phi} + \sin^2 \theta \, \cos 2 \phi 
\, \overline{W}_{TT}^{\cos 2 \phi} \right ) 
\nonumber \\
& & \hspace{0.5cm} 
+ \sin (\phi_{S_1} + \phi_{S_2}) \, (\sin 2 \theta \sin \phi 
\, W_{TT}^{\sin \phi} + \sin^2 \theta \, \sin 2 \phi \, 
W_{TT}^{\sin 2 \phi} ) 
\nonumber \\
& & \hspace{0.5cm} + \left. \left. 
\sin (\phi_{S_1} - \phi_{S_2}) \, (\sin 2 \theta \sin \phi 
\, \overline{W}_{TT}^{\sin \phi} + \sin^2 \theta \, \sin 2 \phi \, 
\overline{W}_{TT}^{\sin 2 \phi} ) 
\right ]   + \ldots \right \}\,.   
\label{fulldy}
\eq
$\phi_{S_1}$ and $\phi_{S_2}$ are the azimuthal angles 
of the spin vectors of hadrons $A$ and $B$, respectively.
This angular structure is valid in any dilepton c.m. frame, 
but  the numerical values of the structure functions 
are frame-dependent.

In the parton model and at leading order in QCD
the two invariants
$  x_1 = Q^2/2P_1{\cdot}q$ and   $x_2 = Q^2/2P_2{\cdot}q$
can be interpreted as the fractions of
the longitudinal momenta of the hadrons $A$ and $B$ carried by the quark and
the antiquark that annihilate into the virtual photon.
In the c.m. frame of the two colliding hadrons, which is 
the most convenient frame to study the partonic structure 
of the hadronic tensor, the photon momentum $q^{\mu}$
can be parametrised as $q^{\mu} = (x_1 P_1^+, x_2 P_2^-, \Vec q_T)$  
and acquires a transverse component $\Vec q_T$.  
Neglecting terms of order $1/Q^2$, one has $Q^2/x_1x_2s = 1$, that
is $\tau \equiv Q^2/s=x_1x_2$. 
The structure functions in eq.~(\ref{fulldy}) can be 
expressed in terms of the four variables $x_1, x_2, 
Q_T\equiv \vert \Vec q_T \vert, Q$. 

Other variables customarily used are the rapidity of the virtual photon,  
$  y \equiv \frac{1}{2} \, \ln (q^+/q^-) =
  \frac{1}{2} \ln (x_1/x_2)$ 
and the Feynman variable $x_F = 2 q_L/\sqrt{s} = x_1 - x_2$. 
In a dilepton c.m. frame, $y=\frac{1}{2}(1+\cos\theta)$. The relation between 
$(x_1, x_2)$ and $(\tau, y)$ is $  x_1 = \sqrt\tau \, \E^y\,, \;
  x_2 = \sqrt\tau \, \E^{-y}$. 
The DY cross-section can be variously reexpressed in terms of these variables: 
\begin{equation}
\frac{\D^4 \sigma}{\D^4 q} = 
\frac{2}{s} 
  \frac{\D ^4 \sigma}{\D x_1  \D x_2  \D^2\Vec{q}_T } 
= 2 \, \frac{\D^4 \sigma}{\D y  \D Q^2 \, \D^2\Vec{q}_T} 
= 2 \, (x_1 + x_2) \frac{\D^4 \sigma}{\D x_F  \D Q^2 \, \D^2\Vec{q}_T}\,. 
  \label{dy16}
\end{equation}

\subsubsection{DY asymmetries in the TMD approach}
\label{dy_tmd}

\begin{figure}[t]
\begin{center}
\includegraphics[width=0.40\textwidth]
{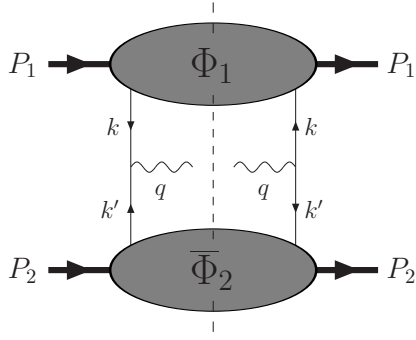}
  \caption{\label{handbag5} The parton-model 
diagram for the DY hadronic tensor.}
\end{center} 
\end{figure}

In the parton model, calling $k_1$ and $k_2$ the momenta of the quark (or
antiquark) coming from hadron $A_1$ and $A_2$ respectively, 
the hadronic tensor shown in Fig.~\ref{handbag5} is
\begin{equation}
  W^{\mu\nu} =
  \frac{1}{3} \, \sum_a e_a^2 \int \! \D^4 k_1
  \int \! \D^4 k_2 \, \delta^4(k_1 + k_2 - q) \,
  {\rm Tr} \,  [\Phi (k_1) \, \gamma^\mu \, \bar \Phi (k_2) \gamma^\nu].
  \label{dy27}
\end{equation}
Here $\Phi$ is the quark correlation matrix for hadron $A_1$,
$\anti\Phi$ is the antiquark correlation matrix for hadron $A_2$,
 and the factor $1/3$ has been added since in $\Phi$ and
$\bar \Phi$ summations over colours are implicit. 
It is understood that, in
order to obtain the complete expression of the hadronic tensor, one must add to
(\ref{dy27}) a term with $\Phi (k_1)$ replaced by 
$\Phi (k_2)$ and $\bar \Phi (k_2)$
replaced by $\bar \Phi (k_1)$, which accounts for the case where a quark is
extracted from $A_2$ and an antiquark is extracted from $A_1$. 
In the following
formulae we shall denote this term symbolically by $[1\leftrightarrow2]$.

Introducing the longitudinal momentum fractions $\xi_1 = k_1^+/P_1^+$ 
and $\xi_2 = k_2^-/P_2^-$, and 
working out the delta function of four-momentum conservation, 
one finds $\xi_1 = x_1, \, \xi_2 = x_2$, and the hadronic tensor becomes
\be
  W^{\mu\nu} = 
  \frac1{3} \, \sum_a e_a^2
  \int \!  \D^2\Vec{k}_{1T}
  \int \!  \D^2\Vec{k}_{2T} \,
  \delta^2(\Vec k_{1T} + \Vec k_{2T} - \Vec{q}_T)
  {\rm Tr} \, 
  [
    \Phi (x_1, \Vec k_{1T}) \, \gamma^\mu \, \anti\Phi (x_2, \Vec k_{2T}) 
   \gamma^\nu
  ]
  +
  [1 \leftrightarrow 2]  .
  \label{dy29}
\ee
Inserting here the explicit partonic expressions of 
the quark correlators, it is not difficult to get  
the parton model expressions of the DY structure functions
 \cite{Tangerman:1994a1,Boer:1999mm,Arnold:2008kf}. 
Only 24 of these 48 structure functions are non vanishing at leading twist.   
The most relevant ones are, in the Collins-Soper frame  
(but the Gottfried-Jackson expressions differ from these 
only by subleading terms $\mathcal{O}(Q_T/Q)$)
\bq
& & W_{UU}^1 = \frac{1}{3} \, \mathcal{C} \, [f_1 \bar f_1]\,, 
\label{wdy1} \\
& & W_{UU}^{\cos 2 \phi} = \frac{1}{3} \, \mathcal{C} \left [ 
\frac{2 (\hat{\Vec h} \cdot \Vec k_{1T}) 
(\hat{\Vec h} \cdot \Vec k_{2T}) - \Vec k_{1T} 
\cdot \Vec k_{2T}}{M_1 M_2} \, h_1^{\perp} \bar h_1^{\perp} 
\right ] \,, 
\label{wdy2} \\
& & W_{TU}^1 = - \frac{1}{3} \, \mathcal{C} \left [\frac{\hat{\Vec h}\cdot 
\Vec k_{1T}}{M_1} \, f_{1T}^{\perp} \bar f_1 \right ] \,, 
\;\;\;
W_{UT}^1 =  \frac{1}{3} \, \mathcal{C} \left [\frac{\hat{\Vec h}\cdot 
\Vec k_{2T}}{M_2} \, f_1 \, \bar {f}_{1T}^{\perp} \right ] \,, 
\label{wdy3} \\
& & W_{TU}^{\sin (2 \phi - \phi_{S_1})} = 
\frac{1}{3} \, \mathcal{C} \left [\frac{\hat{\Vec h}\cdot 
\Vec k_{2T}}{M_2} \, h_1 \bar h_1^{\perp} \right ]\,, 
\;\;\;
W_{UT}^{\sin (2 \phi - \phi_{S_2})} = -  
\frac{1}{3} \, \mathcal{C} \left [\frac{\hat{\Vec h}\cdot 
\Vec k_{1T}}{M_1} \, h_1^{\perp} \bar h_1 \right ]\,, 
\label{wdy4} \\
& & W_{TT}^{\cos (2 \phi - \phi_{S_1} - \phi_{S_2})} =
\frac{1}{3} \mathcal{C} \, [h_1 \bar h_1]\,, 
\label{wdy5} 
\eq
where $\hat{\Vec h} \equiv \Vec q_T/Q_T$ and 
$\mathcal{C}$ denotes the transverse-momentum convolution of 
eq.~(\ref{convol}) with the addition of the $[1 \leftrightarrow 2]$ term. 
In eq.~(\ref{wdy4}) we defined the combinations 
$W_{TU (UT)}^{\sin (2 \phi - \phi_{S_1 (S_2)})} \equiv - \frac{1}{2} 
(W_{TU (UT)}^{\cos 2 \phi} - W_{TU (UT)}^{\sin 2 \phi})$ and 
$W_{TT}^{\cos (2 \phi - \phi_{S_1} - \phi_{S_2})} \equiv 
\frac{1}{2} (W_{TT}^{\cos 2 \phi} + W_{TT}^{\sin 2 \phi})$, 
which correspond to the angular modulations indicated by their 
superscripts. 

Eqs.~(\ref{wdy1}-\ref{wdy5}) contain a series of interesting 
results. First of all, the Boer-Mulders function $h_1^{\perp}$
generates a $\cos 2 \phi$ asymmetry in unpolarised 
DY. At leading twist, the $\nu$ parameter is given  by 
\be
\nu  = 2 \, \frac{W_{UU}^{\cos 2 \phi}}{W_{UU}^1}
= 2 \, \frac{\mathcal{C} \left [ 
(2 (\hat{\Vec h} \cdot \Vec k_{1T}) 
(\hat{\Vec h} \cdot \Vec k_{2T}) - \Vec k_{1T} 
\cdot \Vec k_{2T}) \, h_1^{\perp} \bar h_1^{\perp} 
\right ]}{M_1 M_2 \, \mathcal{C} \, [f_1 \bar f_1]}
\, .
\label{nuparton}
\ee
Since $\nu \neq 0$ and $\lambda = 1$, 
the Lam-Tung relation is violated, and this is one of the 
remarkable consequences of the intrinsic transverse motion of quarks. 

Concerning singly-polarised Drell-Yan processes, the Boer-Mulders 
function combines with the transversity distribution in the 
$\sin (2 \phi - \phi_{S_1})$, or $\sin (2 \phi - \phi_{S_2})$, 
asymmetry, whereas the Sivers function is probed 
via the $F_{TU}^1$ (or $F_{UT}^1$) structure function 
associated with the $\sin \phi_{S_1}$ (or $\sin \phi_{S_2}$) 
asymmetry. Note that upon integration over the lepton angles 
only the Sivers asymmetry is non vanishing.   

\subsubsection {DY double transverse asymmetries}
\label{doubletrans} 

With two transversely polarised 
colliding hadrons, the $\cos (2 \phi - \phi_{S_1} - \phi_{S_2})$ 
term provides a direct access to transversity. Inserting 
eqs.~(\ref{wdy1}, \ref{wdy5}) into eq.~(\ref{fulldy}) 
and integrating the cross section over $\Vec q_T$, one gets 
\bq
  \frac{\D^3\sigma}{\D{x}_1 \D{x}_2 \D \Omega} &=&
  \frac{\alpha_\text{em}^2}{12 Q^2} \, \sum_a \, e_a^2
  \left [
    (1 + \cos^2 \theta)\,  f_1^a(x_1) \, \bar{f}_1^a(x_2) \right.
\nonumber \\
& & + \left. S_{1T} S_{2T} \, \sin^2 \theta \, 
    \cos(2\phi - \phi_{S_1} - \phi_{S_2}) 
h_1^a (x_1) \bar h_1^a(x_2) \right ] 
  + [1\leftrightarrow2] \, .
  \label{dy33}
\eq
This parton-model expression can be generalised 
to QCD by resorting to the collinear factorisation theorem, 
which for the polarised DY process reads \cite{Collins:1992xw}
\be
  \D\sigma =
  \sum_a \sum_{\lambda_1 \lambda_1' \lambda_2 \lambda_2'} 
\int \! \D\xi_1 \int \! \D\xi_2 \;
  \rho_{\lambda_1'\lambda_1}^{(1)} \, f_a(\xi_1, \mu^2) \,
  \rho_{\lambda_2'\lambda_2}^{(2)} \, \bar f_a(\xi_2, \mu^2)
  \, 
    \D\hat\sigma_{\lambda_1 \lambda_1' \lambda_2 \lambda_2'}
(Q^2,\mu^2, \alpha_s(\mu^2))\,, 
  \label{dyfac0}
\ee
where $\xi_1$ and $\xi_2$ are the momentum fractions of the quark (from
hadron $A_1$) and antiquark (from $A_2$), $\rho^{(1)}$ and $\rho^{(2)}$ are the
quark and antiquark spin density matrices, 
$\D\hat\sigma_{\lambda_1 \lambda_1' \lambda_2 \lambda_2'}$ 
is the cross-section matrix  of the elementary subprocesses in
the quark and antiquark helicity space, $\mu$ is the factorisation scale.
At leading order, i.e. $\mathcal{O}(\alpha_s^0)$, 
the only contributing subprocess is $q \bar q \to \ell^+ \ell^-$ 
and $\xi_1 = x_1$, $\xi_2 = x_2$.  In the transversely polarised 
case, one reobtains eq.~(\ref{dy33}), except that 
all distribution functions acquire a $Q^2$-dependence.  
Thus the LO  double transverse asymmetry is 
\be
  A_{TT}^{DY}
= a_{TT} \, 
  \frac{\sum_a e_a^2 \, h_1^a(x_1, Q^2) \bar h_1^a(x_2, Q^2) 
+ [1\leftrightarrow2]}
       {\sum_a e_a^2 \,      f_1^a(x_1, Q^2)    \bar f_1^a(x_2, Q^2) 
+ [1\leftrightarrow2]}\,, 
\label{attdy}
\ee
where 
\be
a_{TT} = 
 \frac{\sin^2\theta}{1 + \cos^2\theta} 
\,   \cos(2\phi - \phi_{S_1} - \phi_{S_2})\,, 
\ee
is the elementary double-spin asymmetry for 
 $q \bar q \to \ell^+ \ell^-$. 
We see that a measurement of $A_{TT}^{DY}$ would directly provide the
product of quark and antiquark transversity distributions, 
with no mixing with other unknown quantities. 
At next-to-leading order (NLO) the DY transverse cross section 
gets  contributions from virtual-gluon (vertex and self-energy) corrections 
and real-gluon emission, which were calculated by 
several authors with different methods 
\cite{Vogelsang:1993jn,Contogouris:1994ws,Kamal:1996as,Vogelsang:1997ak}.
The NLO double transverse asymmetry was investigated in 
Refs.~\cite{Martin:1998rz,Martin:1999mg,Barone:2005cr,Ratcliffe:2004we}. 

\subsubsection{DY azimuthal and spin asymmetries in QCD}
\label{dy_qcd} 

As in the case of SIDIS, perturbative gluon radiation 
can generate a non-zero transverse momentum $Q_T$. 
For instance, the contribution of the quark-antiquark annihilation process 
$q \bar q \to \gamma^* g$ to the unpolarised angular 
distribution in the Collins-Soper frame is \cite{Collins:1978yt}
\be
\frac{1}{N_{\rm tot}} 
\, \frac{\D N}{\D \Omega} = \frac{3}{16 \pi} 
\left ( \frac{Q^2 + \frac{3}{2} Q_T^2}{Q^2 + Q_T^2}
+ \frac{Q^2 - \frac{1}{2} Q_T^2}{Q^2 + \frac{1}{2} Q_T^2} 
\cos^2 \theta +  
 \frac{1}{2} \, \frac{Q_T^2}{Q^2 + Q_T^2} 
\, \sin^2 \theta \, \cos 2 \phi + \ldots \right )\,, 
\label{qcd_dy1}
\ee
where we have omitted the $\sin 2 \theta \, \cos \phi$ 
term which is the only one depending on the quark and antiquark 
distributions.  
From (\ref{qcd_dy1}) one gets
\be
\lambda = \frac{Q^2 - \frac{1}{2} Q_T^2}{Q^2 + \frac{3}{2} Q_T^2} 
\,, 
\;\;\; 
\nu = \frac{Q_T^2}{Q^2 + \frac{3}{2} Q_T^2} \,, 
\label{qcd_dy2} 
\ee
and the Lam-Tung relation is fulfilled. The contribution of 
the $q g \to \gamma^* q$ is more complicated, but also satisfies 
the Lam-Tung relation, which 
holds for the complete leading-order cross section.  

The perturbative QCD approach to the 
DY angular distribution sketched above holds for $Q_T \sim Q$. 
At small $Q_T$, large logarithms of the form $\ln (Q^2/Q_T^2)$ 
appear, which must be resummed.  
This is done in the space conjugate to 
$\Vec q_T$ and gives rise to a Sudakov form factor, 
according to the Collins-Soper procedure \cite{Collins:1981uk}. 
The Sudakov resummation for the structure functions 
of unpolarised DY production, including those
related to azimuthal asymmetries,  has been studied in 
Refs.~\cite{Boer:2006eq,Berger:2007jw}  
and the Lam-Tung relation is found to be unaffected
by the resummation. 
 
The perturbatively generated $\cos 2 \phi$ asymmetry 
is suppressed as $Q_T^2/Q^2$ at small $Q_T$. 
A further contribution to this asymmetry can arise 
in the twist-three approach from the product 
of two quark-gluon correlation functions 
$E_F$, one associated with the quark from hadron $A_1$, 
the other with the antiquark from hadron $A_2$ \cite{Zhou:2009rp}.  

In the singly-polarised DY case the situation 
is again analogous to SIDIS. 
At large $Q_T \sim Q \gg \Lambda_{\rm QCD}$, 
a DY single-spin asymmetry is generated by 
the $G_F$ quark-gluon correlator of  the polarised hadron. 
A smooth transition  from this twist-three mechanism to the Sivers effect 
occurs in the intermediate region $\Lambda_{\rm QCD} 
\ll Q_T \ll Q$, where both the higher-twist and the TMD 
factorisations apply \cite{Ji:2006vf}. 
Another SSA's arises from the chirally-odd 
quark-gluon correlation function $E_F$ of the 
unpolarised hadron coupled to the transversity distribution 
of the transversely polarised hadron, but in the low $Q_T$ 
limit this contribution vanishes (after integration over the lepton angles 
only the Sivers asymmetry survives).  

\subsection{Inclusive hadroproduction}
\label{transvhad}

We finally discuss a third class of reactions that  
probe the transverse-spin and transverse-momentum 
structure of hadrons: inclusive hadroproduction
with one transversely polarised hadron in the initial state, that is 
$  A^\uparrow  +  B  \rightarrow  h  +  X $,   
where an unpolarised (or spinless) hadron $h$  is produced with a 
transverse momentum $\Vec{P}_{T}$ with respect to the collision axis. 
An interesting variation of this process is the production 
of a transversely polarised hadron, i.e., a $\Lambda$ hyperon, 
from unpolarised hadron-hadron scattering, that is 
$A + B \rightarrow h^{\uparrow} + X$ (Section~\ref{lambdahh}). 
We will limit ourselves to a brief description 
of these reactions, referring the reader for more detail 
to some reviews \cite{D'Alesio:2007jt,Liang:2000gz}
and to the original papers. 

We first consider hadroproduction with a 
transversely polarised colliding particle.
The measured quantity is the single-spin asymmetry
\begin{equation}
  A_N =
  \frac{\D\sigma^{\uparrow} - \D\sigma^{\downarrow}}
       {\D\sigma^{\uparrow} + \D\sigma^{\downarrow}} \,, 
  \label{at}
\end{equation}
with the cross sections usually expressed as functions
of $ P_T^2$ and of the Feynman variable $x_F = 2 P_L/\sqrt{s}$ 
($P_L$ being the longitudinal momentum of the produced hadron). 
In terms of the scattering angle $\theta$, 
Feynman's $x$ can be written as $x_F = 2 P_T/\sqrt{s} \tan \theta$.  
Another often used variable is the pseudorapidity $\eta = 
- \ln \tan (\theta/2)$. 

According to the QCD factorisation theorem \cite{Collins:1985ue,Collins:1992xw}
the differential cross-section for hadroproduction at large $ P_T $ 
can be formally written as
\begin{equation}
  \D\sigma =
  \sum_{abc} \sum_{\lambda_a \lambda_a' 
\lambda_c \lambda_c'} \rho^a_{\lambda_a \lambda_a'} \,
  f_a(x_a) \otimes f_b(x_b) \otimes
  \D \hat{\sigma}_{\lambda_a \lambda_a' \lambda_c \lambda_c'} \otimes
  \mathcal{D}^{h/c}_{\lambda_c \lambda_c'}(z) \, .
  \label{hh8}
\end{equation}
Here $f_a$ ($f_b$) is the distribution of parton $a$ ($b$) inside the hadron
$A$ ($B$), $\rho^a_{\lambda_a \lambda_a'}$ is the spin 
density matrix of parton $a$, $\mathcal{D}^{h/c}_{\lambda_c \lambda_c'}$ 
is the fragmentation matrix of parton $c$
into hadron $h$, and $\D\hat\sigma$ is the (perturbatively 
calculable) cross-section
of the elementary process $a + b \to c + \ldots$ (a two-body 
scattering, $a + b \to c + d$, at lowest order). 

If the produced hadron is unpolarised, or spinless, 
only the diagonal elements of $\mathcal{D}^{h/c}_{\lambda_c \lambda_c'}$
are non zero, i.e. $\mathcal{D}^{h/c}_{\lambda_c \lambda_c'} \sim 
\delta_{\lambda_c \lambda_c'}\,D^{h/c}$, where
$D^{h/c}$ is the unpolarised fragmentation function. 
Together with helicity
conservation in the partonic subprocess, this implies $\lambda_a=\lambda_a'$.
Therefore, the  cross section (\ref{hh8}) carries no 
dependence on the spin of hadron $A$ and
all single-spin asymmetries vanish \cite{Kane:1978nd}.
In order to escape such a conclusion one must 
consider either the intrinsic transverse
motion of quarks \cite{Sivers:1989cc,Collins:1992kk,Collins:1993kq}, 
or higher-twist effects
\cite{Efremov:1981sh,Efremov:1983eb,Efremov:1984ip,Ratcliffe:1985mp,Qiu:1991pp,Qiu:1991wg}. 
In the former case, one can probe 
a number of distribution and fragmentation 
functions, including the transversity distribution 
(transversely polarised quarks in hadron 
$A^{\uparrow}$), the Sivers function 
(unpolarised quarks in hadron $A^{\uparrow}$), the Boer-Mulders 
function (transversely polarised quarks in hadron $B$), the Collins function
(transversely polarised quarks fragmenting into hadron $h$). 
The twist-three single-spin asymmetries involve 
various quark-gluon correlators, either 
in the initial state (distribution functions), 
or in the final state (fragmentation functions). 
The main problem is that all TMD or twist-three 
contributions mix up in a single observable, $A_N$, 
which makes the physical interpretation of the results quite unclear. 

\subsubsection{Hadroproduction in the extended parton model}

When the intrinsic transverse motion of quarks is taken into account, the
QCD factorisation theorem for inclusive hadroproduction is not 
proven, and actually is known to be explicitly violated in some 
cases \cite{Collins:2007nk,Rogers:2010dm}.  
Nevertheless, one can write 
a non-collinear factorisation formula in the context of the 
extended parton model, with a tree-level elementary kernel.  
 One must obviously recall that: {\it i)} the generalisation of this kernel 
to higher order in $\alpha_s$ 
is not a legitimate procedure; {\it ii)} the transverse-momentum 
dependent distribution and fragmentation functions appearing 
in the hadroproduction factorisation formula  
are not guaranteed to be universal quantities, 
i.e.,  to be the same functions as in other processes. 

The extended-parton model formula generalising eq.~(\ref{hh8}) 
is (we consider the production of a spinless hadron) \cite{Anselmino:2005sh}
\begin{equation}
  \D\sigma =
  \sum_{abcd} \sum_{\lambda_a \lambda_a' \lambda_c \lambda_c'} 
 \rho^a_{\lambda_a \lambda_a'} \,
  f_a(x_a, \Vec k_{Ta}) \otimes f_b(x_b, \Vec k_{Tb}) \otimes
\D \hat \sigma_{\lambda_a \lambda_a' \lambda_c \lambda_c'} \otimes 
  \mathcal{D}_{\lambda_c \lambda_c'}^{h/c}(z, \Vec p_T) \, .
  \label{hhtrasv1}
\end{equation}
where the convolutions $\otimes$ are now not only on the 
longitudinal momentum fractions $x_a, x_b, z$, but also 
on the transverse momenta $\Vec k_{Ta}, \Vec k_{Tb}, \Vec p_T$. 
Note that even though $h$ is unpolarised, its $\Vec p_T$-dependent 
fragmentation matrix $\mathcal{D}_{\lambda_c \lambda_c'}^{h/c}$ 
is non diagonal. 
The elementary cross sections have the structure 
$\D \hat {\sigma}_{\lambda_a \lambda_a' 
\lambda_c \lambda_c'} \sim \sum_{\lambda_b \lambda_d} 
\hat{M}_{\lambda_c \lambda_d, \lambda_a \lambda_b} 
\hat{M}^*_{\lambda_c' \lambda_d, \lambda_a' \lambda_b}$.   
The amplitudes $\hat{M}$ refer to the 
elementary subprocess $a \, b \to c \, d$ (remember that 
we are considering the tree level only).  
A natural reference frame is the center-of-mass frame of the colliding hadrons.
The collision axis  forms with the direction 
of the produced hadron a plane, that we call the hadronic plane. 
The tricky point about eq.~(\ref{hhtrasv1}) 
is that, due to intrinsic transverse momenta,  
the partonic scattering does not take place in the hadronic plane. 
This non-planar geometry gives rise to some non-trivial  phases 
in the distribution and fragmentation matrices. 
Also, the amplitudes $\hat{M}$ appearing in eq.~(\ref{hhtrasv1}) 
must be Lorentz transformed to the canonical amplitudes 
$\hat{M}^0$ defined in the partonic center-of-mass frame, 
an operation which introduces further  phases. 
This complicated structure has been fully worked out in 
Ref.~\cite{Anselmino:2005sh}, where all the details 
concerning the kinematics and the scattering amplitudes 
can be found. 
Here we limit ourselves to quoting some general results. 
The contribution to the transverse single-spin asymmetry from the 
$q q \to qq$ subprocess schematically reads
\bq
\D \Delta \sigma_{qq \to qq}
&\sim& 
 f_{1T}^{\perp a} \otimes f_1^b \otimes \D \Delta \hat{\sigma}' \otimes 
D_1^c + h_1^a \otimes f_1^b \otimes \D \Delta \hat{\sigma}''
\otimes H_1^{\perp c}  \nonumber \\
& & + 
h_1^a \otimes h_1^{\perp b} \otimes \D \Delta \hat{\sigma}'''
 \otimes D_1^c + 
f_{1T}^{\perp a} \otimes h_1^{\perp b} \otimes 
\D \Delta \hat{\sigma}'''' \otimes H_1^{\perp c}\,. 
\eq
One recognises the Sivers effect (first term), the Collins effect 
(second term), the Boer-Mulders effect (third term) and a 
mixed effect (fourth term).   
The other contributions,  
$q \bar q \to gg$, $q g \to q g$, $ q g \to g q$, $gq \to gq$, 
$g q \to q g$, $gg \to q \bar q$, $ gg \to gg$, 
are explicitly given in Ref.~\cite{Anselmino:2005sh}. 
They contain, besides the distributions and fragmentation functions 
of linearly polarised gluons, the gluon Sivers function 
$f_{1T}^{\perp g}$. 
For instance, the $g q \to g  q$ and $g g \to \bar q q$  contributions are  
\be
\D \Delta \sigma_{gq \to g q} 
\sim f_{1T}^{\perp g} \otimes f_1^b \otimes \D \Delta \hat{\sigma}_I 
\otimes D_1^c + \ldots \,, \;\;\;\;
\D \Delta \sigma_{gg \to \bar q q} 
\sim f_{1T}^{\perp g} \otimes f_1^g \otimes \D \Delta \hat{\sigma}_{II}
 \otimes D_1^c + \ldots \,. 
\ee
Processes that select these terms and allow accessing  
the gluon Sivers function are $D$ meson production 
(which is dominated by the $g g \to \bar c c$ channel) \cite{Anselmino:2004nk}
and  pion production at midrapidity (which probes 
in the RHIC kinematics the small $x_a$ region and   
thus proceeds predominantly via gluonic channels)
\cite{Anselmino:2006yq}.  
Other reactions probing the Sivers functions 
of quarks and/or gluons without contributions
from the fragmentation sector are prompt-photon 
production $A^{\uparrow} + B \rightarrow \gamma + X$ 
\cite{D'Alesio:2004up,Schmidt:2005gv,D'Alesio:2006fp}, 
photon--jet production $A^{\uparrow} + B 
\rightarrow \gamma + {\rm jet} + X$ \cite{Schmidt:2005gv,Bacchetta:2007sz}, 
back-to-back dijet production $A^{\uparrow} + B \rightarrow 
{\rm jet}_1 + {\rm jet}_2 + X$ \cite{Boer:2003tx}. 

On the contrary, the Collins effect can be singled out 
by studying asymmetric azimuthal correlation of hadrons 
inside a jet, that is $A^{\uparrow} + B \rightarrow {\rm jet} + X 
\rightarrow h + X$ \cite{Yuan:2007nd}. 

\subsubsection{Single-spin asymmetries at twist three}
\label{singlespintwist3}

As pointed out in Ref. \cite{Efremov:1981sh,Efremov:1984ip},
non-vanishing single-spin asymmetries can be obtained in
perturbative QCD at higher-twist level. 
A twist-three factorisation theorem was proven 
for direct photon production \cite{Qiu:1991pp,Qiu:1991wg} and
hadron production \cite{Qiu:1998ia}. 
This work has been extended to cover the chirally-odd contributions 
\cite{Kanazawa:2000hz,Kanazawa:2000kp,Koike:2002ti}. 
Here we limit ourselves to quoting the
main general results of these works.
The twist-three phenomenological studies of data is treated in
Section~\ref{sec:phen_hp}.

At twist three the hadroproduction cross section is formally given by 
\begin{eqnarray}
  \D\sigma &=&
  \sum_{abc}
  \left\{
    G_F^a(x_a, x_a') \otimes f_1^b(x_b) \otimes 
    \D\hat\sigma' \otimes D_1^{h/c}(z) + 
  h_1^a(x_a) \otimes E_F^b(x_b, x_b') \otimes 
  \D\hat\sigma'' \otimes D_1^{h/c}(z) \right. 
  \nonumber
\\
  && \hspace{2em} \null +
  \left.
    h_1^a(x_a) \otimes f_1^b(x_b) \otimes 
    \D\hat\sigma''' \otimes \hat{E}_F^c (z, z')
  \right\} ,
  \label{singtwist31}
\end{eqnarray}
where $G_F(x_a,x_a')$ and $E_F(x_a,x_a')$ are the quark--gluon correlation
functions introduced in Section~\ref{ht_tmd},  $\hat{E}_F^c (z, z')$ 
is a quark--gluon correlator in the fragmentation process
and $\D \hat\sigma'$, $\D\hat\sigma''$ and $\D\hat\sigma'''$
 are cross-sections of hard partonic subprocesses.
The first term in (\ref{singtwist31})
corresponds to the chirally-even mechanism considered by Qiu  and Sterman 
\cite{Qiu:1998ia}. 
The second term  is the initial-state chirally-odd
contribution analysed in Ref.~\cite{Kanazawa:2000hz}. 
The third term is the final-state contribution 
studied in Ref.~\cite{Koike:2002gm}. 
The details and the elementary cross-sections
can be found in the original papers.

\subsubsection{$ \Lambda$ production}
\label{lambdahh}

The origin of the large transverse polarisation of hyperons 
measured since the 70's \cite{Bunce:1976yb,Heller:1978ty}
in high-energy unpolarised hadron-hadron scattering 
is a longstanding  problem (for reviews see 
Refs.~\cite{Panagiotou:1989sv,Felix:1999tf}).  
A transverse-momentum mechanism able to produce 
sizable asymmetries in $A + B \rightarrow \Lambda^{\uparrow} + 
X$ involves the polarising fragmentation 
function $D_{1T}^{\perp}$ introduced in Section~\ref{frag}, which 
describes the fragmentation of an unpolarised 
quark in a transversely polarised hadron \cite{Anselmino:2000vs}. 
In the twist-three approach the $\Lambda$ polarisation 
is generated by the chirally-odd spin-independent 
quark-gluon correlation function $E_F$ \cite{Zhou:2008fb}.  

While in  $A + B \rightarrow \Lambda^{\uparrow} + 
X$ the $\Lambda$ polarisation must vanish  
at pseudorapidity $\eta = 0$ for symmetry reasons.  
in $\Lambda$+jet production no such constraint exists. 
The process $A + B \rightarrow {\rm jet} + {\rm jet} + X 
\rightarrow \Lambda + {\rm jet} + X$ has been proposed as 
an alternative way of probing $D_{1T}^{\perp}$ through
the correlation between the transverse momentum 
and spin of the $\Lambda$ with respect to the dijet axis \cite{Boer:2007nh}. 

\subsection{Other processes} 
\label{other} 

We conclude our discussion of the transverse-spin effects in hard processes 
by listing a series of reactions that have been proposed 
as sources of information on transversity and TMD's. 
\begin{itemize} 
\item[-]
Hadron production in transversely polarised lepton-proton 
scattering: $ \ell + p^{\uparrow} \to h + X$. Note that 
the final lepton is not detected, so this process is similar  
to $p + p^{\uparrow} \to h + X$. 
Its transverse SSA has been calculated in the twist-three factorisation 
approach \cite{Koike:2002ti} and in the extended parton 
model \cite{Anselmino:2009pn}.  
\item[-]
Dilepton photoproduction: $\gamma + N^{\uparrow} \to \ell^+ + \ell^- + X$.  
It has been shown \cite{Pire:2009ap} that the transverse SSA of this reaction  
involves the transversity distribution multiplied by 
the chiral-odd distribution amplitude of the photon. 
\item[-]
Exclusive $\pi$ electroproduction: $ e + p^{\uparrow} \to e' + \pi^0 + p'$. 
Using a model for the GPD's and relating $H_T(x, \xi, t)$ to the 
tensor charge, the authors of Ref.~\cite{Ahmad:2008hp}
show that the transverse-spin asymmetry of this process 
can provide information on the tensor charge $\delta u$. 
\item[-]
Photo- and electroproduction of two vector mesons: $ \gamma^{(*)} + 
N^{\uparrow} \to \rho_1 + \rho_2 + N'$. 
If one of the mesons is transversely polarised this reaction probes 
the transversity GPD $H_T(x, \xi, t)$ \cite{Enberg:2006he}. 
\end{itemize}

\section{Experimental results and phenomenological analyses}
\label{sec:results}

In the last decade, many transverse-spin effects have been measured
in SIDIS on transversely polarised targets mainly 
by the HERMES and COMPASS Collaborations, 
in hadron-hadron scattering by the RHIC spin experiments, and in unpolarised 
Drell-Yan processes at Fermilab. 

In this section, we review some of the recent experimental findings 
together with their phenomenological interpretation. 
The selected data are organised according to the physics information 
they provide. 
We start with the measurements aiming
to access the transversity distribution, including the related
measurements which are being performed in $e^+e^-$ collisions.
Then we describe the experimental results related to the $T$-odd TMD's (Sivers 
and Boer-Mulders function) and we conclude this part with a brief 
discussion of some measurements involving the $T$-even TMD's
and higher-twist PDF's.   
The RHIC hadroproduction results cannot easily be fitted in this scheme, 
and are presented in a separate subsection.   

Since most of the results currently used to access transversity and TMD's 
come from the SIDIS experiments, we feel useful
to give in section~\ref{sec:sidis_kinem} 
some details on the kinematical ranges of the present SIDIS
experiments and on the analyses these experiments are doing, 
which are essentially common to all the SSA's extraction.

\vspace{0.5cm} 

\underline{Note}: Although one should in principle distinguish between the 
longitudinal momentum fraction 
$x=k^+/P^+$ and the Bjorken variable $x_B=Q^2/2P\cdot q$, 
we have seen that they coincide as far as $1/Q^2$ corrections 
are neglected. 
For the sake of simplicity, in this section 
we ignore this distinction and write the distributions as functions 
of $x$. Analogously, we take $z=P_h^-/k^-$ to be the the argument of 
fragmentation functions. 
In the data plots, obviously $x$ and $x_B$ always stay for $x_B$
and $z$ and $z_h$ for $z_h=P \cdot P_h / P \cdot q$.\\
Other notations: $k_T \equiv \vert \Vec k_T \vert$ 
is the transverse momentum of the initial quark and $p_T \equiv 
\vert \Vec p_T \vert$ is the transverse momentum of the 
produced hadron with respect to the fragmenting quark.
In SIDIS $P_{h \perp} 
\equiv \vert \Vec P_{h \perp} \vert$ is the transverse momentum of the 
final  hadron with respect to the $\gamma^* N$ axis.
In hadroproduction 
$P_T \equiv \vert \Vec P_T \vert$ is the transverse momentum of the final 
hadron with respect to the collision axis. 

\subsection{SIDIS kinematics and SSA extraction}
\label{sec:sidis_kinem}

The SIDIS events are usually identified with  standard cuts, with
some differences for the different channels and for the various 
experiments due to the different beam energies and thus to the 
different kinematical domain.
The DIS events are selected requiring the photon virtuality $Q^2$
to be larger than 1 GeV$^2$. 
The fractional energy $y$ transfered from the beam lepton to the virtual 
photon has to be larger than 0.1, to remove events affected 
by poor energy resolution, and smaller than 0.9 (or 0.95),
to avoid the  region most affected by radiative corrections.
A minimum value of  invariant mass of the final hadronic state
$W\simeq  2$ GeV is also needed to exclude the resonance region.
Typically, values of $W^2$ larger than 4, 10 and 25 GeV$^2$
are  required in the data analyses of the JLab, HERMES and COMPASS
experiments respectively.

\begin{figure}[tbh]
\begin{center}
\includegraphics[width=0.55\textwidth]
{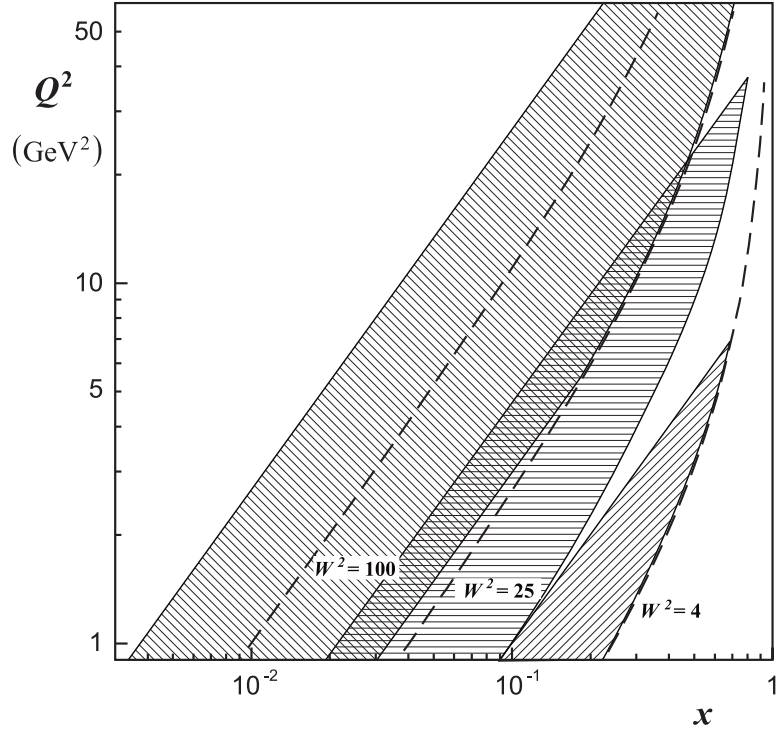}
\end{center}
\caption{The $x-Q^2$ DIS regions with lepton beams of 160 GeV,
27.5 GeV and 6 GeV momenta (left to right).
At large $x$, the region is limited by $W^2 > 25$ GeV$^2$
or $y>0.1$, $W^2 > 10$ GeV$^2$ and $W^2 > 4$ GeV$^2$ for the 
three beam momenta respectively, while at  low $x$ it is limited 
by the requirement $y < 0.9$.
The dashed curves give the $Q^2-x$ correlations
for $W^2$ equal to 100, 25 and 4 GeV$^2$.
}
\label{fig:SIDISkin} 
\end{figure}

The variables $x$, $Q^2$ and $W$, for the selected events, 
cover ranges which strongly depend on the lepton beam energy.
Fig.~\ref{fig:SIDISkin} shows the regions of the ($x$, $Q^2$) plane
kinematically accessible with lepton beams of 160 GeV,
27.5 GeV and 6 GeV momenta, corresponding to the COMPASS,
HERMES and JLab experiments respectively.
In the COMPASS experiment the $x$ range is between
0.004 and 0.3, where the upper limit is given by the  low 
luminosity; for the HERMES experiment $0.02<x<0.4$, while
the JLab experiments  can presently measure with 
high precision at $x>0.1$, in the valence region.
The average $Q^2$ values are also different 
and there is a strong $x-Q^2$ correlation. 
At $x \simeq 0.1$ the mean $Q^2$ value is 6.4 GeV$^2$  at
COMPASS and about 2.5 GeV$^2$ at
HERMES, while at $x\simeq 0.3$ the values are about 20 GeV$^2$ and
6.2 GeV$^2$ respectively.
In the overlap region the ratio of the $Q^2$ mean values
measured in the two experiments
goes from 2 to 3 with increasing $x$, in spite of the similar
mean values when integrating over the whole $x$ range.
The $W^2$ values are between 25 and 200 GeV$^2$ for COMPASS, 
between 10 and 50 GeV$^2$ for HERMES, and
below 10 GeV$^2$ for  the present JLab experiments.
The differences in the covered kinematical regions
make the experiments complementary, and, all together, they guarantee
a very good coverage of the phase space.

In addition to the requirements on the inclusive DIS variables, 
in the data analysis  cuts on the 
energy final state hadrons are applied, which also depend on the experiment
and on the physics channel under consideration.
Particle identification implies momenta above the RICH thresholds, 
which depend on the detector used in the experiment.
In the single hadron analyses, 
the relative energy $z$ of each hadron has to be
between 0.2 and 0.8.
The upper limit, usually not required in the COMPASS analyses, is chosen
to reject exclusively produced hadrons.
The lower limit is used to select hadrons from the current 
fragmentation region.
To do that, a selection based on the hadron rapidity should 
be applied, or, equivalently, the so-called Berger criterion
should be fulfilled.
This criterion~\cite{Berger:1987zu} has been tuned on unpolarised SIDIS data
and allows to relax the request based on the $W$ value alone
by asking for each hadron a large enough $z$.
Thus, if $W>7.4$ GeV, all hadrons belong to the current fragmentation.
Going down to $W\simeq 5$ GeV only hadrons with $z>0.2$ belong to this
region, while if $W\simeq 3$ GeV one should cut at $z>0.5$
to be safe.


The azimuthal asymmetries in SIDIS introduced in Section~\ref{sidis} are
defined as
\be
A^{f(\Phi)}_{XY}=F  ^{f(\Phi)}_{XY}/F_{UU} \, ,
\label{eq:ssa}
\ee
where $f(\Phi)$ is a trigonometric function of a linear combination
$\Phi$ of $\phi_H$ and $\phi_S$, and $XY$ refer to the
beam and target polarisations (U, L, or T).
The asymmetries can be extracted from the measured 
distribution of the final state hadrons in the relevant azimuthal
angle $\Phi$.
In the following, the methods used to evaluate the transverse SSA's 
in the HERMES and in the COMPASS experiments will briefly described.

In principle, since all the trigonometric functions
appearing in the cross-section are orthogonal, the amplitudes 
of the azimuthal modulations (the so-called ``raw asymmetry'' $a$)
can be obtained as twice the mean value of  $f(\Phi)$, or by fitting
the azimuthal distribution with a function of the type
$F(\Phi)=const \cdot [ 1 + a \cdot f(\Phi)]$.
In practice this procedure requires to correct the azimuthal
distribution for possible acceptance effects by means of Monte Carlo 
simulations.
This can be avoided in the case of the SSA's by collecting data
with two opposite spin orientations, indicated with ``+'' and ``-''
in the following.
If the acceptance and the detector efficiencies are the same
for the two sets of data, no Monte Carlo correction is needed  when
fitting the function $F(\Phi)$ on the quantities
\be
A(\Phi)=\frac{N^+(\Phi)-rN^-(\Phi)}{N^+(\Phi)+rN^-(\Phi)} \, .
\label{eq:elssa}
\ee
Here $N^{\pm}$ are the  numbers of events in a given
$\Phi$ bin, $r$ is the normalisation factor between the two sets of data, 
and $\Phi$ is always measured assuming the same orientation of the
target polarisation for  both sets of data.
In the case of the HERMES experiment, the target spin orientation is 
flipped every second, so that acceptance and overall efficiencies can 
safely be assumed to be the same for the two sets of data.
In the case of the COMPASS experiment, in which the target polarisation
can be reversed typically only after a few days of data taking,
the target is divided in cells with opposite polarisation directions.
This allows to minimise the possible systematic effects due to
acceptance variations by using, instead of $A(\Phi)$ the 
so-called ``ratio product'' quantities~\cite{Ageev:2006da} which 
combine the number of events from the different cells and with the 
different target polarisation orientation. 
These quantities do not dependent on the 
beam flux and on the acceptances, under the assumption that the  
relative variations are the same for all the target
cells, and they have a very simple expression in terms of the
azimuthal modulation one wants to extract.

The methods described above are simple and direct, still some
systematic effect can be relevant. 
In particular, the apparatus acceptance can introduce correlations
between the physical asymmetries.
For this reason different and more elaborated methods have been 
developed, which include the binning of the data
in the $(\phi_h,\phi_S)$ plane and the fit with a function
which includes all the modulations which appear in the cross-section.
In the most recent analyses, both 
HERMES~\cite{Diefenthaler:2007rj,Airapetian:2009ti} and 
COMPASS~\cite{Bressan:2009es,Wollny:2009eq,Alekseev:2010rw} have introduced
``unbinned'' maximum-likelihood methods based on maximum-likelihood fits 
with the data unbinned in $\phi_h$ and $\phi_S$.
The probability distributions include all the expected azimuthal
modulations, both for the spin independent and the transverse target spin
dependent parts of the cross-section.
The spin independent part turned out not
to influence the results for the SSA's.
In the COMPASS case, the same is true for the acceptance of the apparatus, 
assumed to have the same relative variations for all the target
cells~\cite{Wollny:2009eq}, and the
SSA's obtained with the ``unbinned'' maximum-likelihood method are in 
very good agreement 
with those extracted from the simpler methods used previously.

In order to obtain the final results, here called SSA's, the
raw asymmetries obtained with such fits have to be divided by the target
(and beam) polarisation.
The result are the HERMES ``moments'' or ``amplitudes'', usually
indicated with $2 <f(\Phi)>$.
In the COMPASS data, the raw asymmetries are divided by the
target polarisation, by its dilution factor $f$, and by the
kinematical $y$-dependent factors, usually dependent on the experimental 
acceptance in $y$. 
As an example, the Collins SSA's published by COMPASS
are obtained by dividing the raw asymmetries $a$ by the target
polarisation, by the dilution factor $f$, and by the mean value of the 
transverse polarisation transfer from the initial to the
final quark in the elementary lepton-quark scattering
$D_{NN} =(1-y)/(1-y+y^2/2)$.
The HERMES results for the Collins asymmetry (the so-called ``lepto-beam
asymmetries'') are only divided by the target polarisation.

Concerning the evaluation of possible systematic effects,
it has to be noted that,
even if the methods described above allow to extract simultaneously
all the SSA's, the systematic uncertainties  have to be evaluated
independently for all the modulations, since they can
be different.
For this reason, the results are not usually published all at the same time.

A final remark concerns the measurements of the asymmetries
as functions of different kinematical variables, typically
$x$, $z$ and $P_{h \perp}$.
Since in the experiments there is a strong $x-Q^2$ correlation, very
few attempts have been done to measure the $Q^2$ dependence of the
asymmetries in the various $x$ bins, which would be better studied by 
comparing the results of the different experiments.
Usually the SSA's are measured binning
the data alternatively in $x$, $z$ or $P_{h \perp}$, and integrating
on the other two variables.
This extraction introduces some correlation between the data,
which should be taken into account when fitting all the results 
in a global analysis.
To avoid this problem, the HERMES Collaborations is doing multi-dimensional
analysis, which are not yet possible in COMPASS, due to the
limited statistics.

\subsection{Accessing transversity}
\label{sec:results_transversity}

Today, the most direct information on transversity 
is coming from SIDIS measurements with transversely 
polarised targets, which are complementary to the 
DY experiments and have the advantage of allowing 
a flavor separation by identification of the final state hadrons. 
  
Among the various SIDIS observables related to transversity, 
the measurements performed so far 
have provided data on three of them: 
the Collins asymmetry, the two-hadron asymmetry,
and the $\Lambda$ polarisation. 
They will be presented in the following 
subsections, after a a brief description of the 
event and hadron selection. 

\subsubsection{Collins asymmetry in SIDIS}

The main source of information on the transversity PDF's
is at present the Collins asymmetry, 
which couples $h_1$ to the Collins fragmentation 
function $H_1^{\perp}$. 
The Collins asymmetry has been measured by the HERMES 
\cite{Airapetian:2004tw,Diefenthaler:2007rj}   
and by the COMPASS 
\cite{Alexakhin:2005iw,Ageev:2006da,Alekseev:2008dn,Alekseev:2010rw} 
Collaborations. 

Before describing these results, it has to be mentioned 
that the asymmetries measured by the two experiments differ for
the already mentioned correction by the $D_{NN}$ factor, 
applied by COMPASS only, and for the sign because of the different 
definition of the Collins angle $\Phi_C$.
In HERMES following the so-called ``Trento 
convention''~\cite{Bacchetta:2004jz}
it is defined as $\Phi_C=\phi_h+\phi_S$, while in COMPASS
the original definition \cite{Collins:1992kk}
$\Phi_C=\phi_h+\phi_S-\pi$ is used, as mentioned in
Section \ref{sidis_parton}.

The first signal for a non-zero Collins asymmetry came from
HERMES in 2005~\cite{Airapetian:2004tw}, when
the results on the data collected with the transversely polarised
target in 2002 were published.
The asymmetry had values clearly different from zero in the valence
region and of opposite sign for positive and negative pions, and this
was the first evidence
that both the transversity and the Collins FF had to be
different from zero.
An interesting feature of the HERMES results
is that the size of the asymmetry turned out to be roughly the same 
for positive and negative pions.
As suggested in Ref.~\cite{Airapetian:2004tw}, this result implied that
the favoured ($u \rightarrow \pi^+, \, d \rightarrow \pi^-$)
and the 
unfavoured ($u \rightarrow \pi^-, \, d \rightarrow \pi^+$)
Collins fragmentation functions $H_1^{\perp, {\rm fav}}$
and $H_1^{\perp, {\rm unf}}$ should be of the same size.
In fact, neglecting the sea contribution (the asymmetry is different
from zero only in the valence region) the flavour
structure of the Collins asymmetry for a proton target
can be written as
\be
 A_{\rm Coll}^{p, \pi^+} \sim 
e_u^2 h_1^u H_1^{\perp, {\rm fav}} + e_d^2 h_1^d H_1^{\perp, {\rm unf}} 
\,, \;\;\;\;\;
 A_{\rm Coll}^{p, \pi^-} \sim 
e_u^2 h_1^u H_1^{\perp, {\rm unf}} + e_d^2 h_1^d H_1^{\perp, {\rm fav}} 
\label{eq:collpm}
\,.  
\ee
Due to the weight factor given by the quark charge,
the measured asymmetries are not sensitive to $h_1^d$,
thus the result
$|A_{\rm Coll}^{p, \pi^+}| \simeq |A_{\rm Coll}^{p, \pi^-}|$
implies that $ H_1^{\perp, {\rm fav}} \simeq - H_1^{\perp, {\rm unf}}$.
This finding, unexpected at the time, can be 
understood~\cite{Airapetian:2004tw} in
the framework of the string model of fragmentation which in its most 
recent version~\cite{Artru:2010st} is described in Section 
\ref{collinsfunction}.
If a favoured pion forms at the string end created by the first 
break, a disfavoured pion from the next break will be opposite in charge
and will inherit transverse momentum from the first break in
opposite direction from that acquired by the first pion.
Also, assuming that the $u$ and $d$ quark contributions 
add up in the asymmetries, in the same model it is expected
that their transversity distributions
have opposite sign.

\begin{figure}[tb]
\begin{center}
\includegraphics[width=0.98\textwidth,bb=0 295 590 550]
{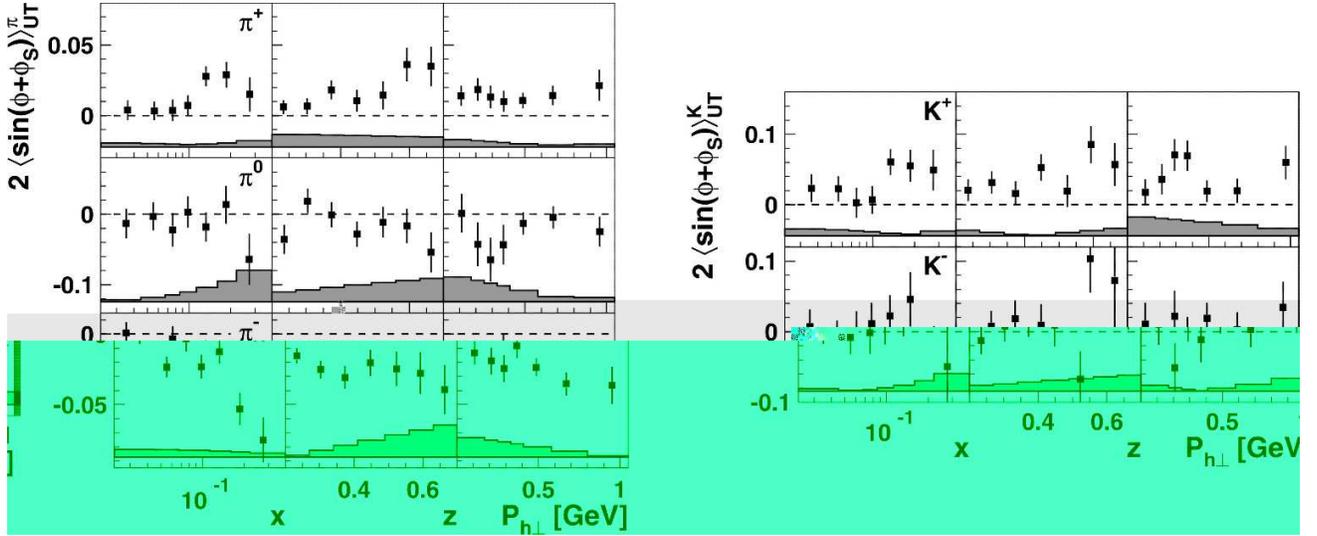}
\hfill
\end{center}
\caption{HERMES  results for the Collins asymmetry
from the 2002-2005 data collected with the transversely polarised 
proton target~\cite{Airapetian:2010ds}.
The asymmetries are shown as function of $x$, $z$ and $P_{h \perp}$.
The left plots show the asymmetries for pions,
the right plots the asymmetries for charged kaons.
}
\label{fig:coll_hermes07} 
\end{figure}

The COMPASS experiment started data  taking with the 
transversely polarised deuteron
target, and the first results~\cite{Alexakhin:2005iw}
were published almost at the same time as the HERMES results.
The measured Collins asymmetries were all compatible with
zero, both for positive hadrons and for negative hadrons.
This result is compatible with the HERMES finding.
Limiting again the analysis to the valence region,
the deuteron asymmetries can be written as 
\be
 A_{\rm Coll}^{d, \pi^+} \sim 
(h_1^u + h_1^d)(e_u^2 H_1^{\perp, {\rm fav}} + e_d^2 H_1^{\perp, {\rm unf}})
\,, \;\;\;\;\;
 A_{\rm Coll}^{d, \pi^-} \sim 
(h_1^u + h_1^d)(e_u^2 H_1^{\perp, {\rm fav}} + e_d^2 H_1^{\perp, {\rm unf}})
\,.  
\ee
The straightforward conclusion from the COMPASS deuteron measurements
is that $h_1^u$ and $h_1^d$ must have roughly the same size and opposite sign,
very much as in the case of the helicity quark distributions.

Both HERMES and COMPASS have continued the measurements with the proton and
the deuteron targets respectively, and have produced
results with considerably better statistics, which have confirmed the
first measurements.

The  HERMES results based on the whole data collected 
from 2002 to 2005~\cite{Airapetian:2010ds} 
are shown in Fig.~\ref{fig:coll_hermes07} for pions (left)
and charged kaons (right).

The applied cuts are
$Q^2> 1$ GeV$^2$, $0.1<y<0.95$, $W^2> 10$ GeV$^2$ and all the hadrons
with 2 $ < P_h <$ 15 GeV, 0.2 $< z <$ 0.7, and a polar angle with
respect to the direction of the virtual photon larger than 0.02 rad
are used in the extraction of the asymmetries.
In the figure the bands represent the 
maximal systematic uncertainty which includes hadron misidentification and
acceptance and detector smearing effects, and is smaller than the statistical 
one. 
The scale uncertainty due to the target polarisation 
has been evaluated to be about 8\%. 
The fraction of charged pions  and charged kaons produced in vector meson 
decay has also been estimated.
As can seen in the figure, the asymmetries for $\pi^+$ and $\pi^-$ 
have opposite sign, increase from very small values at $x\simeq 0.03$
to about 5\% at the highest $x$ values, and have a similar magnitude.  
The $\pi_0$ asymmetries are compatible with 
zero. 

\begin{figure}[tb]
\begin{center}
\includegraphics[width=0.75\textwidth]
{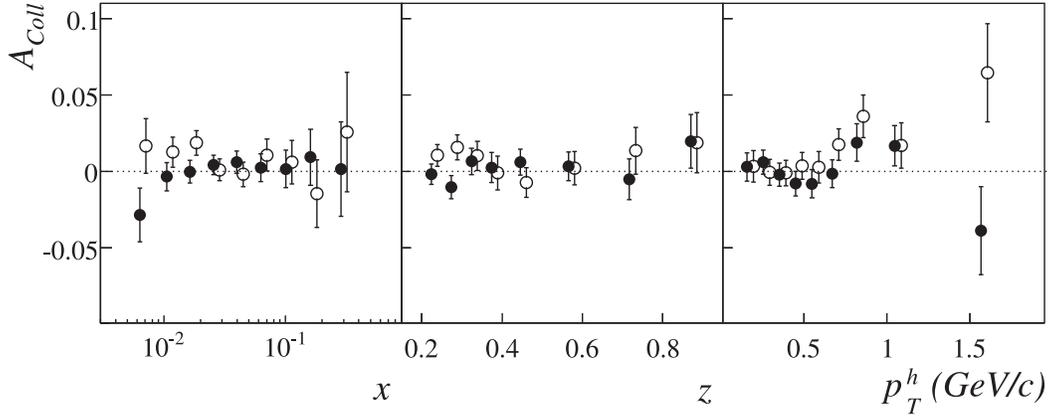}
\caption{COMPASS results for the Collins asymmetry
from the 2002, 2003, and 2004 data collected with the transversely polarised 
deuteron target~\cite{Ageev:2006da}.
The asymmetries are shown as functions of $x$, $z$ and $P_{h \perp}$
for all positive (full circles) and all negative hadrons (open circles).
In  the plots the open circles are slightly shifted horizontally 
with respect to the measured value. 
\label{fig:coll_cmpdah} 
}
\end{center}
\end{figure}

The COMPASS results for the Collins asymmetry from 
all the data collected from 2002 to 2004 with the deuteron target
are shown in Fig.~\ref{fig:coll_cmpdah} for 
charged positive and negative hadrons~\cite{Ageev:2006da}.
The error bars are statistical only.
The systematic errors have been estimated to be negligible with respect
to the statistical precision, and
the overall scale uncertainty is 7.3\% including the uncertainties
on the target polarisation and on its dilution factors.
Here the DIS events are selected requiring
$Q^2> 1$ GeV$^2$, $0.1<y<0.9$, $W^2> 25$ GeV$^2$.
The hadrons used in the analysis have $P_{h \perp} >$ 0.1 GeV
and $z > 0.2$.
Following early suggestions by Collins and Artru~\cite{Artru:2002pua}, 
the asymmetries
have been extracted also for ``leading'' hadrons, selected as
the highest $z$ hadron with $z>0.25$. 
Also in this case, the asymmetries turned out to be
compatible with zero.

\begin{figure}[tb]
\begin{center}
\includegraphics[width=0.75\textwidth]
{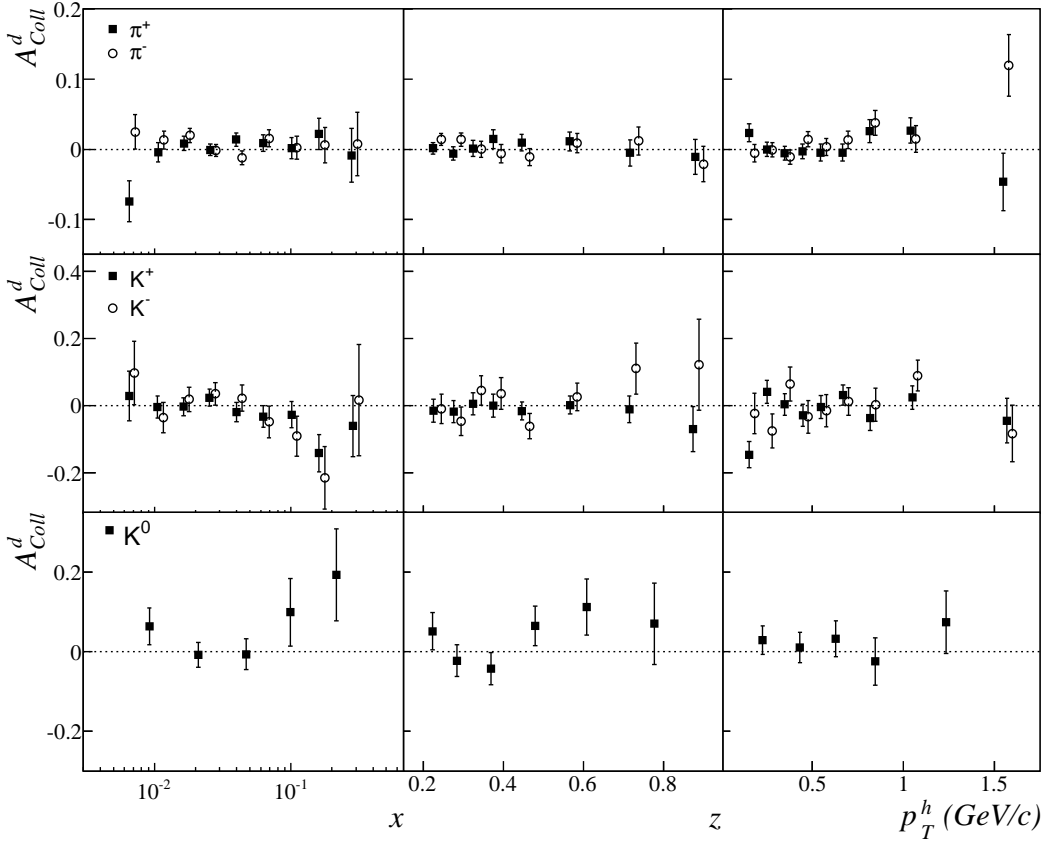}
\caption{COMPASS results for the Collins asymmetry for 
charged pions and kaons, and for neutral kaons
from all the data collected with the transversely polarised 
deuteron target~\cite{Alekseev:2008dn}.
The full and open circles refer to positive and negative hadrons
respectively.
The asymmetries are shown as function of $x$, $z$ and $P_{h \perp}$.
\label{fig:coll_cmp_dpk} 
}
\end{center}
\end{figure}

Again compatible with zero are the asymmetries
measured by COMPASS on deuteron for charged pions and for kaons.
The final results are shown in Fig.~\ref{fig:coll_cmp_dpk}
as functions of $x$, $z$ and $P_{h \perp}$ for charged pions (top),
charged kaons (middle) and neutral kaons (bottom).
They have been obtained using all the 2002-2004 data for K$^o$
and the 2003 and 2004 data for the charged hadrons, since in 2002
the RICH was not working during the transverse target polarisation
data taking.
The kinematical cuts are the same as for the unidentified
charged hadrons,
plus the requirement to have charged pion and kaon momenta
above 3.1 GeV and 10 GeV respectively, and 
below 50 GeV. The lower limit is due to the RICH threshold
and the upper correspond to 1.5 $\sigma$ mass separation between 
the two mass hypotheses.

The charged pion and kaon asymmetries have been corrected
for the purity of the particle identification, which,
anyhow, is quite good~\cite{Alekseev:2008dn}.
The overall systematic errors have been estimated to be negligible
with respect to the statistical errors.

The  quantitative interpretation of 
SIDIS data on Collins asymmetries and the extraction of the transversity 
distributions require external information on the other unknown quantity 
of the process, the Collins fragmentation function. 
This has been recently obtained 
from the inclusive hadron production in $e^+ e^-$  annihilation,
described in the next section.

\subsubsection{Collins effect in $e^+ e^-$ annihilation} 

The first indication of the Collins effect in $e^+ e^-$ annihilation 
came from a study of the DELPHI data on charged hadron production 
at the $Z^0$ pole \cite{Efremov:1998vd}, which gave an estimate 
of about 10 \% for the analysing power $\langle H_1^{\perp} \rangle/
\langle D_1 \rangle$, with a considerably uncertainty.   

More recently, data on azimuthal asymmetries in inclusive production 
of back-to-back hadrons from $e^+ e^-$ annihilation at $s \simeq 110$ 
GeV$^2$ have been 
presented by the Belle Collaboration \cite{Abe:2005zx,Seidl:2008xc}. 
In their analysis they use  
 both reconstruction methods described 
in Section~\ref{epluseminus}. The measured quantities  are
\be
R_{12} \equiv \frac{N (\phi_1 + \phi_2)}{\langle N_{12} \rangle}\,,
\;\;\;\; R_0 \equiv \frac{N (\phi_0)}{\langle N_0 \rangle}\,, 
\label{belle1}
\ee
where $N(\phi_1 + \phi_2)$ and $N(\phi_0)$ are the numbers of hadron pairs 
with $\cos (\phi_1 + \phi_2)$ and  $\cos 2 \phi_0$ 
modulation, respectively, 
and $\langle N_{12} \rangle$, $\langle N_{0} \rangle$ are 
the total average number of pairs. 
In terms of the asymmetries $a_{12}$ and $a_0$ defined 
in eqs.~\ref{a12} and~\ref{a0}, $R_{12}$ and $R_0$ are given by 
\be
R_{12} = 1 + a_{12} \, \cos (\phi_1 + \phi_2)\,, 
\;\;\;
R_0 = 1 + a_0 \, \cos 2 \phi_0\,. 
\label{belle2} 
\ee
In order to eliminate the contribution of gluon radiation which 
is insensitive to the charge of the hadrons and the acceptance effects,
the ratios of the normalised distributions for unlike-sign (U) 
hadron pairs over like-sign (L) hadron pairs are taken,  
$R_{12}^{\rm U}/R_{12}^{\rm L}$ and $R_{0}^{\rm U}/R_{0}^{\rm L}$. 
Focusing on the $\cos (\phi_1 + \phi_2)$ modulation, one finds
\be
\frac{R_{12}^{\rm U}}{R_{12}^{\rm L}} \simeq 1 + \cos (\phi_1 + \phi_2) 
 A_{12}^{\rm UL}(z_1, z_2) \,,
\label{belle4} 
\ee
with the asymmetry parameter $A_{12}^{\rm UL}$ given by 
\be
A_{12}^{\rm UL} = \frac{\sin^2 \theta}{1 + \cos^2 \theta} \, 
\left \{
\frac{\sum_a e_a^2 \left (H_{1}^{\perp, {\rm fav}\, [1]} \bar H_{1}^{\perp, 
{\rm fav}\, [1]} + H_{1}^{\perp, {\rm unf}\, [1]}  \bar {H}_{1}^{\perp, 
{\rm unf}\, [1]} \right )}{\sum_a e_a^2 \left ( D_{1}^{\perp, {\rm fav}\, 
[0]}  \bar D_{1}^{\perp, {\rm fav}\, [0]} 
+ D_{1}^{\perp, {\rm unf}\, [0]}  \bar D_{1}^{\perp, {\rm unf}\, [0]} 
\right) } - 
\frac{\sum_a e_a^2 H_{1}^{\perp, {\rm fav}\, [1]}  \bar H_{1}^{\perp, 
{\rm unf}\, [1]}}{\sum_a e_a^2 D_{1}^{\perp, {\rm fav}\, [0]}  
\bar D_{1}^{\perp, {\rm unf}\, [0]}}
\right \}\,, 
\label{belle5}
\ee
where the superscripts ``fav'' and ``unf'' denote, as usual,  the favored 
and the unfavored FF's, respectively. 
Another independent combination of these functions, given by 
the ratio $R_{12}^{\rm U}/R_{12}^{\rm C}$ of unlike-sign 
pairs over all charged (C) pairs, is also determined, following a suggestion  
of Ref.~\cite{Efremov:2006qm}. 
This quantity can be written as in eq.~(\ref{belle4}) with an 
asymmetry parameter $A_{12}^{\rm UC}$. 

An analysis similar to the one we have just sketched is performed  
with the $\cos 2 \phi_0$ method, leading   
to the ratios  $R_0^{\rm U}/R_0^{\rm L}$ and $R_0^{\rm U}/R_0^{\rm C}$, 
and to the asymmetry parameters $A_0^{\rm UL}$ and $A_0^{\rm UC}$.  
Notice that the $\cos (\phi_1 + \phi_2)$ 
and the $\cos 2 \phi_0$ analyses of the same events are not independent,
and thus cannot be included together in a fit.  

\begin{figure}[tb]
\begin{center}
\includegraphics[width=0.45\textwidth]
{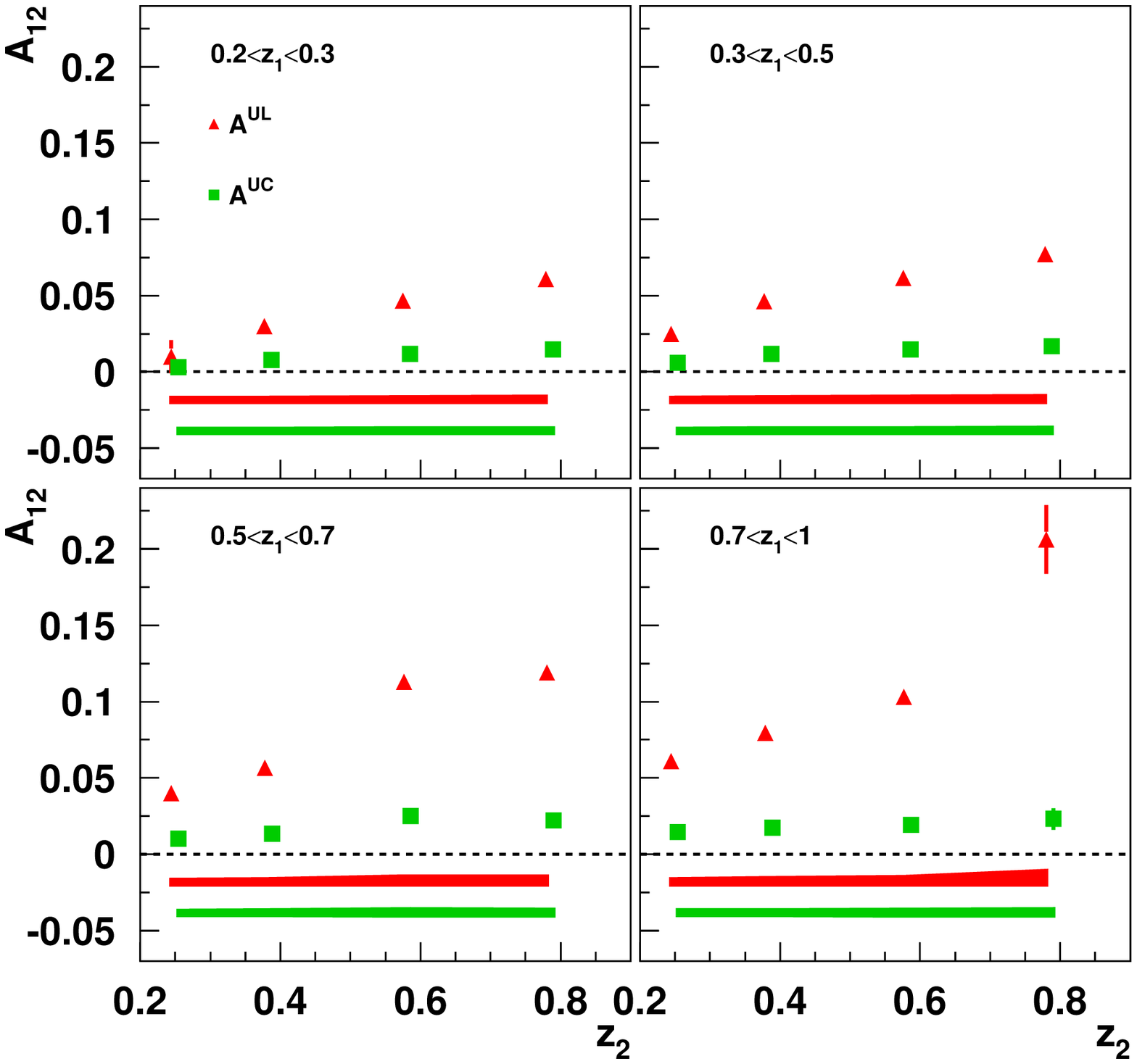}
\hspace{0.5cm}
\includegraphics[width=0.45\textwidth]
{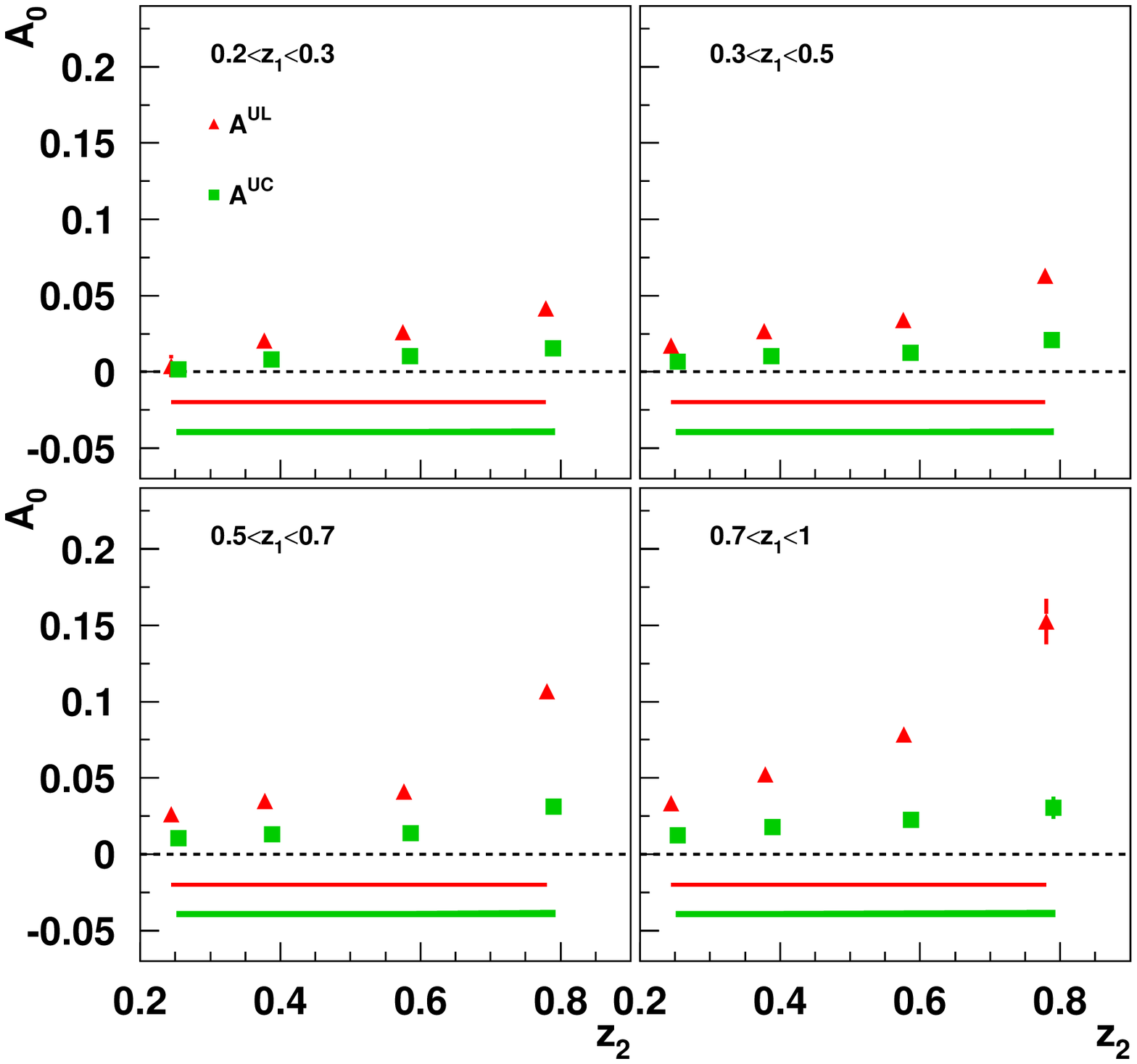}
\end{center}
\caption{\label{fig:belle} $A_0$ (left panel) and $A_{12}$ 
(right panel) as a function of $z_2$ for 
some $z_1$ bins~\cite{Seidl:2008xc}. 
The UL data are represented by triangles and their 
systematic uncertainty is given by 
the upper error band. The UC data are represented 
by the squares and their systematic uncertainty is given 
by the lower error band. 
}
\end{figure}

The Belle results are presented in Fig.~\ref{fig:belle}. 
A clear rising behaviour of the asymmetries with $z_1$ and $z_2$ is visible, 
suggesting a similar trend for the ratio $H_1^{\perp}/D_1$
(recall that the FF's are probed by Belle at the scale $Q^2 = s 
\simeq 110$ GeV$^2$).  
Due to the quadratic nature of the asymmetries in terms  
of the Collins function, the difference between $H_1^{\perp, {\rm fav}}$ and 
$H_1^{\perp, {\rm unf}}$ is poorly determined.
However, combining the Belle constraint on the product
 $H_1^{\perp, {\rm fav}} \cdot H_1^{\perp, {\rm unf}}$  with the 
SIDIS measurements of the Collins asymmetry,
the two Collins functions can be separately determined
\cite{Anselmino:2007fs,Anselmino:2008jk,Efremov:2006qm,Efremov:2008vf}. 

\subsubsection{Phenomenology of the Collins effect and determination 
of transversity}

The first phenomenological analysis of the SIDIS experimental results 
on the Collins asymmetry was performed by Vogelsang and Yuan 
\cite{Vogelsang:2005cs}, who assumed Soffer saturation 
of transversity, i.e., $\vert h_1 \vert = 1/2 (f_1 + g_1)$, 
to extract from the first HERMES  measurement \cite{Airapetian:2004tw}
the favoured and unfavoured Collins functions. Given the poor statistics 
of those data, the uncertainties on $H_1^{\perp, {\rm fav}}$ and 
$H_1^{\perp, {\rm unf}}$ were large. 

Efremov et al.\cite{Efremov:2006qm,Efremov:2008vf}  
analysed the same data with the transversity distributions taken  
from the chiral quark-soliton model \cite{Schweitzer:2001sr}.  
The resulting $H_1^{\perp}$ was shown 
to reproduce satisfactorily also the COMPASS deuteron 
data \cite{Alexakhin:2005iw,Ageev:2006da}, 
and to be compatible with the Collins function determined from 
the Belle data \cite{Abe:2005zx}. 
The main finding about $H_1^{\perp}$ 
supports the HERMES interpretation of their data, namely that
the favoured and unfavoured Collins functions 
are opposite in sign, and that $H_1^{\perp, {\rm unf}}$ 
is surprisingly large, being comparable in magnitude
to $H_1^{\perp, {\rm fav}}$   
at the average $Q^2$ scale (few GeV$^2$) of the HERMES experiment.
Moreover, it explains why the $\pi^0$ asymmetry is nearly zero.

A combined analysis of the first SIDIS data from HERMES and COMPASS, and 
of the $e^+ e^-$ Belle data, was performed by Anselmino et al. 
\cite{Anselmino:2007fs} and led to the first extraction of the $u$ and 
$d$-quarks transversity distributions. 
This analysis has been updated in Ref.~\cite{Anselmino:2008jk} using the 
preliminary HERMES data \cite{Diefenthaler:2007rj}, 
and the COMPASS \cite{Alekseev:2008dn}  
and Belle \cite{Seidl:2008xc} published data.
The fit does not include $K^{\pm}$ and 
$\pi^0$ data, nor the preliminary COMPASS results with the proton target. 
The TMD's are written as factorised functions  of $x$ and $k_T$, and their 
transverse-momentum dependence is assumed to have a Gaussian form. 
These two simplifying assumptions are supported by recent lattice studies 
\cite{Musch:2007ya,Musch:2008jd}. 
Fragmentation functions are parametrised in a similar way. 
Thus the unintegrated unpolarised quantities $f_1(x, k_T^2)$ 
and $D_1 (z, p_T^2)$ are expressed as 
\be
f_1(x, k_T^2) = f_1(x) \, \frac{\E^{- k_T^2/\langle k_T^2 \rangle}}{\pi 
\langle k_T^2 \rangle}\,, \;\;\;\;
D_1(z, p_T^2)  = D_1(z) \, \frac{\E^{- p_T^2/\langle p_T^2 \rangle}}{\pi 
\langle p_T^2 \rangle}\,.  
\label{unp_tmd} 
\ee
The resulting average transverse momentum of the hadron  is 
\be
\langle P_{h \perp} (z) \rangle = \frac{\sqrt{\pi}}{2} 
\, \sqrt{z^2 \langle k_T^2 \rangle + \langle p_T^2 \rangle}\,. 
\label{averageph}
\ee
The widths $\langle k_T^2 \rangle$ and $\langle p_T^2 \rangle$ 
are those obtained in \cite{Anselmino:2005nn}, namely:
$\langle k_T^2 \rangle = 0.25$ GeV$^2$ and 
$\langle p_T^2 \rangle = 0.20$ GeV$^2$.
The parametrisation of the  u- and d- transversity distribution 
and of the Collins function adopted  
in Refs.\cite{Anselmino:2007fs,Anselmino:2008jk} is 
\be
h_{1}^q(x,  k_T^2)  =  
 \mathcal{N}_q \, x^{a} (1 - x)^{b} \, 
 [f_{1}^q(x) + g_{1}^q(x) ] 
\, \frac{\E^{- k_T^2/\langle k_T^2 \rangle}}{\pi 
\langle k_T^2 \rangle}\,, \;\;\;
H_{1}^{\perp q}(z,  p_T^2)  =  
 \mathcal{N}^C_q \, z^{c} (1 - z)^{d} \, D_1(z) 
\, \E^{- p_T^2/\mu_C^2} \,, 
\label{ans12}
\ee
where $N_u, \, N_d, \, N_{fav}^C,  \, N_{unf}^C, \, a, \, b, \, c, \, d, 
\, \mu_C$ are free parameters. 
Antiquark contributions are ignored.
All the data used as input of the combined analysis, i.e. the COMPASS deuteron,
the HERMES proton and the Belle $e^+e^-$ data, are very well fitted.

Note that since the SIDIS and the $e^+e^-$ data are taken 
at very different $Q^2$ (up to 6 and 20 GeV$^2$ 
for HERMES and COMPASS respectively vs.  $\sim 10^2$ GeV$^2$), 
some assumption about the scale dependence of $H_1^{\perp}$  is required. 
The simple hypothesis adopted in all the phenomenological analyses
is that $H_1^{\perp}$ has the same evolution 
as $D_1$, so that the ratio $H_1^{\perp}/D_1$ is the same at all scales.  

\begin{figure}[tb]
\begin{center}
\includegraphics[width=0.40\textwidth,bb= 50 30 570 760,angle=-90]
{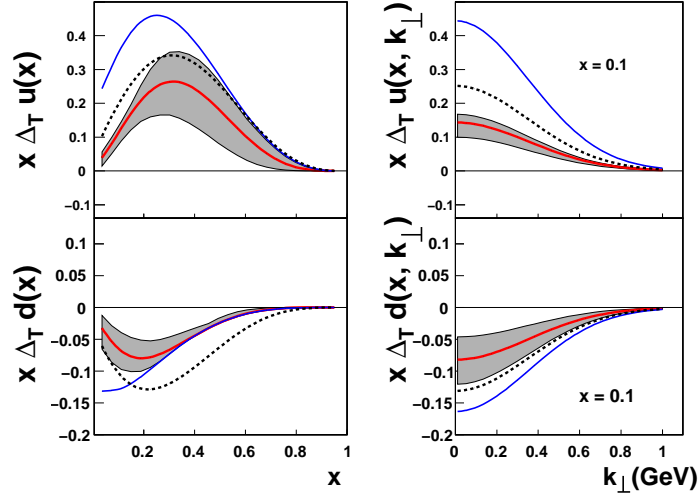} 
\end{center}
\caption{\label{fig_transv} 
The transversity distributions $x \Delta_T u \equiv x h_1^u$
and $x \Delta_T d \equiv x h_1^d$ 
at $Q^2 = 2.4$ GeV$^2$ 
from the fit of Ref.~\cite{Anselmino:2008jk}. The shaded bands 
represent the uncertainty of the fit. 
The solid 
line is the Soffer bound. Also shown are the helicity 
distributions (dashed curves).} 
\end{figure}

In Fig.~\ref{fig_transv} we show the transversity distributions 
extracted from the SIDIS measurements in Ref.~\cite{Anselmino:2008jk}. 
They have opposite sign, with $\vert h_{1}^d \vert$ 
smaller than $\vert h_{1}^u \vert$. 
While the magnitude and the intermediate-$x$ behaviour of $h_1$ 
are reasonably well constrained, its high-$x$ tail is 
not determined by the data.  
%
%
By integration of $h_1^q$, the tensor charges are found to be 
$\delta u = 0.54^{+0.09}_{-0.22}$ and  $\delta d = - 0.23^{+0.09}_{-0.16}$ 
at the reference scale $Q^2 = 0.8$ GeV$^2$.     
The value for $u$ is smaller than the predictions 
of lattice QCD \cite{Gockeler:2006zu} and of most models.  
However,  one should recall that the model scales are very small and 
usually just guessed, so the evolution from these scales to a 
higher $Q^2$ is affected by large uncertainties \cite{Wakamatsu:2008ki}.   

A general caveat about the phenomenological analyses 
of the Collins asymmetry is in order. 
They all ignore the  soft factor appearing in the TMD factorisation formulae. 
At tree level, this factor is equal to 1, but as $Q^2$ 
rises  it increasingly suppresses the asymmetry \cite{Boer:2001he}. 
This effect, which is not taken into account in the present fits, leads to 
underestimate the Collins function extracted from Belle data
and consequently to overestimate the transversity distributions 
obtained by using that function. 

Given the relatively large $Q^2$ values of the COMPASS 
 data in the quark valence region, where the HERMES data showed 
 the largest values of the Collins asymmetry, a comparison 
 of the COMPASS proton data with the HERMES results was regarded
 as very important to establish the leading twist nature of 
 the effect and to check  the robustness 
 of the overall picture and of the phenomenological analysis.

\begin{figure}[tb]
\begin{center}
\includegraphics[width=0.75\textwidth]
{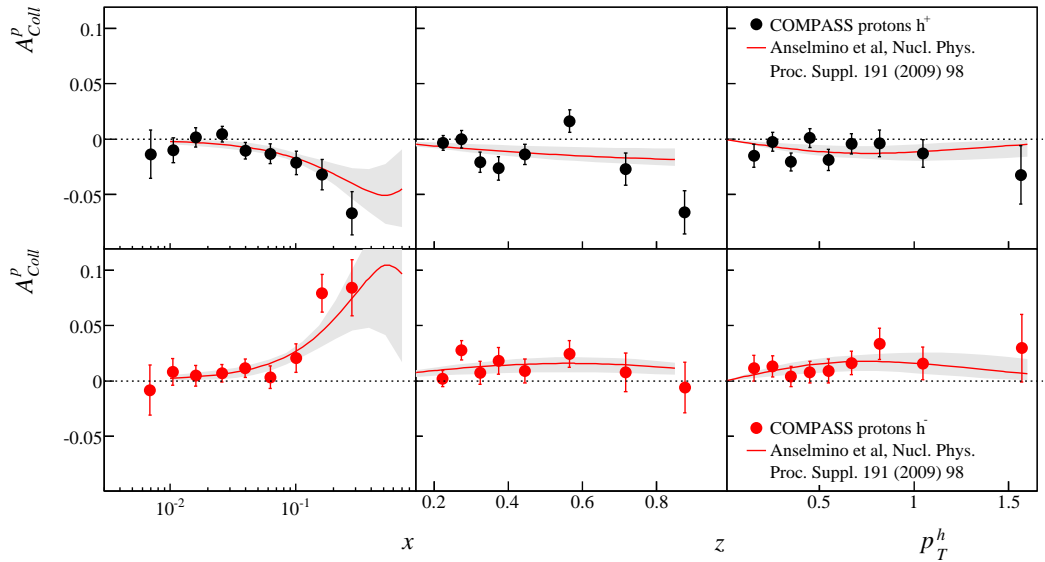}
\end{center}
\caption{COMPASS results for the Collins asymmetry for 
positive (top) and negative (bottom) hadrons
from the data collected with the transversely polarised 
proton target~\cite{Alekseev:2010rw}.
The error bars are statistical only. 
The curves show the calculation by Anselmino et al. based on the global fit
of Ref.~\cite{Anselmino:2008jk}.
}
\label{fig:coll_compass09_p_ah} 
\end{figure}

In 2008 COMPASS produced the first preliminary results~\cite{Levorato:2008tv} 
from part of the data collected in 2007 with the
transversely polarised proton target.
The results obtained using all the available statistics
have been published recently~\cite{Alekseev:2010rw} and are shown in 
Fig.~\ref{fig:coll_compass09_p_ah} for positive
(top) and negative (bottom) hadrons.
The applied kinematical cuts for DIS event and hadron selection are 
the same as for the deuteron data.
The systematic uncertainties have been evaluated 
to be about 0.5 the statistical one, and include a 5\% scale uncertainty 
due to the target polarisation measurement.
At small $x$, in the previously unmeasured region,
the asymmetries are compatible with zero.
At larges $x$ a clear signal develops both for positive and negative hadron.
The results are compatible with the HERMES results in the
overlap region, and in very good agreement 
with the values expected values expected on the basis the global 
fit by Anselmino et al.~\cite{Anselmino:2008jk}, shown by the curves
in fig.~\ref{fig:coll_compass09_p_ah}.
For the first time, the Collins asymmetry has
been measured to be different from zero at $Q^2 \sim 10$ GeV,
and in perspective these data should provide
information on the $Q^2$ evolution of the transversity and the Collins
function.

\subsubsection{Two-hadron asymmetries in SIDIS and $e^+ e^-$ 
annihilation}

The transverse spin asymmetry in the distribution of the azimuthal 
plane of  hadron pairs in the current jet of DIS have been measured 
by HERMES~\cite{Airapetian:2008sk} with the proton target and by 
COMPASS both with the deuteron
\cite{Sozzi:2007zz,Bradamante:2006mt,Martin:2007au,Vossen:2007mh,Massmann:2008zz}
and the proton~\cite{Wollny:2009eq} target.

Also for this asymmetry there are some 
differences between the analysis performed by the two experiments.
The first difference concerns the 
azimuthal angle of the two-hadron production plane:
with reference to the definitions introduced in Section~\ref{twoparticle},
the angle $\phi_R$ is used in the HERMES analysis,
while $\phi_R$ is used in the most recent COMPASS analysis 
(see f.i. Ref.~\cite{Joosten:2007zz}), following the suggestions
of Ref~\cite{Artru:2002pua,Artru:1995zu}. 
The two angles, however, coincide in the $\gamma^* N$ system.
Also, HERMES measures the amplitude of the modulation 
in the angle $\phi_R+\phi_S$, while COMPASS measures the modulation
in $\phi_r+\phi_S+\pi$, in line with the definition of the Collins angle.
Thus the asymmetries measured by the two experiments
are expected to have opposite sign.

Other  differences are 
in the treatment of the kinematical factor $D_{NN}$, as in the Collins
case, and in the kinematical cuts.
Finally, the HERMES extraction of the asymmetries is based on a
Legendre expansion of the dihadron fragmentation functions, as
suggested in Ref.~\cite{Bacchetta:2003vn}.

In HERMES~\cite{Airapetian:2008sk} the events are selected requiring
$Q^2> 1$ GeV$^2$, $0.1<y<0.85$, $W^2> 10$ GeV$^2$ and 
the missing mass larger than 2 GeV to avoid contributions from
exclusive two pion production.
Also a minimum pion momentum of 1 GeV is required for pion identification.
For each event all the possible combinations $\pi^+\pi^-$
have been used, labelling as 1 the positive particle.

If one adopts the partial wave expansion of Ref.~\cite{Bacchetta:2002ux}, 
a convenient observable is the  asymmetry
\bq
A_{UT}^{\sin (\phi_R + \phi_S) \sin \theta} (x_B, y, z_h, M_h^2) 
&\equiv& 2 \, \frac{\int \! \D \cos \theta \, \int \! \D \phi_R \, 
\int \! \D \phi_S \, 
 \sin (\phi_R + \phi_S) \, \D^7 \sigma/\sin \theta}{ \int \! \D \cos \theta \, 
\int \! \D \phi_R \, \int \! \D \phi_S \, \D^7 \sigma} 
\nonumber \\
&=& -   \frac{1}{2} \, \hat{D}_{NN}(y)  \, 
\sqrt{1 - \frac{4 m^2}{M_h^2}} \frac{\sum_a e_a^2 \, h_1^a (x_B) 
H_{1a}^{\newangle,  sp}(z_h, M_h^2)}{\sum_a e_a^2 \, f_1^a (x_B) 
\, D_{1a}^o (z_h, M_h^2)}\,, 
\label{twoasym3}
\eq
which selects the interference fragmentation function $H_1^{\newangle, sp}$. 
Thus, in HERMES, the spin asymmetries defined in eq.~(\ref{eq:elssa}) and
divided by the target polarisation, have been measured
in each ($\phi_R, \, \theta'$) bin, and fitted with the function
$a \sin\phi_R \sin\theta' /[1+b(3 \cos^2 \theta'-1)]$.
Here  $\theta'=||\theta-\pi/2|-\pi/2|$, $a$ is the free
parameter and $b$ has been varied to take into account the unknown
dependence on $\cos^2 \theta$ of the unpolarised dihadron 
fragmentation function.
where $A_{U\perp}^{sin\phi_R^H sin \theta} = a$ is the free
parameter and $b$ is varied to take into account the unknown
dependence on $\cos^2 \theta$ of the unpolarised dihadron 
fragmentation function.
The published asymmetries $A_{U\perp}^{sin\phi_R^H sin \theta}$ are 
the fitted values of $a$.
The final results from the data collected from 2002 to 2005
with the proton target~\cite{Airapetian:2008sk} are shown versus $M_{h}$,
$x$ and $z=z_1+z_2$.
The bottom plots give the average values of the other two variables that were
integrated over.
%
\begin{figure}[tb]
\begin{center}
\includegraphics[width=0.75\textwidth]
{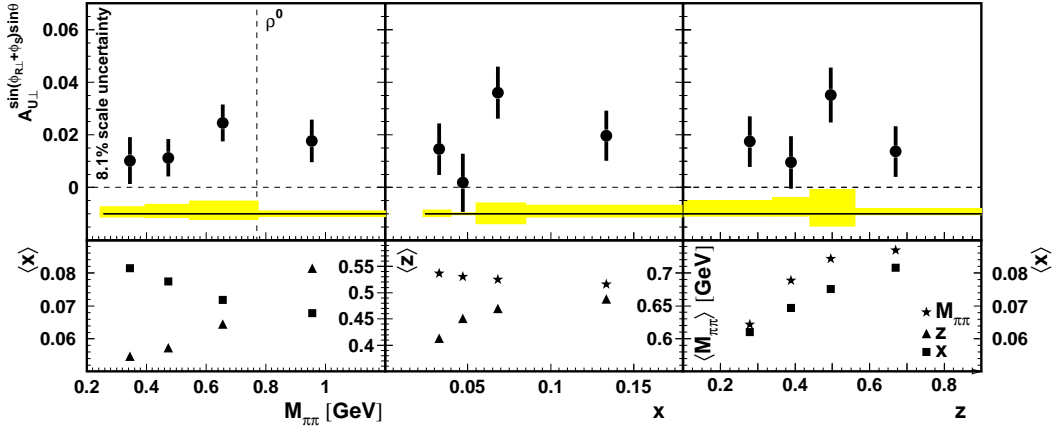}
\end{center}
\caption{The two hadron asymmetries versus $M_{h}$, 
$x$, and $z$
measured by the HERMES experiment with the transversely polarised
proton target. 
The bottom panels show the average values of the variables that were 
integrated over.  
The bands represent the systematic uncertainty.
}
\label{fig:2h_hermes_08} 
\end{figure}
The asymmetries in bins of $x$ and $z$ have been evaluated requiring
0.5 $< M_{h} < 1.0$ GeV.
The systematical errors, which also include the uncertainty due to the
value of $b$ and acceptance effects, are given by the band centred at -0.01.
The scale uncertainty due to the target polarisation uncertainty was about 8\%.
As apparent from the figure, the asymmetries are different from
zero, indicating that the spin-dependent part of the dihadron
fragmentation function is different from zero.
Also, the data show clear trends in each of the 
three kinematical variables, and the signal have the same sign
and smaller values than that of the Collins asymmetry.

In the COMPASS analysis, the event selection is similar to that
described for the single hadron asymmetries, namely only events with
$Q^2> 1$ GeV$^2$, $0.1<y<0.9$, $W^2> 25$ GeV$^2$ are accepted.
Only  hadrons with $z_{1,2}>0.1$ are used in the analysis
and for each pair of hadrons $z<0.9$ is required.
The cuts $x_F>0.1$ and $r_T>7$ MeV are also applied in the more recent 
analysis.
In the extraction of the asymmetries no dependence on $\theta$
is taken into account.
This is justified even in the framework of Ref.~\cite{Bacchetta:2003vn},
since in the COMPASS kinematics the $\sin \theta$
distribution is strongly peaked at one 
($<\sin\theta >=0.94$) and the $\cos \theta$
distribution is symmetric around zero.
The asymmetries are extracted from the azimuthal distributions
using the same methods described for the Collins asymmetry.
In particular the results from the transversely polarised
target data were obtained with the ``ratio product method''.
 
Preliminary results using  the data collected
with the deuteron target have been produced looking at different 
selections for the hadron pair.
The asymmetries have been evaluated using all the combinations of 
the positive and negative selected hadrons~\cite{Sozzi:2007zz};
taking only the two hadrons with higher transverse momentum with
different charge combinations~\cite{Bradamante:2006mt};
taking only the two hadrons with the highest $z_i$, again for
the different charge combinations~\cite{Bradamante:2006mt};
taking all the possible combinations of identified charged pions and 
kaons~\cite{Vossen:2007mh};
taking only the charged pions and kaon with higher 
$z_i$~\cite{Massmann:2008zz}.
All the corresponding asymmetries turned out to be compatible
with zero and no clear signal could be seen.
This result could have been expected on the basis of the
null result on the measured Collins asymmetry.

\begin{figure}[tb]
\begin{center}
\includegraphics[width=0.75\textwidth]
{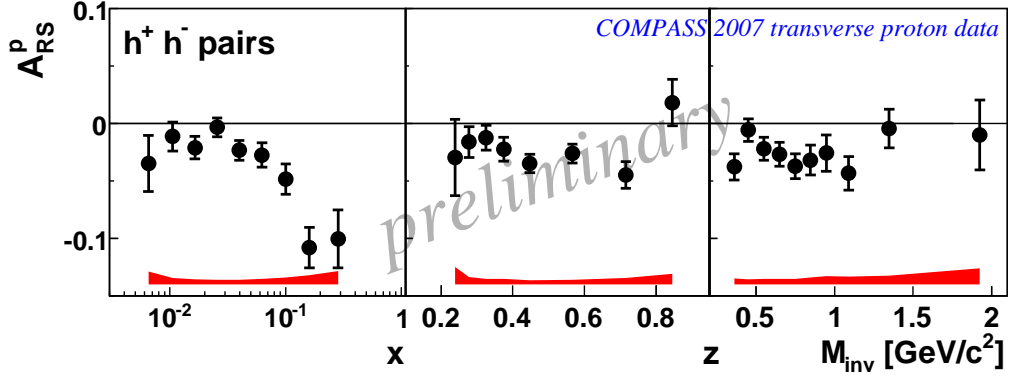}
\end{center}
\caption{The COMPASS proton preliminary results for the two hadron asymmetries 
versus $x$, $z$ and 
$M_{h}$~\cite{Wollny:2009eq}. 
}
\label{fig:2h_compass_p_ch} 
\end{figure}

Also for this SSA there was a strong
interest for the COMPASS measurement with the proton target, since
not much variation was expected
going from the HERMES to the COMPASS energy.
The preliminary results from all the data collected in 
2007~\cite{Wollny:2009eq} are shown in Fig.~\ref{fig:2h_compass_p_ch} 
versus $x$, $z$, and $M_{h}$ for all the combinations
of positive and negative hadrons, selected as in the deuteron case.
The systematic uncertainties have been evaluated to be 
not larger than one half the statistical errors.
%
%
As can be seen, the asymmetries are clearly different
from zero. 
They have the same sign of the Collins asymmetries on proton,
the behaviour in $x$ is very similar and
the absolute values at large $x$ are even larger.
Also, there is no clear indication for a structure, in particular 
as a function of the invariant mass.

Very recently, preliminary results   on the Artru-Collins asymmetries
described in Section~\ref{epluseminus} have
been produced by BELLE~\cite{Vossen:2009xz}.
They show asymmetries different from zero, which depend on $z_h$ and $M_h$.
Both these results and the COMPASS proton results are fresh, and no
attempt to perform global analysis has been done yet.

On the contrary,
The HERMES data on $A_{UT}^{\sin (\phi_R + \phi_S) \sin \theta}$ 
have been recently analysed \cite{Bacchetta:2008wb} in terms 
of the model for $D_1 (z, M_h^2)$ and $H_{1,sp}^{\newangle}$ 
developed in Ref.~\cite{Bacchetta:2006un}. 
It turns out that, in order to get a fair description of the 
HERMES asymmetry, using for the  transversity 
distributions the parametrisation of Ref.~\cite{Anselmino:2008jk},   
the interference fragmentation function $H_{1, sp}^{\newangle}$  
of Ref.\cite{Bacchetta:2006un} must be reduced by a factor 3. 
The predicted $M_h^2$ dependence 
of the analysing power $H_{1, sp}^{\newangle}/D_1$, 
with the typical bumps at the $\omega$ and $\rho$ masses,
is not incompatible with the data, within their large errors.  
However, the interference function 
that fits HERMES data largely undershoot the asymmetry measured 
by COMPASS and, once evolved to high $Q^2$  
by means of the DGLAP equations for dihadron 
fragmentation functions \cite{Ceccopieri:2007ip}, yields values 
of the Artru-Collins asymmetry which are much smaller than those  
found by Belle \cite{Vossen:2007mh}. 
Moreover, the invariant mass behaviour predicted in 
Ref.~\cite{Bacchetta:2008wb} does not match the Belle findings. 
A reconsideration  of interference fragmentation functions in the light 
of the recent HERMES, COMPASS and Belle measurements seems to be 
necessary.

\subsubsection{$\Lambda$ polarisation}

As shown in Section~\ref{transvbar}  detecting a transversely polarised
spin $1/2$ hadron in the final state of a semi-inclusive DIS process
with a transversely polarised target
probes the leading-twist combination $h_1 (x)\, H_1(z)$. 
The typical example of such processes 
is $\Lambda$ (or $\overline{\Lambda}$) production \cite{Jaffe:1996wp}. 
The $\Lambda$ polarisation is in fact easily measured by studying the 
angular distribution of the $\Lambda\to{p}\pi$ decay.

From the phenomenological viewpoint, the
main problem is that, in order to compute the $\Lambda$ polarisation, one
needs to know the fragmentation functions $H_1^{\Lambda/q}(z)$, which are 
completely unknown. 
Predictions for $\mathcal{P}_T^{\Lambda}$ have been presented
by various authors \cite{Anselmino:2000ga,Ma:2001rm,Yang:2001sy} 
and span a wide range of values. 
In the calculation of Ref.~\cite{Anselmino:2000ga}, where 
the transversity distributions are assumed to saturate the Soffer bound, 
$\mathcal{P}_T^{\Lambda}$ lies in the interval $\pm 10 \%$
at $x \sim 0.1$. 
In particular, in the SU(6) non relativistic model, the entire spin of 
the $\Lambda$ is carried by the strange quark and one therefore 
expects a polarisation close to zero, due to the smallness of $h_1^s$.  

\begin{figure}[tb]
\begin{center}
\includegraphics[width=0.65\textwidth,bb=0 310 590 540]
{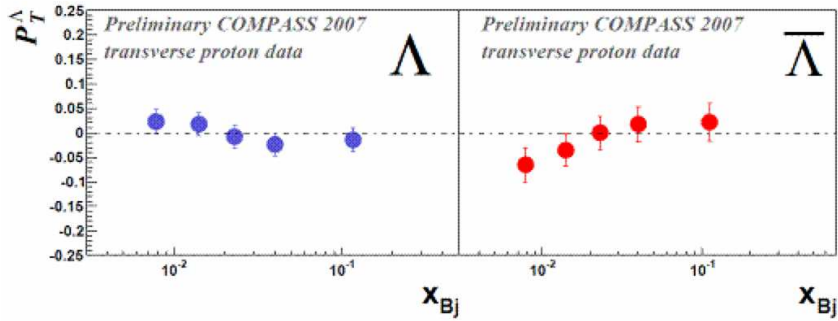}
\end{center}
\caption{COMPASS results for $\Lambda$ and $\bar{ \Lambda}$
polarisation versus $x$ from the data collected with the transversely
polarised proton target~\cite{Negrini:2009zz}.
}
\label{fig:lambda_compasss_p} 
\end{figure}

The only existing results for this experimentally 
difficult channel are from the  COMPASS experiment.
On deuteron, the preliminary analysis gave polarisation values for
$\Lambda$ and $\bar{\Lambda}$ compatible with zero
within the non negligible statistical errors~\cite{Ferrero:2007zz}.
The same indication comes from the polarisation measured with
the proton target.
Fig.~\ref{fig:lambda_compasss_p} shows $\Lambda$ and $\bar{\Lambda}$  
polarisation versus $x$ from part of the 2007 proton 
data~\cite{Negrini:2009zz}.
%

As in the case of the Collins function, independent information on 
$H_1$ can be obtained from $e^+ e^-$ annihilation 
\cite{Artru:1992jg,Chen:1994ar,Contogouris:1995xc}. 
The specific process 
is back-to-back $\Lambda \overline{\Lambda}$ inclusive production, 
$e^+ e^- \to \Lambda \, \overline{\Lambda} \, X$, 
with the hyperon and the anti-hyperon 
decaying into $p \pi^-$ and $\bar p \pi^+$, respectively. 
It was shown in Ref.~\cite{Chen:1994ar} that the unpolarised 
cross section of this process contains an azimuthal 
modulation proportional to $H_1^{\Lambda/q} 
H_1^{\overline{\Lambda}/\bar q} = (H_1^{\Lambda/q})^2$, which is selected 
by the asymmetry between the number of $p \bar p$ pairs 
on the same side of the scattering plane and the number 
of pairs on opposite sides. 
The measurement of such asymmetry was attempted by the ALEPH Collaboration at 
LEP \cite{Buskulic:1996vb}, but the scarce sensitivity of data 
did not allow getting any significant result. 

\subsection{Accessing TMD distributions: T-odd leading twist functions}
\label{sec:results_Todd}

SIDIS represents, at this moment, the best   
source of knowledge on the two $T$-odd distribution 
functions: the Sivers function, involved in transversely polarised 
SIDIS, and the Boer-Mulders function, which generates 
asymmetries in unpolarised SIDIS.  
Some information on the Boer-Mulders distribution comes also from 
unpolarised Drell-Yan processes. 
DY can probe the Sivers function as well, if one of the two
colliding hadrons is transversely polarised, but this 
reaction so far has not been experimentally explored. 

\subsubsection{Sivers effect in SIDIS}
\label{siv_sidis}

As in the case of the transversity PDF's, 
the only measurements which give today a clean access to the Sivers function 
are the SSA's in SIDIS on transversely polarised targets.
The relevant  quantity is the so-called Sivers asymmetry 
$A_Siv=F_{UT}^{\sin(\phi_h-\phi_S)}/F_{UU}$, where 
$F_{UT}^{\sin(\phi_h-\phi_S)}$ and $F_{UU}$ are the structure functions 
introduced in eq.(\ref{sidiscs}).
This SSA couples $f_{1T}^{ \perp}$ to the unpolarised fragmentation 
function $D_1$.
Till now it has been measured only by the COMPASS and the HERMES experiments.
The same data  as for the measurement of the Collins asymmetry have been
used, and the same kind of analysis, described in Section
\ref{sec:sidis_kinem}, has been performed.
In this case the relevant modulation is that
in the azimuthal angle $(\phi_h-\phi_S)$, and, at variance with the 
Collins case, the Sivers asymmetry is defined in the same way and
with the same sign in the COMPASS and in the HERMES experiments.
It has to be noted that in the HERMES papers the Sivers asymmetry 
is indicated as $2< \sin (\phi_h-\phi_S)>$.

The first results on the Sivers asymmetry have been produced by the
HERMES experiment using the data collected in 2002 with the transversely
polarised proton target~\cite{Airapetian:2004tw}, and showed large positive
values for the $\pi^+$ and $K^+$
while for $\pi^-$ and $K^-$ the asymmetries were compatible with zero.
The preliminary results from  the 2002-2005 data~\cite{Diefenthaler:2007rj} 
and the final results published recently~\cite{Airapetian:2009ti}
confirmed with better statistical precision the previous measurement.
%
\begin{figure}[tbh]
\begin{center}
\includegraphics[width=0.48\textwidth]
{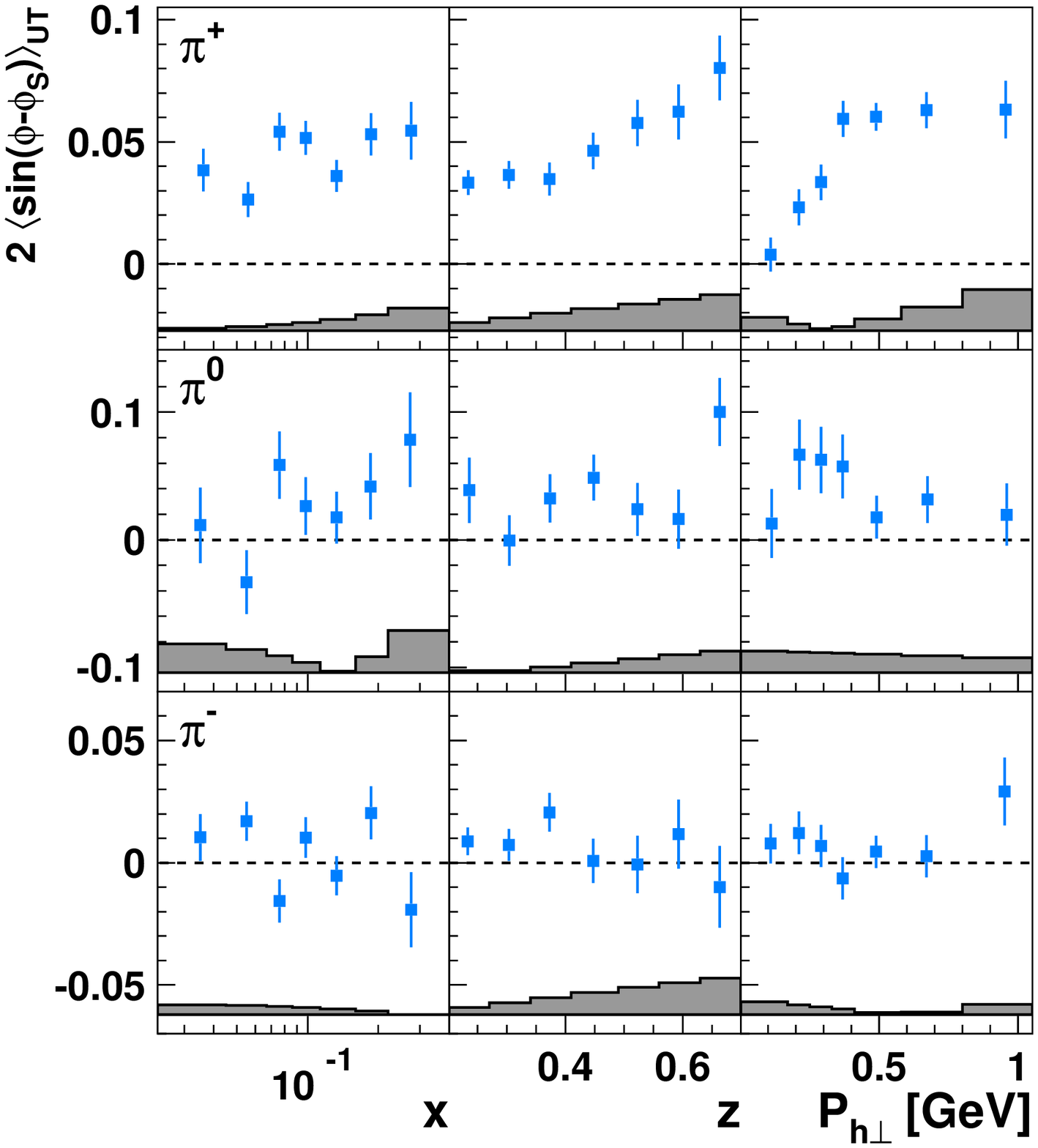}
\includegraphics[width=0.48\textwidth]
{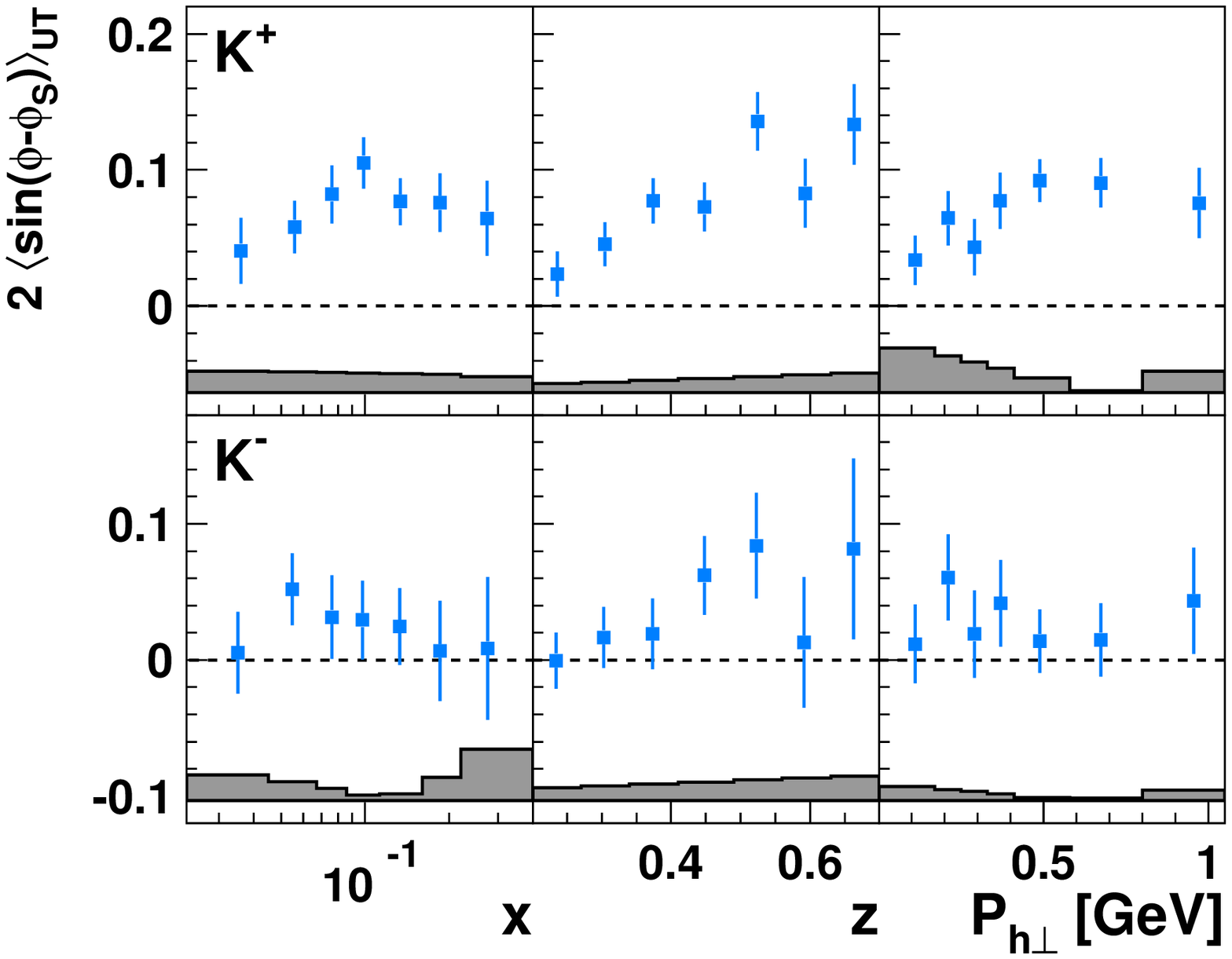}
\hfill
\end{center}
\caption{HERMES results for the Sivers asymmetries on proton from
the 2002-2005 data~\cite{Airapetian:2009ti} as functions
of $x$, $z$ and $P_{h \perp}$, for
pions (left) and charged kaons (right).
}
\label{fig:siv_hermes09_pk} 
\end{figure}
Fig.~\ref{fig:siv_hermes09_pk} shows the final Sivers asymmetries
versus $x$, $z$ and $P_{h \perp}$
for pions and charged kaons.
The error bars are statistical only.
The systematic uncertainty is shown by the bands
and include possible contributions due to the
longitudinal component of the target spin as well as
acceptance  and smearing effects, radiative effects,
and effects due to the hadron identification.
The additional scale error due to the target polarisation uncertainty
is quoted to be 7.3\%.
As can be seen, the $\pi^+$ asymmetry is of the order of 5\% almost over
all the measured $x$ range, compatible with zero
for $\pi^-$, and slightly positive for $\pi^0$. 
Slightly positive signals can be seen also for
$K^-$, while the values for $K^+$ are quite large, up to 10\%.
The pion results can naively be explained  in the framework of the
quark model.
With the assumptions used in eq. (\ref{eq:collpm}), they
can be understood as due to a the $d$-quark Sivers function of roughly 
twice the size of that of the $u$ quark, and of opposite sign.
The $K$ results are more difficult to be explained, and
further studies on the difference between $\pi^+$ and $K^+$
results are quoted in~\cite{Airapetian:2009ti}.
Also, the asymmetry in the  difference of the
distributions of $\pi^+$ and $\pi^-$, which should be
more related to the $u$ and $d$ quark Sivers functions in
the valence region, has been measured.
Finally, a study of the $Q^2$ dependence has  been performed,
without finding clear effects indicating sizable $1/Q^2$ effects.
It has to be reminded, however, that the mean values of $Q^2$ are
all in the range between 1 GeV$^2$ and about 7 GeV$^2$.

\begin{figure}[tbh]
\begin{center}
\includegraphics[width=0.7\textwidth]
{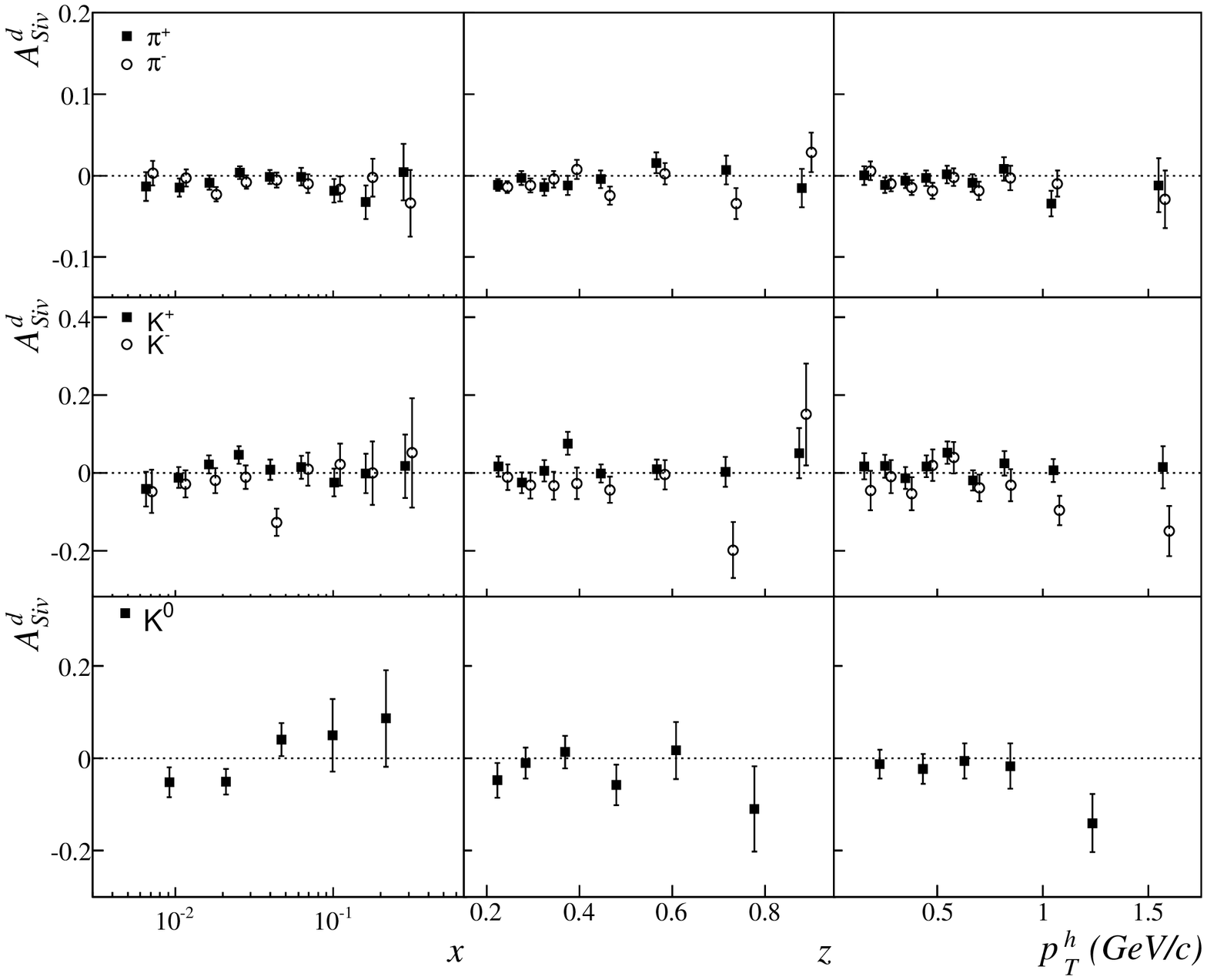}
\end{center}
\caption{
Final COMPASS results for the Sivers asymmetry 
on deuteron against $x$, $z$ and $P_{h \perp}$
for  charged pions and kaons~\cite{Alekseev:2008dn}. 
}
\label{fig:siv_compass09_d_pk} 
\end{figure}

As in the Collins case, COMPASS has  measured for the first time the
Sivers asymmetry on the deuteron.
First results for charged hadrons from the 2002 data were
published in 2005~\cite{Alexakhin:2005iw}, and
later on final results from all the collected
deuteron data have been produced for charged hadrons~\cite{Ageev:2006da}
and for identified pions and kaons~\cite{Alekseev:2008dn}.
The measured Sivers asymmetries for identified hadrons are shown 
in Fig.~\ref{fig:siv_compass09_d_pk}.
%
The errors are statistical only. 
The quoted systematic errors are negligible with respect to the 
statistical ones.
All the measured values are compatible with zero within 
the small statistical errors.
This result for pions can  be interpreted naively in the
framework of the parton model~\cite{Ageev:2006da} as due to
opposite Sivers functions for the $u$ and  $d$ quarks.

After the first phenomenological study \cite{Efremov:2004tp} of 
preliminary HERMES data on the $P_{h \perp}$--weighted Sivers asymmetry 
\cite{Seidl:2004dt}, various theoretical 
groups \cite{Anselmino:2005nn,Vogelsang:2005cs,Collins:2005ie,Collins:2005wb} 
extracted the Sivers distributions (or their moments)  
from the HERMES measurement of $A_{UT}^{\sin (\phi_h - \phi_S)}$ 
\cite{Airapetian:2004tw}. 
The fits were then extended \cite{Anselmino:2005nn,Collins:2005wb}  
to higher-precision HERMES preliminary data \cite{Diefenthaler:2005gx} and 
to the COMPASS deuteron data \cite{Alexakhin:2005iw}. 
A comparison of the results of these analyses \cite{Anselmino:2005an} shows   
a certain qualitative agreement and some common features: 
a negative $f_{1T}^{\perp u}$ and a positive $f_{1T}^{\perp d}$,
as predicted by the the impact-parameter approach \cite{Burkardt:2002ks}, 
with comparable magnitudes, as expected in the large-$N_c$ 
limit \cite{Pobylitsa:2003ty} or in chiral models \cite{Drago:2005gz}. 

\begin{figure}[tbh]
\begin{center}
\includegraphics[width=0.30\textwidth,angle=-90]
{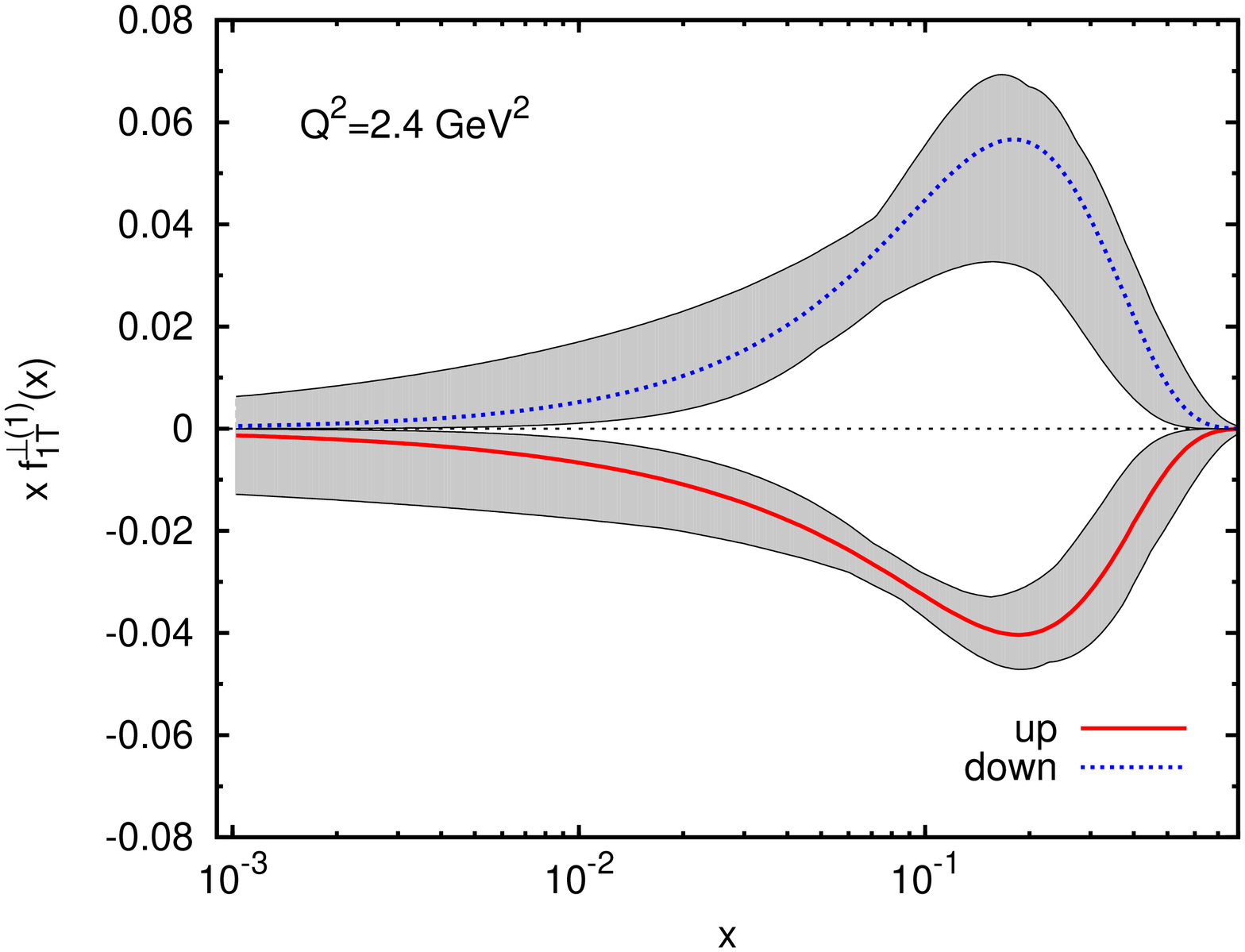}
\end{center}
\caption{\label{fig_ans_siv} 
The first moments of  the Sivers $u$ and $d$ quark distributions 
from the fit of Ref.~\cite{Anselmino:2008sga}. 
}   
\end{figure}

More recent fits \cite{Anselmino:2008sga,Arnold:2008ap} 
take into account the new HERMES \cite{Airapetian:2009ti}   
and COMPASS deuteron data \cite{Alekseev:2008dn}. 
The surprisingly large values of the $K^+$ asymmetry 
and of the $K^+ - \pi^+$ difference call for a careful 
reconsideration of the sea.  
The Sivers functions are factorised in $x$ and $k_T$, with a Gaussian 
dependence on $k_T$. 
Taking as an example the parametrisation  of Ref.~\cite{Anselmino:2008sga}, 
the functional form of $f_{1T}^{\perp q}$ is 
\be
f_{1T}^{\perp q} (x, k_T^2) = N_q 
x^{\alpha} (1 - x)^{\beta} f_1^q (x) \, \E^{- k_T^2/\mu_S^2} \,, 
\label{sivers_ans} 
\ee
where $N_q, \alpha_q, \beta$ and $\mu_S$ are free parameters,
the last two being taken to be the same for all flavors.   
It turns out that the exponent $\beta$ governing 
the high-$x$ tail of the distributions is not well 
constrained by the data which extend up to $x \sim 0.3$. 
The overall quality of the fit is rather good
and the first moments of the extracted Sivers functions for the $u$ 
and $d$ quarks displayed in Fig.~\ref{fig_ans_siv}. 
%
%
As expected, a non negligible strange sea is required to reproduce 
the $K^+$ data. 
The $\bar u$ and $\bar d$ distributions are more uncertain, even in their sign.

The two fits of Ref.~\cite{Anselmino:2008sga} and Ref.~\cite{Arnold:2008ap} 
indicate that the Burkardt sum rule is approximately saturated by 
the quark and antiquark distributions, thus little room is left for the
gluon Sivers component and the orbital motion is restricted to 
valence quarks. 

\begin{figure}[tb]
\begin{center}
\includegraphics[width=0.8\textwidth]
{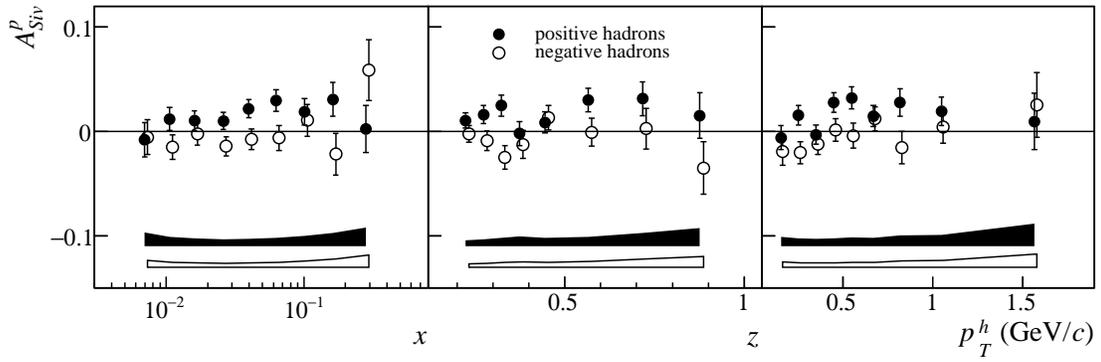}
\end{center}
\caption{COMPASS results for the Sivers asymmetry on proton
against $x$, $z$ and $P_{h \perp}$
for  positive (closed points) and negative hadrons (open
points)~\cite{Alekseev:2010rw}. 
}
\label{fig:siv_compass08_p_ch} 
\end{figure}

In this situation, in which all the existing experimental data
on the Sivers asymmetry could be explained coherently, the
COMPASS measurements with the transversely polarised proton target
came as a surprise.
The preliminary results for charged hadrons from part of the 2007 data 
were released in 2008~\cite{Levorato:2008tv} and 
showed small asymmetries both for positive and negative hadrons,
compatible with zero within the statistical errors.
The analysis of the complete set of data was concluded only 
recently~\cite{Alekseev:2010rw}, and the final results are shown in 
Fig.~\ref{fig:siv_compass08_p_ch}.
%
For positive hadrons the data indicate small positive values,
up to about 3\% in the valence region.
These values are somewhat smaller than but still compatible
with the ones measured by HERMES at smaller $Q^2$.
The systematic errors is estimated to be 0.8$\sigma_{stat}$,
plus a $\pm0.01$ systematic uncertainty in the absolute scale
due to a systematic difference in the mean values of the
asymmetry which was found between the first and the second parts 
of the 2007 run.

Given the importance of the Sivers function the COMPASS
Collaboration has decided to remeasure SSA's on NH$_3$
in 2010 with an improved spectrometer and better statistics.

\subsubsection{Boer Mulders effect in SIDIS}
\label{sec:results_BoerM_SD}

The existence of $\cos \phi_h$ and $\cos 2 \phi_h$ 
asymmetries in unpolarised SIDIS is experimentally well established.  
They have been investigated many years ago by the EMC and ZEUS
experiments \cite{Aubert:1983cz,Arneodo:1986cf,Breitweg:2000qh}
in the large $Q^2$ region where they are dominated by perturbative QCD effects.
It is indeed known that these asymmetries are perturbatively 
generated by gluon radiation 
\cite{Georgi:1977tv,Mendez:1978zx,Konig:1982uk,Chay:1991nh,Nadolsky:1999kb,Nadolsky:2000kz} 
and at high $Q^2$ and high $P_{h \perp}$ they are  dominated by these effects.
The recent HERMES \cite{Giordano:2009hi}, COMPASS \cite{Kafer:2008ud} 
and CLAS \cite{Osipenko:2008rv} results cover a kinematical region 
(small $Q^2$ and $P_{h \perp} < 1$ GeV) where gluon emission is negligible 
\cite{Barone:2008tn} and the asymmetries can be described 
in terms of TMD's and higher-twist contributions.  

The unpolarised azimuthal asymmetries have been measured 
by the COMPASS and HERMES Collaborations using part of the data collected
with polarised targets and combining them in order to cancel
the net target polarisation.

\begin{figure}[tbh]
\begin{center}
\includegraphics[width=0.65\textwidth]
{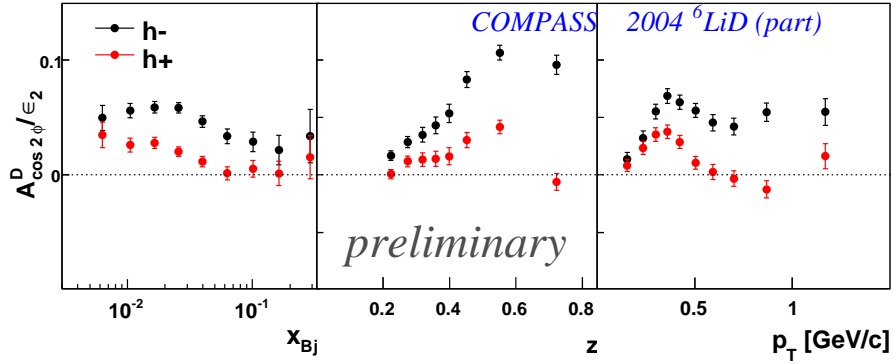}
\end{center}
\caption{
The preliminary results for the $\cos 2\phi$ azimuthal asymmetries 
versus $x$, $z$ and $P_{h \perp}$ for positive
and negative hadrons measured by the COMPASS experiment on $^6$LiD
~\cite{Kafer:2008ud}.
}
\label{fig:bmc_phi} 
\end{figure}

The COMPASS preliminary results~\cite{Kafer:2008ud} have been 
obtained from part of the data collected
in 2004 with the $^6$LiD target polarised both transversely and longitudinally
with respect to the muon beam direction. 
The data have been combining in such a way to cancel
the net target polarisation.
To reduce the acceptance effects, only the events with a vertex
in the downstream target cell have been used, and the statistics for each
polarisation orientation chosen in such a way to have a zero net polarisation.
The events are selected requiring the usual cuts
$Q^2 > 1$ GeV$^2$, mass of the final hadronic state 
$W > 5$ GeV, $0.1 < y < 0.9$.
For the final state hadrons it is  required that $ 0.2 < z  < 0.85$ and
$0.1 < P_{h \perp} < 1.5$ GeV.
The data have been binned alternatively in $x$, $z$ and $P_{h \perp}$
and in each bin the measured $\phi_h$
distribution has been corrected for the acceptance of the spectrometer,
evaluated with a Monte Carlo simulation.
Since the COMPASS beam is longitudinally polarised,
a $\sin \phi_h$ modulation is also possible, so 
the corrected $\phi_h$ distributions have been fitted
with a function containing the $\cos \phi_h$, the $\cos 2\phi_h$ 
and the $\sin \phi_h$ modulations.
This last amplitude turned out to be always compatible with zero.
The preliminary results for the $\cos 2\phi_h$ 
asymmetries are shown in Fig.~\ref{fig:bmc_phi} as functions of $x$,
$z$ and $P_{h \perp}$ for positive and negative hadrons.
The measured asymmetries have been corrected for the corresponding
$y$ dependent kinematical factor appearing in the cross-section.
%
The errors are statistical only.
The systematic errors are of the order of 2\% in both cases and are largely 
dominated by the acceptance correction.
Both the $\cos \phi_h$ and the $\cos 2\phi_h$ asymmetries are quite large, 
with a strong dependence on the kinematical variables.
Also, for the first time,  the asymmetries have been produced separately
for positive and negative hadrons and the measured
difference points to different contributions 
of the $u$ and $d$ quarks to the underlying mechanisms.

\begin{figure}[tbh]
\begin{center}
\includegraphics[width=0.65\textwidth]
{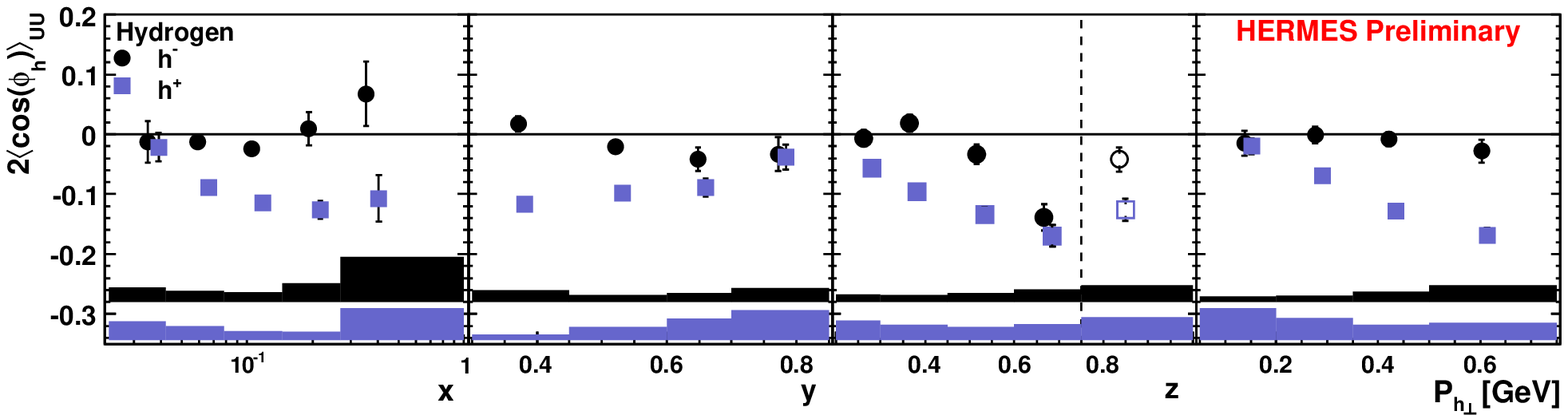}\\
\includegraphics[width=0.65\textwidth]
{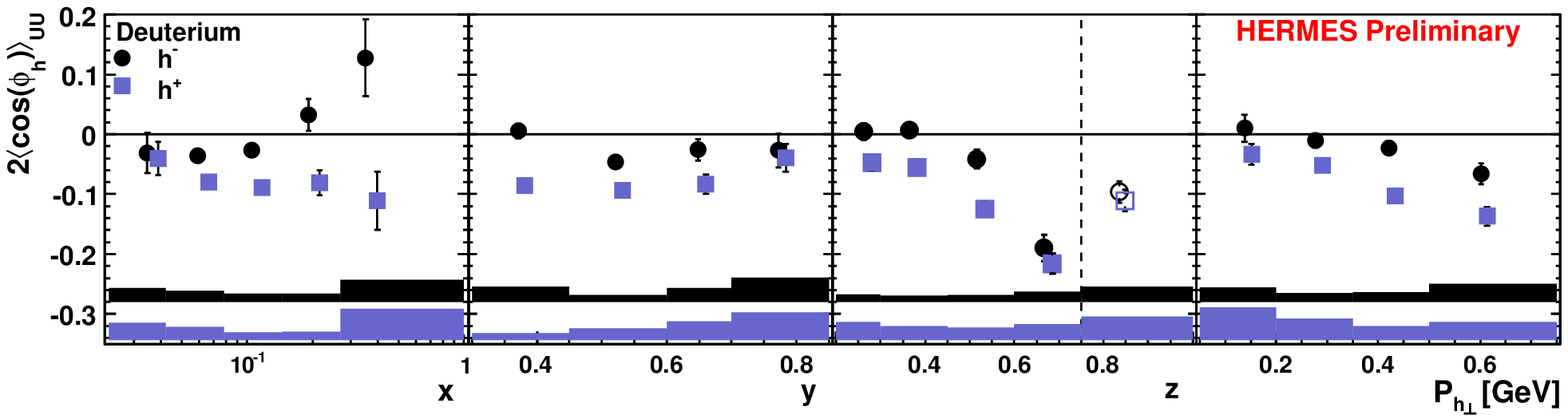}\\
\end{center}
\caption{
The preliminary results for the spin-independent $\cos \phi_h$ 
azimuthal asymmetries versus $x$, $y$, $z$ and 
$P_{h \perp}$ for positive
and negative hadrons measured by the HERMES experiment 
with the 28 GeV electron beam~\cite{Giordano:2009hi}
on p (top) and on d (bottom). 
}
\label{fig:bm_hermes} 
\end{figure}

The HERMES experiment has produced  results
for the  $\cos \phi_h$ and the $\cos 2\phi_h$ asymmetries
in hydrogen and deuterium using the data collected in 2000, 2005, and  2006.
The cuts applied in the event and hadron selection are:
$x> 0.023$, $1<Q^2 <20$ GeV$^2$, $10 <W^2 <45$ GeV$^2$, 
$0.3 < y < 0.85$, $x_F > 0.2$,
$ 0.2 < z  < 0.75$, and $0.05 < P_ {h \perp}< 0.75$ GeV.
To correct for acceptance of the spectrometer,
detector smearing, and QED radiative effects, 
the data were analysed in a 5-dimensional grid in the variables
$x$, $y$, $z$, $P_{h \perp}$ and $\phi_h$.
A 10-dimensional smearing matrix was populated by Monte Carlo
simulation and incorporated into the fitting procedure, which
has been extensively tested with Monte Carlo data.
The released asymmetries are one dimensional projections of the 
asymmetries in which the other four variables have been integrated over.
The measured $\cos \phi_h$  asymmetries for
protons and deuterons are shown in Fig. \ref{fig:bm_hermes}
for positive and negative hadrons versus $x$, $y$, $z$, and $P_{h \perp}$.
%
The error bars are statistical, and the bands show the systematic error.
As can be seen in the figure, the proton and the deuteron
asymmetries are very similar, giving a hint for alike sign for the
$u$ and $d$ Boer Mulders functions.
The difference between positive and 
negative hadrons is remarkable, as in the case of the COMPASS 
results. 
At variance with the COMPASS data shown here, the HERMES
results incorporate the $y$-dependent kinematical
factor appearing in the cross section.

The CLAS results at JLab for the $\cos 2\phi_h$ asymmetry for $\pi^+$
\cite{Osipenko:2008rv} agree with the HERMES measurements
at large $z$.
A striking feature of their data is the large values measured for $z<0.2$,
which increase with $P_{h \perp}$ and for which there is no
comparison with other experiments.

Phenomenologically, the Cahn contribution to the $\cos \phi_h$ asymmetry
was studied by Anselmino and coworkers \cite{Anselmino:2005nn}.
Using the EMC data \cite{Aubert:1983cz,Arneodo:1986cf} in
the region $P_{h \perp} \leq 1$ GeV, they extracted
the average values of quark momenta: $\langle k_T^2 \rangle
= 0.25$ GeV$^2$ in the distribution functions, $\langle
p_T^2 \rangle = 0.20$ GeV$^2$ in the fragmentation functions.

\begin{figure}[tb]
\begin{center}
\includegraphics[angle=270,width=0.45\textwidth]
{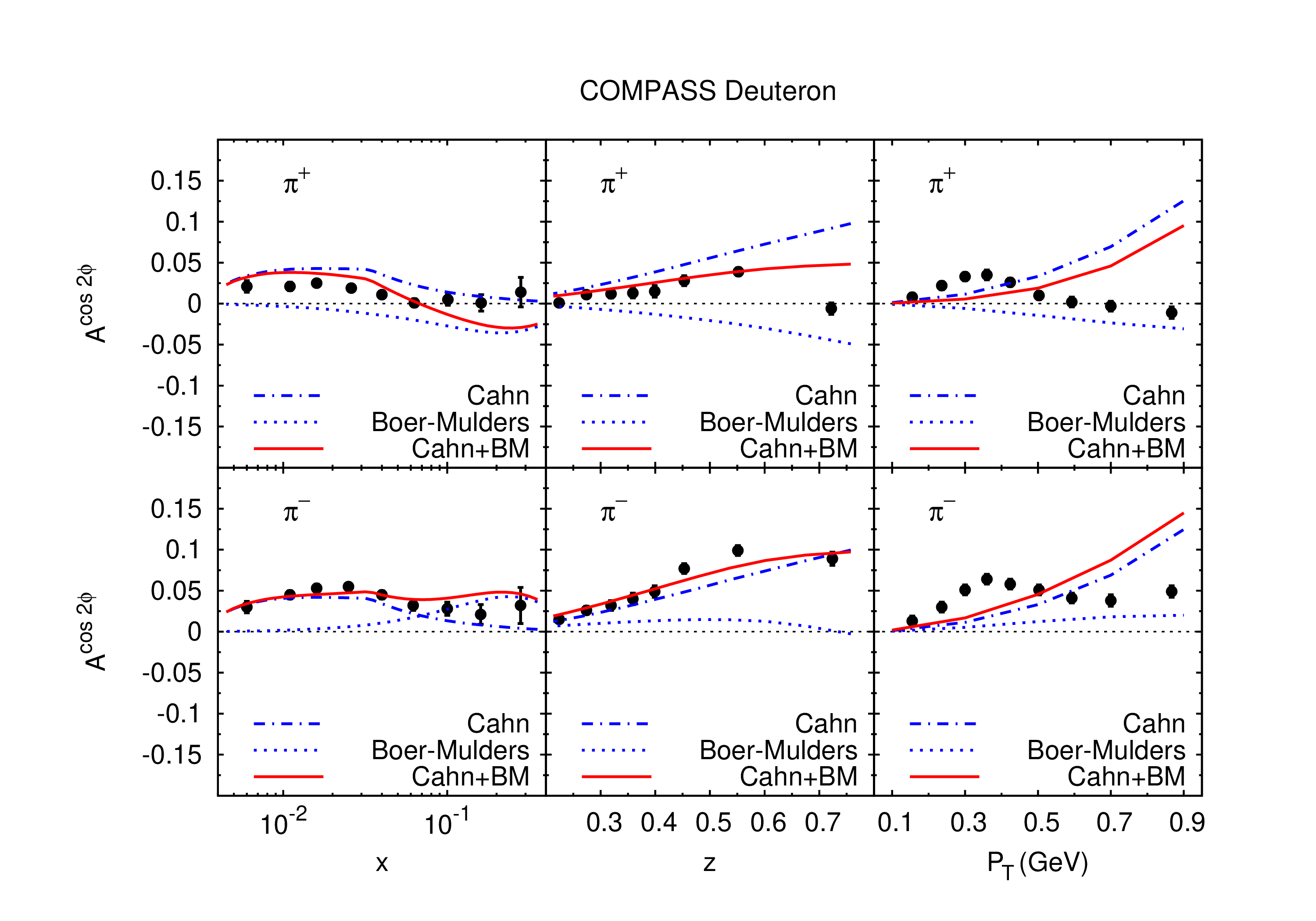}
\includegraphics[angle=270,width=0.45\textwidth]
{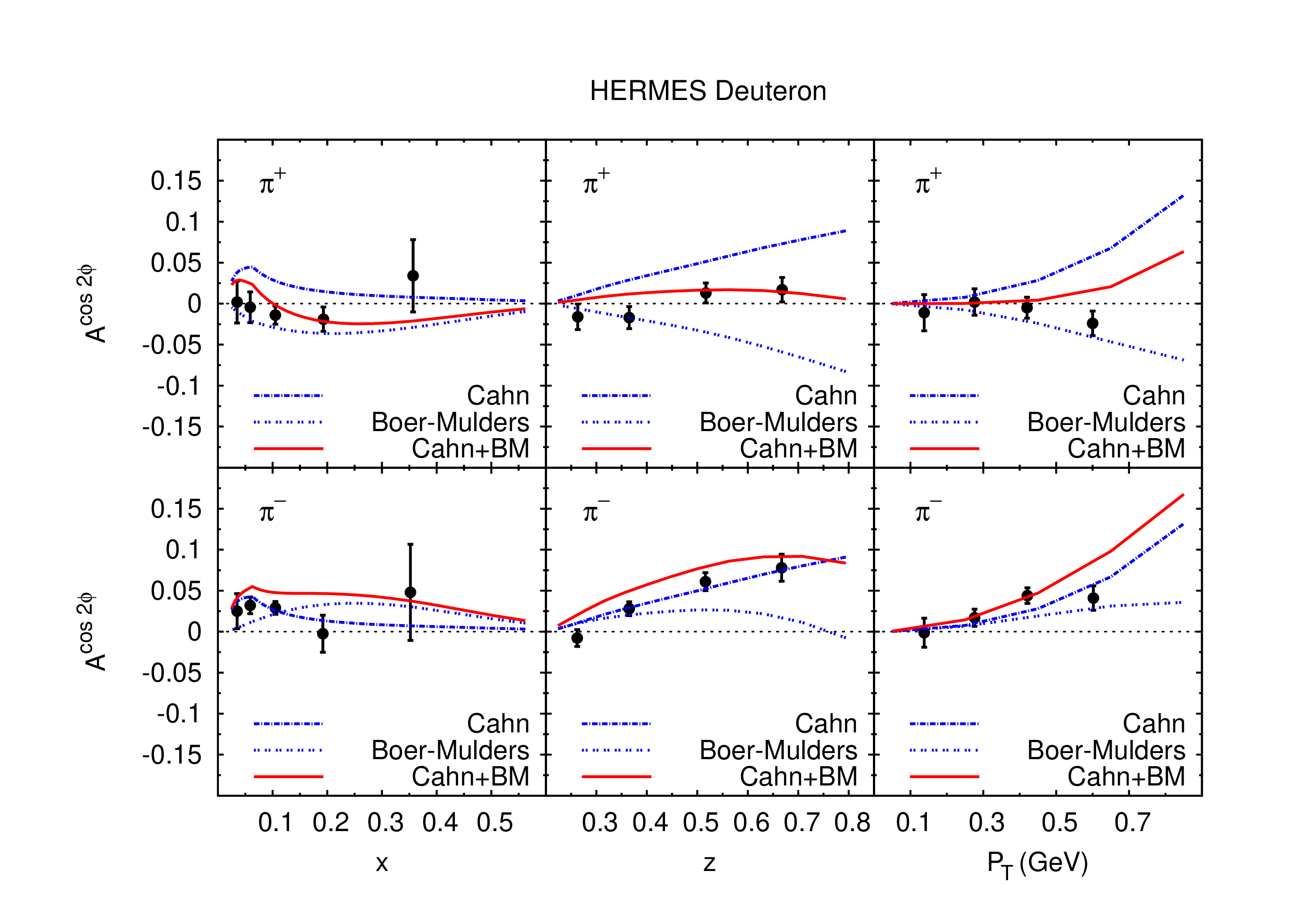}
\end{center}
\caption{
The preliminary results for the
$\cos 2\phi$ spin-independent azimuthal asymmetries for deuteron
from COMPASS (left panel) and HERMES (right) as functions of 
$x$, $z$ and $P_{h \perp}$ compared with
the fits to the data of~\cite{Barone:2009hw}.
}
\label{fig:bm_fit_cos2phi} 
\end{figure}

The $\cos 2 \phi_h$ asymmetry has been 
analysed in detail in Refs.\cite{Barone:2008tn,Barone:2009hw,Zhang:2008ez}. 
In Ref.~\cite{Barone:2008tn}, 
which anticipated the measurements, it was predicted 
that $A_{UU}^{\cos 2 \phi_h}$ is of the order of at most $5 \%$, 
and that the $\pi^-$ asymmetry should be larger than the $\pi^+$ 
asymmetry, as a consequence of the Boer-Mulders effect. 
These predictions have been substantially confirmed by 
the experimental results. 
A fit to the HERMES and COMPASS preliminary data has been recently 
presented in Ref.~\cite{Barone:2009hw}. 
It assumes that $A_{UU}^{\cos 2 \phi_h}$ can be 
described by the leading-twist Boer-Mulders component 
and by the twist-4 Cahn term (which is however only part 
of the full twist-4 contribution, still unknown). 
The available data do not allow a complete determination  
of the $x$ and $k_T$ dependence of $h_1^{\perp}$. 
Thus, the Boer-Mulders functions are simply taken to be proportional to 
the Sivers functions of Ref.~\cite{Anselmino:2008sga}, $h_1^{\perp q} 
= \lambda_q f_{1T}^{\perp q}$, and the parameters $\lambda_q$ 
are obtained from the fit 
(the Boer-Mulders sea is not constrained by the data and is taken 
to be equal in magnitude to the Sivers sea). 
The result is 
\be
h_{1}^{\perp u} = 2.0 \, f_{1T}^{\perp u}\,, 
\;\;\;
h_1^{\perp d} = - 1.1 \, f_{1T}^{\perp d}\,,
\label{bm_barone}
\ee
and the comparison with the data is shown in Fig.~\ref{fig:bm_fit_cos2phi}
%
%
Since $f_{1T}^{\perp u}$ is negative and $f_{1T}^{\perp d}$ 
is positive, the $u$ and $d$ Boer-Mulders distributions 
are both negative. This is what one expects in 
large-$N_c$ QCD \cite{Pobylitsa:2003ty} and in some other models
\cite{Bacchetta:2008af,Courtoy:2009pc}. The results are also 
consistent with the predictions of the 
impact-parameter picture \cite{Burkardt:2005hp} combined 
with lattice calculations \cite{Gockeler:2006zu}, 
which indicate a $u$ component of $h_1^{\perp}$ larger 
in magnitude than the corresponding Sivers component, and the 
$d$ distributions with the same magnitude and opposite sign. 
A very recent model calculation of $h_1^{\perp}$ \cite{Pasquini:2010af} 
is in good agreement with the findings of Ref.~\cite{Barone:2009hw}.  

\subsubsection{Boer-Mulders effect in DY production}
\label{sec:results_BoerM_DY}

As shown is Section \ref{dy_kin} the cross-section for the DY process 
on unpolarised nucleon is given by eq.~\ref{dycross2} where the quantities
$\lambda$ and $\nu$ are related by the  Lam-Tung relation  
$\lambda + 2 \nu = 1$.

The NA10 data on $\pi^- N$ DY off a tungsten target show that the angular 
distribution for the DY events do not show any c.m. energy dependence 
nor any nuclear dependence. 
The data show that the value for $\lambda$ is close to 1, as expected by the 
naive parton model (for massless quarks and no intrinsic quark momentum 
the angular distribution should just be  ($1 +  \cos^2\theta$). 
Also, the value for $\mu$ is close to expectation: it is essentially 
consistent with zero, indicating that the annihilating partons 
contribute equally to the transverse momentum of the muon-pair. 
Both the values of $\lambda$  and of $\mu$  are essentially independent 
of any kinematical variable. 
But the most striking result from this experiment is the large value they 
obtain for $\nu$ and the strong dependence of $\nu$  on the dimuon transverse 
momentum. 

\begin{figure}[tbh]
\begin{center}
\includegraphics[width=0.40\textwidth]
{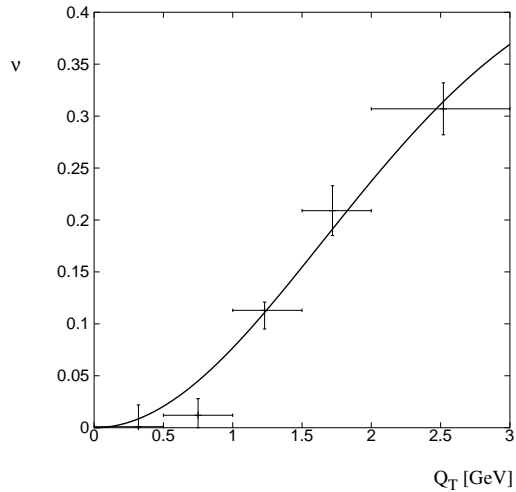}
\end{center}
\caption{\label{fig_boer_nu} 
The amplitude $\nu$ of the $\cos 2 \phi$ modulation in $\pi^- N$ Drell-Yan 
process \cite{Falciano:1986wk}. The Collins-Soper 
frame is adopted. The curve is the prediction of 
Ref.~\cite{Boer:1999mm}. 
}   
\end{figure}

Similar results have been obtained at FermiLab by the experiment E615
which investigated the same  $\pi^- N \rightarrow \mu^+ \mu^- X$ DY 
process.
As shown in Fig.~\ref{fig_boer_nu} $\nu$ was found to be as large 
as 30\%, and steeply rising with $Q_T$, an effect
which is not explained by pQCD and was interpreted 
as a manifestation of the Boer-Mulders mechanism \cite{Boer:1999mm}.
Also, by projecting the data points on the $x_{\pi}$ axis, the 
pion valence structure function  $F_{\pi}(x_{\pi})$ could be precisely 
determined in the $x_{\pi}$ range 
from 0.21 to 1, and found in good agreement with the extraction of 
NA3 and NA10. 

\begin{figure}[tbh]
\begin{center}
\includegraphics[width=0.40\textwidth]
{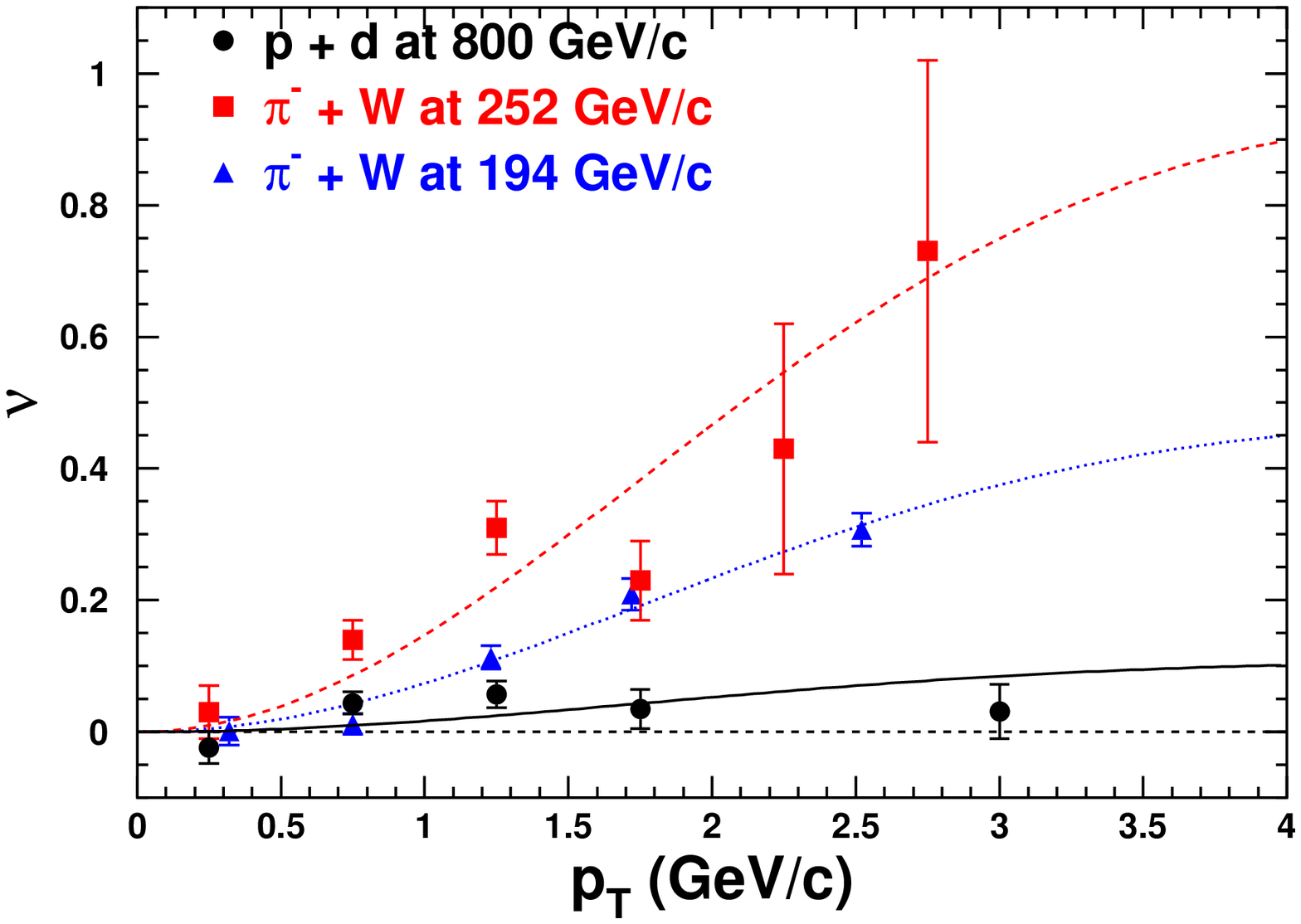}
\hspace{0.2cm}
\includegraphics[width=0.45\textwidth]
{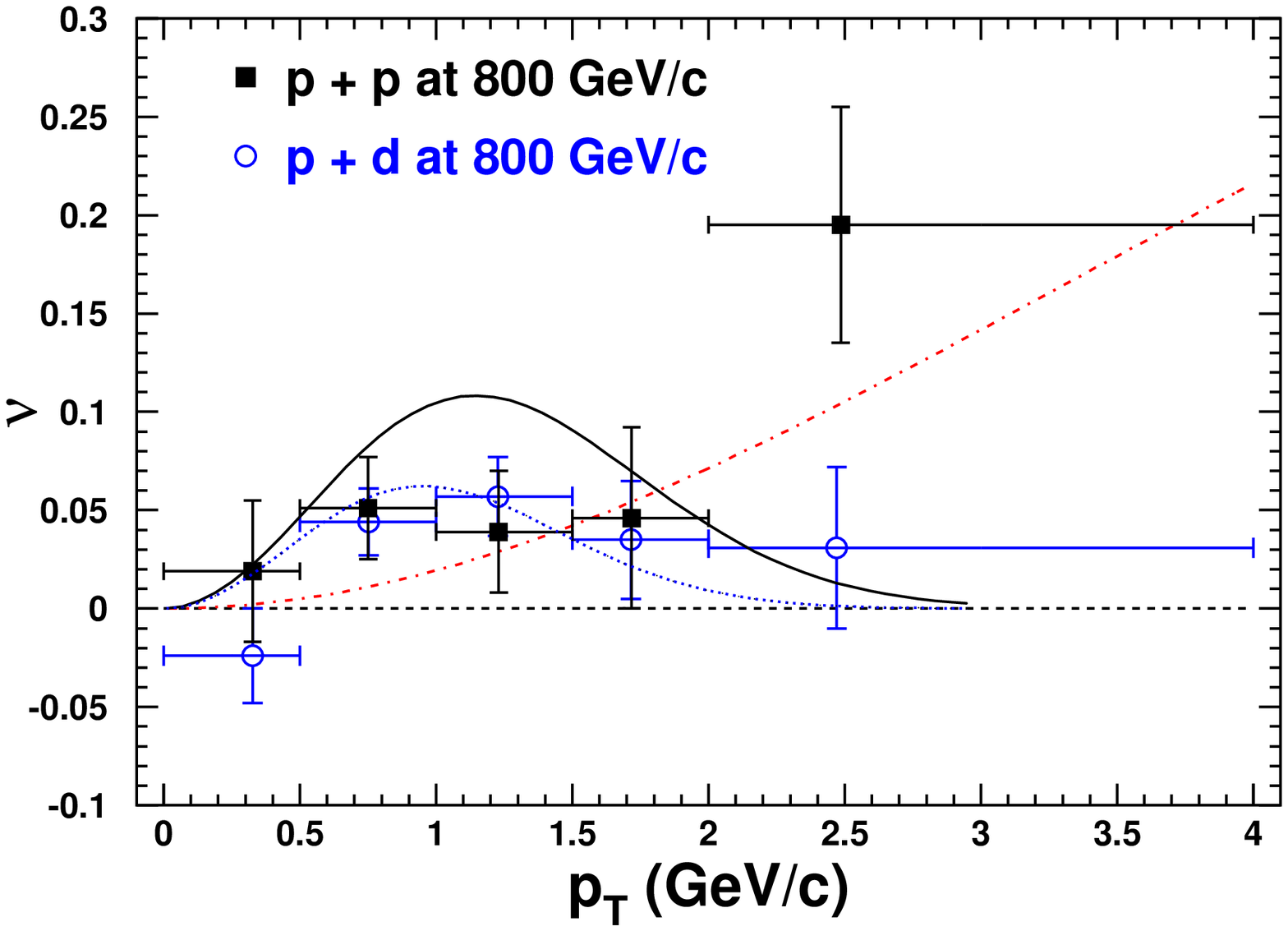}
\end{center}
\caption{\label{fig_pp_pd_dy} 
Left: the $\nu$ coefficient in the Collins-Soper frame 
for three DY measurements (dots: E866/NuSea \cite{Zhu:2006gx}; 
squares: E615 \cite{Conway:1989fs}; triangles: NA10 \cite{Falciano:1986wk}). 
The curves are fits similar to that of Ref.~\cite{Boer:1999mm}. 
Right: the $\nu$ coefficient for $pd$ \cite{Zhu:2006gx} 
and $pd$ DY \cite{Zhu:2008sj}. The dot-dashed curve is 
the perturbative QCD contribution. The solid and dotted curves
are the calculations of Ref.~\cite{Zhang:2008nu}
for $pp$ and $pd$, respectively, based 
on the Boer-Mulders effect. 
}   
\end{figure}

Recently, the E866/NuSea Collaboration at 
FNAL has presented data on the angular distributions of 
dimuon production in $pd$ \cite{Zhu:2006gx} and $pp$ \cite{Zhu:2008sj}
collisions. Much smaller $\nu$ values (less than 0.05) 
than in $\pi N$ DY are found,
as shown in Fig.~\ref{fig_pp_pd_dy}, and the Lam-Tung relation is satisfied. 
These data could in principle give some information about the antiquark 
Boer-Mulders distributions \cite{Zhang:2008nu,Lu:2009ip}. 
However, in the $P_T$ region above 
1-1.5 GeV  they are expected to be described by pQCD \cite{Collins:1978yt}. 
Thus the only points that have likely  to do  
with the Boer-Mulders effect are those below $P_T \sim 1.5$ GeV. 

We conclude this section by mentioning that a 
theoretical study of the DY azimuthal asymmetries 
at small and moderate $P_T$ in the context of twist-three factorisation 
has been performed in Ref.~\cite{Zhou:2009rp}. 

\subsection{Accessing the TMD distributions: leading-twist $T$-even functions
and higher-twist functions}
\label{access_teven}

At leading twist, besides the Collins, Sivers and Boer--Mulders 
terms, there are other three angular modulations in the SIDIS 
cross section, which probe $T$-even TMD distributions: 
\begin{itemize} 
\item[-]
Unpolarised beam and longitudinally polarised target: 
$\sin 2 \phi_h$ modulation, involving $h_{1L}^{\perp}$. 
\item[-]
Unpolarised beam and transversely polarised target: 
$\sin (3 \phi_h - \phi_S)$ modulation, involving $h_{1T}^{\perp}$. 
\item[-]
Longitudinally polarised beam and transversely polarised 
target: $\cos (\phi_h - \phi_S)$ modulation, involving $g_{1T}$. 
\end{itemize} 

From the data collected with the deuteron target,
the COMPASS experiment has produced preliminary results
on these asymmetries for charged hadrons~\cite{Kotzinian:2007uv,Savin:2010du}.
All in all, it is hard to find a signal in any of
these observables. It will be interesting to look 
at the corresponding results with the proton target.

The HERMES collaboration has  measured the 
$\sin 2\phi_h$ moment, both with a  proton \cite{Airapetian:1999tv}
and a deuteron target~\cite{Airapetian:2002mf}, finding it 
to be compatible with zero. 
Signals of a non vanishing $\sin 2 \phi_h$ asymmetry 
have been recently reported by the CLAS collaboration 
\cite{Avakian:2010ae}.

For the moment, all these data can only be confronted with model predictions. 
Focusing on $h_{1T}^{\perp}$, which has attracted 
some attention for its interesting physical content,
a calculation of $A_{UT}^{\sin (3 \phi_h - \phi_S)}$ based on positivity bounds
\cite{Avakian:2007mv} gives limits of about $\pm 0.03$ 
for the asymmetry on a deuteron target and 
a slightly larger value for a proton target. 
The COMPASS deuteron data lie within these bounds. 

The light-cone constituent quark model of Ref.~\cite{Boffi:2009sh} predicts 
an asymmetry $A_{UT}^{\sin (3 \phi_h - \phi_S)}$ smaller than 0.01, also 
consistent with the COMPASS findings. 
All the asymmetries related to $T$-even TMD's 
are calculated in Ref.~\cite{Boffi:2009sh} and found to be generally close 
to zero, hence compatible with the COMPASS findings.  

At subleading twist, that is at order $1/Q$, the situation 
is much more involved. 
The SIDIS structure functions of eq.~(\ref{sidiscs}) 
have have in fact the general form: 
\[
F \sim {\rm l.t.} \; {\rm TMD} \otimes {\rm h.t.} 
\; {\rm FF} \; + \;  
{\rm h.t.} \; {\rm TMD} \otimes {\rm l.t.} 
\; {\rm FF}\, 
\]
where ``l.t.'' = leading twist, and  ``h.t'' = higher twist.
They contain both leading-twist and twist-three 
TMD distributions and fragmentation functions. 
Thus, the phenomenological interpretation 
of these observables is rather intricate. 
In SIDIS with unpolarised (U) and/or longitudinally (L)
and transversely (T) polarised beams and targets
there are 8 twist-three modulations:   
$$
UU: \; \cos \phi_h; \;\;LU: \; \sin \phi_h; \;\; 
UL: \; \sin \phi_h, \;\sin 2 \phi_h; 
LL: \; \cos \phi_h;  \;\; UT: \; \sin \phi_S; \;\; 
$$
$$
UT:  \; \sin (2 \phi_h - \phi_S); \;\; LT: \; \cos \phi_S,\;  
\cos (2 \phi_h - \phi_S) \, .
$$ 
COMPASS has measured all these quantities on a 
deuteron target \cite{Kotzinian:2007uv,Kafer:2008ud,Savin:2010du}
and found them all to be consistent with zero. 
HERMES presented results on the $\sin 2 \phi_h$
and $\sin \phi_h$ modulations with a longitudinally polarised 
proton \cite{Airapetian:1999tv} and deuteron target \cite{Airapetian:2002mf}. 
While the $\sin 2 \phi_h$ asymmetry was found to vanish, 
the $\sin \phi_h$ asymmetry showed a  large signal (up to 4\%
for proton and 2\% in deuteron), incompatible with zero for positive and
neutral pions, originally interpreted in terms of large transversity PDF's.

A remark about the definition of the target polarisation is 
now in order. 
Experimentally, the target polarisation is defined with respect to the a 
longitudinal (transverse) polarisation with respect to the beam axis has 
a transverse (longitudinal) component with respect to the virtual photon axis, 
which is kinematically suppressed by a factor $1/Q$~\cite{Barone:2003fy}.
Thus, any measured $w( \phi_h, \phi_S)$ modulation 
with a ``longitudinally'' (``transversely'') polarised target 
is mixed with a transverse (longitudinal) 
modulation of the same type, suppressed by $1/Q$. 
This may be relevant in some cases. 
For instance, the $\sin \phi_h$ asymmetry measured with 
a target longitudinally polarised with respect to the beam axis
gets a twist-three contribution from $A_{UL}^{\sin \phi_h}$, 
but receives also contributions from the leading-twist Collins and Sivers 
asymmetries, $A_{UT}^{\sin (\phi_h + \phi_S)}$ and 
$A_{UT}^{\sin (\phi_h - \phi_S)}$, which are multiplied 
by a kinematical factor $\sim 1/Q$. 
In principle, all these contributions might be equally relevant. 
The HERMES analysis of the $\sin \phi_h$ asymmetry   
\cite{Airapetian:2005jc} has shown that the Collins and Sivers 
contributions to $\langle \sin \phi_h \rangle$ 
are small, and therefore the large measured asymmetry 
is a genuine subleading-twist effect. 

\begin{figure}[tbh]
\begin{center}
\includegraphics[width=0.5\textwidth,bb=0 200 590 650]
{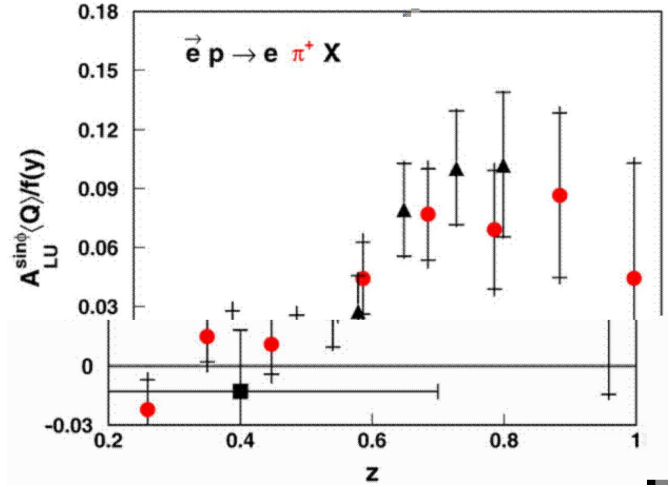}
\end{center}
\caption{Comparison of the kinematically rescaled $\sin \phi_h$ asymmetries
between the HERMES (circles) and CLAS (triangles)
measurements. The full square represents a previous HERMES measurement
averaged over the range $0.2 < z<0.7$. The
outer error bars represent the quadratic sum of the systematic uncertainty and
the statistical uncertainty (inner error bars) \cite{Airapetian:2006rx}.
}
\label{fig:ht_hermes_p_sinphi}.
\end{figure}

Another indication of the relevance of  twist-three effects comes from 
the beam-spin asymmetry $A_{LU}^{\sin \phi_h}$, which  has been measured  
by CLAS at 4.3 GeV for positive pions \cite{Avakian:2003pk} 
and by HERMES at 27.6 GeV for charged and neutral 
pions \cite{Airapetian:2006rx}.
The asymmetry for positive pions is large and positive. 
The CLAS and HERMES  results nicely agree if one rescales
the HERMES data by the mean value of $Q^2$ and the $y$ dependent
kinematical factor, as shown in Fig.~\ref{fig:ht_hermes_p_sinphi}.
%

\subsection{Inclusive hadroproduction}
\label{sec:phen_hp}

Sizable SSA's in polarised inclusive hadroproduction have been reported 
since the 70s \cite{Dick:1975ty,Klem:1976ui,Dragoset:1978gg,Antille:1980th},  
for center-of mass energies in the range 5-10 GeV. 
In the same years, Fermilab experiments discovered that $\Lambda$ hyperons 
produced in unpolarised $pp$ collisions have a large 
transverse polarisation with respect to the production plane 
\cite{Bunce:1976yb,Heller:1978ty}. 
These findings provoked a certain theoretical interest, as it was widely 
held that large transverse polarisation effects could
not be reproduced in the framework of perturbative QCD
\cite{Kane:1978nd}. 
On the other hand, the small values of $\sqrt{s}$ and $P_T$ explored by 
those experiments left the door open to interpretations based on
soft physics.  

In 1991 the E581/E704 experiment at Fermilab extended the investigation of 
transversely polarised hadroproduction to higher energies and found 
remarkably  large transverse SSA's  in the forward 
region \cite{Adams:1991cs,Adams:1991ru}. 
These results were confirmed by the RHIC measurements, which moved the 
energy frontier one order of magnitude upwards. 
On the $\Lambda$ polarisation front, the striking effect 
discovered in the early experiments was observed also  at higher $P_T$ 
values \cite{Heller:1983ia,Lundberg:1989hw,Ramberg:1994tk}.  
Thus, it has become clear that there must be some hard mechanism 
behind the transverse polarisation phenomena observed in 
hadroproduction.

\subsubsection{SSA's in inclusive hadroproduction}

As seen in Section \ref{sec:sidis_kinem}, 
the left-right asymmetries associated to
an azimuthal modulation of a cross-section, are best measured 
by comparing data taken with up- and down-polarised beam or target.
Integrating over $\phi$, a global left-right asymmetry $A_N$ 
can then be defined as 
\be
A_N = \frac{1}{\mathcal{P}} \, \frac{N_L - N_R}{N_L + N_R} \,, 
\ee
with $\mathcal{P}$  the beam or target polarisation.

The modern era of experimental (and theoretical) work 
on SSA's in hadroproduction was inaugurated
by the E704 investigation of $pN$ and $\bar p N$ 
collisions with transversely polarised secondary proton and antiproton beams 
at the c.m. energy $\sqrt{s} \simeq 19.4$ GeV 
\cite{Adams:1991cs,Adams:1991ru}. 
Two kinematical regions were covered:  
1) the  beam fragmentation (or forward) region, $0.2 \leq x_F \leq 0.6$, 
with  $P_T$ in the range $0.2 - 2.0$ GeV;
2) the central rapidity region, $\vert x_F \vert \leq 0.15$, with $P_T$ up to  
4 GeV. 
In pion production, large SSA's were found at high $x_F$
\cite{Adams:1991cs,Adams:1991rw,Adams:1991ru,Bravar:1996ki}: 
the asymmetries are nearly zero up to $x_F \sim 0.3$ 
and then start rising with $x_F$, reaching  
15 \% for $\pi^0$ and 30-40 \% for $\pi^{\pm}$. In $p^{\uparrow} N$
collisions $A_N$ is positive   
for $\pi^+$ and negative for $\pi^-$, with about the same size. 
Signs are reversed in $\bar p^{\uparrow} N$ scattering.  
The asymmetry for $\pi^0$ is roughly half of that of charged 
pions and always positive. 
As for the $P_T$ dependence,  $A_N$ is zero below $P_T \sim 0.5$ GeV, 
and above this value increases  in magnitude with $P_T$. 
In the central rapidity region, where larger values 
of $P_T$ are reached, the asymmetries turn out to be consistent 
with zero \cite{Adams:1994yu}. 
E704 has also measured sizable SSA's in inclusive $\Lambda$ 
and $\eta$ production \cite{Bravar:1995fw,Adams:1997dp} 
but the rather low transverse momentum of the $\Lambda$'s does not allow a  
safe perturbative QCD analysis.  
 
The E704 findings have been substantially confirmed by 
other fixed-target experiments at lower energies, at IHEP (Protvino) 
\cite{Apokin:1990ik,Abramov:1996vp}
and at the BNL--AGS \cite{Krueger:1998hz,Allgower:2002qi}. 

On the phenomenological side,the E704 pion asymmetries have 
been interpreted in terms of the Sivers effect 
\cite{Anselmino:1994tv,Anselmino:1998yz,D'Alesio:2004up} and of 
the Collins effect \cite{Artru:1995bh,Anselmino:1999pw,Boglione:1999dq}. 
A recent reassessment of the situation  \cite{Anselmino:2008uy} 
has shown that, contrary to a previous  prediction 
of a strong suppression of the Collins effect \cite{Anselmino:2004ky}, 
both Collins and Sivers mechanisms may give sizable 
contributions to the SSA's. 
The E704 results have 
also been studied in the context of twist-3 factorisation, 
considering quark-gluon correlations in the initial state   
\cite{Qiu:1991pp,Qiu:1998ia,Kanazawa:2000hz,Kanazawa:2000kp,Kouvaris:2006zy} 
and in the final state \cite{Koike:2002ti}. 
All these approaches are able to reproduce at least qualitatively   
the data,  thus showing that many different physical mechanisms 
may be at work in polarised hadroproduction. 
Thus, it is  impossible for the moment to draw definite conclusions as to 
the dynamical source of single-spin transverse asymmetries.

\begin{figure}[tbh]
\begin{center}
\includegraphics[width=0.40\textwidth]
{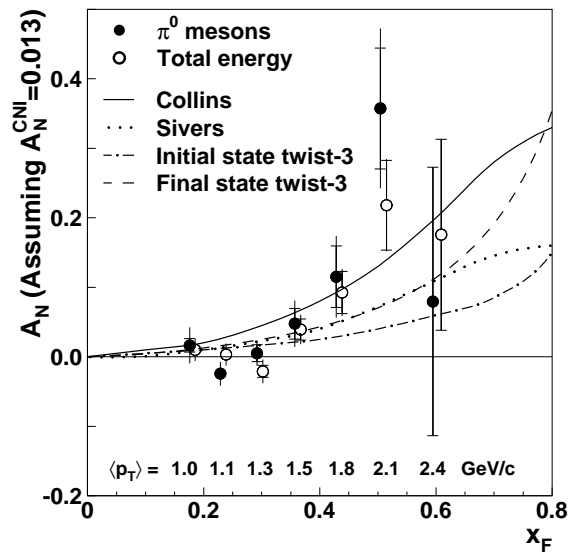}
\end{center}
\caption{\label{fig:star_2004} The asymmetry 
$A_N ( p^{\uparrow} p \to \pi^0 X)$ measured by STAR \cite{Adams:2003fx}
at $\sqrt{s} = 200$ GeV. The curves are the predictions of 
Ref.~\cite{Anselmino:1999pw} (solid), 
Ref.~\cite{Anselmino:1998yz} (dotted), 
Ref.~\cite{Qiu:1998ia} (dot-dashed), Ref.~\cite{Koike:2002ti} (dashed).}
\end{figure}

\begin{figure}[tbh]
\begin{center}
\includegraphics[width=0.4\textwidth]
{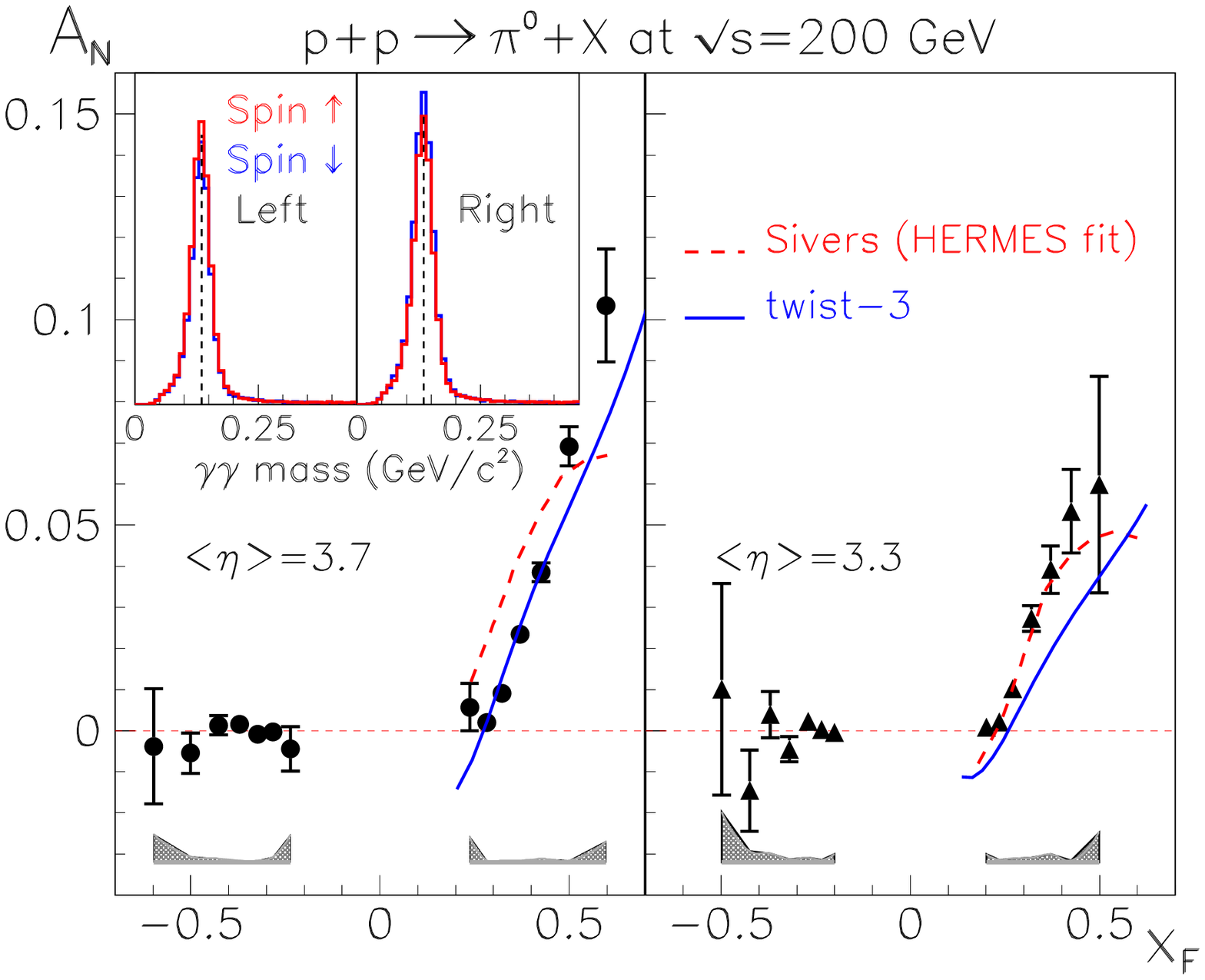}
\hspace{0.2cm} 
\includegraphics[width=0.4\textwidth]
{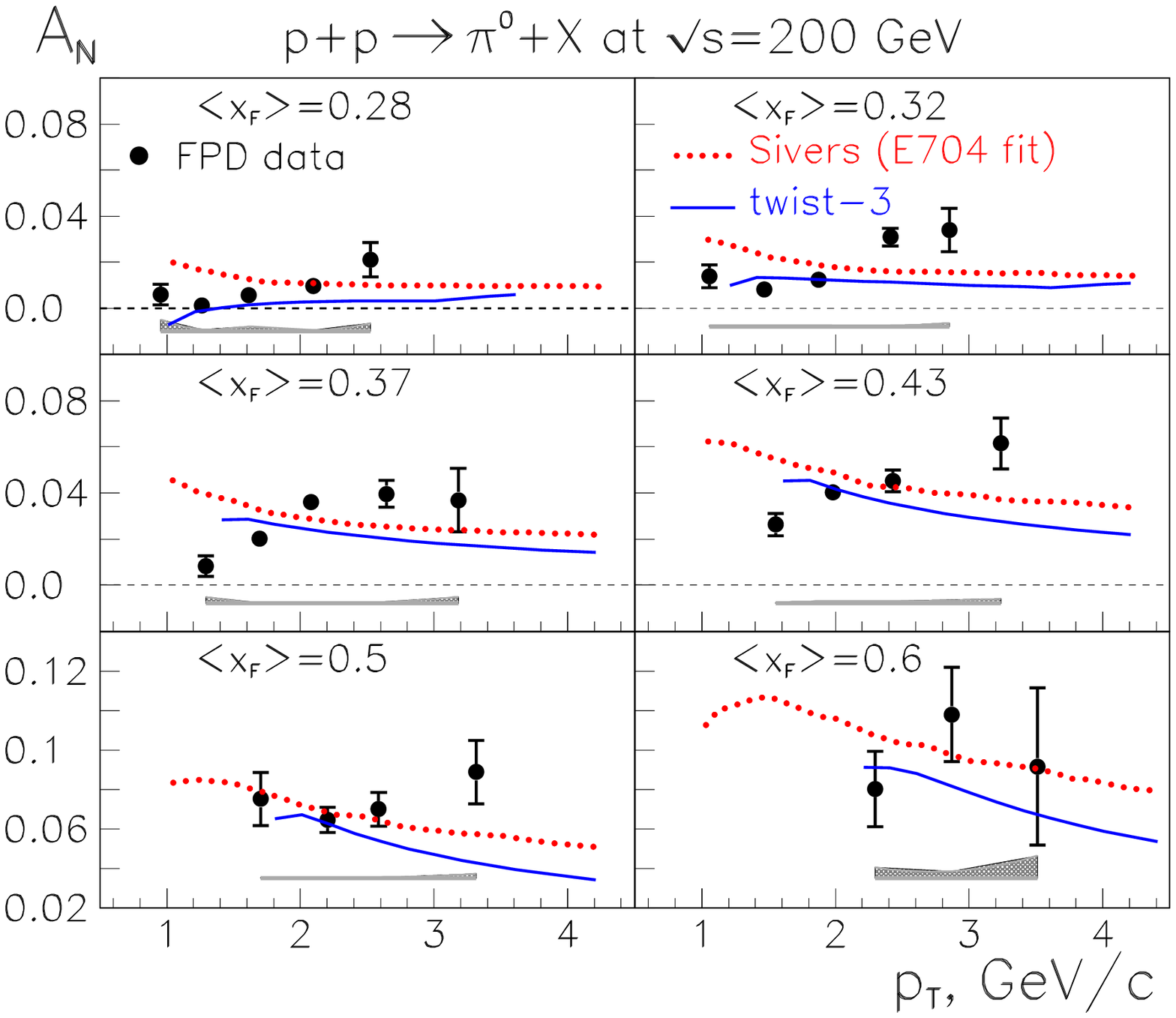}
\end{center}
\caption{\label{fig:star2008} The asymmetry 
$A_N ( p^{\uparrow} p \to \pi^0 X)$ measured by STAR \cite{Abelev:2008qb}
at $\sqrt{s} = 200$ GeV, as a function of $x_F$ 
for two different values of the average pseudorapidity $\langle \eta \rangle$
(left) and the same asymmetry as a function of $P_T$ at fixed $x_F$ 
(right).
The curves are the predictions of 
Ref.~\cite{D'Alesio:2004up,Boglione:2007dm} (dashed line) 
and of Ref.~\cite{Kouvaris:2006zy} (solid line).} 
\end{figure}

The E704 measurement might have left  the doubt that 
transverse SSA's would disappear at collider energies.  
Studying the $p^{\uparrow} \, p \to \pi^0 \, X$ reaction
at $\sqrt{s} = 200$ GeV in the first polarised collisions at RHIC, 
the STAR Collaboration showed that this is not the case: 
the large effects found by E704 
persist at an order of magnitude  higher energy \cite{Adams:2003fx}. 
As shown in Fig.~\ref{fig:star_2004} STAR measured a large positive $A_N$ above
$x_F \sim 0.3$ in the transverse-momentum range $1.0 < \langle P_T \rangle
< 2.4$ GeV. 
More recently, the negative $x_F$ region has been explored, finding an 
asymmetry consistent with zero, and the $P_T$ dependence of the SSA has been 
determined \cite{Abelev:2008qb}. 
In Fig.~\ref{fig:star2008} one sees that
the rise of the SSA's at large $x_F$ is fairly well reproduced  by the 
Sivers mechanism in the generalised parton model with the Sivers function 
extracted from the HERMES SIDIS data \cite{D'Alesio:2004up,Boglione:2007dm}
and by the twist-3 factorisation scheme \cite{Kouvaris:2006zy}.
On the contrary the $P_T$ behaviour of the data, showing a clear tendency to 
increase at fixed $x_F$, contradicts the theoretical expectations, which 
predict a decrease of $A_N$ with $P_T$.   

Concerning the description of hadroproduction SSA's 
in terms of TMD's taken from SIDIS analysis,  
one should recall that RHIC asymmetries scan the parton distributions 
over a wide range of the Bjorken variable, including the large-$x$ region, 
whereas the SIDIS data are limited to $x < 0.3$ 
and do not constrain the tails of the transversity distribution
and of the Sivers function. 
Thus, the generalised parton model predictions 
of the SSA's are quite uncertain at high $x_F$ \cite{Anselmino:2009hk}. 

\begin{figure}[tbh]
\begin{center}
\includegraphics[width=0.40\textwidth]
{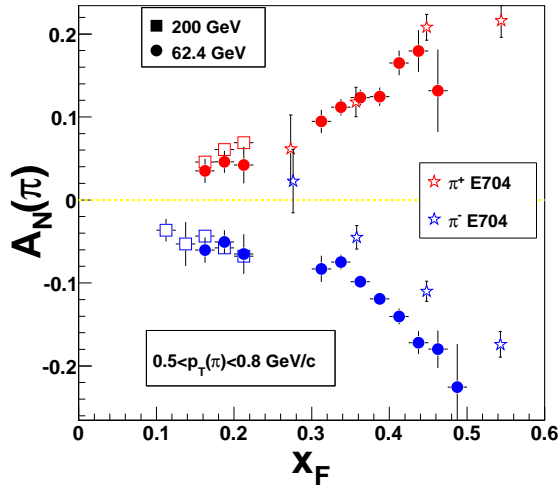}
\end{center}
\caption{\label{fig:aidala_compilation} Charged pion asymmetries 
measured by BRAHMS at $\sqrt{s} = 200$ GeV \cite{Lee:2009ck} 
and $\sqrt{s} = 62.4$ GeV \cite{Arsene:2008mi}, 
and by E704 at $\sqrt{s} = 19.4$ GeV \cite{Adams:1991cs}.  
}
\end{figure}

\begin{figure}[tbh]
\begin{center}
\includegraphics[width=0.80\textwidth]
{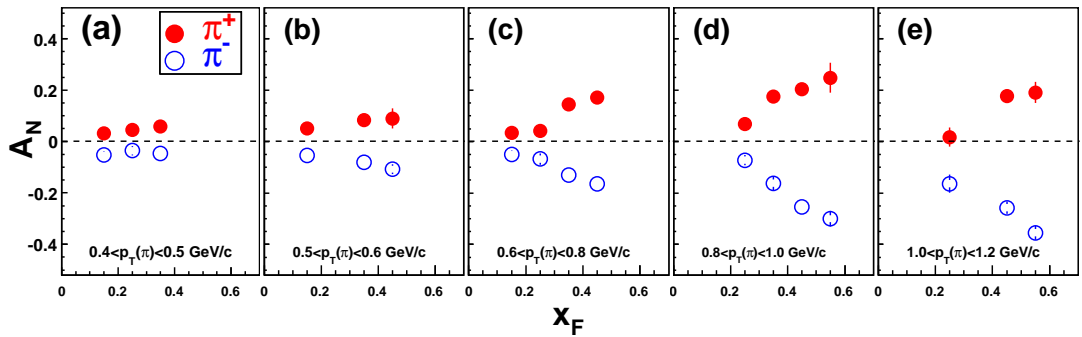}
\end{center}
\caption{\label{fig:brahms_pt} Transverse SSA of charged pions 
at $\sqrt{s} = 62.4$ GeV in bins of $P_T$, measured by BRAHMS 
\cite{Arsene:2008mi}. 
}
\end{figure}

Measurements of forward charged pions by BRAHMS at $\sqrt{s} = 200$ GeV 
\cite{Lee:2009ck} and $\sqrt{s} = 62.4$ GeV \cite{Arsene:2008mi} shown 
in Fig.~\ref{fig:aidala_compilation} confirm the asymmetry pattern observed 
by E704, with the mirror effect of $\pi^+$ and $\pi^-$ 
and large absolute values of $A_N$. 
The SSA's for $\pi^+$ and $\pi^-$ at $\sqrt{s} = 62.4$ GeV are plotted 
in bins of $P_T$ in Fig.~\ref{fig:brahms_pt}. 
A clear rise with $P_T$ is visible up to a transverse momentum of about 1 GeV, 
where the asymmetries reach magnitudes of about 0.3. 
BRAHMS has also measured kaon production at $\sqrt{s} = 62.4$ 
finding that the $K^+$ asymmetry has the same sign and 
approximately the same magnitude as the $K^-$ asymmetry \cite{Arsene:2008mi}. 
The results are fairly well reproduced both by the twist-3 approach 
and by the Sivers effect in the generalised parton model 

In the midrapidity region ($\vert \eta \vert < 0.35$). the PHENIX measurements 
of neutral pion and charged hadron production at $\sqrt{s} 
= 200$ GeV  \cite{Adler:2005in} at $\sqrt{s} = 200$ GeV, 
showed transverse SSA's consistent with zero. 
This result is in agreement with the fixed-target E704 finding and extends it 
to higher $P_T$, up to 5 GeV. 

\subsubsection{Spin-averaged hadroproduction cross sections}

The QCD-based descriptions of the E704 and RHIC 
results on hadroproduction SSA's have been criticised \cite{Bourrely:2003bw} 
on the ground that, whereas the spin-averaged cross sections 
at $\sqrt{s} = 200$ GeV \cite{Adams:2006uz,Adare:2007dg,Arsene:2007jd} 
are well described by next-to-leading 
order perturbative QCD \cite{deFlorian:2002az}, the cross sections 
at lower energies (for instance at $\sqrt{s} = 19.4$ GeV 
\cite{Adams:1994yu}) are not. 
The conclusion of Ref.~\cite{Bourrely:2003bw} is that 
the single-spin asymmetries discovered by E704 and those 
found at RHIC  are different physical phenomena. 
However, it has been recently shown that the 
resummation of large logarithms arising from soft gluon radiation
in the limit $x_T \equiv 2 P_T/\sqrt{s} \to 1$  
significantly improves the agreement of perturbative QCD calculations (limited
however to the cross sections integrated over rapidity) with  data 
at low and intermediate energies \cite{deFlorian:2005yj,deFlorian:2007ty}.   
Predictions for GSI-FAIR and J-PARC kinematics are given 
in Ref.~\cite{deFlorian:2008wt}. 
The inclusion of intrinsic transverse momentum effects 
has been also shown to reduce the discrepancy 
between QCD-parton model predictions and cross section data 
in the fixed-target regime \cite{D'Alesio:2004up}. 
  
Thus,  the hope to achieve a consistent QCD picture
of both polarised and spin-averaged hadroproduction 
phenomena seems to be well founded.

\section{Future measurements}
\label{sec:future}

In this section account is given of near future and more distant future 
SIDIS and DY experiments.
The main goal of the SIDIS experiments is
to measure the SSA's over a detailed grid 
of the kinematic variables $x$, $P_{h \perp}$, and $z$, facilitating the extraction 
of the DF's and of the FF's. 
In addition, it should be possible to study sub-leading twist effects by 
probing their $1/Q$ dependence, and to explore the transition from 
non-perturbative small transverse
momentum, typically lower than 1 GeV, to the transverse momentum large 
regime.
In the case of the DY measurements, the main goal is to perform
for the first time experiments with polarised nucleon targets
and/or polarised beams, to test the predicted test of sign of the
T-odd DF's.

\subsection{SIDIS experiments} 
%
{\bf COMPASS at CERN:}
The analysis of the 2007 transversely polarised proton data is still ongoing, 
and several results on SSA's have still to be obtained.
Many more data on the 
same target (NH$_3$) will be collected in the long 2010 run, so that  in 
the near future a large amount of data is expected.\\
For what concerns a more distant future,
the COMPASS Collaboration is presently preparing a proposal for 
measurements aiming to study chiral perturbation theory, 
generalised parton distributions via Deeply Virtual Compton Scattering
(DVCS), and TMD parton distributions via Drell-Yan processes.
The DVCS measurements will be performed using a liquid hydrogen target
and a 190 GeV muon beam.
In parallel with these measurements, SIDIS data will be collected
to extract with high precision the unpolarised $\cos \phi_h$ and
$\cos 2 \phi_h$ azimuthal asymmetries as well as the beam helicity dependent 
$\sin \phi_h$ asymmetry.
Such information cannot be extracted from the data collected 
with the transversely polarised target because of the complications
of using a nuclear target.
\par \noindent
{\bf JLab experiments:}
In the near future also the CLAS Collaboration in Hall B will take SIDIS data 
on a transversely polarised target. For transverse running, the use 
of a novel HD-ice target is planned, which in a frozen-spin state requires 
only small holding fields. 
The use of the  HD-ice 
target by the E08-015 Collaboration has many advantages: being a solid 
target, it can be short, a few cm, and thanks to the smallness of the holding 
field it can be located in the centre of the detector, thus increasing the 
acceptance of the spectrometer. In addition, the HD target has almost no 
dilution, which maximises the figure-of-merit, and being of low atomic number, 
comparatively few bremsstrahlung photons will be produced in the target. The 
experiment should run in the second half of 2011, and an upgraded version of 
the detector has already been proposed for JLab12.  \\
In a medium term range,
Jlab12 GeV upgrade could meet the requirements to study TMD's in the 
valence region, thus covering a complementary kinematic region with respect 
to COMPASS.
The Clas12 experiment in Hall-B is designed to achieve a very broad 
kinematic coverage while increasing by a factor 10 the luminosity with 
respect to the current 6 GeV setup. 
In particular, the forward spectrometer comprises a 2 T toroid with 
improved geometry to minimise the not-active azimuthal coverage and 
a RICH detector is under study to extend the hadron identification over 
the full energy range of the experiment. 
The spectrometer
is complemented by a central detector embedded in a 5 T solenoid. \\
Also, an upgraded version of experiment E06-010 (PR09-018) has already 
been proposed and conditionally approved to run in Hall A.
The experiment aims to measure the SSA's of the SIDIS process 
$e^+n \rightarrow e'h X$, where $h$ is either a $\pi$ or a $K$.
The experiment will use the large-solid-angle Super BigBite Spectrometer
as hadron arm, the BigBite Spectrometer as electron arm, and a novel 
polarised $^3$He target that includes alkali-hybrid optical pumping 
and convection flow to achieve very high luminosity.
Thanks to the large acceptances of the electron and hadron arms, 
an electron - polarised nucleon luminosity at the level of 
$4 \times 10^{36}$ cm$^{-2}$s$^{-1}$, and a target polarisation of 65\%, 
the experiment should collect in a two-month run about 100 times more 
statistics than that obtained by the past experiments. 
\par \noindent
{\bf $e-N$ and $e-A$ future colliders:}
In depth studies of hadron structure can be best performed at a high energy
polarised electron-polarised proton collider.
Large Collaborations at BNL and JLab are 
elaborating proposals which are well advanced and are being encouraged 
by the USA agencies. As written in the NSAC 2007 Long Range Plan, 
"the allocation of resources are recommended to develop accelerator 
and detector technology necessary to lay the foundation for a polarised 
Electron Ion Collider (EIC). 
The EIC would explore the new QCD frontier of strong color 
fields in nuclei and precisely image the gluons in the proton".   
To carry out a rich and diversified physics program the 
recommended energies for the electrons are between 3 and 10 GeV, 
for the protons between 25 and 250 GeV, and for the heavy ions 
between 25 and 100 GeV. 
The luminosity in the case of the $e-p$ collider should be $10^{33}-10^{34}$ 
cm$^{-2}$s$^{-1}$, i.e. about 100 times the luminosity of the HERA collider. 
Recently preliminary ideas for a polarised  electron-nucleon collider 
(ENC) at GSI, Darmstadt, have been discussed mostly amongst the German 
community. \\
The advantage of the Collider configuration over fixed target experiments 
are manifold:\\
- it provides a large range of $Q^2$, $x$, $W$ and $P_{h \perp}$.\\ 
- the figure of merit for asymmetry measurements is 
very much better. 
For ammonia (NH$_3$) $f\simeq 0.15$, thus the figure of merit when 
scattering on a pure proton beam is better by a factor $\simeq 50$. 
Needless to say, the comparison is done assuming the same number of 
collected events, so a high luminosity for the EIC is a prerequisite.\\
- It provides access to the interaction region, so that modern vertex 
tracking systems can identify short living particles, like $D^0$ produced 
in the interaction.\\
Having access to the interaction region, exclusive reactions are at reach. 
This opens up the whole field of GPDs, which to-day are the only way to 
quantify how the orbital motion of quarks in the nucleon contributes to 
the nucleon spin. 
Also, it allows  measurements in the
target  fragmentation region, which is presently poorly known due to
the difficulty of measuring slowly moving hadrons in fixed-target
experiments, opening a window to
the study of spin-independent and spin-dependent "fracture" functions. \\
The US groups of RHIC and JLab are proceeding jointly to the formulation of 
two different proposals for two different colliders, eRHIC and ELIC, based 
in the two different laboratories. 
The RHIC project clearly foresees the use of the highly polarised proton 
and nuclear existing beams.  
Two accelerator design options are being worked upon, both aiming at  high 
brightness 10 GeV electron beams.
A Ring-Ring option, which requires a new electron storage ring for 
polarised electron or positron beams, is technologically more mature, 
and could provide a peak luminosity of $0.5\times 10^{33}$ cm$^{-2}$s$^{-1}$.
The second option is a Linac-Ring option, which  offers higher 
luminosity (by a factor of 5) and possibly higher energy, 
but requires intensive R\&D for the high-current polarised electron source.\\
The starting point of the JLab project is the availability of the 
12 GeV electron beam from the upgraded CEBAF. 
The proton complex has to be built from scratch, so it
is being designed taking full advantage of the 
expertise matured at RHIC and other laboratories on acceleration and 
storage of polarised proton beams. 
The design goal for the collider luminosity is very ambitious,   
$3 \times 10^{34}$ cm$^{-2}$s$^{-1}$ for beam energies of 10 and
250 GeV for the electron and protons respectively.\\
Quite recently, in the summer of 2008, discussions 
started\footnote{Private communication from D. von Harrach.}, 
about a possible low-cost realisation of  an ENC at GSI.
The central idea is to use the 15 GeV high energy storage ring HESR, 
which is planned to store an antiproton beam for the PANDA experiment 
(and possibly PAX) as the ring where to store the polarised proton beam.  
By constructing a 3-3.5 GeV electron ring, a "low energy" ENC could be 
realised. The cm energy would be ~14 GeV, i.e. in between the HERMES and 
the COMPASS energies.
To inject polarised protons in HESR a new 70 MeV $p$-linac will be needed. 
The protons would then be injected into the existing SIS18 ring, accelerated 
up to 1.4 GeV, and transferred then into HESR. New hardware for the spin 
manipulations will be needed in SIS 18 and in HESR, but it is the same 
which will be necessary for the PAX experiment.
The electron complex has to be constructed from scratch. An $e$-linac and 
an electron synchrotron will accelerate the electron beam, which will 
be stored into a new storage ring of about the same length as HESR, 
and housed in the same tunnel.
Preliminary machine studies  indicate that a luminosity of at least $10^{32}$  
could be achieved, as well as large polarisations (~80\%) for the two beams. 
To further reduce the cost of the project, it is proposed to use the PANDA 
detector, and to operate the collider in time sharing with the PANDA 
Collaboration.

\subsection{Drell-Yan} 

As described in section \ref{sec:dy}, the DY process 
in transversely polarised hadron scattering is theoretically a very
clean and safe way to access transversity.
The original suggestion of measuring DY in $p^{\uparrow} p^{\uparrow}$
scattering, which could be done at RHIC, turned out to be difficult 
because of the small value expected for the asymmetry,of the order of $1-2$\% 
\cite{Barone:1997mj,Martin:1998rz,Martin:1999mg}.  
The measurement will also require external input to disentangle the  
quark and the antiquark distributions.
It will be done when the RHIC luminosity will be
increased.

These problems can be circumvented by studying 
 DY production with polarised antiprotons at moderate energies, which 
is the ideal process to observe a sizable double transverse asymmetry 
\cite{Barone:1997mj,Anselmino:2004ki,Efremov:2004qs,Barone:2005cr}, 
dominated by the valence distributions. 
Such a measurement  
has been proposed by the PAX Collaboration \cite{Barone:2005pu} 
at the FAIR complex to be built at GSI.
Since the production rate 
for relatively large dilepton masses $M$ ($> 4$ GeV)  
might be too small to allow an easy measurement $A_{TT}^{DY}$, 
it has been proposed \cite{Anselmino:2004ki} to exploit the $J/\psi$ 
peak to measure the asymmetry.  

Finally, we recall that 
the Sivers effect can also be observed in Drell-Yan processes 
with a transversely polarised proton, where it gives rise 
to a $\sin (\phi_{\gamma} - \phi_S)$ asymmetry. 
No measurement has been made so far, but many 
experimental collaborations worldwide plan to investigate 
this class of reactions in the near future. 

\subsubsection{The proposed experiments}

In the following we describe very briefly the DY
proposed experiments.
\par
\begin{figure}[t]
\begin{center}
\includegraphics[width=0.35\textwidth,angle=-90]
{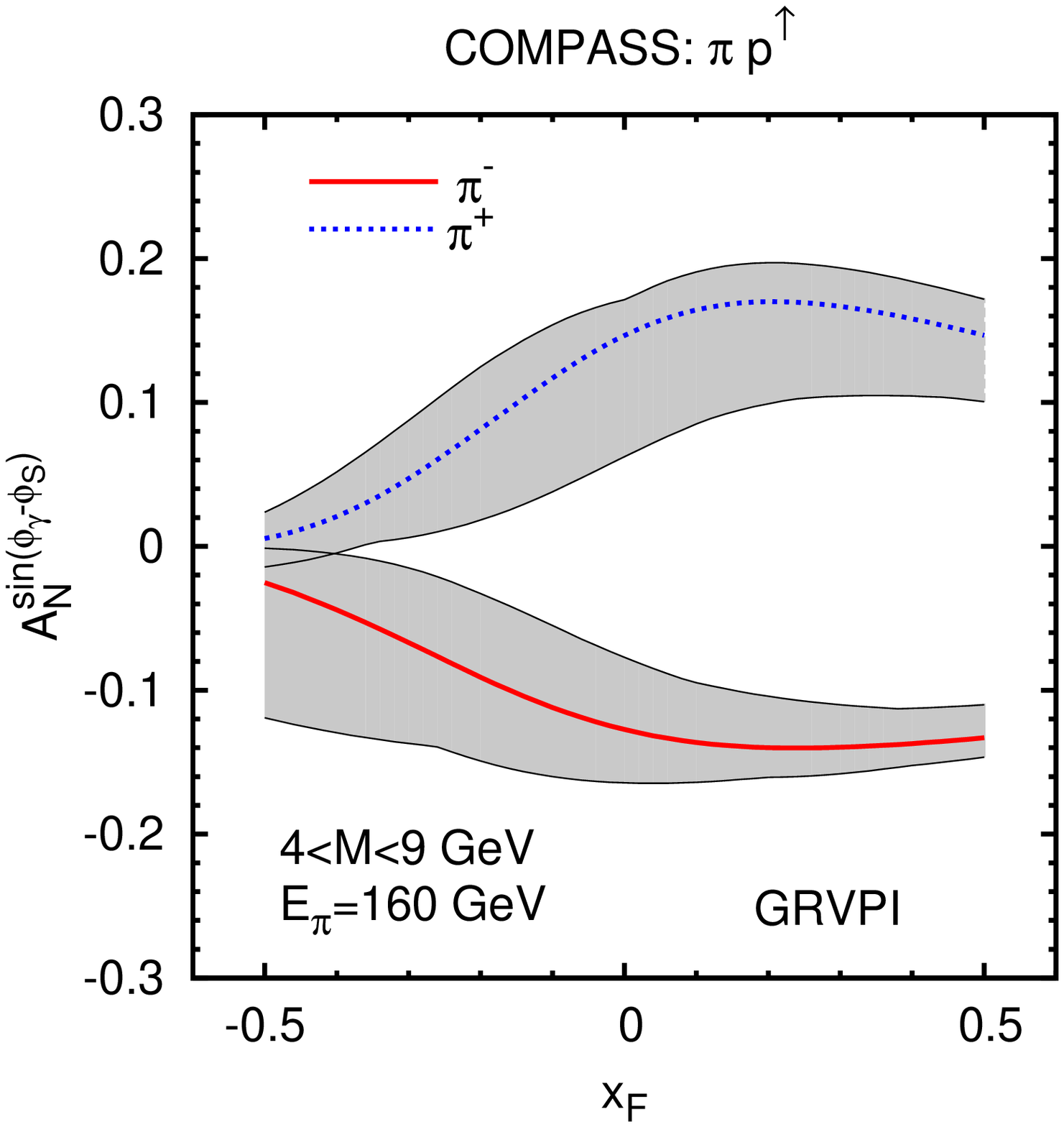}
\includegraphics[width=0.34\textwidth,angle=-90]
{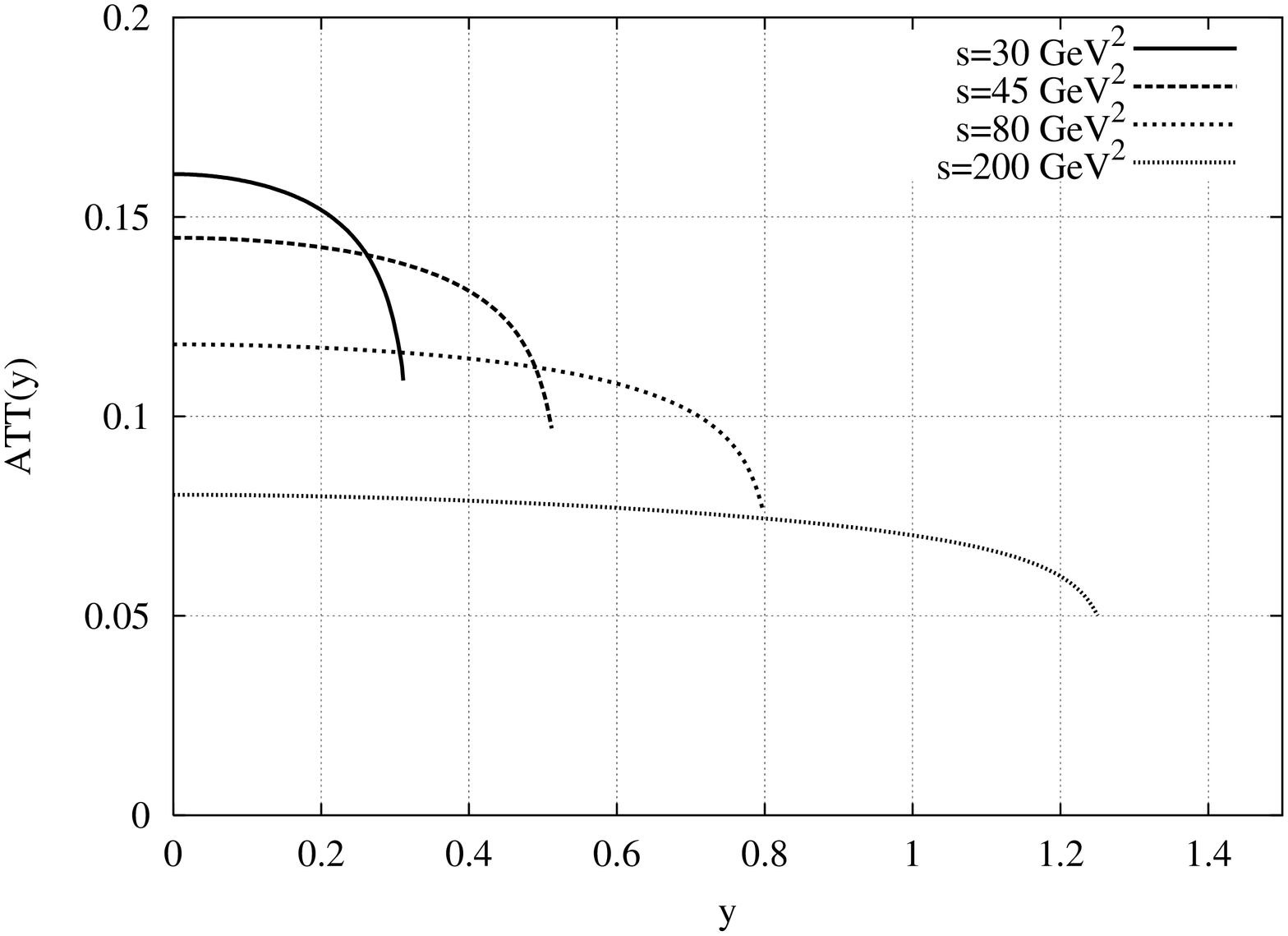}
\hfill
\end{center}
\caption{ 
Left: the prediction of Ref.~\cite{Anselmino:2009st} for the Drell-Yan 
Sivers asymmetry to be measured by COMPASS in $\pi p$ collisions. 
Right: the double spin asymmetry integrated over $M^2$ 
in transversely polarised 
proton-antiproton DY for different values of $\sqrt{s}$ 
\cite{Barone:2005cr}. 
}   
\label{fig_att_DY}
\end{figure}

 \noindent
{\bf COMPASS:}
Among the proposed measurements for a second phase of the COMPASS 
experiment an important issue is the possibility to investigate for the first 
time a $\pi^-$ 
induced DY process on a transversely polarised proton target. 
The high mass of 
the COMPASS target (about 1 kg of NH$_3$) and the excellent performance of the 
COMPASS spectrometer make this measurement feasible, and the number of 
events collected in two years of running would allow to check the expected 
change of sign of the Sivers function. 
Assuming for the magnitude of the Sivers 
function the value extracted from the HERMES measurements in SIDIS
(see Fig.~\ref{fig_att_DY}, left), 
the significance of the 
measurement is expected to be 3 to 4 $\sigma$.  
\par \noindent
{\bf PAX and PANDA at GSI:}
As already mentioned, the PAX Collaboration has proposed to measure DY 
processes in $\bar{p}p$ scattering at FAIR. 
An asymmetric collider is proposed, consisting of 
HESR, where polarised protons will be stored, and of a 
new storage ring for the polarised antiprotons, which could be the 
existing COSY Storage Ring, suitably modified. 
The predictions for $A_{TT}^{DY}$ in $\bar p^{\uparrow} p^{\uparrow}$ 
collisions at GSI-FAIR are shown in Fig.~\ref{fig_att_DY} (right). 
One sees that asymmetries 
of the order of 10-15 \% can be expected at PAX.     
Polarisation of the stored antiproton beam will be done using the 
``spin filtering''  technique \cite{Lenisa:2009zz}. 
The antiprotons beam traverses a polarised proton storage cell, and a 
beam polarisation builds up by repeatedly passing through the cell 
as long as the cross section for parallel spin is different from that of 
antiparallel spins. 
The method has been proven to work for proton beams, but in the case of 
antiprotons the spin dependent cross-sections are not known and corresponding 
measurements have been proposed at the Antiproton Decelerator at CERN.\\ 
The PANDA collaboration also envisage the measurement of DY process in 
$\bar{p} p$ scattering, where both particles are unpolarised. 
The option to put a transversely polarised proton target in the PANDA 
detector is very interesting, but it is technically very difficult and has 
presently been discarded.   
\par \noindent
{\bf Fermilab Experiment E906:}
This experiment is scheduled to run in 2010 for 2 years of data collection.
It will extend DY measurements of E866 (which were done with 800 GeV 
protons) using an upgraded spectrometer and the 120 GeV proton beam from 
the main injector. 
The use of the lower beam energy gives a factor 50 
improvement of luminosity with respect to E866. 
To cut down costs, it will use many components from E866, and data will 
be taken with Hydrogen, Deuterium and Nuclear Targets.
The main goal of the experiment is the study of the structure of the 
nucleon, in particular the $\bar{d}/\bar{u}$ ratio at high $x$. \\
In the future, there are plans to measure SSA's on a transversely 
polarised target, 
and check the change of sign of the Sivers function with respect to SIDIS.
\par \noindent
{\bf J--PARC:}
Two proposals for DY experiments have been submitted:\\
- P04: measurement of high-mass dimuon production at the 50 GeV
proton synchrotron;\\
- P24: polarised proton acceleration.\\
The advantage of an experiment at JAPRC is the high proton beam intensity, and 
consequently the high luminosity. 
The disadvantage is that at the same invariant 
mass the cross section is smaller at lower energy. 
The transverse polarisation 
program is best carried on by P24, but is clearly conditioned by the 
realisation of the polarised proton beam, which is not yet approved.
\par \noindent
{\bf STAR and PHENIX at RHIC:}
According to the present accelerator schedule,  which foresees a long 
longitudinal run for W-physics, a  DY program with transverse spin at 
PHENIX and STAR will not start before 2015.  However, since in the 
intersection regions IP-2 and IP-10, where the PHOBOS and the BRAHMS 
experiments were installed, there are no spin rotators, the polarisation of 
the beams is always transverse and ideas are being put forward to prepare 
both a collider experiment and a fixed target experiment for DY 
measurements.
\par \noindent  
{\bf RHIC internal target:} 
Quite recently ideas have been put forward for a DY experiment at RHIC 
scattering one beam off an internal target. 
With the 250 GeV beam, the 
kinematic range explored would be  $x_1 = 0.25 - 0.4$ 
($x_2 = 0.1 - 0.2$), which would be
ideal to investigate the change of sign of the Sivers function. 
To achieve the necessary luminosity, the use of a pellet target is 
being investigated, in two different scenarios according to the 
target thickness 
(parasitic running or dedicated experiment).
\par \noindent
{\bf NICA:}
To investigate the hadron structure a Nuclotron-based Ion Collider fAcility  
(NICA) is being planned at the JINR in Dubna, based on the existing proton 
synchrotron Nuclotron. The accelerator complex will require new ion and 
polarised proton sources, a new linear accelerator, a new booster 
synchrotron and the two new superconducting storage rings of the collider.
Both polarised proton and polarised deuteron beams should be available in 
the two rings. The main physics objectives will be the study of elastic 
processes and of Drell-Yan processes.

\subsubsection{Summary}
The relevant parameters for all these projects are summarised in 
Table~\ref{tab:coll}.
\begin{table}[tb]
\begin{center}
\begin{tabular}{lccccc}
\hline
Experiment  & particles & energy & $\sqrt{s}$ & $x_1$ or $x_2$  & luminosity     \\ 
\hline        
COMPASS     & $\pi^{\pm}+p^{\uparrow}$ & 160 GeV & 17.4 GeV & $x_2 = 0.2 - 0.3$ &  $2 \times 10^{33}$ cm$^{-2}$s$^{-1}$ \\             
\hline        
PAX         & $p^{\uparrow}+\bar{p}$   & collider &  14 GeV &  $x_1 = 0.1 - 0.9$ & $2 \times 10^{30}$ cm$^{-2}$s$^{-1}$ \\ 
\hline        
PANDA       & $\bar{p}+p^{\uparrow}$    &  15 GeV & 5.5 GeV & $x_2 = 0.2 - 0.4$ &  $2 \times 10^{32}$ cm$^{-2}$s$^{-1}$ \\ 
\hline        
J--PARC      &  $p^{\uparrow}+p$    &  50 GeV  &   10 GeV     &  $x_1 = 0.5 - 0.9$ & $ 10^{35}$ cm$^{-2}$s$^{-1}$ \\ 
\hline        
NICA        & $p^{\uparrow}+p$    &  collider &  20 GeV  &  $x_1 = 0.1 - 0.8$ & $10^{30}$ cm$^{-2}$s$^{-1}$ \\ 
\hline        
RHIC        & $p^{\uparrow}+p$    &  collider &  500 GeV &  $x_1 = 0.05 - 0.1$ & $2 \times 10^{32}$ cm$^{-2}$s$^{-1}$ \\ 
\hline        
RHIC IT phase 1  & $p^{\uparrow}+p$    &   250 GeV   &  22 GeV &  $x_1 = 0.25 - 0.4$ & $2 \times 10^{33}$ cm$^{-2}$s$^{-1}$ \\ 
\hline        
RHIC IT phase 2  & $p^{\uparrow}+p$    &   250 GeV   &  22 GeV &  $x_1 = 0.25 - 0.4$ & $6 \times 10^{34}$ cm$^{-2}$s$^{-1}$ \\ 
\hline        
\end{tabular}
\caption{Compilation of the relevant parameters for the
future planned DY experiments.
 For RHIC, IT stays for Internal Target}
\label{tab:coll}
\end{center}
\end{table}

Theoretical predictions for COMPASS ($\pi p^{\uparrow}$), 
PAX ($\bar p p^{\uparrow}$), RHIC ($p^{\uparrow} p$ at 
$\sqrt{s} = 200$ GeV) and J--PARC ($p^{\uparrow} p$ at 
$\sqrt{s} \simeq 10$ GeV) have been presented by various authors 
\cite{Efremov:2004tp,Collins:2005rq,Bianconi:2005bd,Bianconi:2005yj,Bianconi:2006hc,Anselmino:2005ea,Anselmino:2009st,Sissakian:2005yp,Sissakian:2008th}. 
The conclusion one can draw from these analyses 
is that the future experiments will largely be complementary to each other.

As for $p^{\uparrow}p$ DY, the RHIC data in the negative $x_F$ 
region will probe the contribution of the sea Sivers function, 
while experiments at lower energies, like J--PARC (operating 
at $\sqrt{s} \simeq 10$ GeV) will provide information 
on the large-$x$ behaviour of $f_{1T}^{\perp}$. A comprehensive 
discussion of all future DY measurements of Sivers asymmetries 
can be found in Ref.~\cite{Anselmino:2009st}.

\section{Conclusions and perspectives}

The original finding of the EMC collaboration, that the quark spin 
does not account for the total spin of the proton, has been a strong 
motivation for in-depth studies of the QCD structure of the nucleon 
and for a new generation of experimental investigation of hard scattering 
processes on polarised nucleons. The growing interest in the contribution 
of the quark and gluon orbital angular momentum to the nucleon spin 
naturally led to an increased attention to transverse spin and transverse 
momentum phenomena. 

In this context, the most important experimental finding has been the 
discovery that there is a correlation between the spin of a transversely 
polarised quarks and the $P_T$  of the hadrons created in the quark 
hadronisation process. Convincing evidence for this correlation has been 
provided by both SIDIS processes on transversely polarised nucleons and
high energy $e^+e^-$ annihilations into hadrons. 
Thanks to this correlation, 
it is now possible to measure the transversity distribution function. 
Global analysis of the existing SIDIS data and of the 
$e^+e^-$  data have already provided first rough information of this
two new function.  

A second important discovery of the recent years is that there is also a 
non-zero  correlation between the spin of a transversely polarised nucleon 
and the intrinsic transverse momentum of the quarks.  

In polarised proton-proton scattering the most impressive 
result in transverse spin physics is the confirmation that large SSA's for 
inclusive mesons production persist at centre of mass energies which by 
now are more than one order of magnitude greater than those of the previous 
fixed target experiments. 

On the theoretical side, the main achievement has been
the discovery that the Wilson line structure of parton distributions,
which is necessary to enforce gauge invariance, has also
striking observable consequences, allowing for
single-spin asymmetries that would otherwise be forbidden
by time-reversal symmetry.
At leading order the most general descriptions of 
SIDIS and Drell-Yan processes have been revisited and a number of structure 
functions have been introduced to take into account all possible correlations 
among the transverse momentum and spin of the quarks and the spin of the 
nucleon. 
Many QCD studies have been performed to understand the properties 
and the gauge structure of these unintegrated distribution functions, which 
have been named transverse momentum dependent distribution functions. 
Non-collinear factorisation schemes have been developed and extended to 
polarised processes. In an alternative approach, twist-three effects have 
been evaluated and compared to the TMD description in the intermediate $P_T$ 
region. A third line of attack which is vigorously being pursued is QCD 
computation on the lattice. Recent  refined lattice QCD results shed light 
on the fine structure of TMD's. This approach is particularly interesting 
for transversity, because the tensor charge, the first moment of the 
transversity distribution, which is an all-valence object, is believed to 
have been evaluated with good accuracy on the lattice, thus a good 
measurement of this quantity could provide a good test of the correctness 
of the calculation.  

In spite of these achievements, the amount of work which is still needed 
is not small. 

On the experimental side, the available SIDIS data are only a glimpse to 
a new territory, and many more data are needed to obtain the  $P_T$ and 
$Q^2$ dependence of the asymmetries in the different x-bins, a prerequisite 
to a model independent  extraction of the  TMD functions. The situation is 
worse however in the DY sector, since no polarised DY data exist at all. 
The existing unpolarised DY data allow only to access the Boer-Mulders 
function, but the full exploitation of the DY potential is not even at 
the horizon.

On the theoretical side, the present fits to the data do not make use of 
the $Q^2$ evolution schemes which are already now available for the TMD's 
and for the quark-gluon correlators, and which clearly must be integrated 
in the calculations. 
Also , a better understanding of higher-twist  
contributions to SIDIS observables (which are significant since 
$\langle Q^2 \rangle$ is rather small) requires more insight and more effort. 

In the near future more data will be collected in the SIDIS sector by 
the COMPASS and by the JLab experiments at quite different energies.
In the proton-proton sector, the luminosity of the RHIC collider should 
be enough to allow for measurements of DY processes.

In a more distant future, COMPASS should providing new measurements of 
azimuthal asymmetries in SIDIS 
on a liquid hydrogen target and a the first measurement of polarised 
DY process in $\pi^- p^{\uparrow}$ scattering.
Also, higher energy SIDIS data will come from JLab upgraded at 12 GeV. 

In an even more distant future, many projects have been proposed. 
The PAX experiment at FAIR aims to investigate DY pairs in polarised 
antiproton-polarised proton scattering, a very clean way to address the 
transversity functions, and a terribly difficult experiment. 
Fixed target DY experiments scattering polarised protons on polarised 
protons are being planned at JPARC, in Japan, and at NICA, in Dubna. 
It is fair to say, however, that the future of the field will depend 
in a crucial way on the pending decisions to construct a polarised 
electron-polarised proton collider. 
Ambitious projects are being pursued at BNL and at JLab. 
In Europe, the electron-proton collider 
has a long story, which needs not to be summarised here, we only mention 
the most recent proposal for a polarised collider, which is tailored to 
the new accelerators complex FAIR presently being realised at GSI. 
All these projects are based on existing laboratories and 
existing infrastructures so that optimism is mandatory.

\section{Acknowledgments}

This review on transverse spin and transverse momentum phenomena in hard
processes has been possible thanks to the friendly exchange of ideas and
information which characterizes the "spin community", and we would like to
thank all our colleagues for their invaluable contribution in elucidating
the many facets of this new and rapidly growing field.
This work has been partly supported by the Italian Ministry for Education,
University and Research (MIUR) as a Research Project of National Interest
(PRIN).

\bibliographystyle{h-physrev3}

{
\bibliography{biblio_rev,biblio_1_rev}
}

\end{document}